\documentclass[english,12pt]{article}
\usepackage[T1]{fontenc}
\usepackage[utf8]{inputenc}
\usepackage{textcomp}
\usepackage{amsmath}
\usepackage{cases}
\usepackage{amsthm}
\usepackage{amssymb}
\usepackage{graphicx}
\usepackage{eurosym}
\usepackage{color}
\usepackage{float}
\usepackage{todonotes}
\usepackage{algorithm,algcompatible,amsmath}
\usepackage[shortlabels]{enumitem}
\usepackage[toc,page]{appendix}
\makeatletter
  \theoremstyle{remark}
  
  \theoremstyle{plain}
  
  \theoremstyle{plain}
  
  \theoremstyle{plain}
  \newtheorem*{theo*}{\protect\theoremname}
  \theoremstyle{definition}
  
  \theoremstyle{plain}
  
\theoremstyle{plain}

\usepackage {a4wide}
\usepackage{mathrsfs}
\usepackage{bbm}
\makeatother

\usepackage{babel}
  \providecommand{\definitionname}{Definition}
  \providecommand{\propositionname}{Proposition}
  \providecommand{\theoremname}{Theorem}
  \providecommand{\remarkname}{Remark}
\providecommand{\corollaryname}{Corollary}
\providecommand{\lemmaname}{Lemma}
\newcommand*\samethanks[1][\value{footnote}]{\footnotemark[#1]}

\algnewcommand\INPUT{\item[\textbf{Input:}]}
\algnewcommand\OUTPUT{\item[\textbf{Output:}]}

\begin{document}
\title{Deep reinforcement learning for market making in corporate bonds: beating the curse of dimensionality\thanks{This research has been conducted with the support of J.P. Morgan and under the aegis of the Institut Louis Bachelier. However, the ideas presented in this paper do not necessarily reflect the views or practices at J.P. Morgan. The authors would like to thank Morten Andersen (J.P. Morgan), Fabien Azoulay (J.P. Morgan), Pascal Tomecek (J.P. Morgan), and Sihan Xiong (J.P. Morgan) for the numerous and insightful discussions they had on the subject.}}
\author{Olivier Guéant\thanks{Université Paris 1 Panthéon-Sorbonne. Centre d'Economie de la Sorbonne. 106, Boulevard de l'Hôpital, 75013 Paris. France.}, Iuliia Manziuk\samethanks[2]}
\date{}
\maketitle

\abstract{In corporate bond markets, which are mainly OTC markets, market makers play a central role by providing bid and ask prices for a large number of bonds to asset managers from all around the globe. Determining the optimal bid and ask quotes that a market maker should set for a given universe of bonds is a complex task. Useful models exist, most of them inspired by that of Avellaneda and Stoikov. These models describe the complex optimization problem faced by market makers: proposing bid and ask prices in an optimal way for making money out of the difference between bid and ask prices while mitigating the market risk associated with holding inventory. While most of the models only tackle one-asset market making, they can often be generalized to a multi-asset framework. However, the problem of solving numerically the equations characterizing the optimal bid and ask quotes is seldom tackled in the literature, especially in high dimension. In this paper, our goal is to propose a numerical method for approximating the optimal bid and ask quotes over a large universe of bonds in a model \emph{à la} Avellaneda-Stoikov. Because we aim at considering a large universe of bonds, classical finite difference methods as those discussed in the literature cannot be used and we present therefore a discrete-time method inspired by reinforcement learning techniques. More precisely, the approach we propose is a model-based actor-critic-like algorithm involving deep neural networks.}

\vspace{5mm}

 \vspace{5mm}

\section{Introduction}

On both sides of the Atlantic ocean, the corporate bond markets traditionally operate through market makers (also called dealers) providing liquidity on the one hand, and clients consuming this liquidity on the other.\footnote{Of course, there are also inter-dealer broker markets but the focus of the paper will not be on them.}\\

These mainly OTC markets have undergone a large number of changes since the subprime crisis. First, central banks imposed a low-interest rate environment and purchased numerous securities including corporate bonds. This has resulted in a large number of bond issuances by firms which traditionally borrowed cash through the corporate bond market and by riskier newcomers as investors were looking for higher returns. Subsequently, we have seen a very important increase in the size of corporate bond markets. For instance, the Bank for International Settlement estimates that the European corporate bond market has seen its size multiplied by 3 in the 10 years following the start of the subprime crisis. Secondly, concerning the secondary market, numerous reforms were introduced in the years following the crisis both in the US and in Europe: Basel III, Volcker rule, Dodd-Frank Act, etc. The deleveraging phenomenon in almost all banks after the crisis was therefore
reinforced by regulations, reducing the intermediation capacities of many market makers (see the famous note by Goldman Sachs \cite{gs}). Meanwhile, the corporate bond markets on both sides of the Atlantic ocean have undergone electronification in a process specific to these markets (see for instance \cite{mck}). This electronification process is dominated by Multi-dealer-to-client (MD2C) platforms operated by companies like Bloomberg, Tradeweb, MarketAxess, etc. They allow clients to send the same request for quote (RFQ) to several dealers
simultaneously and therefore instantly put dealers into competition.\footnote{Despite the many efforts of entrepreneurs and major groups like BlackRock, there has been no paradigm shift in corporate bond markets from OTC to all-to-all trading -- i.e. no shift towards (central) limit order books.}\\

Regarding how market makers operate, electronification is also in progress with major market participants replacing traders by algorithms to be able to (i) stream quotes for a large universe of bonds and (ii) automatically answer a large proportion of the RFQs they receive.\\

Determining the optimal bid and ask quotes that a market maker should set for a given universe of bonds is a complex task. Useful models exist, most of them inspired by that of Avellaneda and Stoikov \cite{avellaneda} who revived an old paper by Ho and Stoll \cite{hostoll}; e.g. Cartea, Jaimungal, and Ricci \cite{carteajaimungal}, Guéant, Lehalle, and Fernandez-Tapia \cite{glft}, Guéant \cite{gueant}, etc. These models describe the complex optimization problem faced by market makers: proposing bid and ask prices in an optimal way for making money out of the difference between bid and ask prices while mitigating the market risk associated with holding inventory.\\

While most of the models only tackle one-asset market making, they can often be generalized to tackle multi-asset market making (see for instance \cite{gueant}). However, the problem of solving numerically the equations characterizing the optimal bid and ask quotes is seldom tackled in the literature, especially as far as high-dimensional cases are concerned.\footnote{One important exception is the recent paper \cite{bg} by Bergault and Gu\'eant who proposed a factorial approach to approximate the optimal quotes.}\\

In this paper, our goal is to propose a numerical method for approximating the optimal bid and ask quotes over a large universe of bonds in a model inspired by~\cite{gueant}. Because we want our method to scale to high-dimensional cases, classical finite difference methods as those discussed in the literature (see also the appendix) are not appropriate. We propose instead a discrete-time method inspired by the reinforcement learning (RL) literature. More precisely, the approach we propose is a model-based actor-critic-like algorithm involving deep neural networks.\\

The recent advances on strategy board games and video games, especially the research works published by research teams at Google DeepMind (see for instance~\cite{atari} and \cite{go}), have shed a renewed light on the power of reinforcement learning techniques for solving problems in the field of dynamic optimization. In finance, several efforts have been made to use ideas from RL or approximate dynamic programming in order to solve classical and less classical problems. The most famous paper is the paper ``Deep hedging'' \cite{dh} co-written by researchers from J.P. Morgan in which the authors use deep recurrent neural networks to find the optimal way to hedge a European payoff. Beyond European payoffs, the case of American options (i.e. the addition of an optimal stopping problem) has been tackled, for instance in \cite{cheridito}. Theoretical pricing bounds have been proposed in the interesting paper \cite{phl} dealing with XVA problems. General methods have also been proposed by Pham \emph{et al.} \cite{pham2, pham1, pham}, themselves inspired by the BSDE solver of Jentzen \emph{et al.} \cite{jentzen} who approximate the solution of some specific nonlinear partial differential equations in dimension up to~100. Although not written with the vocabulary of reinforcement learning, most of these papers share ideas with the methods discussed in the RL community.\\

The craze around RL-flavored techniques has impacted several fields in finance and it is interesting to notice that market making has not yet been tackled. There are several reasons for this. The first one is obvious: pricing and hedging derivatives are the main topics of quantitative finance. There is consequently no surprise that these topics (with a focus on option hedging rather than option pricing in fact) were addressed first. But the nature of the different problems that could be tackled are also very important. Market making models are models in which the dynamics of the state process is described by that of point processes and the control of point processes has always been less studied than that of continuous semi-martingales (see~\cite{controlgraph} for a general paper on the control of point processes). Moreover, point processes prevent the massive use of automatic differentiation in a recursive way as in papers like ``Deep hedging'' \cite{dh}. However, unlike many problems in finance, the problem of a corporate bond market maker has no natural time horizon and should be regarded as a stationary or ergodic problem; and RL techniques are well suited to tackle infinite horizon problems.\\

In Section 2 we introduce notation and present the market making model in continuous time. In Section 3 we show how the problem can be transformed into a discrete-time one and we present our actor-critic-like methodology for approximating the value function and the optimal strategies (here the optimal quotes). In Section~4, we present numerical examples in low dimension to compare with the proceed of a finite difference method and examples in high dimension for which finite difference methods cannot be used. In the appendix, we recall the finite difference method classically used to solve the Hamilton-Jacobi-Bellman equations associated with our model.\\

\section{The classical model in continuous time}

\subsection{Notation}

Let $\left(\Omega,\left(\mathcal{F}_{t}\right)_{t\in\mathbb{R}_{+}},\mathbb{P}\right)$
be a filtered probability space, with $\left(\mathcal{F}_{t}\right)_{t\in\mathbb{R}_{+}}$
satisfying the usual conditions. We assume that all the stochastic processes introduced in this paper are defined on $\Omega$ and adapted to the filtration $\left(\mathcal{F}_{t}\right)_{t\in\mathbb{R}_{+}}$.\\

We consider a market maker in charge of $d$ bonds. For each bond $i \in \{ 1, \ldots, d\}$, we consider that there exists a reference price for bond $i$ at any time $t$, denoted by~$S_t^i$  -- it could be based on the Composite Bloomberg Bond Trader (CBBT) mid-price but other options are possible. We assume that the dynamics of prices is given by
  \begin{equation*}
  dS^i_t = \sigma^i dW^i_t,
  \end{equation*}
  where $\left(W^1, \ldots, W^d\right)$ is a $d$-dimensional Brownian motion with a correlation matrix denoted by $\left(\rho^{i,j}\right)_{1 \le i,j \le d}$. We denote by $\Sigma$ the covariance matrix $\left(\rho^{i,j} \sigma^i\sigma^j\right)_{1 \le i,j \le d}$.\\

For each bond $i \in \{ 1, \ldots, d\}$, we assume that the number of RFQs arriving at the bid (respectively at the ask) is a Poisson process with intensity $\lambda_{\textrm{RFQ}}^{i,b}$ (resp. $\lambda_{\textrm{RFQ}}^{i,a}$). We assume that the Poisson processes are independent and independent from the Brownian motions driving reference prices.\\

When a buy RFQ\footnote{In a buy RFQ, the market maker is proposed to buy the bond.} on bond $i$ is received at time $t$ by the market maker, we assume that the size of the request is given by $\Delta^i$ (one size for each bond)\footnote{This hypothesis can be relaxed very easily (see for instance \cite{bg}).}. If the market maker answers a price to that request, that price is denoted by $S^{i,b}_t = S^i_t - \delta_t^{i,b}$ where~$\delta_t^{i,b}$ is assumed to be $\mathcal{F}_{t-}$-measurable. Similarly, for a sell RFQ on bond $i$, we assume that the size of the request is given by $\Delta^i$ and, if the market maker answers a price to that request, that price is denoted by $S^{i,a}_t = S^i_t  + \delta_t^{i,a}$~where $\delta_t^{i,a}$ is assumed to be $\mathcal{F}_{t-}$-measurable. The probability to trade following a buy (resp. sell) RFQ on bond~$i$ given a bid quote $S^i - \delta^{i,b}$ (resp. ask quote $S^i + \delta^{i,a}$) is denoted by $f^{i,b}\left(\delta^{i,b}\right)$ (resp. $f^{i,a}\left(\delta^{i,a}\right)$).\\

Throughout this paper, and in line with classical hypotheses in market making models (see \cite{gueant,glft}), we assume that, for all $i\in \{1, \ldots, d\}$:
      \begin{itemize}
  \item $f^{i,b}$ and $f^{i,a}$ are twice continuously differentiable,
  \item $f^{i,b}$ and $f^{i,a}$ are decreasing, with $\forall \delta \in \mathbb{R}$, ${f^{i,b}}'(\delta) <0$ and ${f^{i,a}}'(\delta) <0$,
  \item $\lim_{\delta \to +\infty} f^{i,b}(\delta) = \lim_{\delta \to +\infty} f^{i,a}(\delta) = 0$,
    \item $f^{i,b}$ and $f^{i,a}$ satisfy
    \begin{equation}
    \sup_{\delta}\frac{f^{i,b}(\delta){f^{i,b}}''(\delta)}{\left({f^{i,b}}'(\delta)\right)^2} <  2 \quad \text{and} \quad \sup_{\delta} \frac{f^{i,a}(\delta){f^{i,a}}''(\delta)}{\left({f^{i,a}}'(\delta)\right)^2} <  2. \label{sup2}
    \end{equation}
\end{itemize}
For each bond $i$, we denote by $N^{i,b}$ and $N^{i,a}$ the point processes associated with the number of transactions at the bid and at the ask. In particular, the inventory in bond~$i$ at time $t$, denoted by $q^i_t$, has the following dynamics:
  \begin{equation*}
dq^i_t = \Delta^i dN^{i,b}_t - \Delta^i dN^{i,a}_t.
\end{equation*}
In what follows, we assume that risk limits are imposed to the market maker in the form of inventory limits $-Q^i$ and $Q^i$ (multiple of $\Delta^i$) for each bond $i$. Therefore, the intensity of the trade processes $N^{i,b}$ and $N^{i,a}$ at time $t$, denoted by $\lambda_t^{i,b}$ and $\lambda_t^{i,a}$, are:
\begin{eqnarray*}
 \lambda_t^{i,b} &=&  \lambda_{\textrm{RFQ}}^{i,b} f^{i,b}(\delta_t^{i,b}) 1_{q^i_{t-}<Q^i},\\
\lambda_t^{i,a} &=&  \lambda_{\textrm{RFQ}}^{i,a} f^{i,a}(\delta_t^{i,a}) 1_{q^i_{t-}>-Q^i}.
\end{eqnarray*}
In particular, $\forall t, q_t \in \mathcal{Q} = \prod_{i=1}^{d} \{-Q^i, -Q^i + \Delta^i, \ldots, Q^i - \Delta^i, Q^i\}$.\\

The resulting cash process of the market maker is $\left(X_t\right)_t$ where
\begin{eqnarray*}
dX_t &=& \sum_{i=1}^d -S^{i,b}_t \Delta^i dN^{i,b}_t + S^{i,a}_t \Delta^i dN^{i,a}_t\\
 &=&  \sum_{i=1}^d -\left(S^i_t - \delta^{i,b}_t\right) \Delta^i dN^{i,b}_t + \left(S^i_t +
\delta^{i,a}_t\right) \Delta^i dN^{i,a}_t\\
&=& \sum_{i=1}^d \left(\delta^{i,b}_t \Delta^i dN^{i,b}_t +
\delta^{i,a}_t \Delta^i dN^{i,a}_t\right)  - \sum_{i=1}^d dq^i_t S^i_t
\end{eqnarray*}
The Mark-to-Market (MtM) value of the portfolio -- hereafter the profit and loss (PnL) -- at time $t$ is therefore $\textrm{PnL}_t = X_t + \sum_{i=1}^d q^i_t S^i_t$. It has the following dynamics:
  \begin{eqnarray*}
d\textrm{PnL}_t &=& \sum_{i=1}^d \delta^{i,b}_t \Delta^i dN^{i,b}_t +
\delta^{i,a}_t \Delta^i dN^{i,a}_t + q^i_t \sigma^i dW^i_t.
\end{eqnarray*}

\subsection{Optimization problem and Bellman equations}

Several objective functions have been proposed in the market making literature. Avellaneda and Stoikov initially used in \cite{avellaneda} an expected utility objective function. Cartea~\emph{et~al.} used in most of their papers (see \cite{carteabook} for an overview) a risk-adjusted expected value of the PnL over a finite time horizon. Guéant showed in \cite{gueant} that these two kinds of objective function lead to similar optimal quoting strategies. In this paper, we start with an infinite-horizon version of the classical risk-adjusted expected PnL objective function with a positive discount rate denoted by $r$.\footnote{As discussed in Section \ref{disc} below, we are in fact interested in the ergodic problem (maximization of the average reward) but the introduction of a discount rate is necessary for mathematical and numerical reasons.}\\

More precisely, the problem consists in maximizing
\begin{equation*}
\mathbb{E}\left[\int_0^{+\infty} e^{-r t}\left(d\textrm{PnL}_t - \psi(q_t)dt\right)\right],
\end{equation*}
where $\psi: \mathbb{R}^d \to \mathbb{R}_+$ penalizes inventory. A classical penalty function is proportional to the instantaneous variance of the MtM value of the portfolio, i.e. $$\psi: q \in \mathbb{R}^d \mapsto \frac 12 \gamma q' \Sigma q,$$ where $\gamma >0$ is the risk aversion parameter of the market maker. It is in line with most papers presented in \cite{carteabook} and consistent with the initial model of Avellaneda and Stoikov and its resolution by Guéant \emph{et al.} in \cite{glft}. A financially relevant alternative consists in considering a penalty proportional to the standard deviation of the MtM value of the portfolio, or equivalently to the Value at Risk in the Gaussian case that we consider. In that case, we consider a penalty function of the form $$\psi: q \mapsto \frac 12 \gamma \sqrt{q' \Sigma q},$$ where $\gamma$ weighs the risk adjustment to be made to the expected PnL.\\

The stochastic optimal control problem is therefore the following:
\begin{eqnarray*}
\sup_{\left(\delta^{i,b}, \delta^{i,a}\right)_{1 \le i \le d} \in \mathcal{A}^{2d}}&&\mathbb{E}\left[\int_0^{+\infty} e^{-r t}\left(\sum_{i=1}^d \left(\delta^{i,b}_t \Delta^i dN^{i,b}_t +
\delta^{i,a}_t \Delta^i dN^{i,a}_t\right) - \psi(q_t) dt\right)\right]\\
=\sup_{\left(\delta^{i,b}, \delta^{i,a}\right)_{1 \le i \le d} \in \mathcal{A}^{2d}}&& \mathbb{E}\Bigg[\int_0^{+\infty} e^{-r t}\Bigg(\sum_{i=1}^d \left(\delta^{i,b}_t \Delta^i \lambda_{\textrm{RFQ}}^{i,b} f^{i,b}(\delta_t^{i,b}) 1_{q^i_{t-}>-Q^i}\right.\nonumber\\
&& \left.{}+ \delta^{i,a}_t \Delta^i \lambda_{\textrm{RFQ}}^{i,a} f^{i,a}(\delta_t^{i,a}) 1_{q^i_{t-}<Q^i}\right) - \psi(q_t)\Bigg)dt\Bigg],
\end{eqnarray*}
where $\mathcal{A}$ is the set of predictable processes.\\

We know from classical stochastic optimal control theory that we can restrict ourselves to closed-loop controls, also called policies in the reinforcement learning community. In other words, with a slight abuse of notations, we can look for a family of functions $(q \in \mathcal{Q} \mapsto \delta^{i,b}(q), q \in \mathcal{Q} \mapsto \delta^{i,a}(q))_{1 \le i \le d}$ and control processes such that $$\forall (i,s) \in \{1, \ldots, d\}\times\{b,a\}, \forall t, \delta^{i,s}_t = \delta^{i,s}(q_{t-}).$$

For a given policy $\delta = \left(\delta^{i,b}, \delta^{i,a}\right)_{1 \le i \le d}$, we can define the value function associated with that policy: $\tilde{\theta}^\delta_{r}: q \in \mathcal{Q} \mapsto \tilde{\theta}^\delta_{r}(q)$. This value function is the unique solution of the following linear Bellman equation:
\begin{eqnarray}
&&- r \tilde{\theta}^\delta_{r}(q) - \psi(q) + \sum_{i=1}^d 1_{q^i<Q^i} \lambda_{\textrm{RFQ}}^{i,b} f^{i,b}(\delta^{i,b}(q))\left( \delta^{i,b}(q)\Delta^i  +  \tilde{\theta}^\delta_{r}\left(q+\Delta^ie^i\right) - \tilde{\theta}^\delta_{r}(q)\right)\nonumber\\
 &&+ \sum_{i=1}^d 1_{q^i>-Q^i} \lambda_{\textrm{RFQ}}^{i,a} f^{i,a}(\delta^{i,a}(q))\left(\delta^{i,a}(q)\Delta^i  +  \tilde{\theta}^\delta_{r}\left(q-\Delta^ie^i\right) - \tilde{\theta}^\delta_{r}(q)\right) = 0.\label{theta_delta}
\end{eqnarray}

We also associate with this stochastic optimal control problem a Hamilton-Jacobi-Bellman equation, satisfied by the optimal value function $\tilde{\theta}^*_{r}: q \in \mathcal{Q} \mapsto \tilde{\theta}^*_{r}(q)$:
\begin{eqnarray}
0 = - r \tilde{\theta}^*_{r}(q) - \psi(q) &+& \sum_{i=1}^d 1_{q^i<Q^i} H^{i,b}\left(\frac{\tilde{\theta}^*_{r}(q) - \tilde{\theta}^*_{r}\left(q+\Delta^ie^i\right)}{\Delta^i}\right)\nonumber\\
 &+& \sum_{i=1}^d 1_{q^i>-Q^i} H^{i,a}\left(\frac{\tilde{\theta}^*_{r}(q) - \tilde{\theta}^*_{r}\left(q-\Delta^ie^i\right)}{\Delta^i}\right),\label{theta}
\end{eqnarray}
where the Hamiltonian functions are
$$H^{i,b}(p) = \Delta^i \lambda_{\textrm{RFQ}}^{i,b} \sup_{\delta \in \mathbb{R}} f^{i,b}(\delta)(\delta - p),$$
$$H^{i,a}(p) = \Delta^i \lambda_{\textrm{RFQ}}^{i,a} \sup_{\delta \in \mathbb{R}} f^{i,a}(\delta)(\delta - p),$$
and where $\left(e^1, \ldots, e^d\right)$ are the vectors of the canonical basis of $\mathbb{R}^d$.\\

In \cite{gueant}, the author studied similar equations in the finite-horizon case. The general analysis of this type of equations, in both the finite-horizon and infinite-horizon cases, is carried out in  the paper \cite{controlgraph} about optimal control on graphs.\footnote{For market making purposes the finite-horizon case can be of interest if the market maker wants to penalize his inventory at a specific time, for instance to diminish the exposure to a specific event (a central bank announcement, an election, etc.). However, before that final time, and in practice just a few hours before that final time (see \cite{gueant}), the influence of the final condition is so small that the optimal behavior is almost the same as in the infinite-horizon problem (see \cite{controlgraph} for a proof of convergence). We believe that a meaningful approach for taking into account a specific event or an increased level of uncertainty is to see the value of the risk aversion parameter $\gamma$ as a variable that can be modified.} We know in particular that the optimal value function is the unique solution to Eq. \eqref{theta}.\\

Moreover, using a verification argument as in \cite{gueant}, we can easily prove that the optimal bid and ask quotes are given by the policy $\delta^*=\left(\delta^{i,b*},\delta^{i,a*}\right)_{1\le i \le d}$, where
\begin{eqnarray}
\delta^{i,b*}(q) &=& {f^{i,b}}^{-1}\left(- \frac{{H^{i,b}}'\left(\frac{\tilde{\theta}^*_{r}(q) - \tilde{\theta}^*_{r}\left(q+\Delta^i e^i\right)}{\Delta^i}\right)}{\Delta^i \lambda_{\textrm{RFQ}}^{i,b}}\right), \label{optimalb}\\
\delta^{i,a*}(q) &=& {f^{i,a}}^{-1}\left(- \frac{{H^{i,a}}'\left(\frac{\tilde{\theta}^*_{r}(q) - \tilde{\theta}^*_{r}\left(q-\Delta^i e^i\right)}{\Delta^i}\right)}{\Delta^i \lambda_{\textrm{RFQ}}^{i,a}}\right), \label{optimala}
\end{eqnarray}
and of course that $\tilde{\theta}^{\delta^*}_{r} = \tilde{\theta}^*_{r}$.\\

In practice, a numerical approximation of the optimal control is usually obtained by first solving (numerically) the Hamilton-Jacobi-Bellman equation $\eqref{theta}$ and then applying Eqs.~\eqref{optimalb} and \eqref{optimala}. Since $\tilde{\theta}^*_{r}$ cannot be obtained in closed form, one classically use a finite difference numerical scheme on a grid to obtain an approximation of the optimal value function (see the appendix for a classical approach). However, as the size of the grid increases exponentially with the number of bonds $d$, the use of finite difference schemes is not possible for $d$ greater than 4, 5, or 6 depending on the values of the parameters and the computer infrastructure. Our main goal is therefore to propose alternatives to methods based on grids.\\

\subsection{Remarks on the parameters}
\label{disc}
Before we propose a numerical method which is not based on grids, let us comment on the parameters because there are two kinds of parameters in the above market making model.\\

First, there are parameters associated with the bonds and the market: the arrival rates of RFQs for each bond and side (i.e. $(\lambda_{\textrm{RFQ}}^{i,b}, \lambda_{\textrm{RFQ}}^{i,a})_i$), the size of trades  (i.e.~$(\Delta^i)_i$) -- assumed constant here but it could be a distribution (see \cite{bg}) --, the probability to trade given a quote for each bond and side (i.e. $(f^{i,b}(\cdot), f^{i,a}(\cdot))_i$), and the covariance between bond price changes (i.e. the matrix $\Sigma$). These parameters are always estimated when used as inputs of Hamilton-Jacobi-Bellman equations.\\

With the craze around reinforcement learning techniques, people sometimes have in mind that optimization can now be carried out without an initial estimation of the model parameters, that is, without a model. However, this is true only when one has gigantic data sets. Is it the case in finance? Perhaps in the case of limit order book data on stock markets if one considers a problem involving a few stocks. But it is clearly not the case when it comes to OTC markets such as corporate bond markets, all the more when one wants to consider a large universe of assets.\footnote{One can simply think of the numerous methods (shrinkage, factor models, Marchenko-Pastur or other approaches from random matrix theory, etc.) that have been proposed to clean objects as simple as correlation matrices, just because financial time series are not long enough or not stationary enough.} In fact, most of the recent scientific advances using reinforcement learning (Go, Atari video games, etc.) used simulators that could provide as many samples as desired. There is of course no such simulator of real financial data, and one has instead to build a simulator using a model, hence the need to estimate parameters.\\

Second, there are parameters that need to be chosen: the magnitude of the risk-adjustment in the objective function (i.e. the risk aversion parameter $\gamma$), and the discount rate $r$. Regarding the magnitude of the risk adjustment, it has to be chosen in line with the risk aversion of the market maker or that of the bank. A classical method consists in testing different values for $\gamma$ and looking for the PnL profile (distribution) that corresponds to the ``desired'' / ``best acceptable'' one. Regarding the discount rate, the situation is different. In fact, we do not really care about maximizing the objective function starting from a given inventory, but we would like instead to maximize the objective function starting from any reasonable inventory. If we consider the stationary probability measure $m^{\delta}$ on $\mathcal{Q}$ associated with the Markov chain $(q_t)_t$ driven by a policy~$\delta$, then a natural objective function to maximize is
\begin{equation*}\sum_{q \in \mathcal{Q}} \tilde{\theta}^\delta_{r}(q) m^\delta(q).\end{equation*} As it is well known in the reinforcement learning community (see Chapter 10 of~\cite{sb}), maximizing such an objective function leads to optimal strategies that are independent of $r$. This is due to the fact that
\begin{eqnarray*}
\sum_{q \in \mathcal{Q}} r \tilde{\theta}^\delta_{r}(q) m^\delta(q) &=& - \sum_{q \in \mathcal{Q}} \psi(q) m^{\delta}(q)\\
&& + \sum_{q \in \mathcal{Q}} \sum_{i=1}^d 1_{q^i<Q^i} \lambda_{\textrm{RFQ}}^{i,b} f^{i,b}(\delta^{i,b}(q))\delta^{i,b}(q)\Delta^i m^{\delta}(q)\\
&&  + \sum_{q \in \mathcal{Q}} \sum_{i=1}^d 1_{q^i>-Q^i} \lambda_{\textrm{RFQ}}^{i,a} f^{i,a}(\delta^{i,a}(q)) \delta^{i,a}(q)\Delta^i m^{\delta}(q)
\end{eqnarray*}
is indeed independent of $r$.\\

In other words, maximizing this type of objective function is equivalent to maximizing the ergodic constant (i.e. the average reward per unit of time) and we should therefore choose $r$ small to be close to the ergodic case (see \cite{controlgraph} for the link between stationary problems, ergodic problems, and the long-term behavior of controlled processes on graphs) but not too small as value functions scale proportionally to $\frac 1r$ and therefore too small values of $r$ can thus lead to numerical problems.\footnote{The ergodic case corresponding to the limit $r \to 0$ cannot be considered directly because the value function is only unique up to a constant in that case.} In particular, the discount rate should not be chosen in line with a financial interest rate: it is instead a purely numerical tool.\\

\section{Going beyond grids}
\label{algo}
\subsection{A discrete-time reformulation of the problem}

The equations \eqref{theta_delta} and \eqref{theta} characterizing the value and optimal value functions are based on infinitesimal (and therefore continuous-time) reasoning. However, the problem of the market maker is to propose a bid or an ask quote upon receiving a request for quote: it is a discrete-time problem. What happens between two requests is just the ``payment'' of the running penalty $\psi(q)$ for holding inventory. We can therefore reformulate the problem in discrete time, by focusing on the decision-making process of market makers.\\

For a given policy $\delta$, we can define two types of value function. The first one, consistent with the above definition is the value function $\tilde{\theta}_{r}^\delta$ at any non-RFQ time. The second one is the value function $\theta^\delta_r$ at the time of an RFQ just before the market maker knows the bond and the side of the RFQ.\\

If we consider an inventory $q$ at an arbitrary time, and define by $\tau$ the duration before the arrival of the next RFQ, it is clear that the value functions $\tilde{\theta}_r^\delta$ and $\theta_r^\delta$ associated with $\delta$ satisfy for all $q \in \mathcal{Q}$ the following equations:
\begin{eqnarray}
\tilde{\theta}_r^\delta(q) &=& \mathbb{E}\left[ \int_0^\tau - e^{-rt}\psi(q) dt +  e^{-r\tau} \theta_r^\delta(q)\right]\nonumber\\
 &=& -\frac{\psi(q)}{r + \sum_{i=1}^d \left(\lambda_{\textrm{RFQ}}^{i,b} + \lambda_{\textrm{RFQ}}^{i,a}\right)} + \gamma_{RL} \theta_r^\delta(q),\label{theta12}
\end{eqnarray}
where
$$\gamma_{RL} = \frac{\sum_{i=1}^d \left(\lambda_{\textrm{RFQ}}^{i,b} + \lambda_{\textrm{RFQ}}^{i,a}\right)}{r + \sum_{i=1}^d \left(\lambda_{\textrm{RFQ}}^{i,b} + \lambda_{\textrm{RFQ}}^{i,a}\right)}$$ is a discount rate adapted to the discrete-time reinforcement learning reformulation of the problem.
Furthermore, for all $q \in \mathcal{Q}$,
\begin{eqnarray}
\theta_r^\delta(q)\!\!  &=& \!\! \mathbb{E}\left[f^{I,s}(\delta^{I,s}(q)) \left(\Delta^I \delta^{I,s}(q) + \tilde{\theta}_r^\delta\left(q + \left(1_{s=b} - 1_{s=a}\right) \Delta^I e^I\right)\right)\right.\nonumber\\
&&\left. + \left(1-f^{I,s}(\delta^{I,s}(q))\right) \tilde{\theta}_r^\delta(q)  \right],\label{theta21}
\end{eqnarray} where the expectation is over the couple of random variables $(I,s)$ distributed in $\{1, \ldots, d\}\times\{b,a\}$ according to the probabilities $$\mathbb{P}((I,s)=(i,b)) = \frac{\lambda_{\textrm{RFQ}}^{i,b}}{\sum_{j=1}^d \left(\lambda_{\textrm{RFQ}}^{j,b} + \lambda_{\textrm{RFQ}}^{j,a}\right)}$$ and $$\mathbb{P}((I,s)=(i,a)) = \frac{\lambda_{\textrm{RFQ}}^{i,a}}{\sum_{j=1}^d \left(\lambda_{\textrm{RFQ}}^{j,b} + \lambda_{\textrm{RFQ}}^{j,a}\right)}.$$

In other words, the discounted running penalty (corresponding to the current inventory and the frequency of RFQs) is paid first. Then a bond and a side are drawn and the quoting strategy for that bond and side is applied. Finally, the client decides to or not to transact.\\

In what follows, we are going to work with $\theta_r^\delta$ rather than $\tilde{\theta}_r^\delta$ but Eqs. \eqref{theta12} and~\eqref{theta21} allow to go from one to the other.\\

It is noteworthy that we can also consider optimal value functions $\tilde{\theta}_r^*$ and $\theta_r^*$, which satisfy for all $q \in \mathcal{Q}$ the following equations:
\begin{eqnarray}
\tilde{\theta}_r^*(q) &=& -\frac{\psi(q)}{r + \sum_{i=1}^d \left(\lambda_{\textrm{RFQ}}^{i,b} + \lambda_{\textrm{RFQ}}^{i,a}\right)} + \gamma_{RL} \theta_r^*(q),\label{opttheta12}
\end{eqnarray}
and
\begin{eqnarray}
\theta_r^*(q)  &=& \mathbb{E}\left[\sup_{\delta \in \mathbb{R}} \left(f^{I,s}(\delta) \left(\Delta^I \delta + \tilde{\theta}_r^*\left(q + \left(1_{s=b} - 1_{s=a}\right) \Delta^I e^I\right)\right)\right.\right.\nonumber\\
&&+ \left. \left. \vphantom{\mathbb{E}\left[\sup_{\delta \in \mathbb{R}} \left(f^{I,s}(\delta) \left(\Delta^I \delta + \tilde{\theta}_r^*(q + \left(1_{s=b} - 1_{s=a}\right) \Delta^I e^I)\right)\right.\right.} \left(1-f^{I,s}(\delta)\right) \tilde{\theta}_r^*(q)\right)  \right]. \label{opttheta21}
\end{eqnarray}

\subsection{Actor-critic approach}
\label{ac_approach}
Now that our problem is written in discrete time, we can use some of the ideas of the reinforcement learning literature in order to approximate the optimal policy.\\

In a nutshell, there are two families of approaches in reinforcement learning for approximating the optimal policy: value iteration and policy iteration.\\

Value iteration consists in approximating $\theta_r^*$ or equivalently $\tilde{\theta}_r^*$ by using a fixed point algorithm on Eqs. \eqref{opttheta12} and \eqref{opttheta21}. More precisely, we define for a function $\theta : \mathcal{Q} \to \mathbb{R}$ the operators

$$\Gamma_1 : \theta \mapsto \left(q \mapsto -\frac{\psi(q)}{r + \sum_{i=1}^d \left(\lambda_{\textrm{RFQ}}^{i,b} + \lambda_{\textrm{RFQ}}^{i,a}\right)} + \gamma_{RL} \theta(q)\right)$$
and
$$\Gamma_2 : \theta \mapsto \left(q \mapsto \mathbb{E}\left[\sup_{\delta \in \mathbb{R}} \left(f^{I,s}(\delta) \left(\Delta^I \delta + \theta\left(q + \left(1_{s=b} - 1_{s=a}\right) \Delta^I e^I\right)\right)\right.\right.\right.$$$$\left.\left. \vphantom{\left(q \mapsto \mathbb{E}\left[\sup_{\delta \in \mathbb{R}} \left(f^{I,s}(\delta) \left(\Delta^I \delta + \theta\left(q + \left(1_{s=b} - 1_{s=a}\right) \Delta^I e^I\right)\right)\right.\right.\right.} + \left.\left(1-f^{I,s}(\delta)\right) \theta(q)\right)  \right] \right),$$
and we easily see that $\theta^*_r$ is a fixed point of $\Gamma_2 \circ \Gamma_1$.\\

Because $\Gamma_1$ and $\Gamma_2$ are Lipschitz functions (with respect to the $\|\cdot\|_\infty$ norm) with Lipschitz constants equal respectively to $\gamma_{RL}<1$ and~$1$, $\Gamma_2 \circ \Gamma_1$ is a contraction mapping. Therefore, using Banach fixed-point theorem, $\theta^*_r$ can be approximated using a classical iterative algorithm of the form $\theta_{r} \leftarrow \Gamma_2 \circ \Gamma_1(\theta_{r})$, with any initial function $\theta_{r}$ -- hence the name value iteration.\\

Nevertheless, value iteration has three important drawbacks. First, in the high-dimensional case, value functions cannot be tabulated and need instead to be approximated. However, nothing guarantees the convergence of value iteration with approximation, as the resulting operator has no reason anymore to be a contraction mapping (approximation is indeed rarely a 1-Lipschitz operator with respect to the $\|\cdot\|_\infty$ norm). Second, value iteration is interested in approximating the optimal value function but not directly the optimal strategy. This is not a problem in itself, and it is, in fact, the case of most methods using Hamilton-Jacobi-Bellman equations. However, because we usually approximate the value functions using neural networks, it is not clear that a good approximation of the value function will provide a good approximation of finite differences of that value function -- this is what we need to approximate strategies (see Eqs. \eqref{optimalb} and \eqref{optimala}). Third, when doing value iteration, one requires Hamiltonian functions (because of the supremum in the definition of~$\Gamma_2$) and then their first derivative to compute the optimal strategies (see again Eqs.~\eqref{optimalb} and~\eqref{optimala}). However, the Hamiltonian functions are often not known in closed form and approximating them raises difficulties in the multi-bond case because, for each bond and each side, we do not know in advance the relevant interval of approximation and the required granularity.\\

Policy iteration consists instead in starting from a given policy $\delta$, represented for instance by a neural network with the state as an input (here the inventory in each bond), and updating it in order to increase, for instance, $\sum_{q \in \mathcal{Q}} \theta^\delta_{r}(q) m^\delta(q)$. The methodology we propose is of this type, or more exactly of the actor-critic type, because we use an approximation of the value function in order to guide the update of the strategy.\\

The approach we propose oscillates between phases of TD learning\footnote{TD stands for Temporal Difference. TD learning is a classical reinforcement technique for evaluating value functions. See for instance \cite{sb}.} for evaluating the value function -- the critic -- associated with a given quoting strategy and phases during which the quoting strategy -- the actor -- is updated.\\

Both the value function and the quoting strategy for each bond are represented by feedforward neural networks. It is noteworthy that, in order to normalize the learning process across bonds, the final layer of the neural network(s) modeling the quoting strategy is not a quote but rather a probability to trade. Of course, in order to go from probabilities to quotes and vice versa, we need the functions $\left(f^{i,s}\right)_{i \in \{1, \ldots, d \}, s \in \{b,a\}}$ and the associated inverse functions. In this paper, we have chosen, for each bond $i$, the same function for $f^{i,b}$ and $f^{i,a}$ in the form of a 4-parameter function
\begin{equation}
\label{SU}f^{i} : \delta \mapsto 1 - \Phi\left(\alpha_i + \beta_i \sinh^{-1}\left(\frac{\delta - \mu_i}{\sigma_i}\right)\right)
\end{equation}
with $\beta_i > 0$ (SU Johnson parametrization).\footnote{This parametrization does not satisfy $\sup_{\delta}\frac{f^{i}(\delta){f^{i}}''(\delta)}{\left({f^{i}}'(\delta)\right)^2} <  2$ for all values of the four parameters. However, in the examples of the next section, the above inequality is satisfied (see Figure \ref{h} in Section \ref{dataset}).} Another related symmetry assumption made throughout the paper is $\forall i \in \{1, \ldots, d\}, \lambda^{i,b} = \lambda^{i,a}$.\\

Our algorithm in the case of $d$ bonds can be implemented in two ways. In the first variant, we need $d+1$ neural networks: one for the value function and $d$ for the $d$ quoting functions -- in what follows we only model the quoting strategy at the bid and assume that $\delta^{i,a}(q) = \delta^{i,b}(-q)$\footnote{This assumption is compatible with the hypothesis made in all the examples of this paper that for all $i \in \{1, \ldots, d\}$ we have $f^{i,b} = f^{i,a}$ and $\lambda^{i,b} = \lambda^{i,a}$.} -- written in terms of probability to trade. In this case, the input of all neural networks is the current inventory in all bonds (normalized by the size of RFQs for each bond). In the second variant, we need only two neural networks: one for the value function with the (normalized) current inventory in all bonds as input and one for the quoting strategy with an extended set of inputs consisting of the current inventory in all bonds and a $d$-dimensional one-hot vector specifying the bond for which the network should output a quote.\\

In what follows, we focus on the former variant. However, we present numerical examples involving the second variant in Section \ref{one_nn}.\\

We denote by $q \in \mathcal Q \mapsto  \theta[\omega^0](q)$ the neural network for the value function and by $q \in \mathcal Q \mapsto p^{i}[\omega^i](q)$, for $i \in \{1, \ldots, d\}$, the $d$ neural networks for the probabilities to trade, where $\omega^0, \ldots, \omega^d$ are the weights of the different neural networks.\\

As far as the starting point of the learning process is concerned, we considered two types of initial strategies for each bond. The first possible initial point is a naive strategy that we call the myopic strategy. For bond $i \in \{1, \ldots, d\}$ and side $s \in \{b,a\}$, the myopic strategy consists in the quote $\delta_{\textrm{myopic}}^{i,s}$ maximizing $\delta^{i,s} \mapsto \delta^{i,s}f^{i,s}(\delta^{i,s})$. This strategy maximizes the PnL without taking into account the risk associated with the inventory. In other words, it corresponds to the optimal strategy when the risk penalty function $\psi$ is equal to $0$. This strategy only takes into account the very short run, hence the name myopic. This first initial guess is typically used as an initial point when our algorithm is used in the single-bond case. Using our algorithm in dimension $1$ allows to obtain approximations of the optimal quotes in the single-bond case. These optimal quotes in the single-bond case constitute our second possible initial point for the multi-bond case. This second possible initial point would be optimal in the case of independent bonds, i.e. if the covariance matrix was diagonal. By starting the learning process from that second initial point, it remains to learn how to take account of the correlations between bonds.\\

For either choice of initial strategy, we pre-train $p^1[\omega^1], \ldots, p^d[\omega^d]$ using classical supervised learning techniques (in our case, mini-batches with random inputs in $\mathcal{Q}$ and Adam optimizer on a mean squared error) to output probabilities close to those associated with the desired initial quotes.\\

As far as the neural network $\theta[\omega^0]$ is concerned, we pre-train it so that the initial value function is close to that associated with the initial strategy $\delta$. In dimension 1, we use the linear equations $\eqref{theta12}$ and $\eqref{theta21}$ to obtain $\theta^\delta_r$. In the multi-bond case, we start from a strategy $\delta$ that does not take the correlations into account. Subsequently, the value function $\theta^\delta_r$ is separable in the sense that $$\forall q = (q^1, \ldots, q^d) \in \mathcal{Q}, \theta^\delta_r(q) = \sum_{i=1}^d \theta^{\delta^i}_r(q^i),$$ where $(\theta^{\delta^i}_r)_i$ are the value functions in the single-bond case. In practice, the functions~$(\theta^{\delta^i}_r)_i$ are, as above, obtained by solving the linear equations $\eqref{theta12}$ and~$\eqref{theta21}$.\\

Let us now come to the algorithm itself. As discussed above, it alternates phases of TD learning to learn the value function associated with the current strategy and phases of policy iteration to improve the current strategy.\\

To carry out learning, we consider rollouts of the market making model. Each rollout is a Monte-Carlo simulation starting from a given initial inventory. In other words, at each step of the simulation we first draw a bond and a side according to the probabilities of occurrence of RFQs for each bond and side. Then, we draw whether or not a trade occurs according to the probability associated with our quoting strategy. Finally, we change the inventory according to the occurrence of a trade, if a trade occurred, and go to the next step of the Monte-Carlo simulation.\\

In our algorithm, we consider long rollouts starting from a flat inventory and short rollouts starting from any admissible inventory, called additional rollouts. The former rollouts are used in particular to obtain the average reward per RFQs ($R_{\text{mean}}$) that will be used in the TD learning process. The latter rollouts are used to make more robust the estimation of the value function. It is important to understand that we can simulate our market making model from any given starting inventory. In particular, exploration can be carried out with simulations starting from different points that we choose and not only by choosing different actions as in most reinforcement learning algorithms.\\

After shuffling the rollout data to avoid the well-known autocorrelation bias, we proceed with a classical TD learning of the value function (up to a translation by~$R_{\text{mean}}$) with mini-batches. Interestingly, here, we know the model and we can therefore work with expected rewards and not simply with rewards. In other words, we carry out a mini-batched stochastic semi-gradient descent (where $(I,s)$ and $q$ are stochastic) on the weights $\omega^0$ of the neural network $\theta[\omega^0]$ for minimizing
\begin{eqnarray*}
\mathbb{E}\left[\left(f^{I,s}(\delta^{I,s}(q)) \left(\Delta^I \delta^{I,s}(q) -\frac{\psi\left(q + \left(1_{s=b} - 1_{s=a}\right) \Delta^I e^I\right)}{r + \sum_{i=1}^d \left(\lambda_{\textrm{RFQ}}^{i,b} + \lambda_{\textrm{RFQ}}^{i,a}\right)}\right.\right.\right.
\end{eqnarray*}
\begin{eqnarray*}
&&\left.\vphantom{\left(f^{I,s}(\delta^{I,s}(q)) \left(\Delta^I \delta^{I,s}(q) -\frac{\psi\left(q + \left(1_{s=b} - 1_{s=a}\right) \Delta^I e^I\right)}{r + \sum_{i=1}^d \left(\lambda_{\textrm{RFQ}}^{i,b} + \lambda_{\textrm{RFQ}}^{i,a}\right)}\right.\right.}
 + \gamma_{RL} \theta[\omega^0]\left(q + \left(1_{s=b} - 1_{s=a}\right) \Delta^I e^I\right)\right)\\
&&+ \left(1-f^{I,s}(\delta^{I,s}(q))\right)\left(-\frac{\psi(q)}{r + \sum_{i=1}^d \left(\lambda_{\textrm{RFQ}}^{i,b} + \lambda_{\textrm{RFQ}}^{i,a}\right)} + \gamma_{RL} \theta[\omega^0](q)\right) \vphantom{f^{I,s}(\delta^{I,s}(q)) \left(\Delta^I \delta^{I,s}(q) -\frac{\psi\left(q + \left(1_{s=b} - 1_{s=a}\right) \Delta^I e^I\right)}{r + \sum_{i=1}^d \left(\lambda_{\textrm{RFQ}}^{i,b} + \lambda_{\textrm{RFQ}}^{i,a}\right)}\right.}\\
&&\left.\left. \vphantom{f^{I,s}(\delta^{I,s}(q)) \left(\Delta^I \delta^{I,s}(q) -\frac{\psi\left(q + \left(1_{s=b} - 1_{s=a}\right) \Delta^I e^I\right)}{r + \sum_{i=1}^d \left(\lambda_{\textrm{RFQ}}^{i,b} + \lambda_{\textrm{RFQ}}^{i,a}\right)}\right.} - R_{\text{mean}} - \theta[\omega^0](q)  \right)^2\right],
\end{eqnarray*}
where $\delta^{I,s}(q)$ is computed using the neural network $p^I[\omega^I]$, and where the gradient is taken only on the last $\omega^0$, hence the term semi-gradient.\\

More precisely, if $N$ is the number of mini-batches and $K$ the size of each mini-batch, this means that given a shuffled sequence of inventories, quotes, bonds and sides $(q_{k,n},\delta_{k,n},I_{k,n},s_{k,n})_{1 \le k \le K, 1 \le n \le N}$ coming from one or several rollouts, we carry out for each $n \in \{1, \ldots, N\}$ the following gradient descent with learning rate $\eta$:
\begin{equation*}
\omega^0 \leftarrow \omega^0+ \eta \frac 1 K \sum_{k=1}^K \nabla_{\omega^0} \theta[\omega^0](q_{k,n}) (\widehat{\theta}_{k,n} - \theta[\omega^0](q_{k,n})),\end{equation*}
where
\begin{eqnarray*}
\widehat{\theta}_{k,n}&=& f^{I_{k,n},s_{k,n}}(\delta_{k,n}) \left(\Delta^{I_{k,n}} \delta_{k,n} - \frac{\psi\left(q_{k,n} + \left(1_{s_{k,n}=b} - 1_{s_{k,n}=a}\right) \Delta^{I_{k,n}} e^{I_{k,n}}\right)}{r + \sum_{i=1}^d \left(\lambda_{\textrm{RFQ}}^{i,b} + \lambda_{\textrm{RFQ}}^{i,a}\right)}\right.\\
&&\left.\vphantom{f^{I_{k,n},s_{k,n}}(\delta_{k,n}) \left(\Delta^{I_n} \delta_{k,n} - \frac{\psi\left(q_{k,n} + \left(1_{s_{k,n}=b} - 1_{s_{k,n}=a}\right) \Delta^{I_{k,n}} e^{I_{k,n}}\right)}{r + \sum_{i=1}^d \left(\lambda_{\textrm{RFQ}}^{i,b} + \lambda_{\textrm{RFQ}}^{i,a}\right)}\right.}+ \gamma_{RL} \theta[\omega^0]\left(q_{k,n} + \left(1_{s_{k,n}=b} - 1_{s_{k,n}=a}\right) \Delta^{I_{k,n}} e^{I_{k,n}}\right)\right)- R_{\text{mean}}\\
&& +\left(1-f^{I_{k,n},s_{k,n}}(\delta_{k,n})\right) \left(- \frac{\psi(q_{k,n})}{r + \sum_{i=1}^d \left(\lambda_{\textrm{RFQ}}^{i,b} + \lambda_{\textrm{RFQ}}^{i,a}\right)} + \gamma_{RL} \theta[\omega^0](q_{k,n} )\right).\\
\end{eqnarray*}

It is noteworthy that the subtraction of $R_{\text{mean}}$ in the above equations permits to center the value functions and avoid very large values (positive or negative). This is particularly important as the value functions scale in $\frac 1r$ and as we focus on $r$ small to be close to the ergodic case.\\

Once each TD learning phase is over, the quoting strategy is going to be improved. For that purpose, we employ a method that is at the frontier between random search and gradient policy. In fact, during each of the rollouts, not only the current strategy~$\delta$ was played but also a randomized version of it, for exploration.\footnote{We only explore locally in the sense that the list of states in the rollout is only that given by the use of the strategy $\delta$. The randomized version of the strategy only helps answering the question: ``what if we did differently for that state?''.} More precisely, we consider for each $(I,s)$ a slightly modified policy $$q \mapsto \delta_\epsilon^{I,s}(q) = {f^{I,s}}^{-1}\left(\nu \vee \left(f^{I,s}(\delta^{I,s}(q)) + \epsilon\right) \wedge (1 - \nu)\right),$$ where $\nu$ is a fixed small number in $(0,1)$, and where $\epsilon$ is a centered random variable (exploration noise) with a distribution to be chosen. Then, we consider that exploration goes into the right or wrong direction depending on whether
$$\!f^{I,s}(\delta_\epsilon^{I,s}(q))\!\!\left(\Delta^I \delta_\epsilon^{I,s}(q)\! - \!\frac{\psi(q\! +\! \left(1_{s=b} \! - \! 1_{s=a}\right) \Delta^I e^I)}{r\! +\! \sum_{i=1}^d \left(\lambda_{\textrm{RFQ}}^{i,b}\! +\! \lambda_{\textrm{RFQ}}^{i,a}\right)}\!+\! \gamma_{RL} \theta[\omega^0](q \! + \! \left(1_{s=b}\! -\! 1_{s=a}\right) \Delta^I e^I)\right)$$
\begin{equation}\label{withnoise}+ \left(1-f^{I,s}(\delta_\epsilon^{I,s}(q))\right)\left(-\frac{\psi(q)}{r + \sum_{i=1}^d \left(\lambda_{\textrm{RFQ}}^{i,b} + \lambda_{\textrm{RFQ}}^{i,a}\right)} + \gamma_{RL} \theta[\omega^0](q)\right)\end{equation}
is above or below
$$\!f^{I,s}(\delta^{I,s}(q))\!\!\left(\Delta^I \delta^{I,s}(q)\! - \!\frac{\psi(q\! +\! \left(1_{s=b} \! - \! 1_{s=a}\right) \Delta^I e^I)}{r\! +\! \sum_{i=1}^d \left(\lambda_{\textrm{RFQ}}^{i,b}\! +\! \lambda_{\textrm{RFQ}}^{i,a}\right)}\!+\! \gamma_{RL} \theta[\omega^0](q \! + \! \left(1_{s=b}\! -\! 1_{s=a}\right) \Delta^I e^I)\right)$$
\begin{equation}\label{withoutnoise}+ \left(1-f^{I,s}(\delta^{I,s}(q))\right)\left(-\frac{\psi(q)}{r + \sum_{i=1}^d \left(\lambda_{\textrm{RFQ}}^{i,b} + \lambda_{\textrm{RFQ}}^{i,a}\right)} + \gamma_{RL} \theta[\omega^0](q)\right).\end{equation}
\normalsize
More precisely, we consider the same rollouts as above and we split the resulting data set into $d$ data sets corresponding to RFQs over each of the $d$ bonds. For each of these $d$ data sets, we compute the difference between \eqref{withnoise} and \eqref{withoutnoise} and normalize it by dividing by the standard deviation over that data set. We then change the weights of the neural network defining each actor so as to go in the right direction, with a gradient proportional to the above normalized difference and proportional to $f^{I,s}(\delta_\epsilon^{I,s}(q)) - f^{I,s}(\delta^{I,s}(q))$.\\

In mathematical terms, for each bond $i$ we have a sequence of inventories, normalized differences between \eqref{withnoise} and \eqref{withoutnoise}, and perturbations in the probability space denoted by $(q_{l,m}, dV_{l,m},dp_{l,m})_{1 \le l \le L , 1 \le m \le M}$, where $M$ is the number of mini-batches, each being of size $L$. The corresponding updates for bond $i$ consist for all $m \in \{1, \ldots, M\}$ in the gradient ascent
\begin{equation*}
\omega^i \leftarrow \omega^i+ \tilde{\eta} \frac 1 L \sum_{l=1}^L \nabla_{\omega^i} p^i[\omega^i](q_{l,m}) dV_{l,m} dp_{l,m},\end{equation*} where $\tilde{\eta}$ is a unique learning rate for all bonds.\\

It is noteworthy that this method is slightly different from a gradient ascent towards the greedy policy associated with a given value function. The difference lies in the fact that random noises are used here whereas a pure gradient ascent would use a deterministic automatic differentiation to decide upon the direction of improvement of each actor. Using random noises enables to explore more and, besides, it allows a normalization that makes easier the choice of hyperparameters.\footnote{The risk with high-dimensional problems is indeed to design a method that requires the choice of a number of hyperparameters that increases with the dimension.}\\

Another important remark regarding our methodology is that we progressively increase the boundaries of the problem. In other words, periodically, after a given number of updates of the critic and the actor, we increase each risk limit by $\Delta^i$ in order to finally reach $Q^i$. This means that the initial rollouts are carried out over a small state space and that the state space expands progressively. This idea, which we call \emph{reverse Matryoshka dolls principle}, is an idea that cannot be used with grids but that perfectly fits our deep learning framework.\\
\vspace{-5mm}
\begin{algorithm*}[h!]
    \caption{Market Making RL Algorithm}
  \begin{algorithmic}[1]
    \INPUT A market making model with $d$ bonds, maximal risk limits, and initial quotes for each bond
    \STATE \textbf{Initialization}
    \FOR{$i \in \{1, \ldots, d\}$}
    \STATE Pre-train actor $i$'s neural network to the initial quote for bond $i$
    \ENDFOR
    \STATE Compute the 1-dimensional value function for each bond
    \STATE  Pre-train the critic's neural network to the resulting value function in the zero-correlation case
    \STATE \textbf{Updates}
    \FOR{$j=1..$\texttt{NB STEPS}}
      \IF{$j$ \textrm{modulo} \texttt{NB STEPS BETWEEN INCREASE} = 0 and  maximal risk limits are not reached}
      \STATE Increase the risk limits
      \ENDIF
      \STATE Carry out rollouts (standard and additional) with the current policy
      \STATE Update the critic using TD learning
      \STATE Update the policy by updating each of the $d$ neural networks of the actor
    \ENDFOR
    \OUTPUT $d$ neural networks for the $d$ quote functions (1 for each bond).
  \end{algorithmic}
\end{algorithm*}

\newpage

\section{Numerical results}

\subsection{A set of 20 corporate bonds}
\label{dataset}
In order to illustrate our reinforcement learning method, we consider a set of 20~European corporate bonds. The characteristics of the bonds and the associated RFQs are documented in the following tables.\footnote{For confidentiality reasons, the bonds are identified by numbers and not by their Bloomberg or Reuters identifier.}\\

We start with Table \ref{lambdadelta} in which we document the characteristics of the RFQs for each  bond. Regarding the size of RFQs, it is indicated in numéraire, but the size has to be in number of bonds to be used in our model. We decided to divide the size by $100$ in order to go from numéraire to number of bonds. This means that we make the assumption that bonds are at par. Of course, we can divide by the current price instead of $100$ and update the model periodically, but given the bonds that we consider, dividing by $100$ is satisfactory and is a second-order assumption compared to the assumption of constant sizes.

\begin{table}[h]
\centering
\begin{tabular}{|c|c|c|}
  \hline
  Bond& Arrival rate & Average size of RFQs \\
  identifier & $\lambda_{\textrm{RFQ}}^{b}= \lambda_{\textrm{RFQ}}^{a}$ & in numéraire \\
  \hline
BOND.1&0.275&700000\\
BOND.2&0.175&300000\\
BOND.3&0.1&900000\\
BOND.4&0.15&1200000\\
BOND.5&0.025&1000000\\
BOND.6&0.1&600000\\
BOND.7&0.05&800000\\
BOND.8&0.175&900000\\
BOND.9&0.4&200000\\
BOND.10&0.125&1300000\\
BOND.11&0.4&500000\\
BOND.12&0.425&1000000\\
BOND.13&0.575&500000\\
BOND.14&0.325&700000\\
BOND.15&0.4&500000\\
BOND.16&0.2&1000000\\
BOND.17&0.125&500000\\
BOND.18&0.125&800000\\
BOND.19&0.3&1100000\\
BOND.20&0.325&1200000\\
  \hline
  \end{tabular}
\caption{Characteristics of the RFQs received by the market maker.}
\label{lambdadelta}
\end{table}
\newpage

Table \ref{SU_table} below documents the probability to trade following an answer to an RFQ. The parametrization is that of Eq. \eqref{SU} (SU Johnson).\\

\begin{table}[h]
\centering
\begin{tabular}{|c|c|c|c|c|}
  \hline
  Bond identifier& $\alpha$ & $\beta$ & $\mu$ & $\sigma$ \\
  \hline
BOND.1&0.4&0.6&0.096&0.086\\
BOND.2&0.4&0.6&0.0576&0.0516\\
BOND.3&0.4&0.6&0.1728&0.1548\\
BOND.4&0.4&0.6&0.0528&0.0473\\
BOND.5&0.4&0.6&0.3408&0.3053\\
BOND.6&0.4&0.6&0.1008&0.0903\\
BOND.7&0.4&0.6&0.1872&0.1677\\
BOND.8&0.4&0.6&0.2496&0.2236\\
BOND.9&0.4&0.6&0.1248&0.1118\\
BOND.10&0.4&0.6&0.0096&0.0086\\
BOND.11&0.4&0.6&0.096&0.086\\
BOND.12&0.4&0.6&0.192&0.172\\
BOND.13&0.4&0.6&0.048&0.043\\
BOND.14&0.4&0.6&0.3312&0.2967\\
BOND.15&0.4&0.6&0.144&0.129\\
BOND.16&0.4&0.6&0.0576&0.0516\\
BOND.17&0.4&0.6&0.0672&0.0602\\
BOND.18&0.4&0.6&0.2832&0.2537\\
BOND.19&0.4&0.6&0.1584&0.1419\\
BOND.20&0.4&0.6&0.12&0.1075\\
  \hline
\end{tabular}
\caption{Characteristics of the probability to trade (SU Johnson, see Eq. \eqref{SU}).}
\label{SU_table}
\end{table}

\begin{figure}[H]
  \centering
  \includegraphics[width=0.44\textwidth]{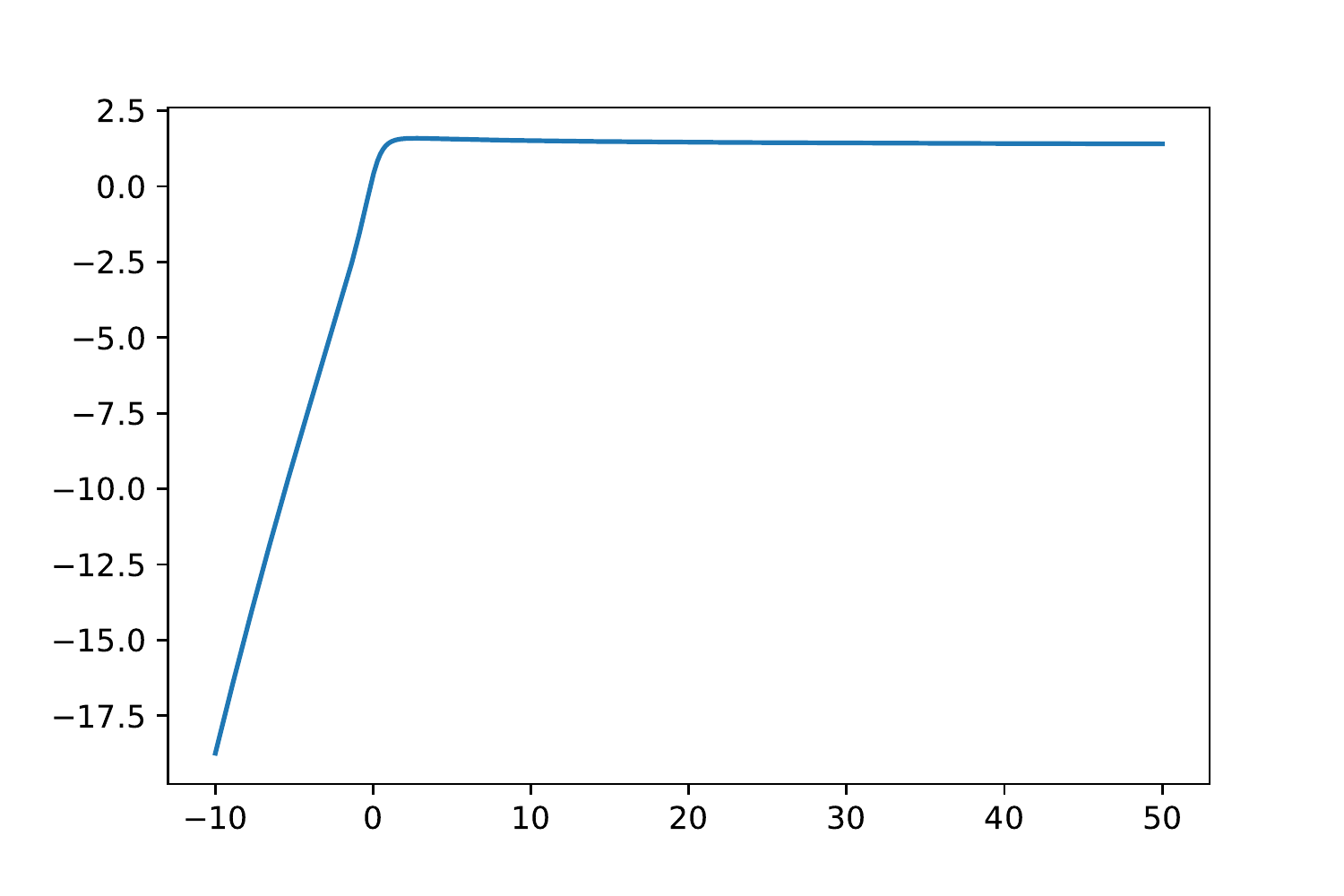}
  \includegraphics[width=0.44\textwidth]{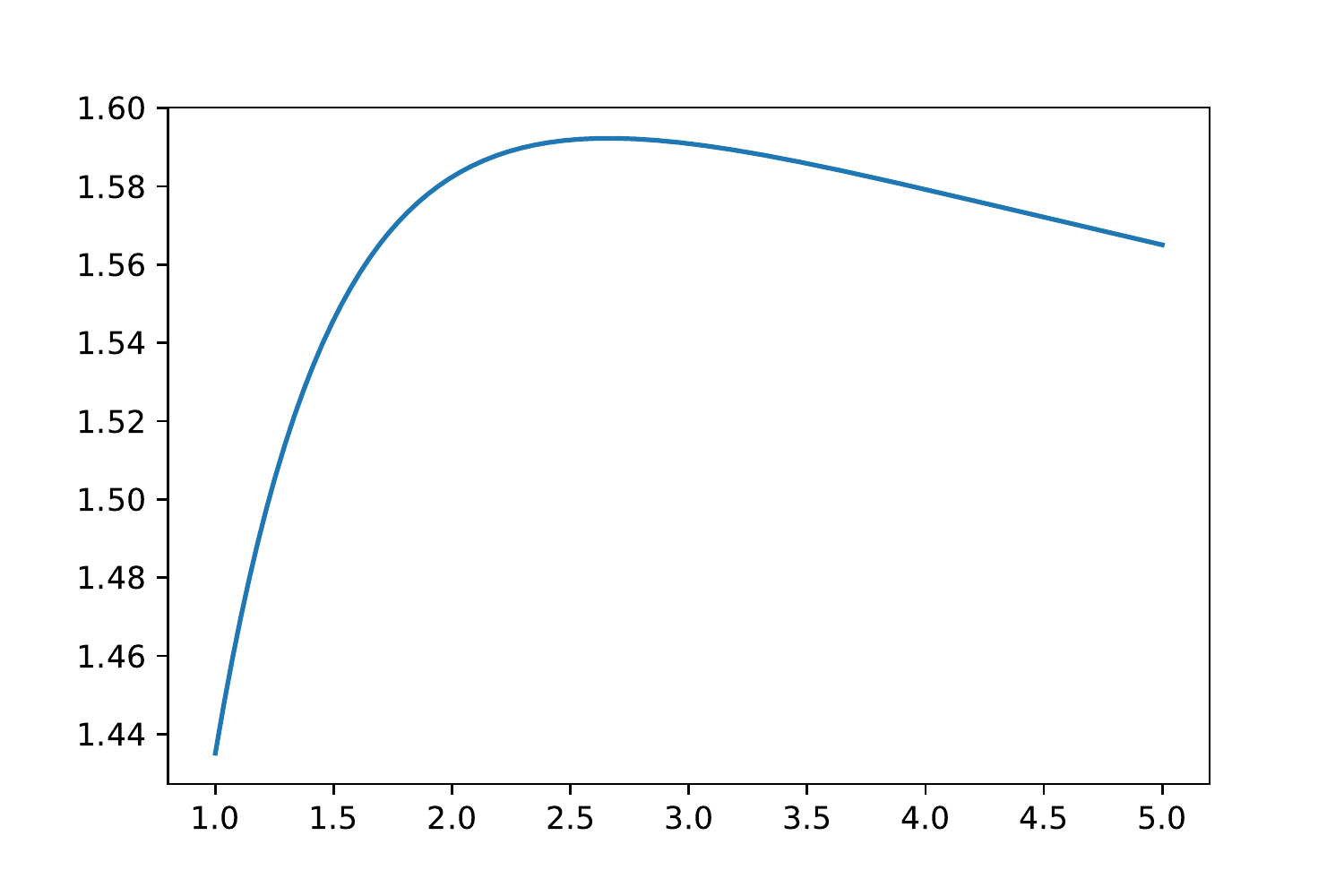}
  \caption{Check of the conditions \eqref{sup2}. Left: large scale. Right: zoom on the maximum.}\label{h}
\end{figure}

Defining $h : z \mapsto 1 - \Phi\left(0.4 + 0.6 \sinh^{-1}\left(z\right)\right)$, the function $z \mapsto \frac{h(z)h''(z)}{h'(z)^2}$ is always below $2$ as it can be seen in Figure \ref{h}. Therefore, we are in the conditions where all the theoretical results of the literature apply.

\newpage

The covariance matrix between the price variations of the 20 bonds is documented in Table \ref{cov_table} below.\\

\begin{table}[H]
\centering
{\scriptsize
\rotatebox[origin=c]{90}{
\begin{tabular}{|c@{\hskip 1mm}c@{\hskip 1mm}c@{\hskip 1mm}c@{\hskip 1mm}c@{\hskip 1mm}c@{\hskip 1mm}c@{\hskip 1mm}c@{\hskip 1mm}c@{\hskip 1mm}c@{\hskip 1mm}c@{\hskip 1mm}c@{\hskip 1mm}c@{\hskip 1mm}c@{\hskip 1mm}c@{\hskip 1mm}c@{\hskip 1mm}c@{\hskip 1mm}c@{\hskip 1mm}c@{\hskip 1mm}c@{\hskip 1mm}|c@{\hskip 1mm}|}
  \hline
\rotatebox[origin=c]{-90}{\ BOND.1\ }&\rotatebox[origin=c]{-90}{\ BOND.2\ }&\rotatebox[origin=c]{-90}{\ BOND.3\ }&\rotatebox[origin=c]{-90}{\ BOND.4\ }&\rotatebox[origin=c]{-90}{\ BOND.5\ }&\rotatebox[origin=c]{-90}{\ BOND.6\ }&\rotatebox[origin=c]{-90}{\ BOND.7\ }&\rotatebox[origin=c]{-90}{\ BOND.8\ }&\rotatebox[origin=c]{-90}{\ BOND.9\ }&\rotatebox[origin=c]{-90}{\ BOND.10\ }&\rotatebox[origin=c]{-90}{\ BOND.11\ }&\rotatebox[origin=c]{-90}{\ BOND.12\ }&\rotatebox[origin=c]{-90}{\ BOND.13\ }&\rotatebox[origin=c]{-90}{\ BOND.14\ }&\rotatebox[origin=c]{-90}{\ BOND.15\ }&\rotatebox[origin=c]{-90}{\ BOND.16\ }&\rotatebox[origin=c]{-90}{\ BOND.17\ }&\rotatebox[origin=c]{-90}{\ BOND.18\ }&\rotatebox[origin=c]{-90}{\ BOND.19\ }&\rotatebox[origin=c]{-90}{\ BOND.20\ }&\\
\hline
.0049&.0014&.0112&.0021&.0203&.0056&.014&.0182&.0084&0&.0063&.0161&0&.0217&.0119&.0021&.0035&.0196&.014&0&BOND.1\\
.0014&.0005&.0031&.0006&.0053&.0016&.0038&.0046&.0023&0&.0017&.0041&0&.0055&.0031&.0007&.001&.0048&.0036&0&BOND.2\\
.0112&.0031&.0282&.0048&.0544&.0133&.0362&.0487&.0213&0&.0155&.0438&0&.0603&.0315&.0047&.008&.0541&.0374&0&BOND.3\\
.0021&.0006&.0048&.001&.0088&.0024&.006&.0078&.0036&0&.0027&.0069&0&.0093&.0051&.0009&.0015&.0085&.0059&.0001&BOND.4\\
.0203&.0053&.0544&.0088&.1381&.0247&.0744&.1098&.0413&0&.0296&.0989&0&.1472&.0662&.0082&.0145&.1295&.0834&.0001&BOND.5\\
.0056&.0016&.0133&.0024&.0247&.0066&.0169&.0222&.0101&0&.0075&.0197&0&.0267&.0145&.0024&.004&.0242&.0171&0&BOND.6\\
.014&.0038&.0362&.006&.0744&.0169&.0482&.0657&.0275&0&.0199&.0598&0&.083&.0421&.0058&.01&.0741&.0508&0&BOND.7\\
.0182&.0046&.0487&.0078&.1098&.0222&.0657&.0971&.0371&0&.0273&.0871&0&.1265&.0589&.0072&.013&.1127&.0734&0&BOND.8\\
.0084&.0023&.0213&.0036&.0413&.0101&.0275&.0371&.0166&0&.0118&.0331&0&.0456&.0242&.0035&.006&.0409&.0285&0&BOND.9\\
0&0&0&0&0&0&0&0&0&0&0&0&0&0&0&0&0&0&0&0&BOND.10\\
.0063&.0017&.0155&.0027&.0296&.0075&.0199&.0273&.0118&0&.0092&.0244&0&.0332&.0174&.0026&.0045&.0303&.0208&0&BOND.11\\
.0161&.0041&.0438&.0069&.0989&.0197&.0598&.0871&.0331&0&.0244&.0829&0&.1182&.0545&.0064&.0115&.1043&.0676&.0006&BOND.12\\
0&0&0&0&0&0&0&0&0&0&0&0&.0001&.0002&0&0&0&.0002&0&.0001&BOND.13\\
.0217&.0055&.0603&.0093&.1472&.0267&.083&.1265&.0456&0&.0332&.1182&.0002&.1873&.0763&.0086&.0155&.1602&.0978&.0009&BOND.14\\
.0119&.0031&.0315&.0051&.0662&.0145&.0421&.0589&.0242&0&.0174&.0545&0&.0763&.0386&.0048&.0085&.0679&.0459&.0002&BOND.15\\
.0021&.0007&.0047&.0009&.0082&.0024&.0058&.0072&.0035&0&.0026&.0064&0&.0086&.0048&.0011&.0015&.0074&.0057&0&BOND.16\\
.0035&.001&.008&.0015&.0145&.004&.01&.013&.006&0&.0045&.0115&0&.0155&.0085&.0015&.0026&.0141&.01&0&BOND.17\\
.0196&.0048&.0541&.0085&.1295&.0242&.0741&.1127&.0409&0&.0303&.1043&.0002&.1602&.0679&.0074&.0141&.1452&.0868&.0009&BOND.18\\
.014&.0036&.0374&.0059&.0834&.0171&.0508&.0734&.0285&0&.0208&.0676&0&.0978&.0459&.0057&.01&.0868&.0578&.0001&BOND.19\\
0&0&0&.0001&.0001&0&0&0&0&0&0&.0006&.0001&.0009&.0002&0&0&.0009&.0001&.0013&BOND.20\\
 \hline
\end{tabular}}}
\caption{Covariance matrix.}
\label{cov_table}
\end{table}

\normalsize

\newpage

\subsection{Results in the case $\psi(q) = \frac 12 \gamma \sqrt{q'\Sigma q}$}
\label{sqrt}
We first start with examples corresponding to a penalty function of the form $$\psi(q) = \frac 12 \gamma \sqrt{q'\Sigma q}.$$ This penalty is proportional to the instantaneous standard deviation of the MtM value of the portfolio. It penalizes inventory in a way proportional to the volatility of bond prices in the single-bond case, and takes account of the correlation between bonds in the multi-bond case. This penalty is less severe than the quadratic penalty, proportional to the variance of the MtM value of the portfolio, used in Section \ref{sqr} and in some papers of the literature (see for instance \cite{carteabook} and \cite{gueant}).\\

\subsubsection{Single-bond cases}

We start with examples of single-bond market making. Our focus is on the comparison between results obtained by using the finite difference method presented in the appendix (hereafter PDE method\footnote{This method is indeed inspired by partial differential equation techniques.}) and results obtained by using our reinforcement learning algorithm. In what follows we consider $\gamma = 5\cdot10^{-2}$ and $r = 10^{-4}$. The risk limits were set to 5 times the size of RFQs.\\

For the method using a finite difference scheme, we first had to choose how to interpolate the Hamiltonian functions. Our methodology was to try several intervals and choices of granularity for each Hamiltonian function. If an approximation of the optimal value function could be obtained (i.e. if the interval of interpolation was large enough), then we computed the associated optimal quotes and compared the value function associated with these optimal quotes to the approximation of the optimal value function. We validated our interpolation and approximation when the two were almost equal.\\

\begin{figure}
  \centering
  \includegraphics[width=0.88\textwidth]{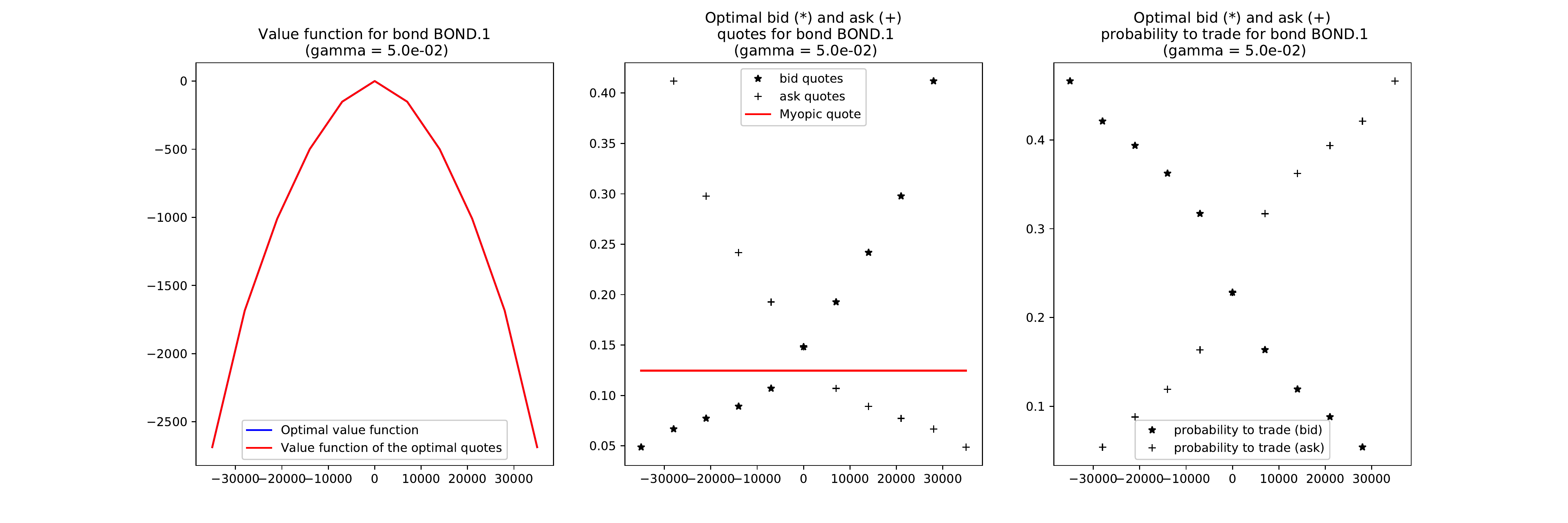}\\
  \includegraphics[width=0.88\textwidth]{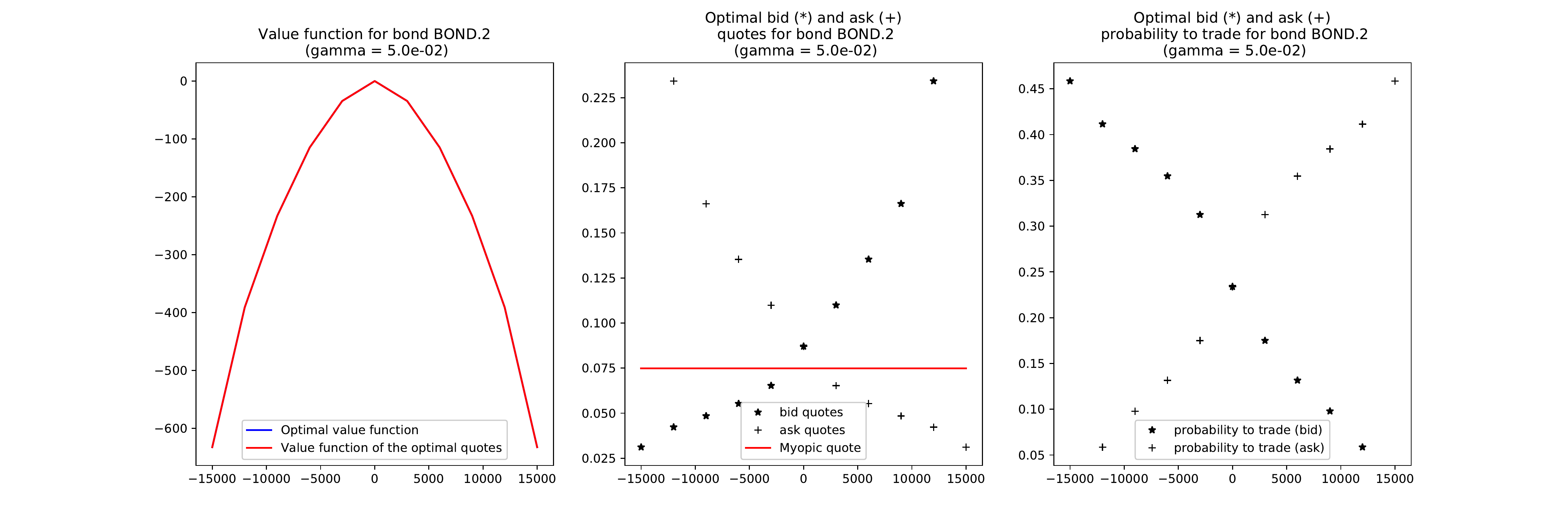}\\
  \includegraphics[width=0.88\textwidth]{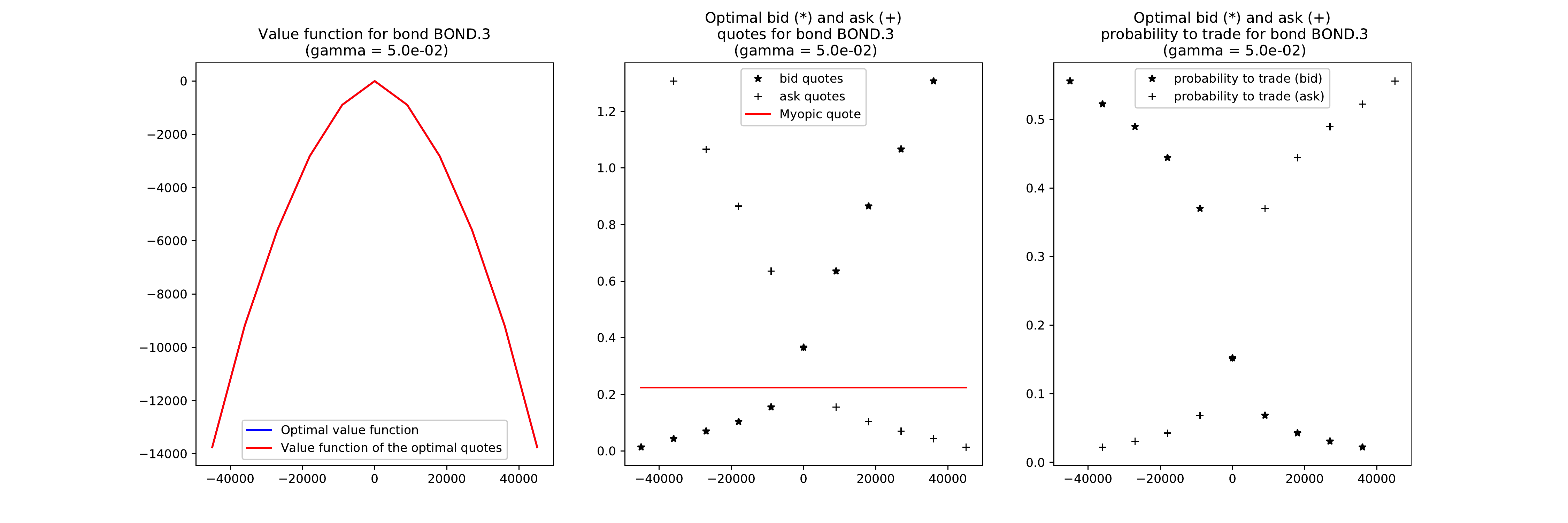}\\
  \includegraphics[width=0.88\textwidth]{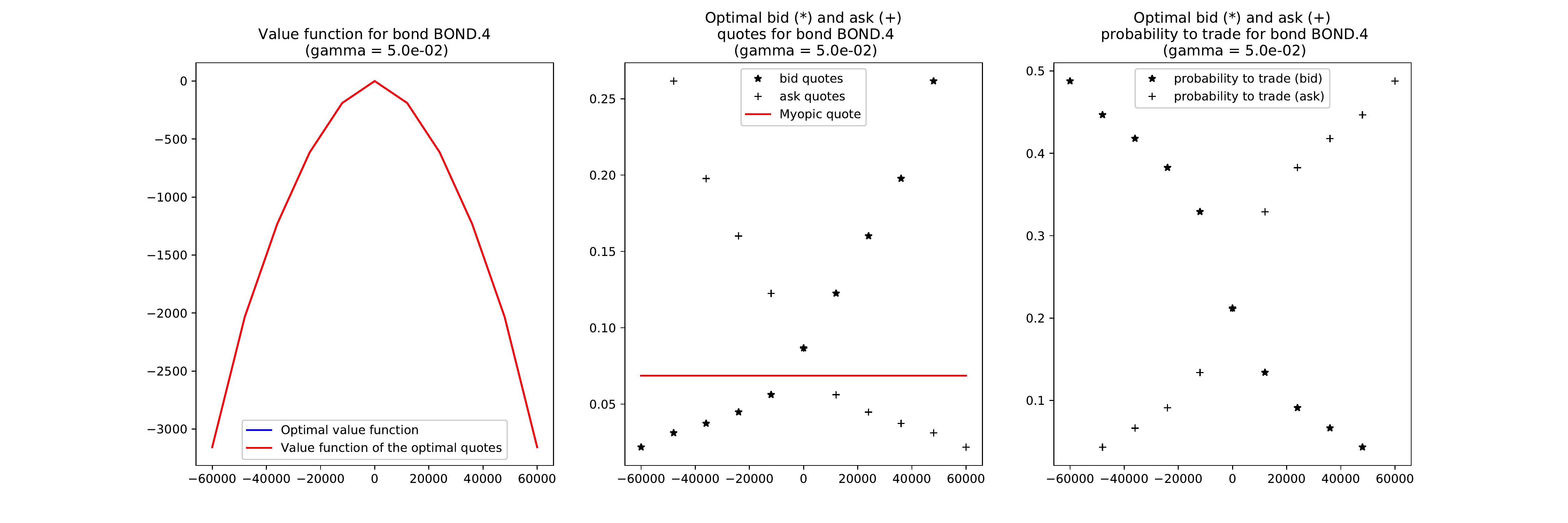}
  \caption{Value functions, optimal quotes, and optimal probabilities to trade with the finite difference approach.}
  \label{pde_sqrt_1}
\end{figure}

\begin{figure}
  \centering
  \includegraphics[width=0.88\textwidth]{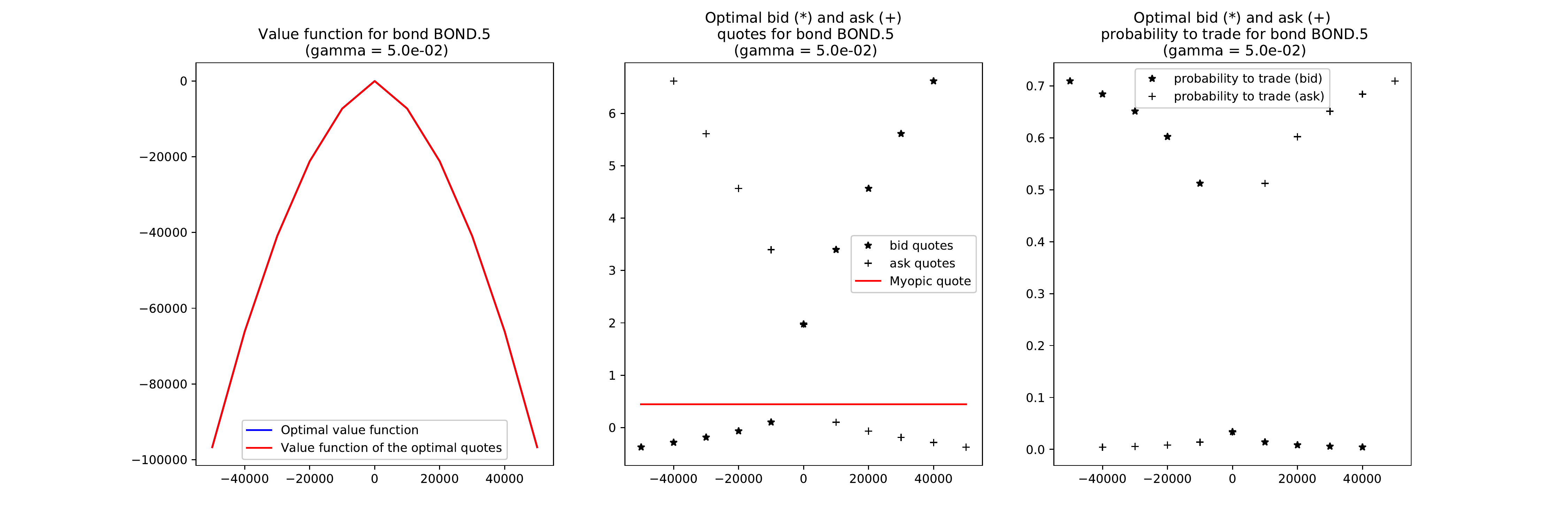}\\
  \includegraphics[width=0.88\textwidth]{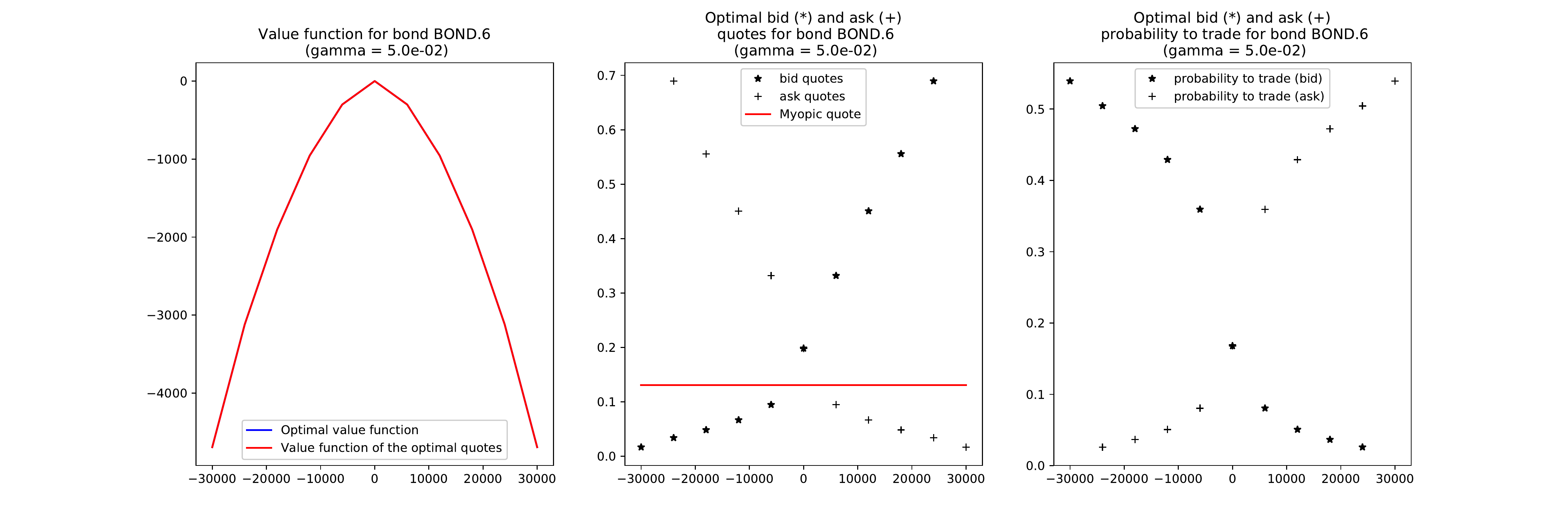}\\
  \includegraphics[width=0.88\textwidth]{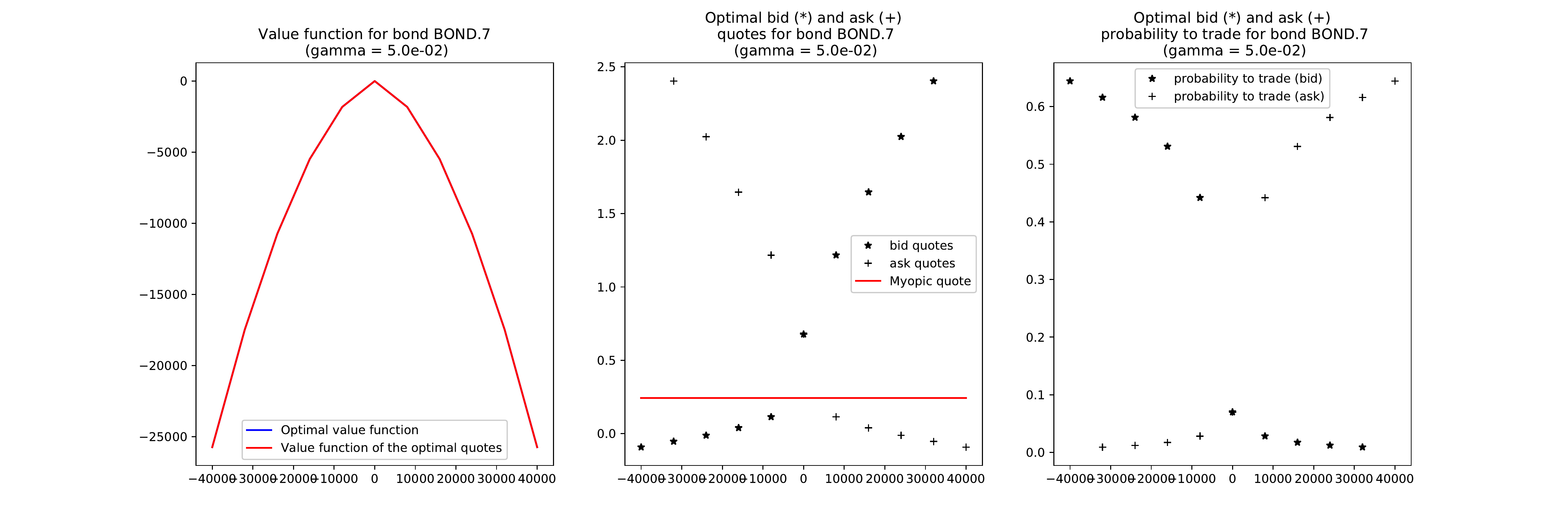}\\
  \includegraphics[width=0.88\textwidth]{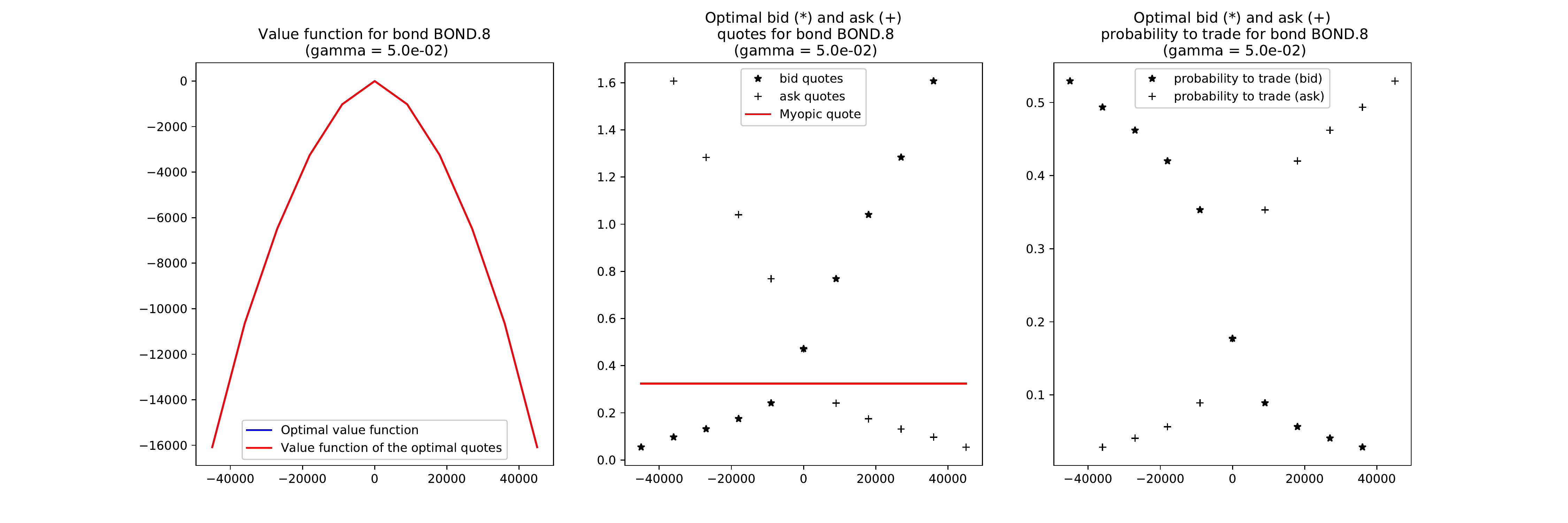}
  \caption{Value functions, optimal quotes, and optimal probabilities to trade with the finite difference approach.}
  \label{pde_sqrt_2}
\end{figure}

\begin{figure}
  \centering
  \includegraphics[width=0.88\textwidth]{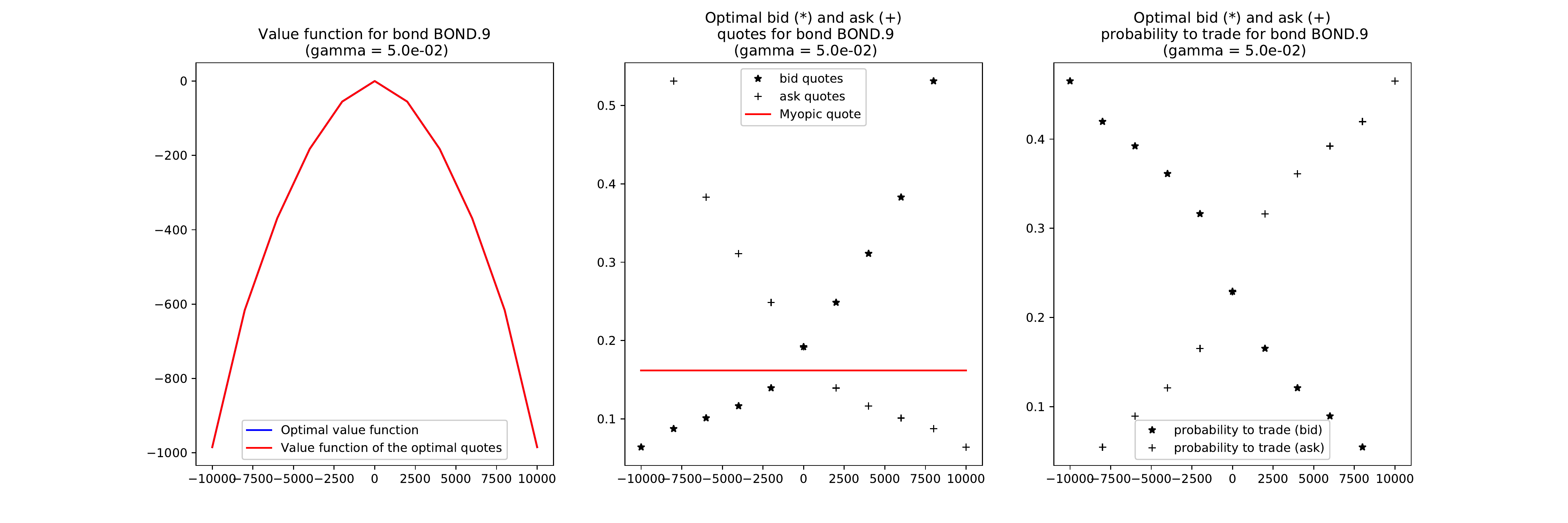}\\
  \includegraphics[width=0.88\textwidth]{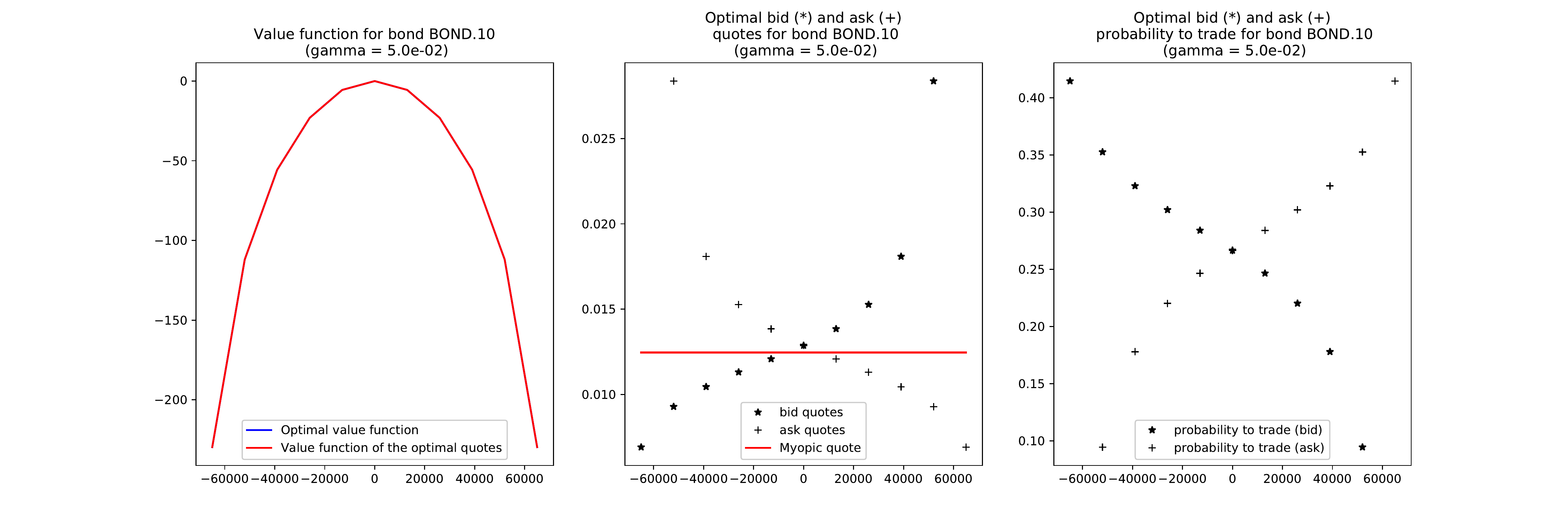}\\
  \includegraphics[width=0.88\textwidth]{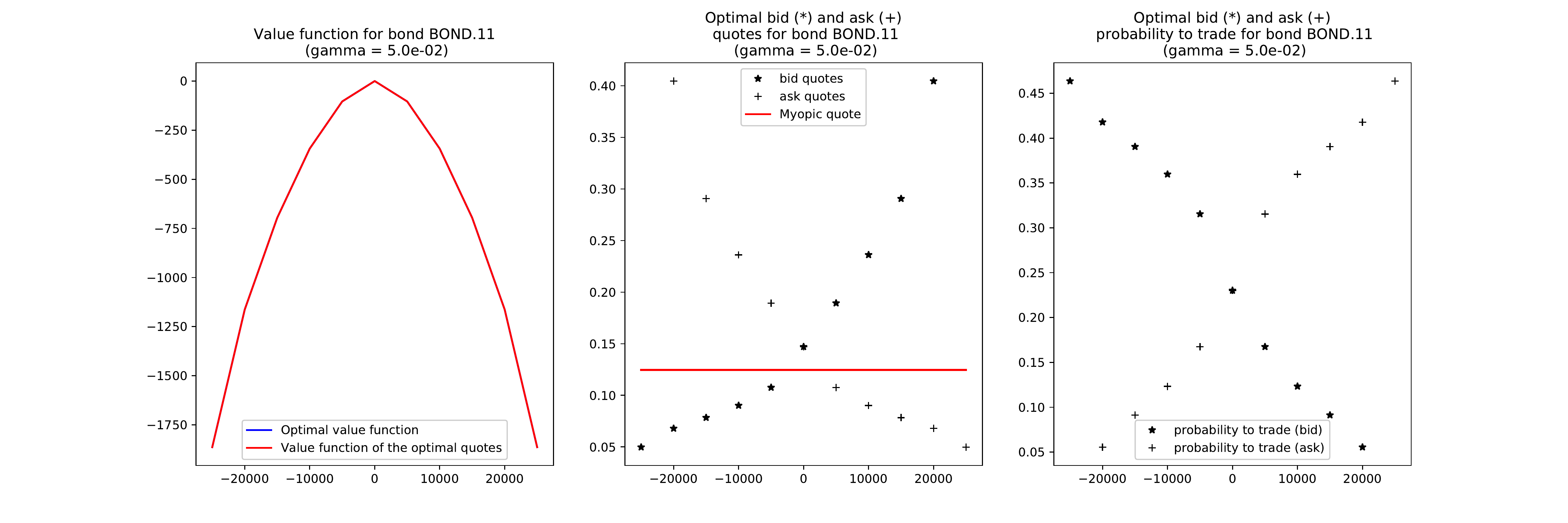}\\
  \includegraphics[width=0.88\textwidth]{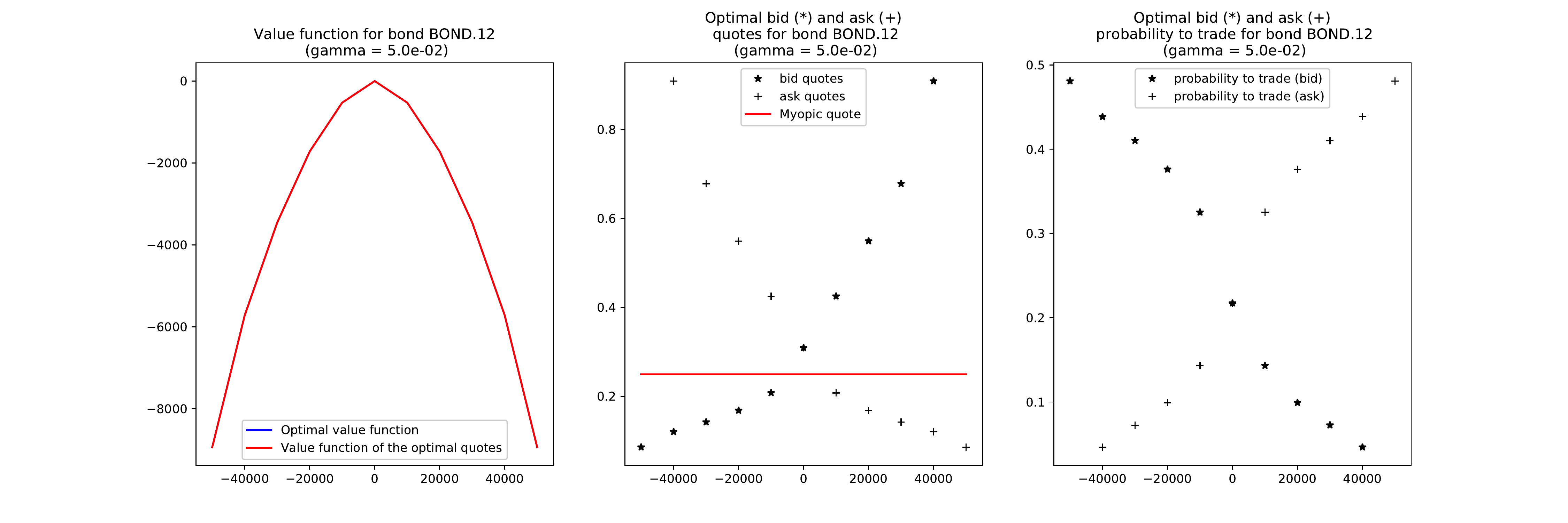}
  \caption{Value functions, optimal quotes, and optimal probabilities to trade with the finite difference approach.}
  \label{pde_sqrt_3}
\end{figure}

\begin{figure}
  \centering
  \includegraphics[width=0.88\textwidth]{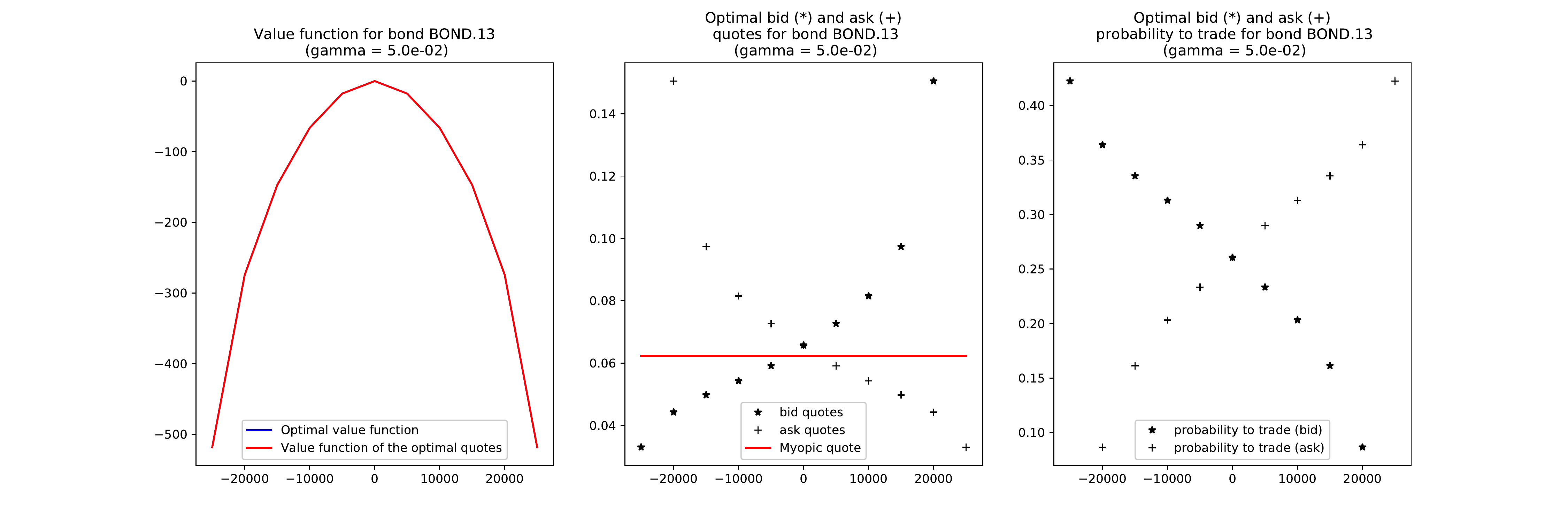}\\
  \includegraphics[width=0.88\textwidth]{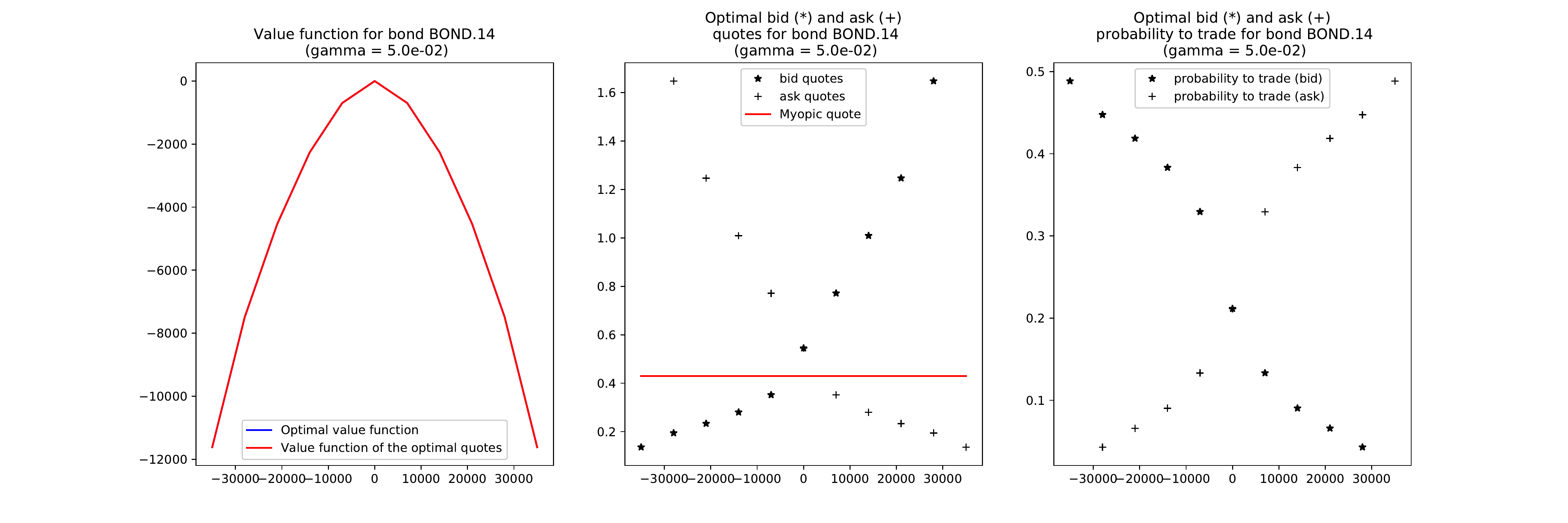}\\
  \includegraphics[width=0.88\textwidth]{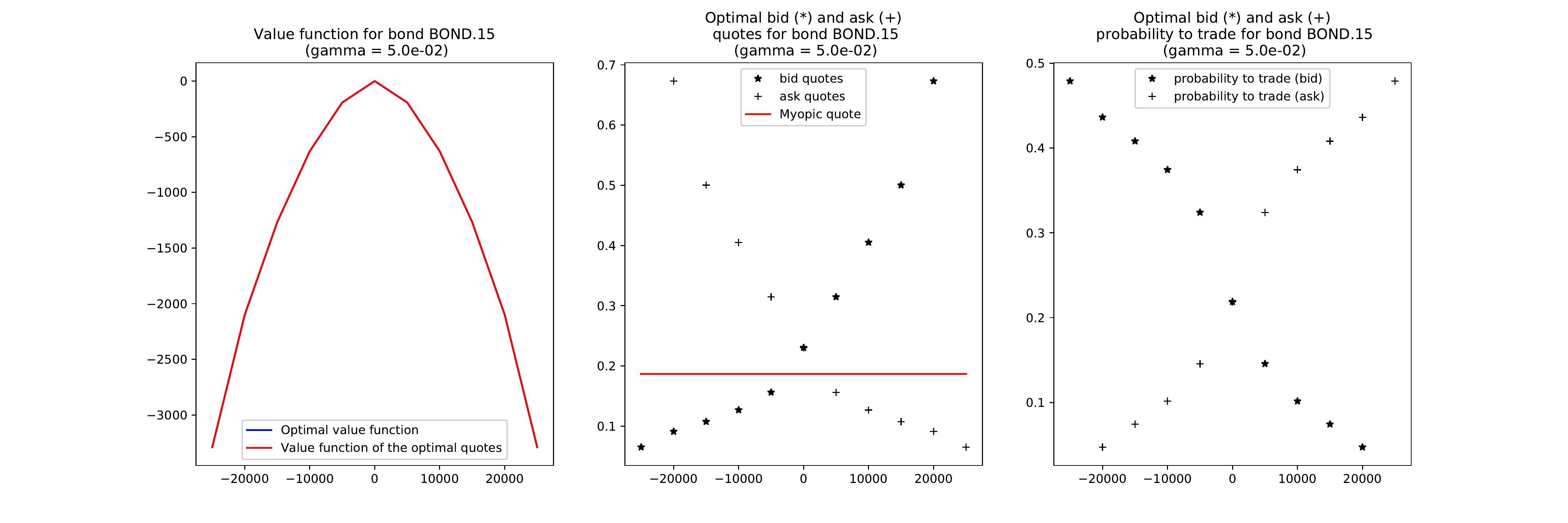}\\
  \includegraphics[width=0.88\textwidth]{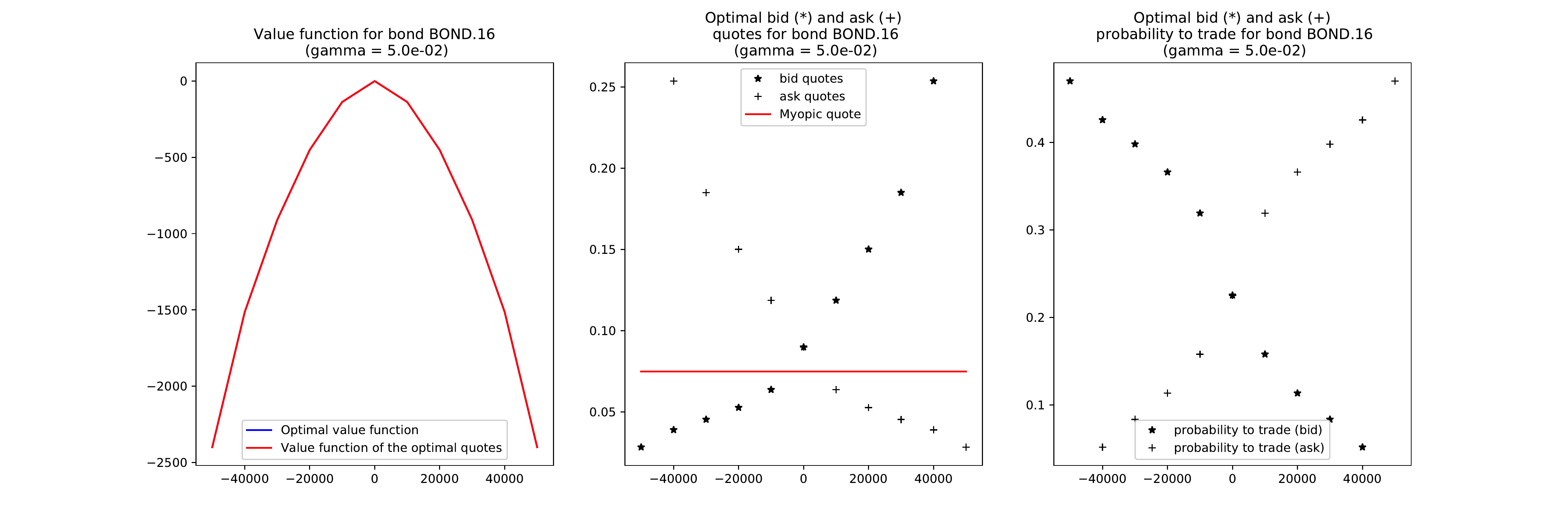}
  \caption{Value functions, optimal quotes, and optimal probabilities to trade with the finite difference approach.}
  \label{pde_sqrt_4}
\end{figure}

In Table \ref{R_mean_table} and Figures \ref{pde_sqrt_1}, \ref{pde_sqrt_2}, \ref{pde_sqrt_3}, \ref{pde_sqrt_4}, and \ref{pde_sqrt_5} we document the results we obtained with the finite difference scheme of the appendix.\\

On the left panels of the figures, we plotted the optimal value function and the value function associated with the optimal quotes (which should be exactly the same).\footnote{We subtracted the maximum of each function before plotting.} The value functions are those of the continuous-time model (i.e. at any time, not specifically at an RFQ time). On the central panels we plotted the optimal quotes along with the myopic quotes. Eventually, on the right panels, we plotted the associated probability to trade.\\

The average rewards per RFQ obtained when using the optimal quotes computed with our PDE method are reported in Table 4. The reported figures were obtained with Monte-Carlo simulations of 3000 RFQs.\\

\begin{table}[h!]
\centering
\begin{tabular}{|c|c|}
  \hline
  Bond identifier& Average reward per RFQ \\
  \hline
BOND.1&199.1\\
BOND.2&53.3\\
BOND.3&354.4\\
BOND.4&180\\
BOND.5&391.6\\
BOND.6&155.2\\
BOND.7&240\\
BOND.8&569.3\\
BOND.9&75.5\\
BOND.10&43.4\\
BOND.11&145.8\\
BOND.12&552.2\\
BOND.13&81.3\\
BOND.14&653.8\\
BOND.15&208.5\\
BOND.16&171.4\\
BOND.17&90.2\\
BOND.18&527.7\\
BOND.l9&469.4\\
BOND.20&473.7\\
  \hline
\end{tabular}
\caption{Average rewards per RFQ for the optimal quotes computed with the finite difference method (Monte-Carlo simulation).}
\label{R_mean_table}
\end{table}

We see in Figures \ref{pde_sqrt_1}, \ref{pde_sqrt_2}, \ref{pde_sqrt_3}, \ref{pde_sqrt_4}, and \ref{pde_sqrt_5} that the optimal quote function at the bid $q \mapsto \delta^{b*}(q)$  is an increasing function and that the optimal quote function at the ask $q \mapsto \delta^{a*}(q)$  is a decreasing function. This means, as expected, that a market maker with a long (resp. short) inventory wants to decrease (resp. increase) the probability to buy and increase (resp. decrease) the probability to sell. The plots of probabilities (right panels) are quite instructive as the shape of the functions highly depends on the liquidity and the volatility of the bond. For instance, for the two illiquid and volatile bonds BOND.5 and BOND.7 (see Tables 1 and 3 for the value of $\lambda_{\textrm{RFQ}}^{i,b} = \lambda_{\textrm{RFQ}}^{i,a}$ and $\sigma^i$) the optimal quotes are associated with very low probabilities ($<10\%$) to buy when the inventory is positive or even equal to $0$ and those probabilities jump to more that $40\%$ when the inventory is negative (in fact less than $-\Delta$).\\

\begin{figure}
  \centering
  \includegraphics[width=0.88\textwidth]{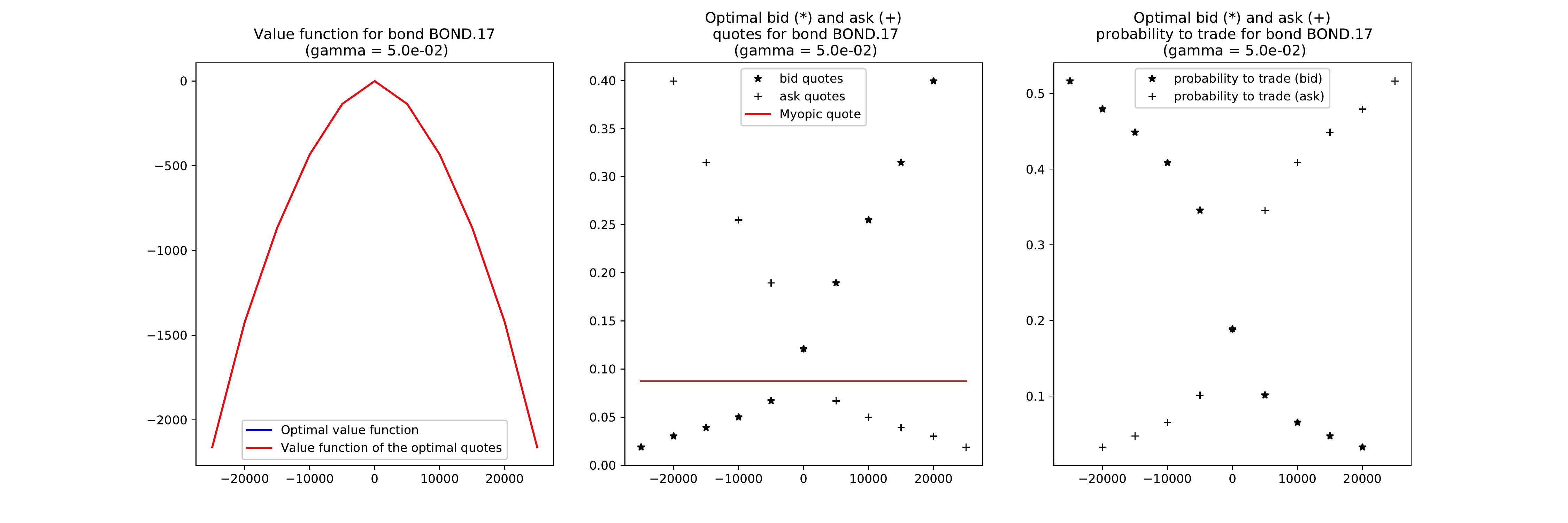}\\
  \includegraphics[width=0.88\textwidth]{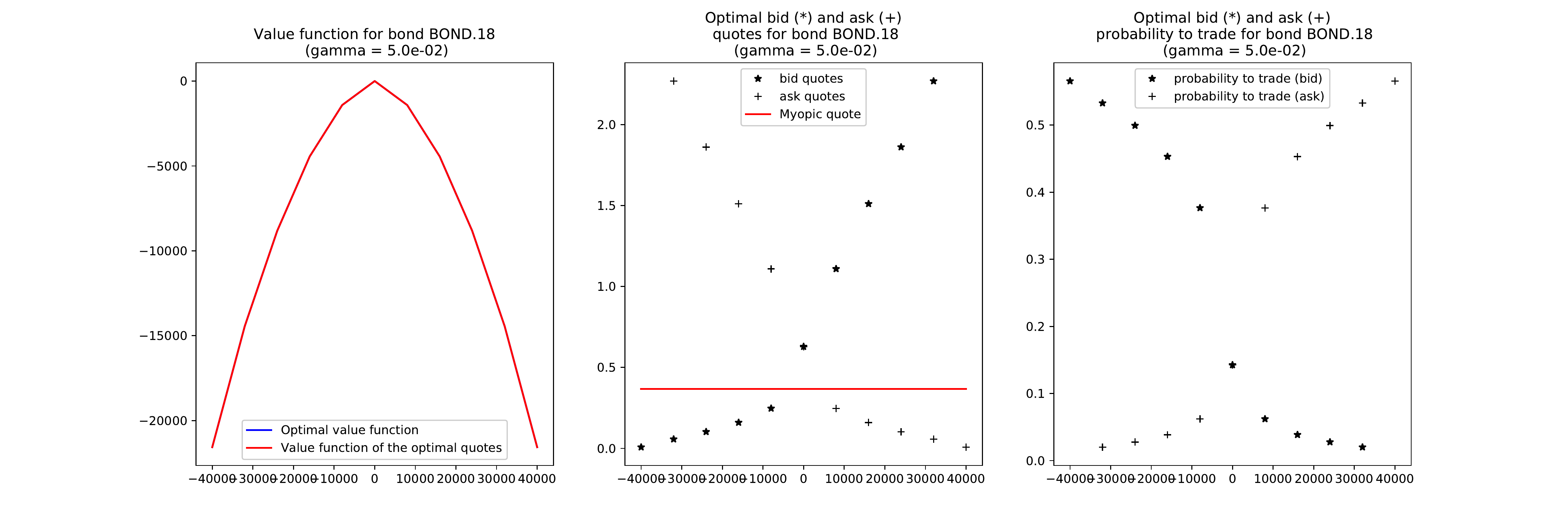}\\
  \includegraphics[width=0.88\textwidth]{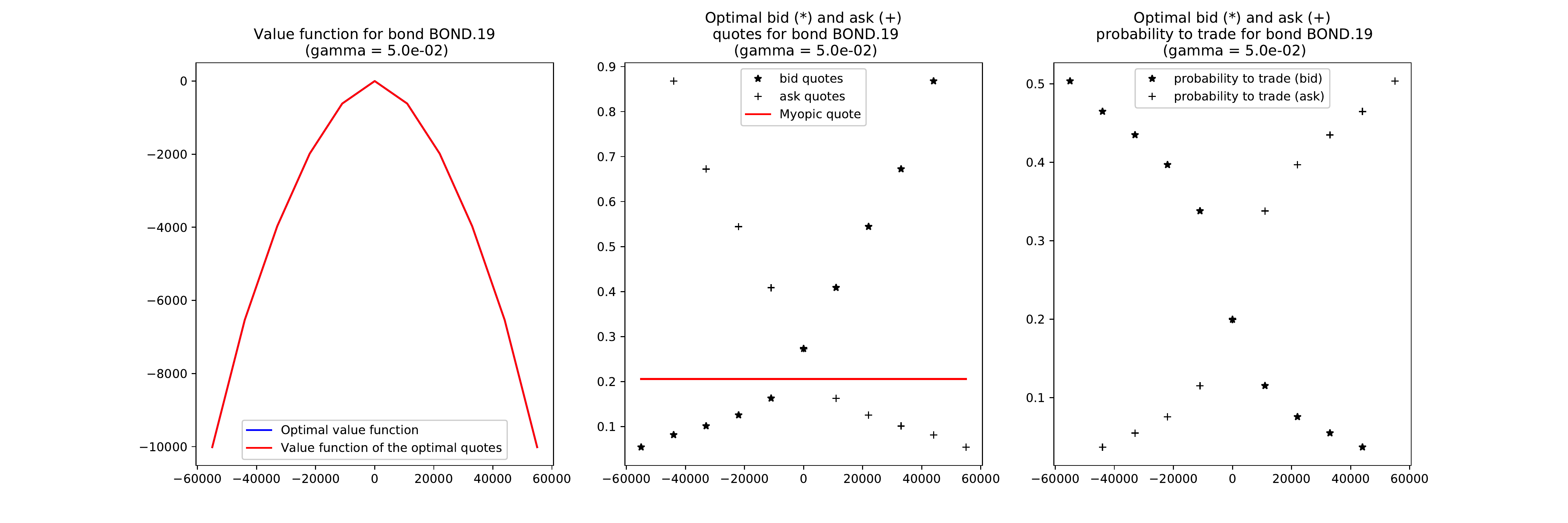}\\
  \includegraphics[width=0.88\textwidth]{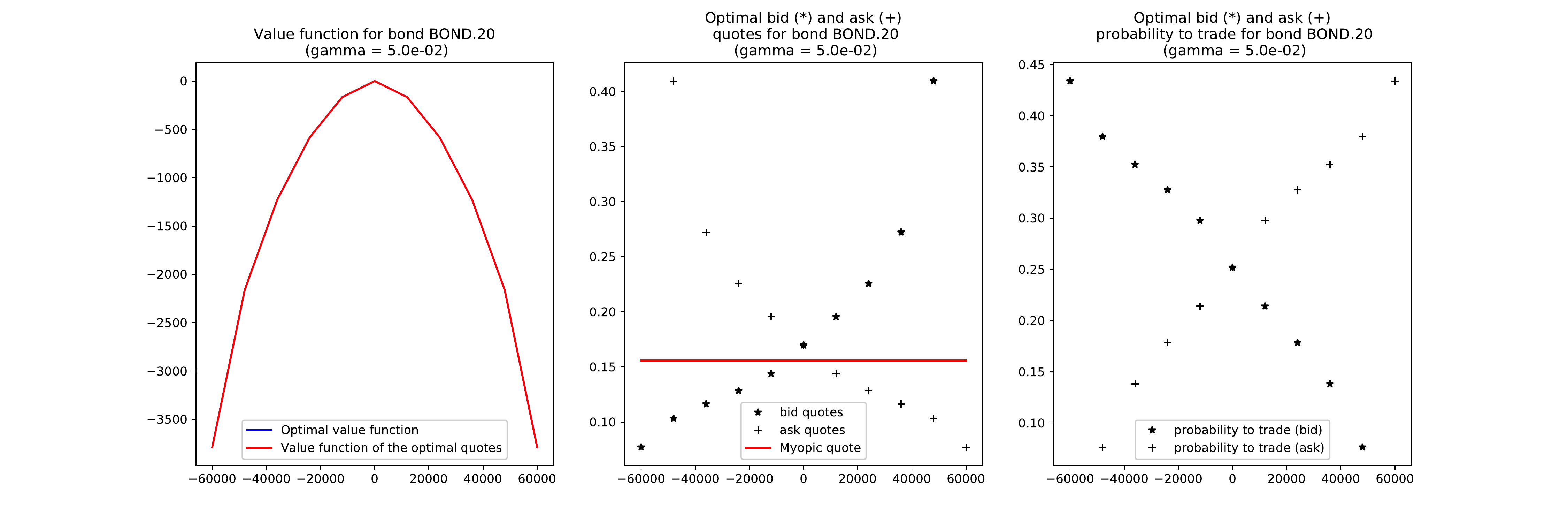}
  \caption{Value functions, optimal quotes, and optimal probabilities to trade with the finite difference approach.}
  \label{pde_sqrt_5}
\end{figure}

For our RL algorithm, we considered again $\gamma = 5\cdot10^{-2}$ and $r = 10^{-4}$. Regarding risk limits, it is noteworthy that in the single-bond case, we did not use our reverse Matryoshka dolls principle and considered risk limits equal to 5 times the RFQ size. For both the critic and the actor, we considered neural networks with 2 hidden layers and 10 nodes in each of these layers with ReLU activation functions. The final layer of each neural network contains one node and the activation function is affine in the case of the critic and sigmoid in the case of the actor. This is of course a large network compared to the task to be carried out but there is no problem of overfitting here. As far as pre-training is concerned, we used myopic quotes as described in Section \ref{algo}. Regarding the learning phase, for most bonds we considered 50 steps of the algorithm, i.e. 50 steps of TD learning and 50 steps of policy improvement. For 5 bonds we decided to increase this number of steps to either 100 or 200. At each step, we carried out 1 rollout of length 10000 starting from a zero inventory and 100 additional rollouts of length 100 starting from a random inventory. The noise $\epsilon$ in each rollout is distributed uniformly in $[-0.05, 0.05]$ and we chose the probability limit $\nu = 0.005$. The learning rate for the critic is $\eta = 5\cdot 10^{-8}$ and we used mini-batches of size $70$. The learning rate for the actor is $\tilde{\eta} = 0.01$ and we used mini-batches of size $50$.\\

\begin{figure}[H]
  \centering
  \includegraphics[width=0.8\textwidth]{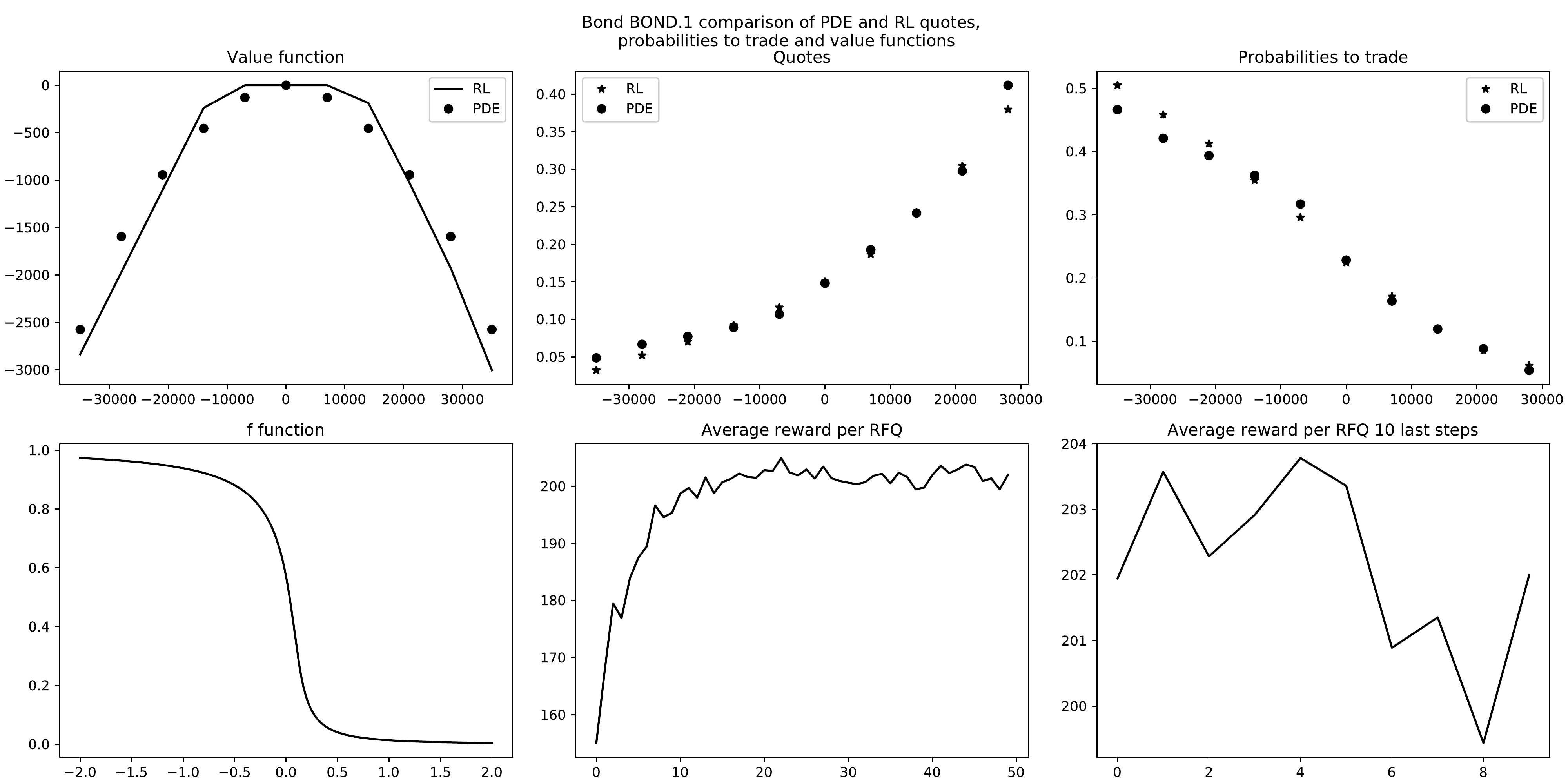}\\
  \caption{Comparison between the two numerical methods.}
  \label{comp_sqrt_1}
\end{figure}

\begin{figure}[H]
  \centering
   \includegraphics[width=0.8\textwidth]{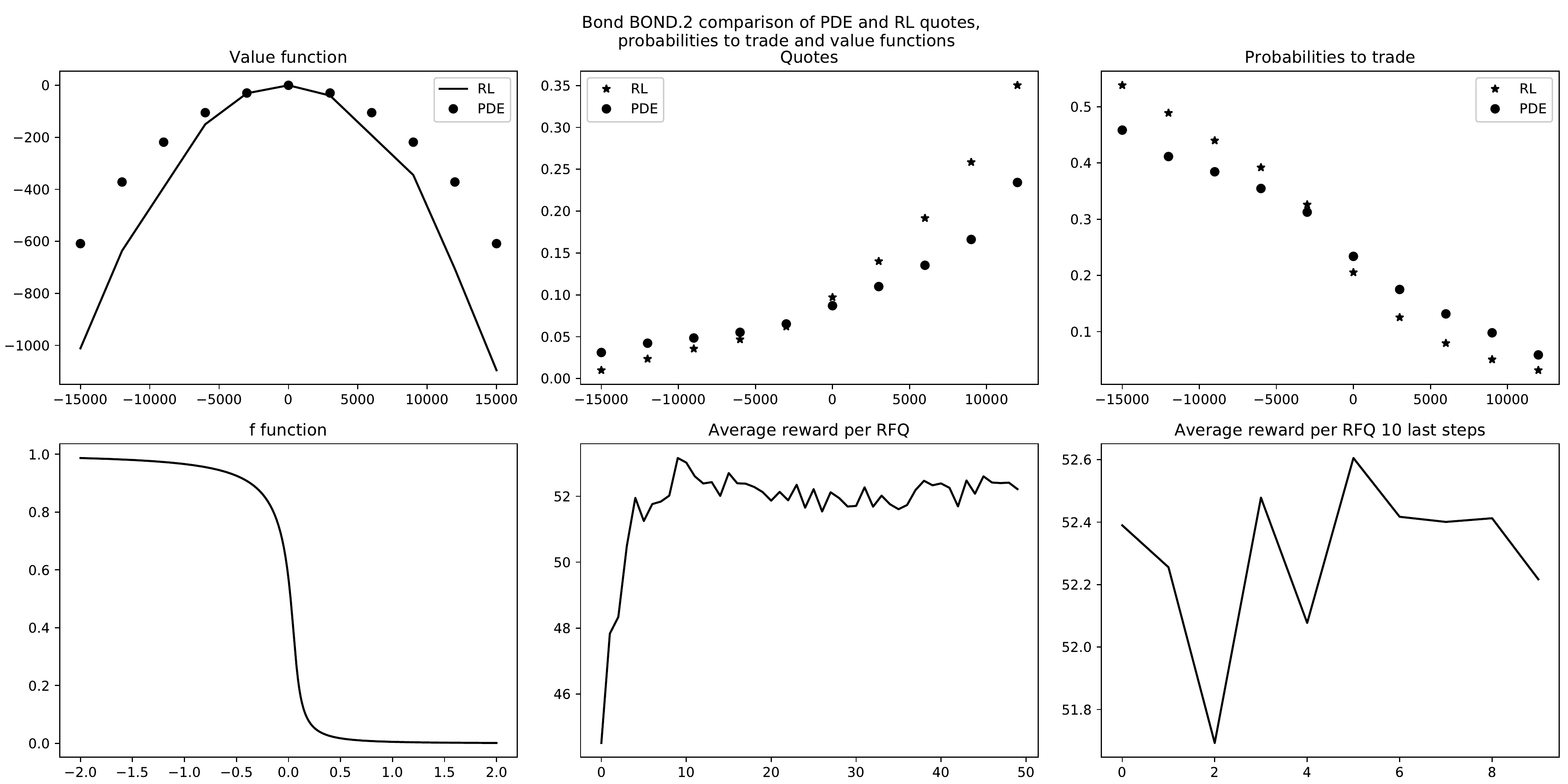}\\
  \includegraphics[width=0.8\textwidth]{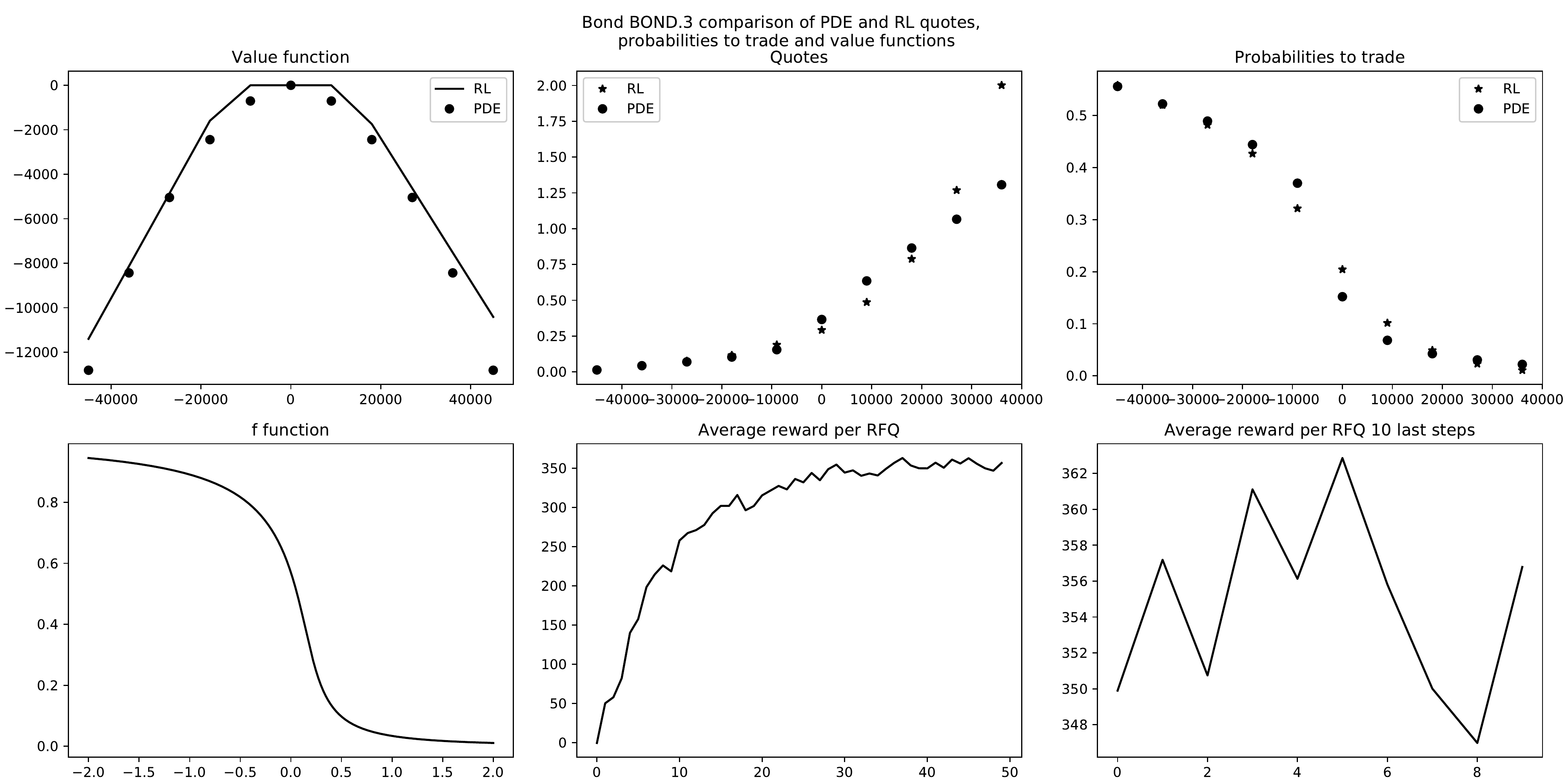}\\
  \includegraphics[width=0.8\textwidth]{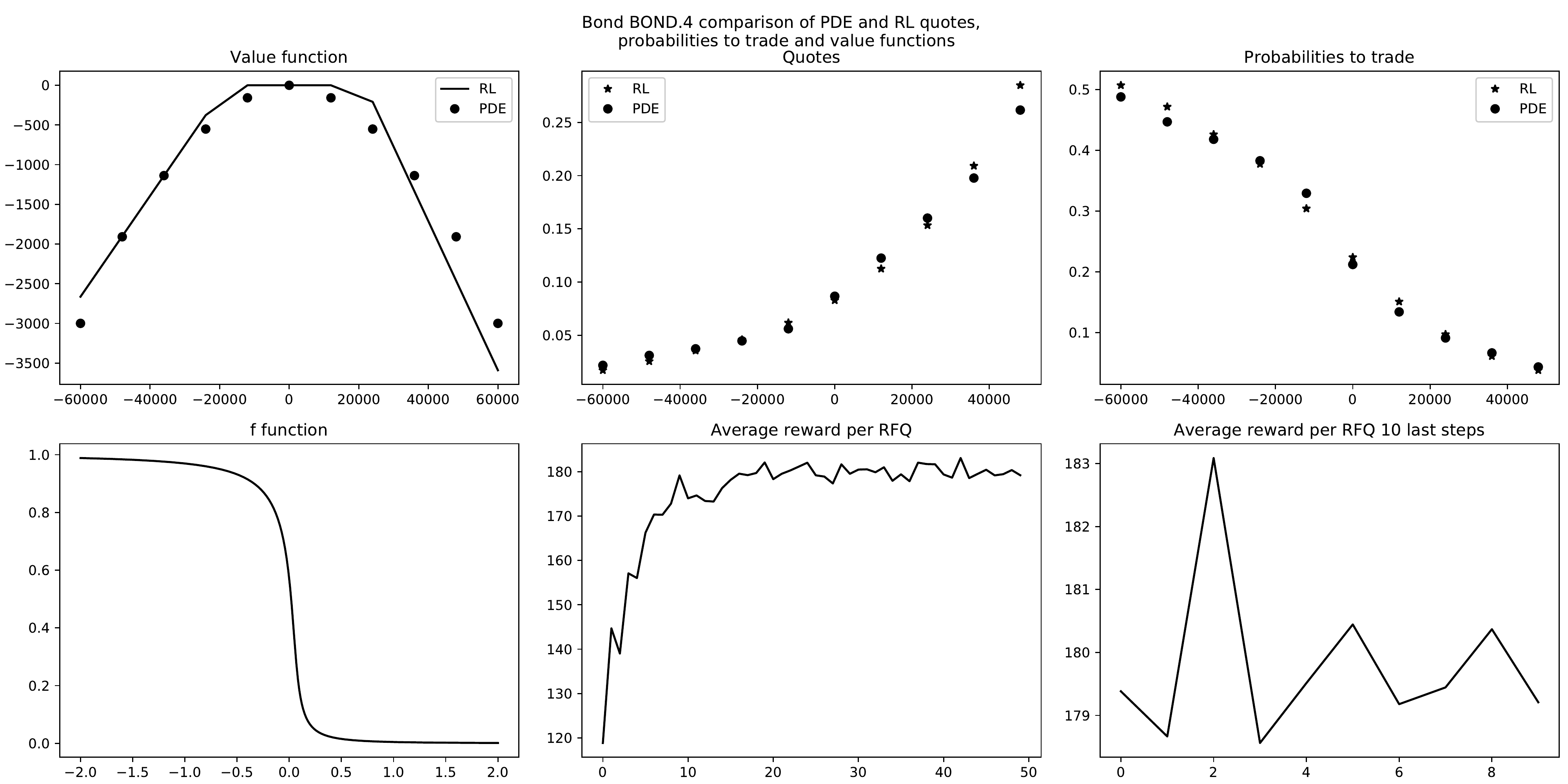}\\
  \caption{Comparison between the two numerical methods.}
  \label{comp_sqrt_2}
\end{figure}

\begin{figure}[H]
  \centering
    \includegraphics[width=0.8\textwidth]{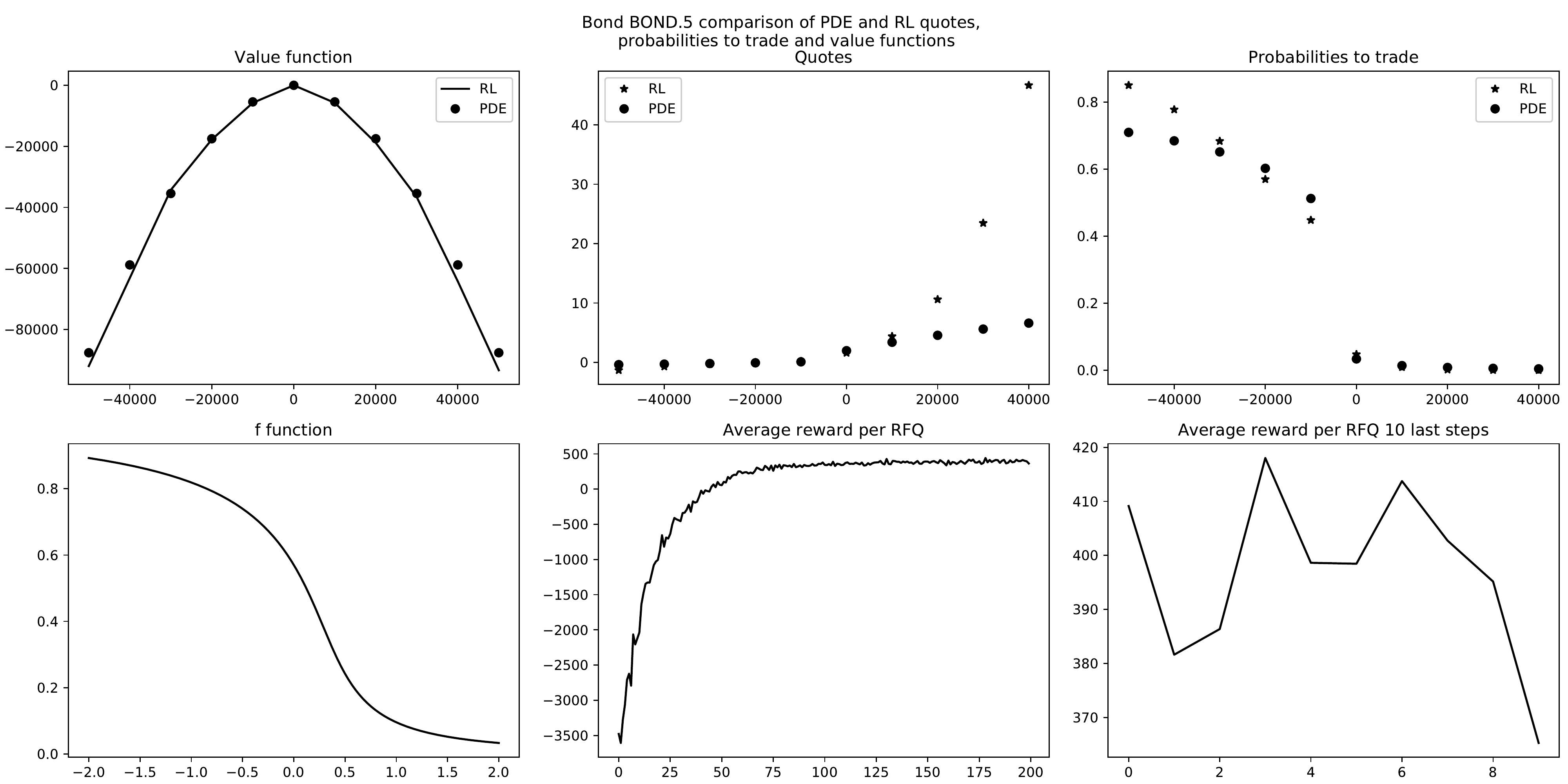}\\
  \includegraphics[width=0.8\textwidth]{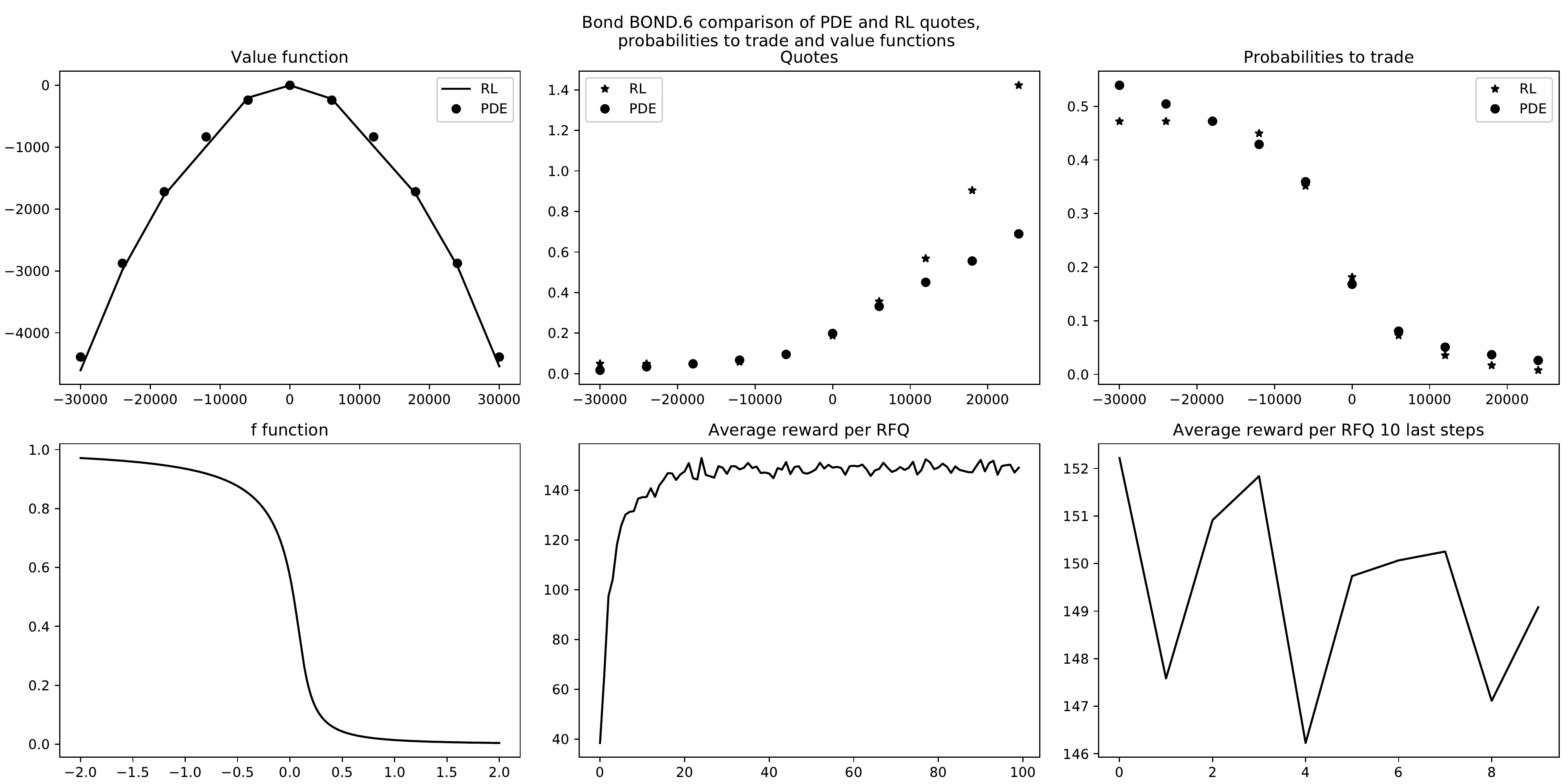}\\
  \includegraphics[width=0.8\textwidth]{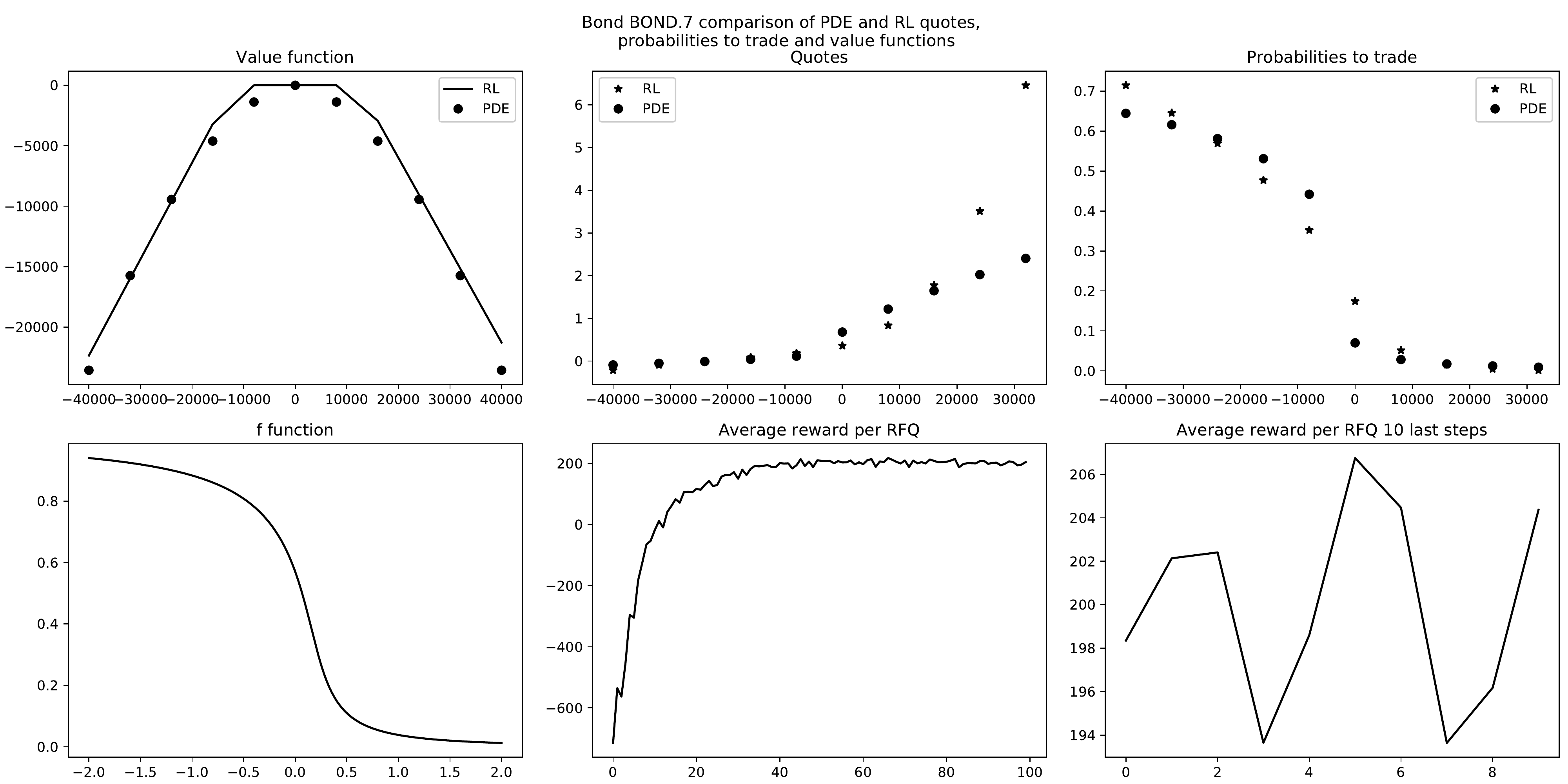}\\
  \caption{Comparison between the two numerical methods.}
  \label{comp_sqrt_3}
\end{figure}

\begin{figure}[H]
  \centering
    \includegraphics[width=0.8\textwidth]{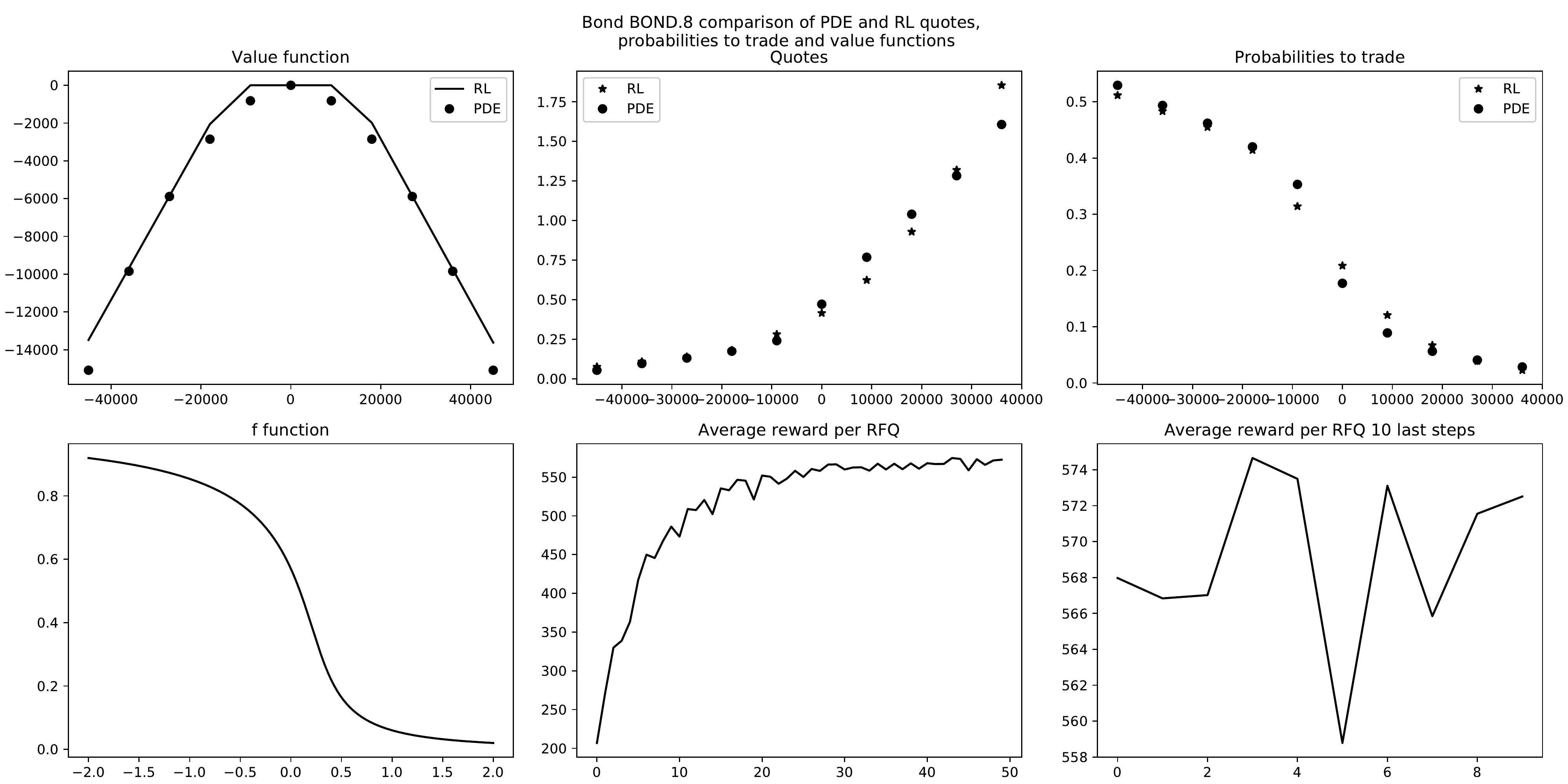}\\
  \includegraphics[width=0.8\textwidth]{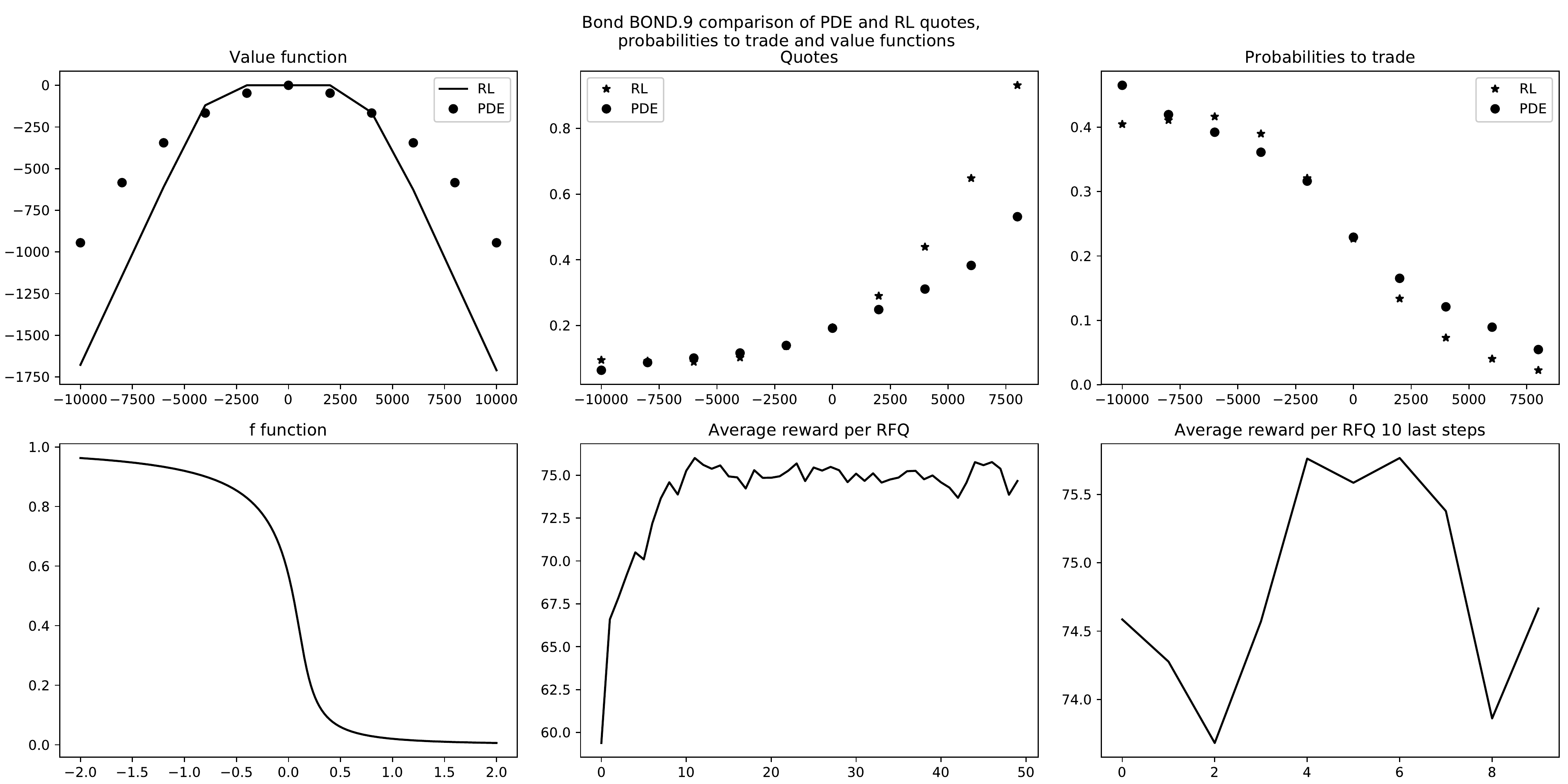}\\
  \includegraphics[width=0.8\textwidth]{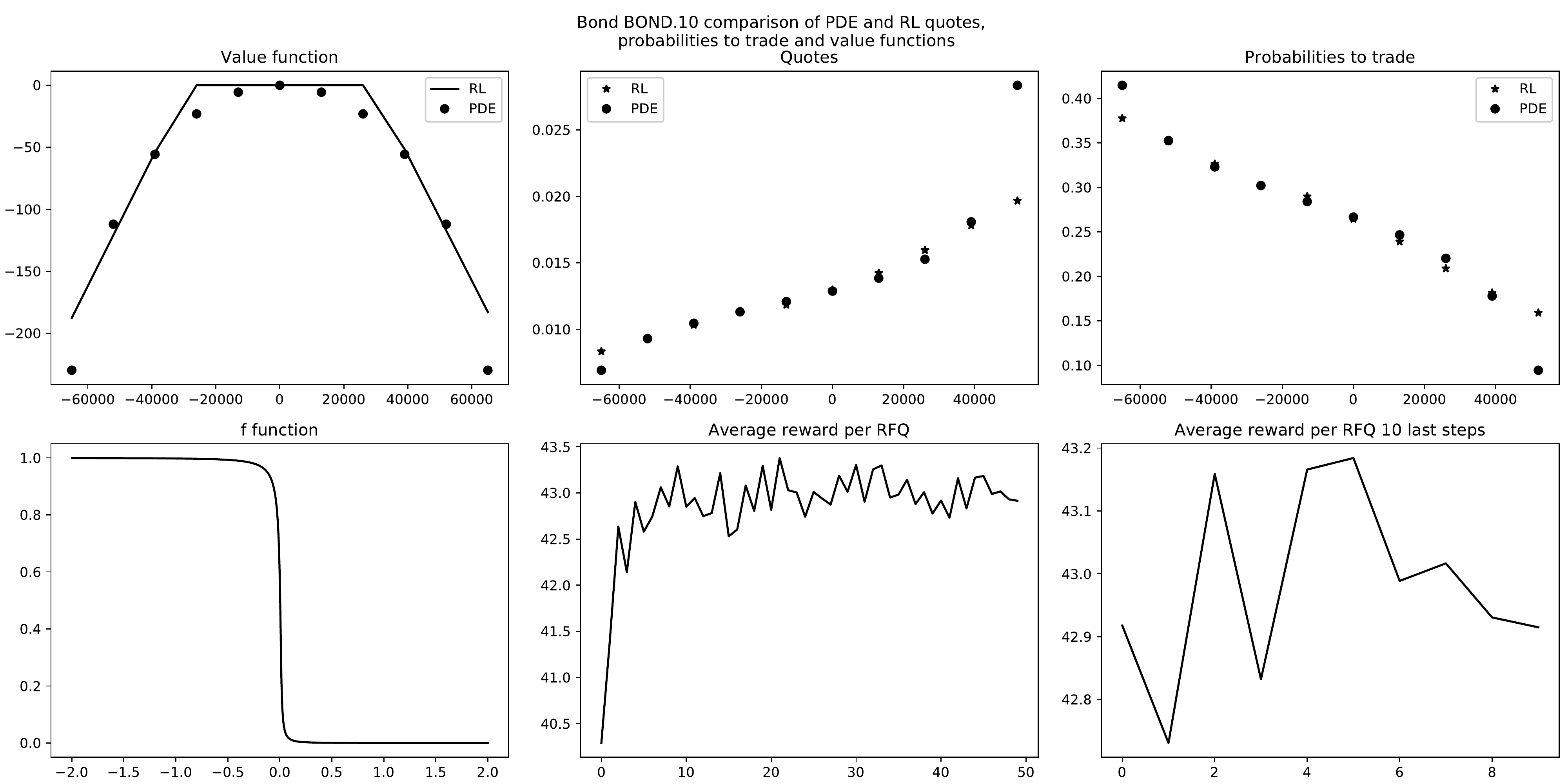}\\
  \caption{Comparison between the two numerical methods.}
  \label{comp_sqrt_4}
\end{figure}

\begin{figure}[H]
  \centering
    \includegraphics[width=0.8\textwidth]{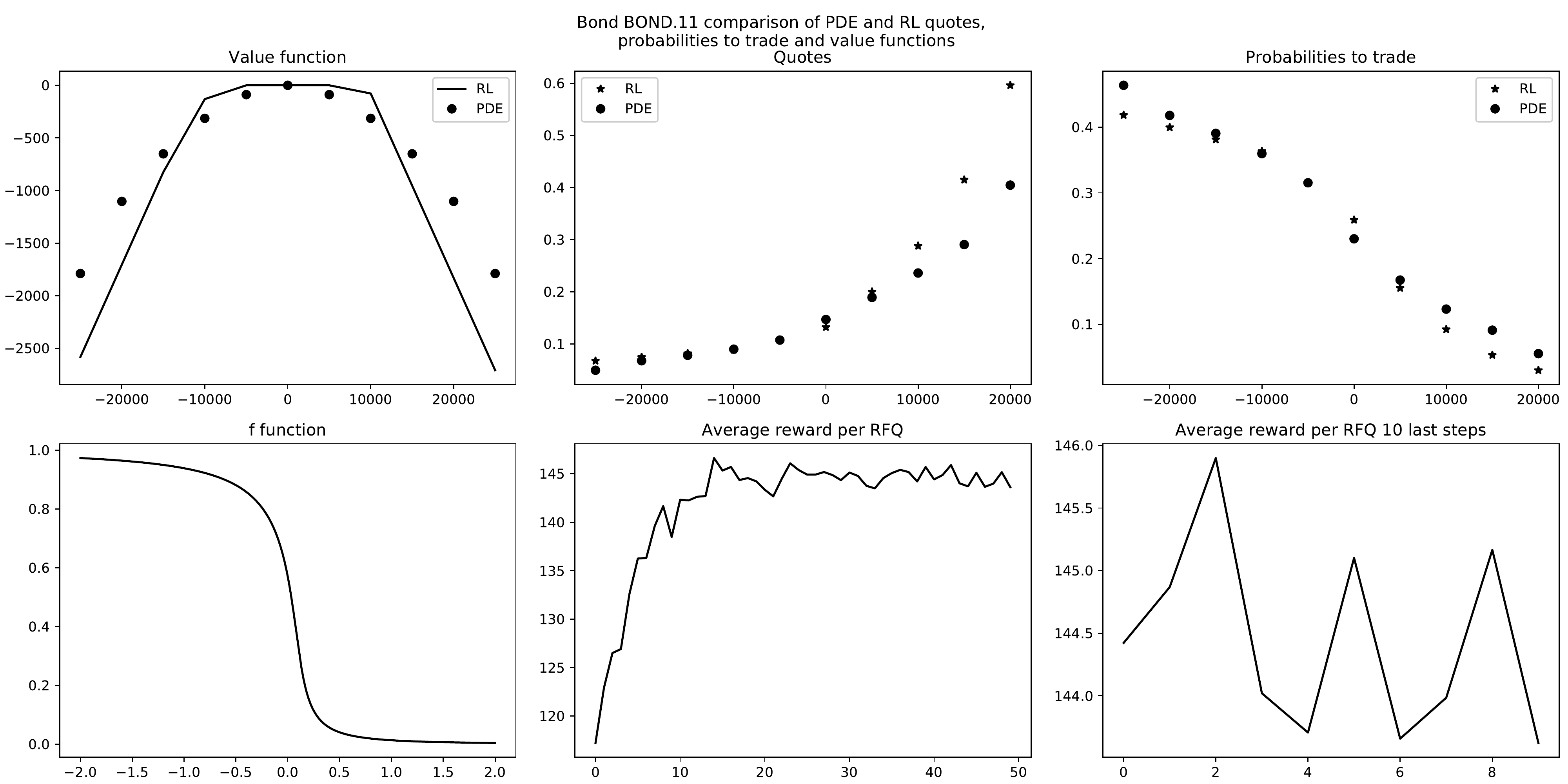}\\
  \includegraphics[width=0.8\textwidth]{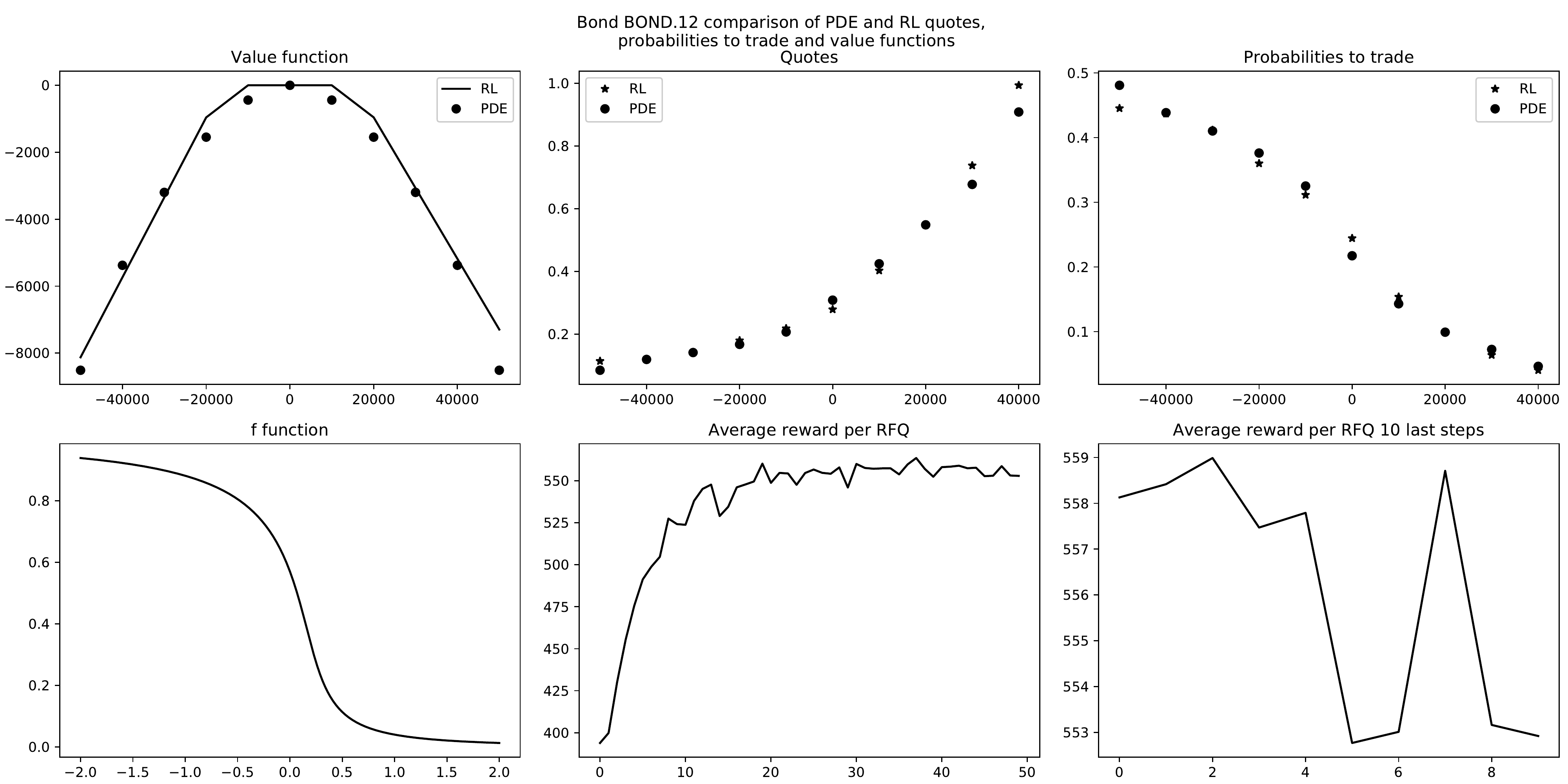}\\
  \includegraphics[width=0.8\textwidth]{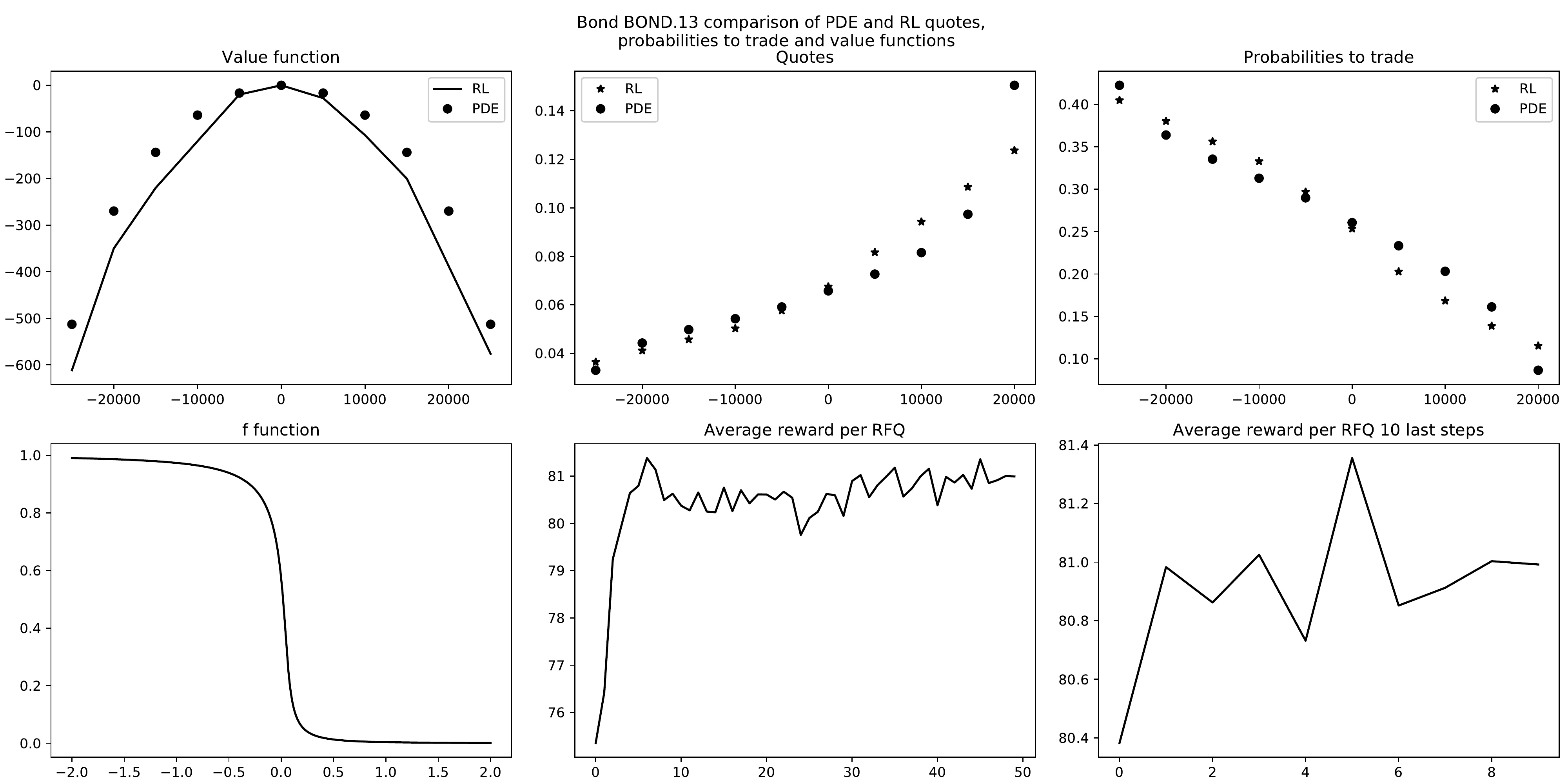}\\
  \caption{Comparison between the two numerical methods.}
  \label{comp_sqrt_5}
\end{figure}

\begin{figure}[H]
  \centering
    \includegraphics[width=0.8\textwidth]{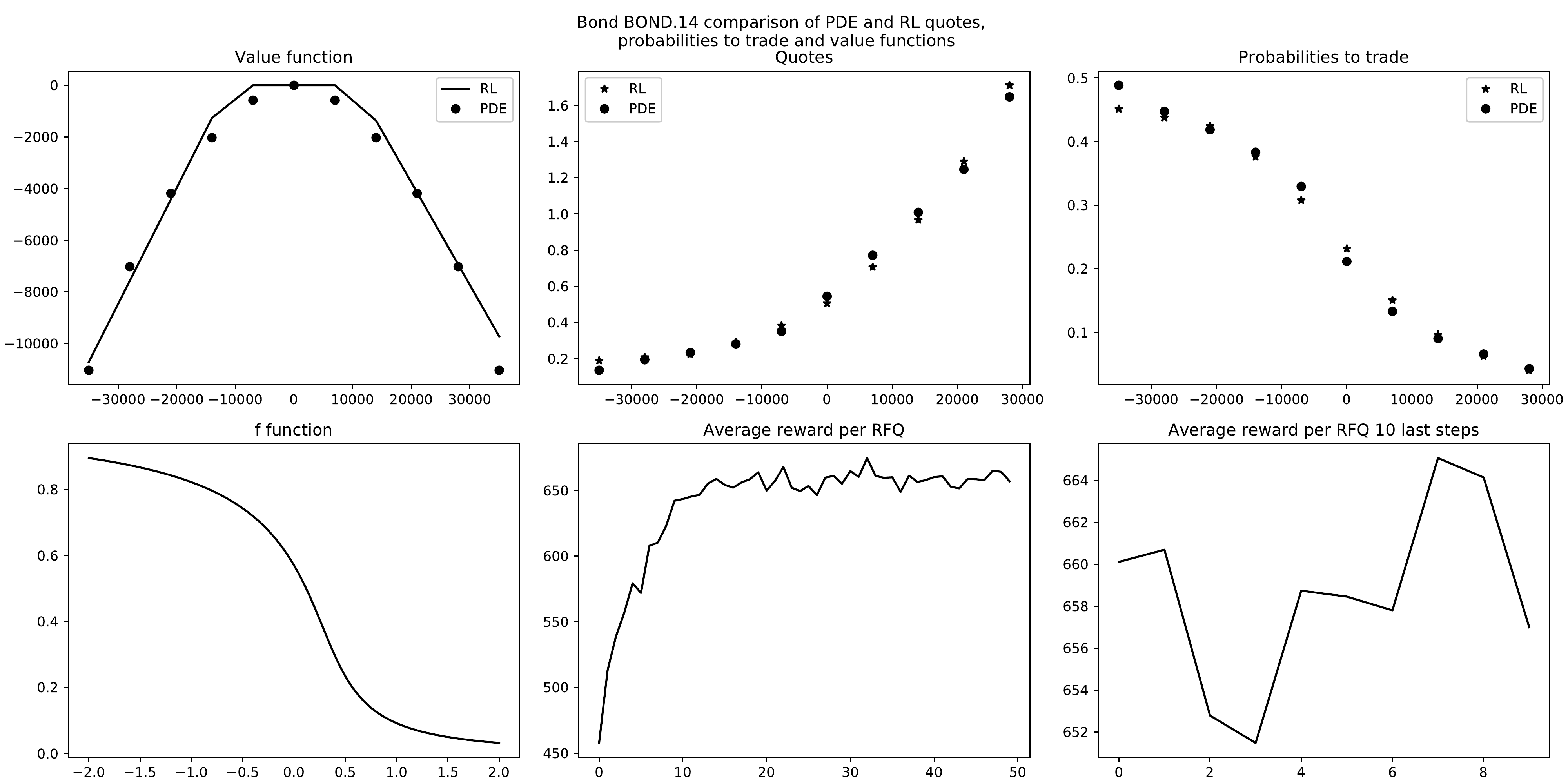}\\
  \includegraphics[width=0.8\textwidth]{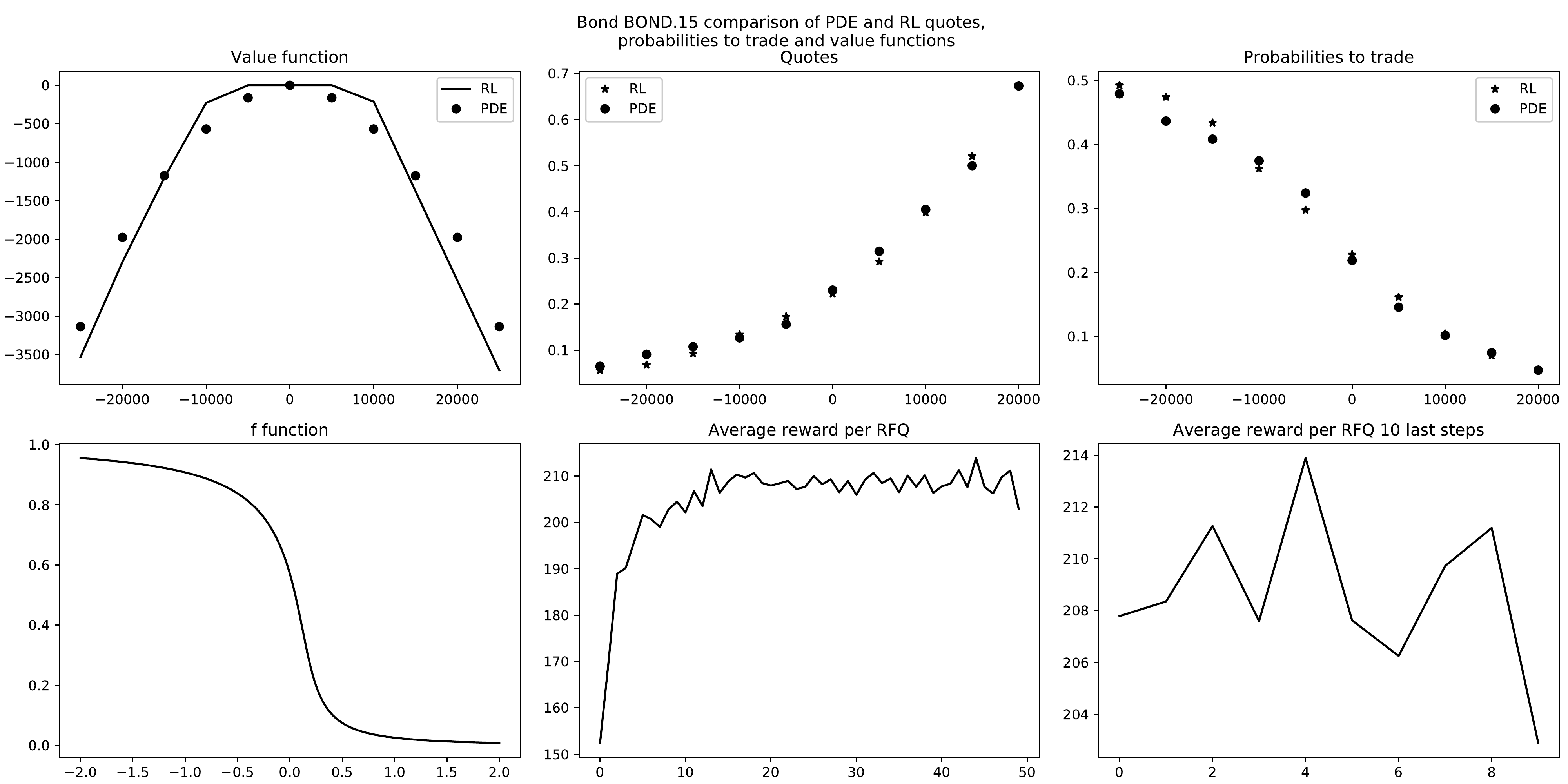}\\
  \includegraphics[width=0.8\textwidth]{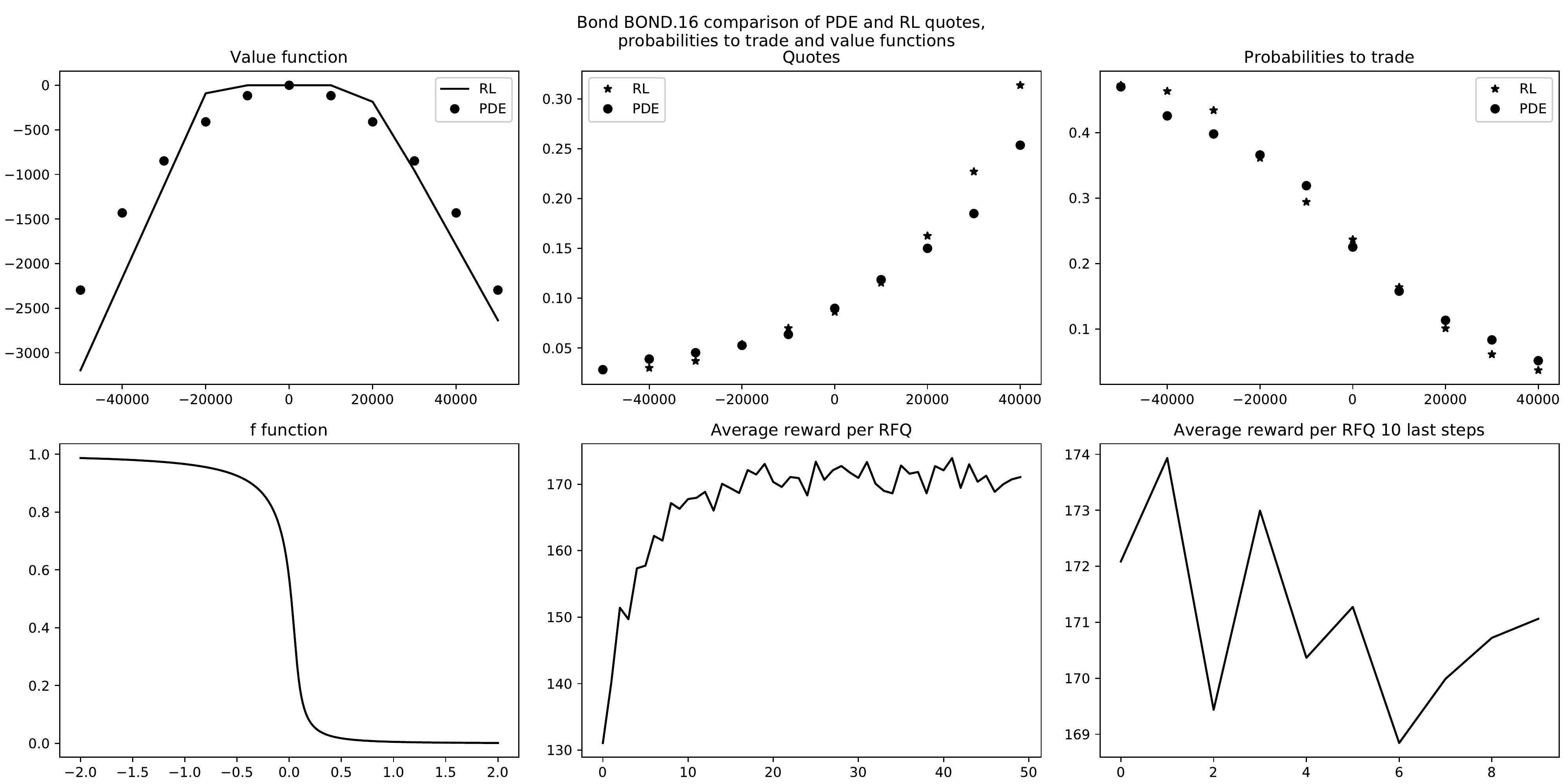}\\
  \caption{Comparison between the two numerical methods.}
  \label{comp_sqrt_6}
\end{figure}

\begin{figure}[H]
  \centering
    \includegraphics[width=0.8\textwidth]{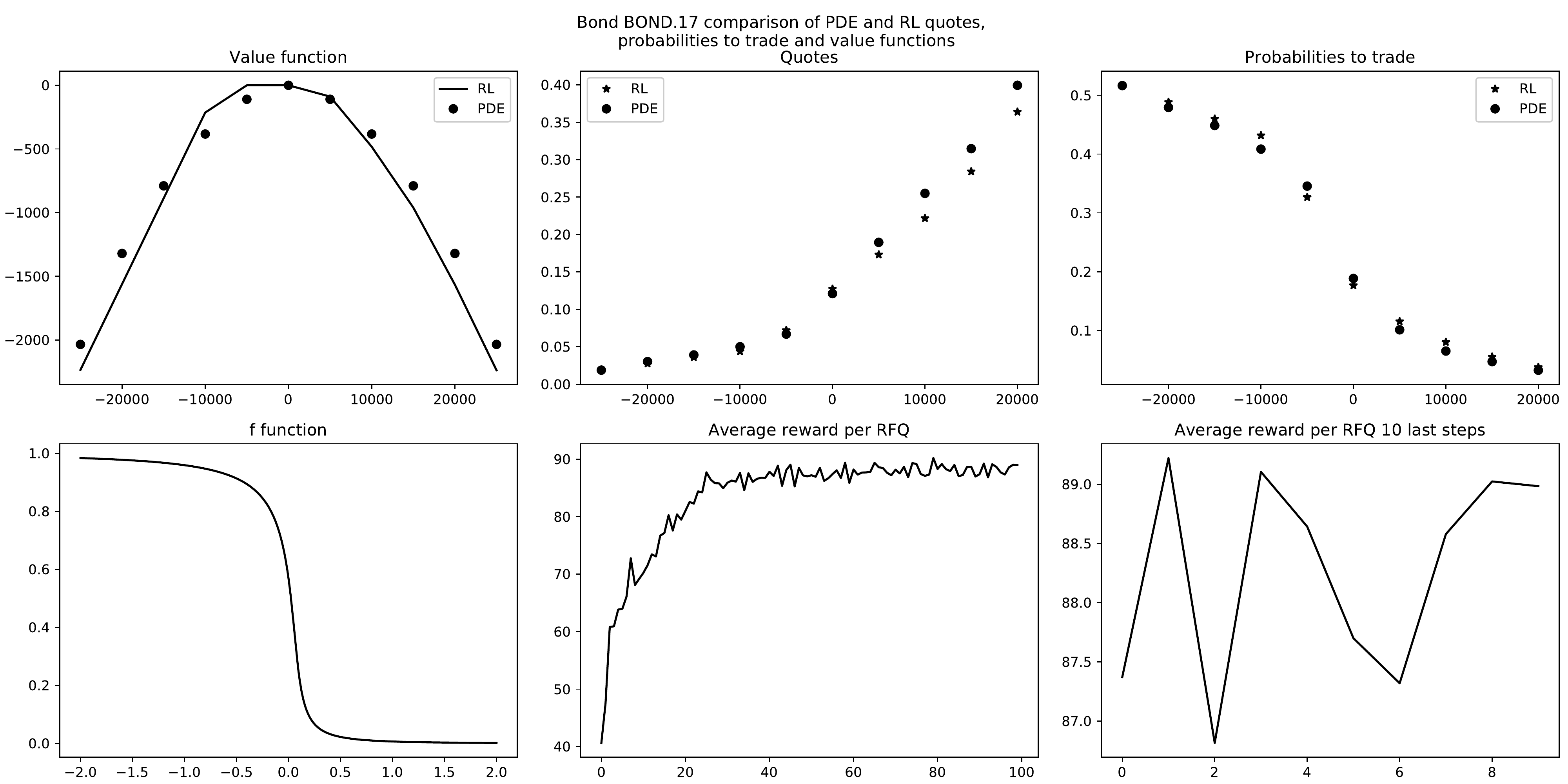}\\
  \includegraphics[width=0.8\textwidth]{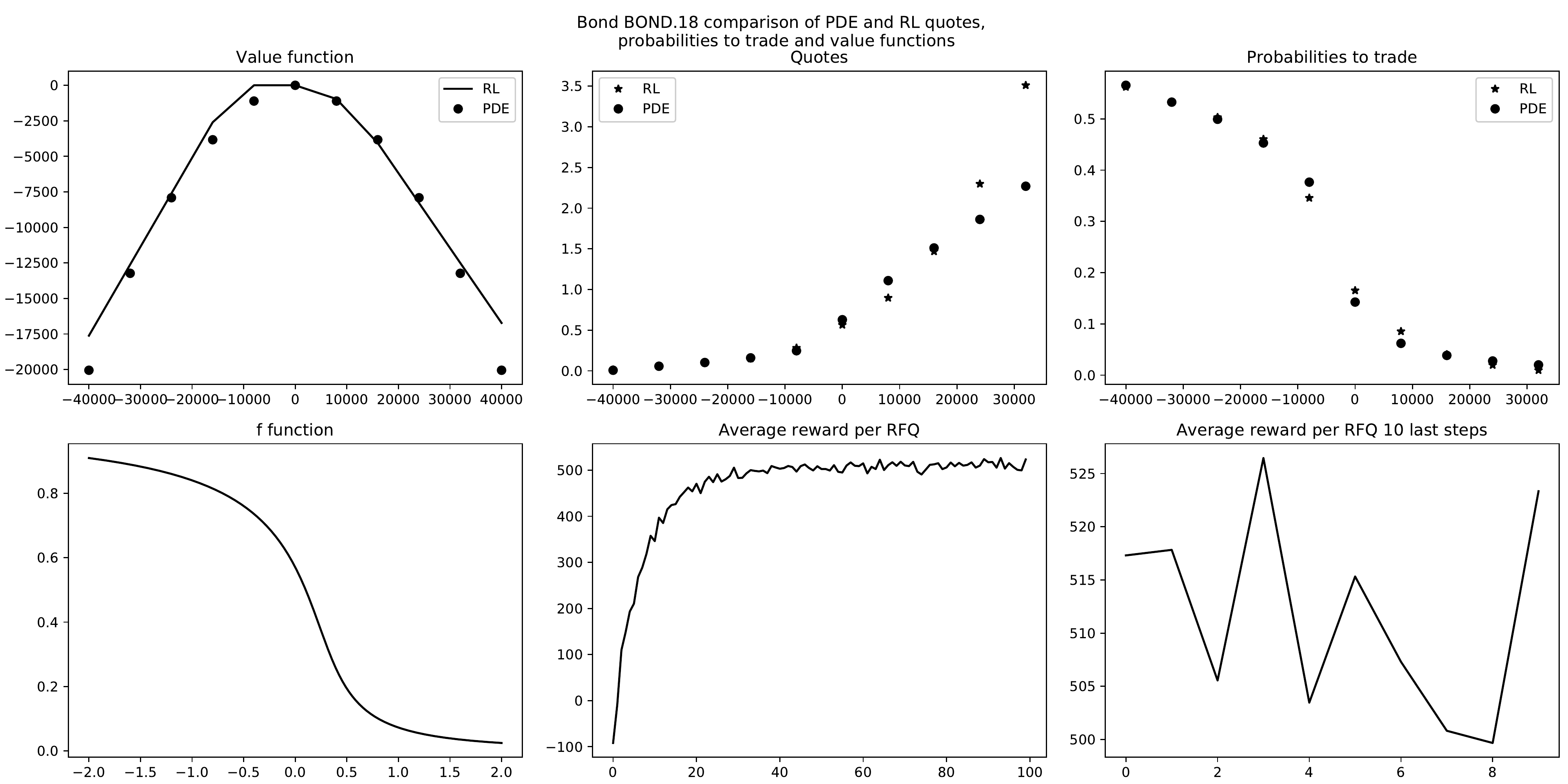}\\
  \includegraphics[width=0.8\textwidth]{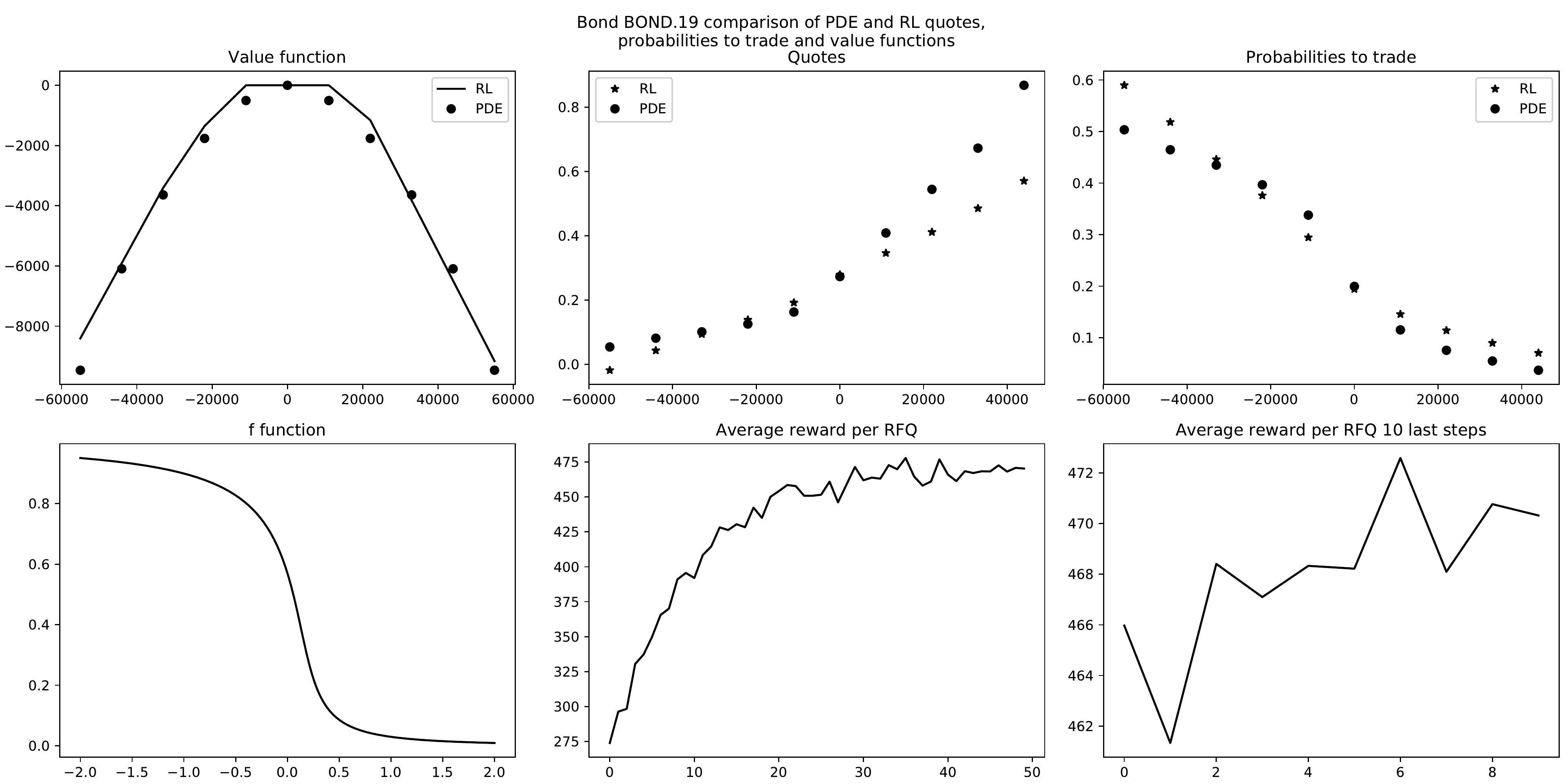}\\
  \caption{Comparison between the two numerical methods.}
  \label{comp_sqrt_7}
\end{figure}

\begin{figure}[H]
  \centering
    \includegraphics[width=0.8\textwidth]{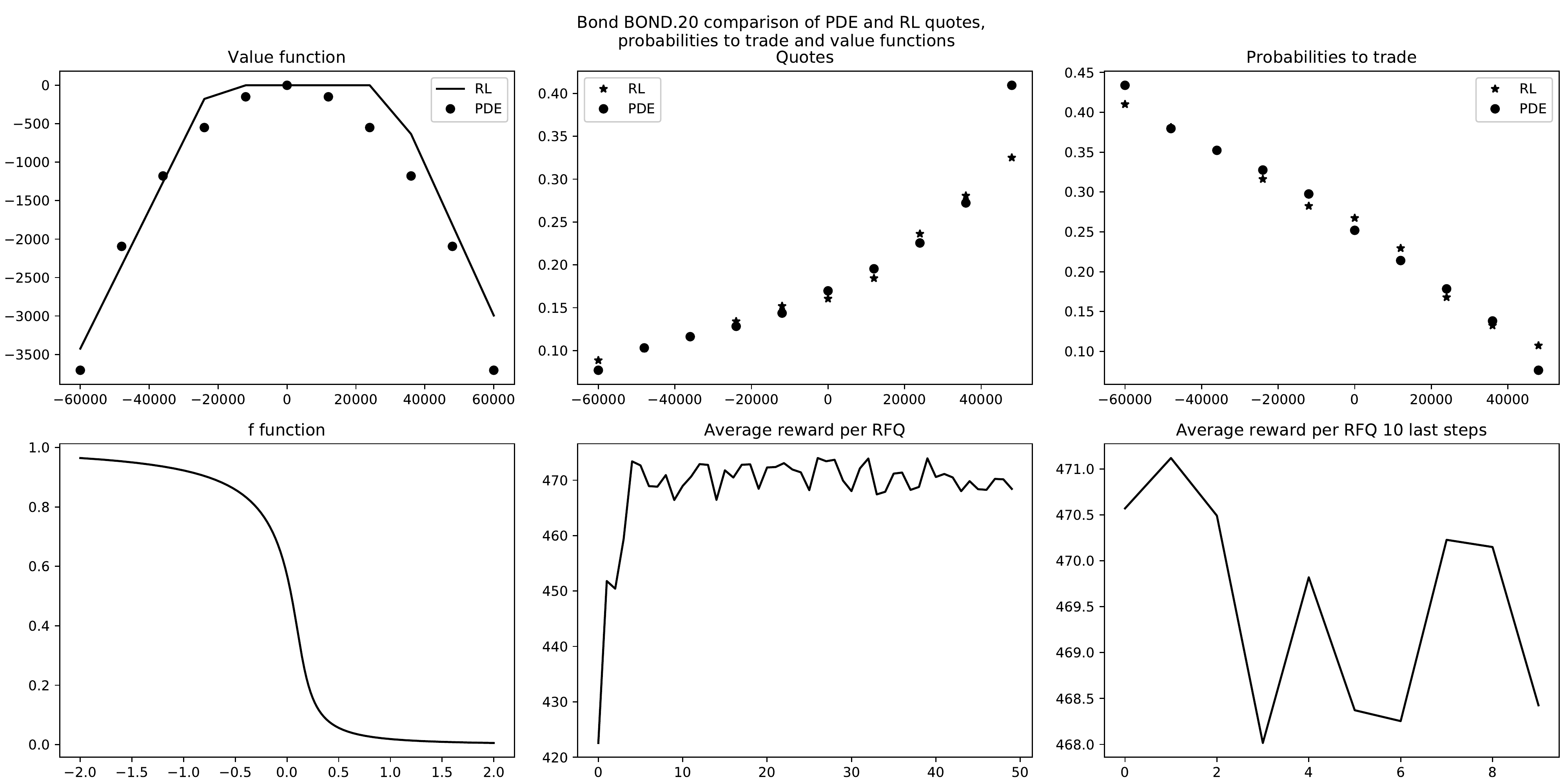}\\
  \caption{Comparison between the two numerical methods.}
  \label{comp_sqrt_8}
\end{figure}

The results are shown in Figures \ref{comp_sqrt_1}, \ref{comp_sqrt_2}, \ref{comp_sqrt_3}, \ref{comp_sqrt_4}, \ref{comp_sqrt_5}, \ref{comp_sqrt_6}, \ref{comp_sqrt_7}, and \ref{comp_sqrt_8} along with the comparison between the PDE method and our RL algorithm.\footnote{It is noteworthy that the value function obtained with the finite difference scheme is an approximation of $\tilde{\theta}_r^*$ and we had therefore to use Eq. \eqref{theta12} to be able to plot $\theta_r^*$ and only then carry out the comparison to the value function obtained with TD learning.}

As it can be easily seen, the results with the two methods are often very close in terms of probability to trade (though sometimes not in terms of quotes) and they are always comparable when it comes to the average reward per RFQ (see the values in the bottom right panel of the figures and the values in Table~4). In particular, in terms of performance, our algorithm is competitive for the single-bond case. The question is now to see how it scales to the multi-bond case.

\subsubsection{2-bond cases}

In order to validate our RL approach in the multi-dimensional case, we first consider several cases with two bonds. With two bonds, we can indeed compute the optimal quotes with the PDE method and the RL approach and compare the results as we did in the single-bond case. In particular, in the 2-bond case, we can see the influence of the correlation between bond prices through the shape of the value function and that of the optimal quote functions, and check whether our RL method manages, as with the PDE method, to account for correlation.\\

We first start with the case of BOND.1 and BOND.6. Our choice is motivated by the high correlation ($98\%$) between the price variations of these two bonds.\footnote{This correlation can easily be computed from the figures of Table 3.}\\

For our RL algorithm, we considered $\gamma = 5\cdot10^{-2}$ and $r = 10^{-4}$. We considered risk limits equal to $5$ times the RFQ size and kept them unchanged during learning. For the critic and for the actor, we considered neural networks with $2$ hidden layers and $12$ nodes in each of these layers with ReLU activation functions. Again, the final layer of each neural network contains one node and the activation function is affine in the case of the critic and sigmoid in the case of the actor. For the pre-training phase, we used for each bond the quotes obtained by our RL algorithm in the single-bond case.  For the learning phase we considered $500$ steps, i.e. $500$ steps of TD learning and $500$ steps of policy improvement. At each step we carried out $1$ rollout of length $10000$ starting from a zero inventory and $100$ additional rollouts of length $100$ starting from a random inventory. The noise $\epsilon$ in each rollout is distributed uniformly in $[-0.05, 0.05]$ and we chose the probability limit $\nu = 0.005$. The learning rate for the critic is $\eta=5\cdot10^{-8}$ and we used mini-batches of size $70$. The learning rate for the actor is  $\tilde{\eta} = 0.01$ and we used mini-batches of size $50$. \\

\begin{figure}[H]
  \centering
  \includegraphics[width=0.4\textwidth]{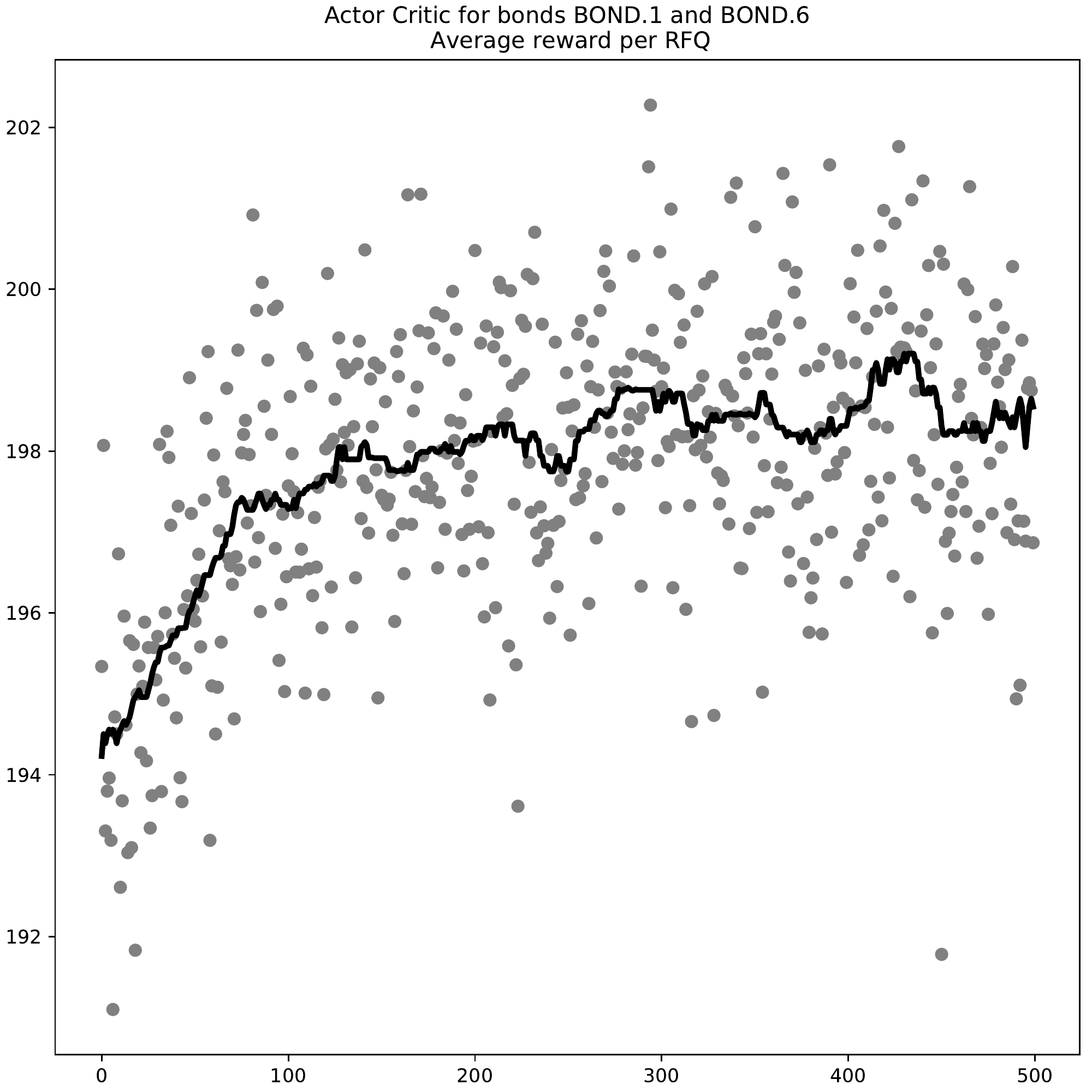}\\
  \caption{Average reward per RFQ -- Learning process for the 2-bond case.}
  \label{rl_2d}
\end{figure}

The learning curve of the algorithm is plotted in Figure \ref{rl_2d}.\footnote{We plotted the moving median over the last 40 points (or less if less points were available).}

We see that the average reward per RFQ increases during the learning phase. This means that our algorithm takes account of the correlation between the two bonds, although the improvement in terms of average reward per RFQ is rather small (from $\simeq194$ to $\simeq198$). Interestingly, the average reward per RFQ obtained when using the optimal quotes computed with the finite difference method is $197.9$ and there is therefore no discrepancy.\\

\begin{figure}[H]
  \centering
  \includegraphics[width=0.48\textwidth]{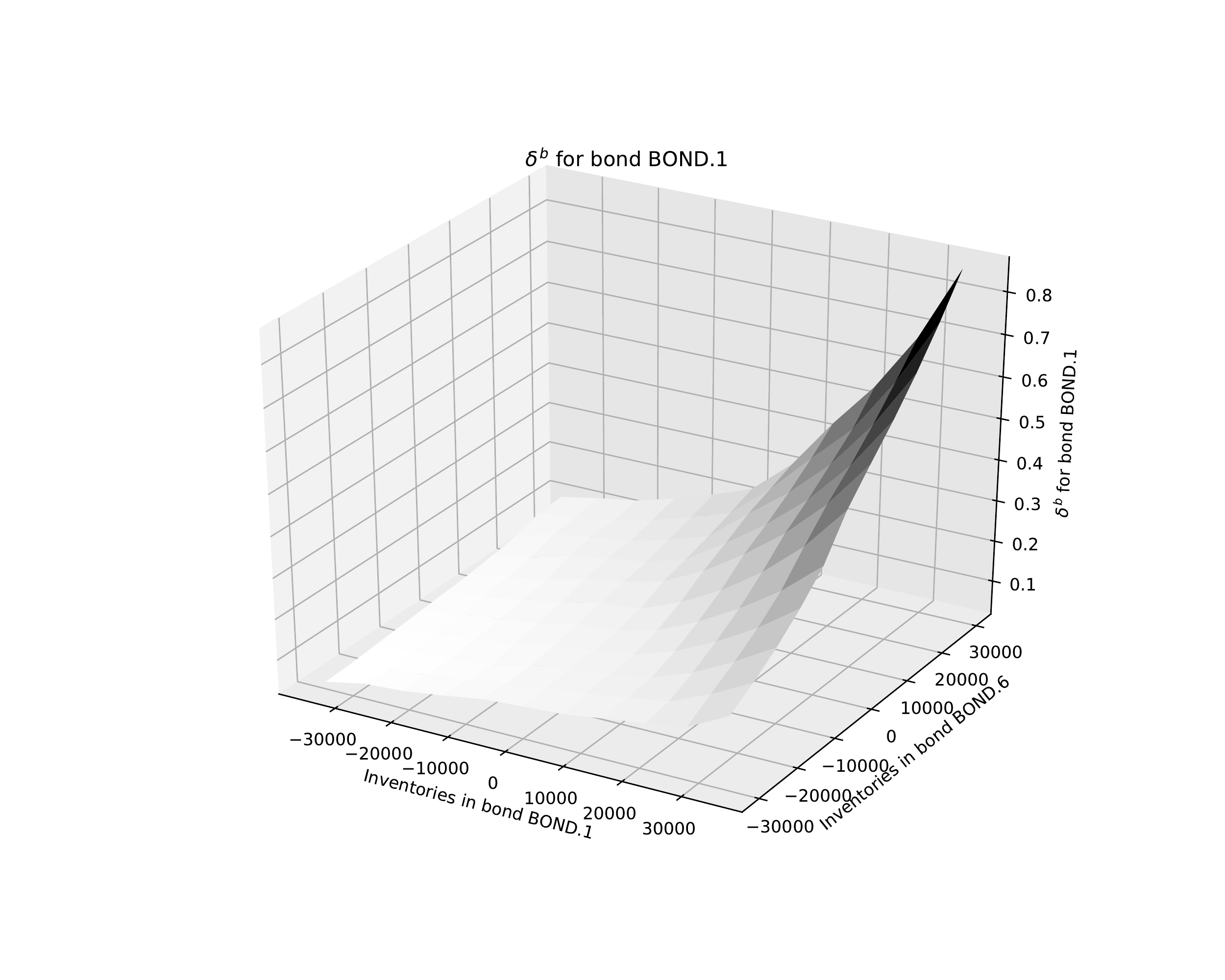}
  \includegraphics[width=0.48\textwidth]{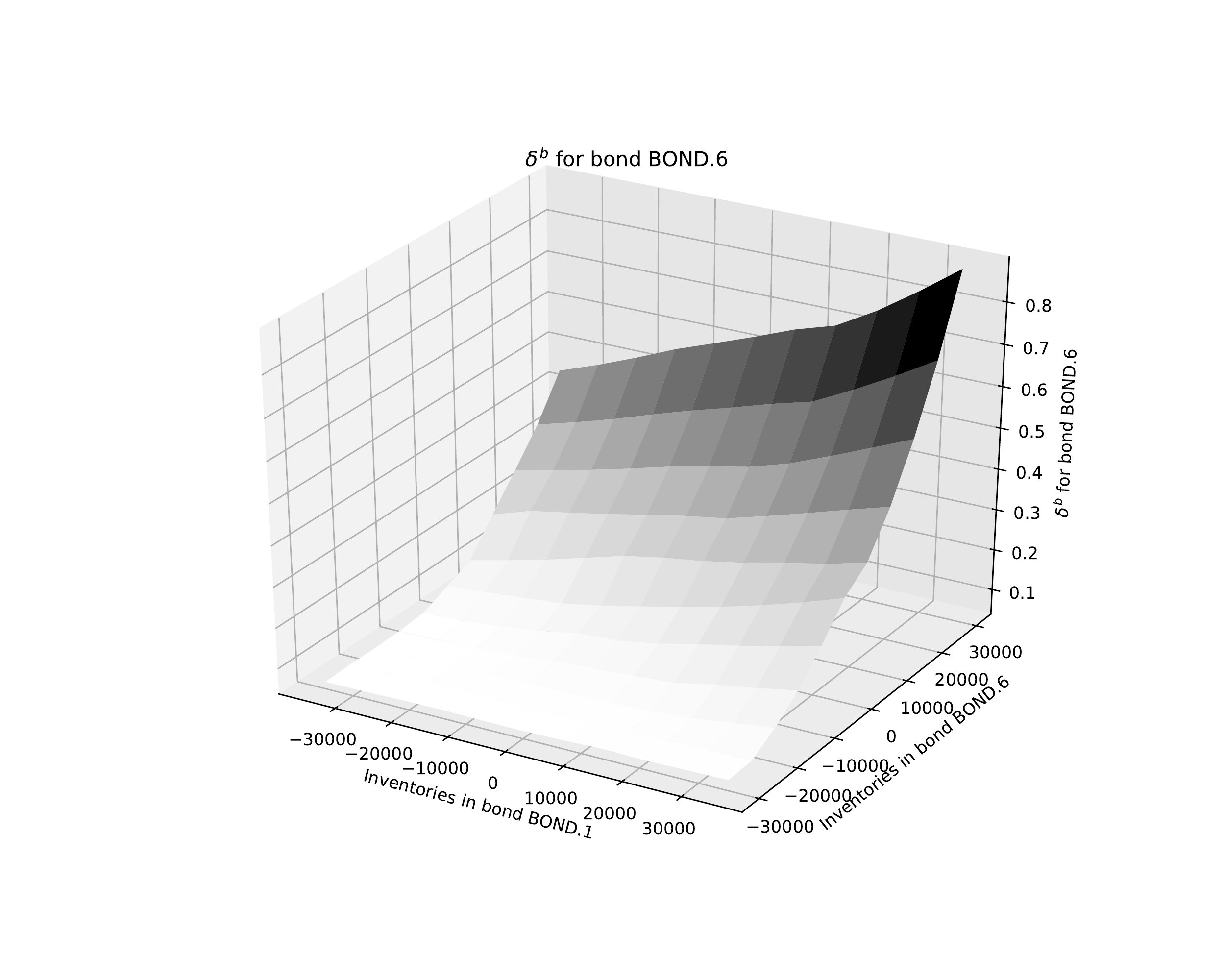}
  \caption{Optimal bid quotes obtained with our RL algorithm.}
  \label{rl_2d_deltas}
\end{figure}

\begin{figure}[H]
  \centering
  \includegraphics[width=0.48\textwidth]{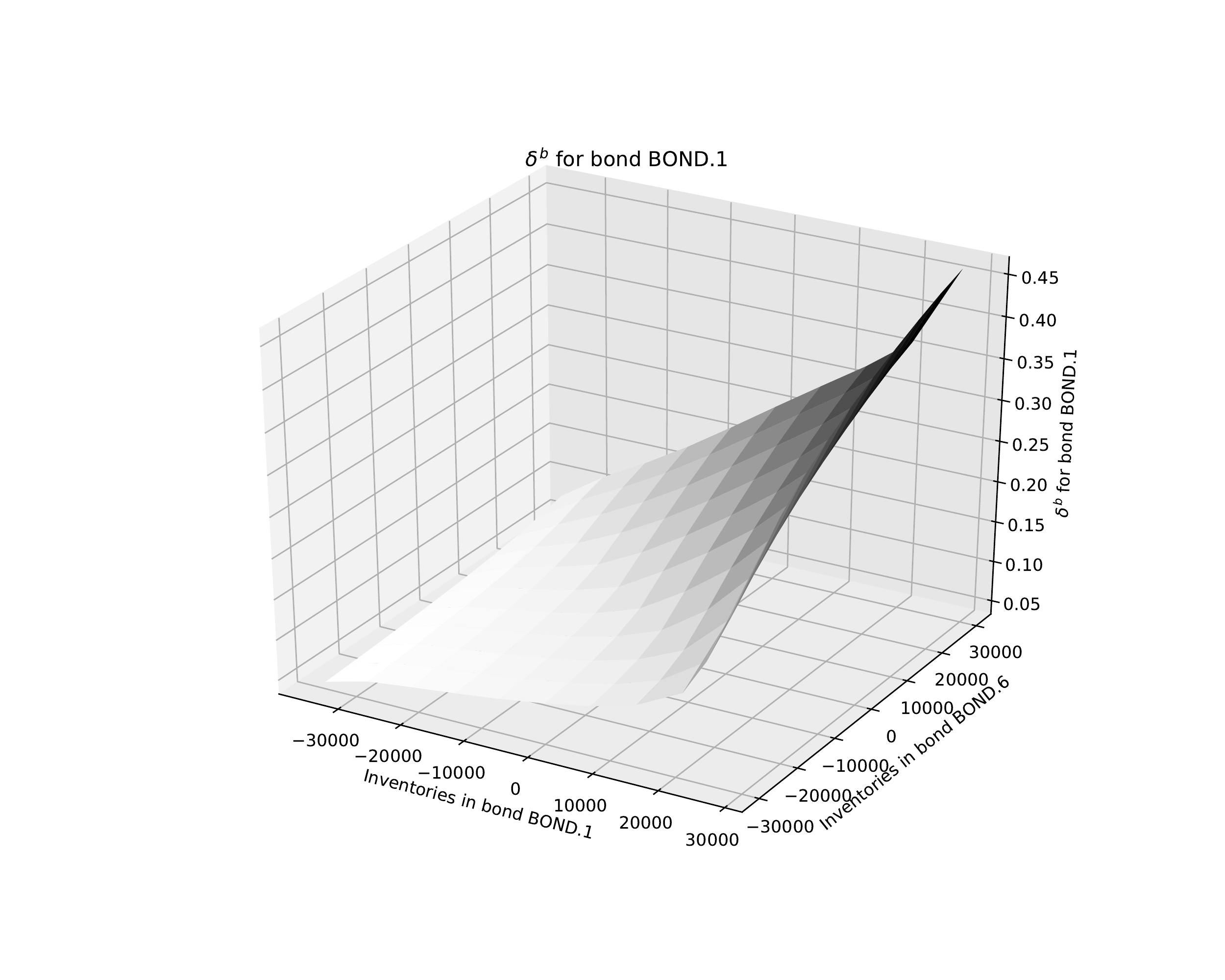}
  \includegraphics[width=0.48\textwidth]{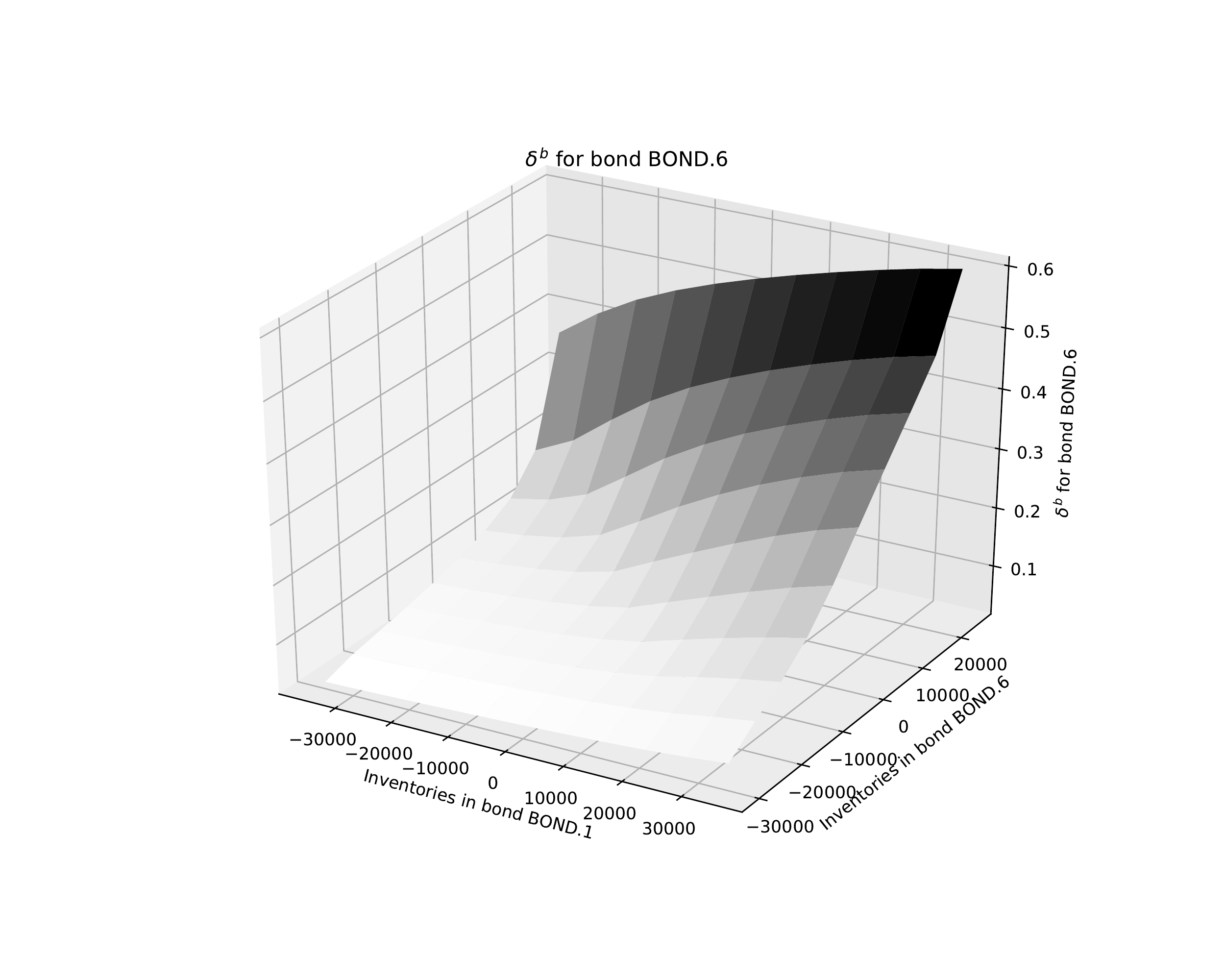}
  \caption{Optimal bid quotes obtained with our finite difference method.}
  \label{th_2d_deltas}
\end{figure}

\begin{figure}[H]
  \centering
  \includegraphics[width=0.48\textwidth]{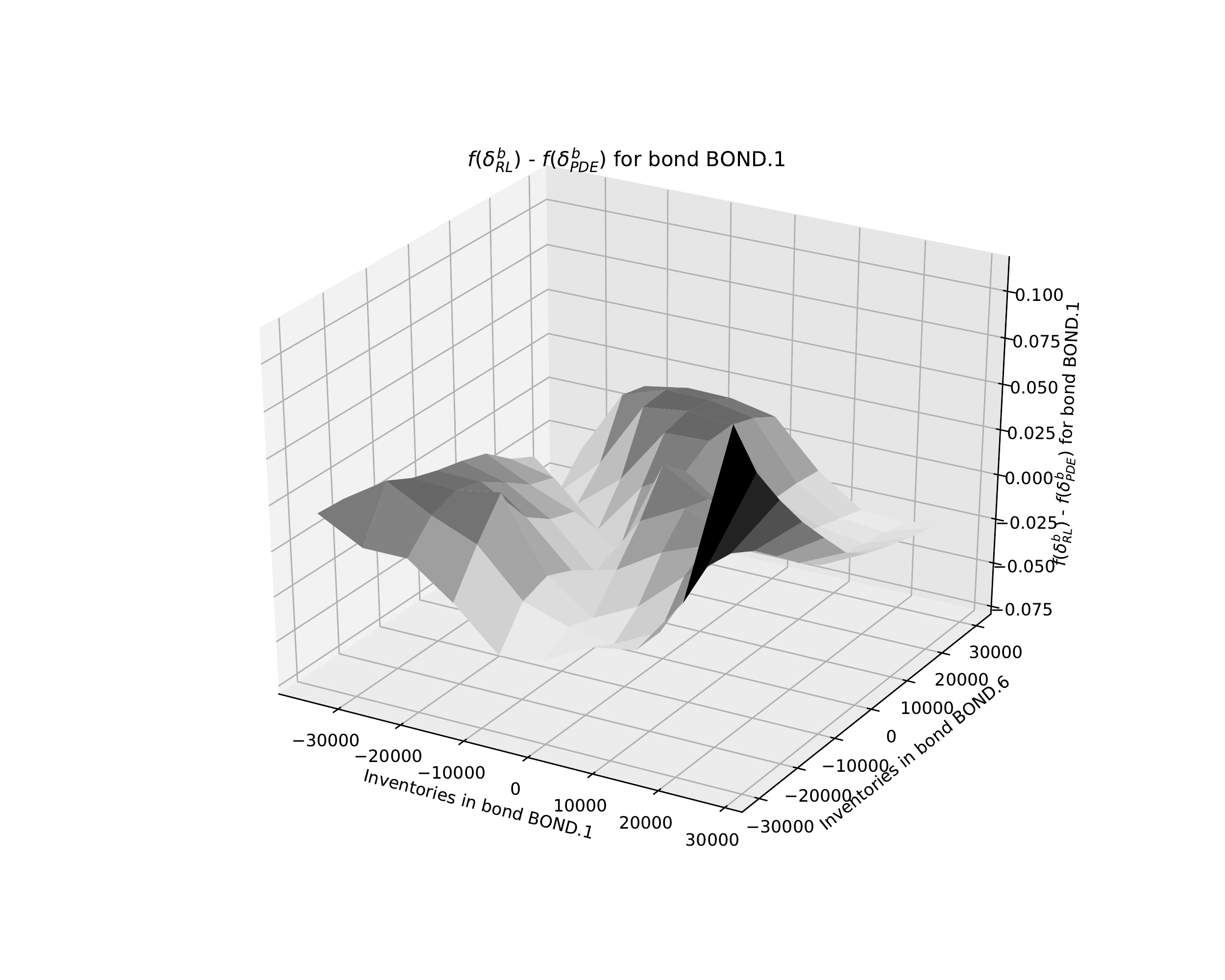}
  \includegraphics[width=0.48\textwidth]{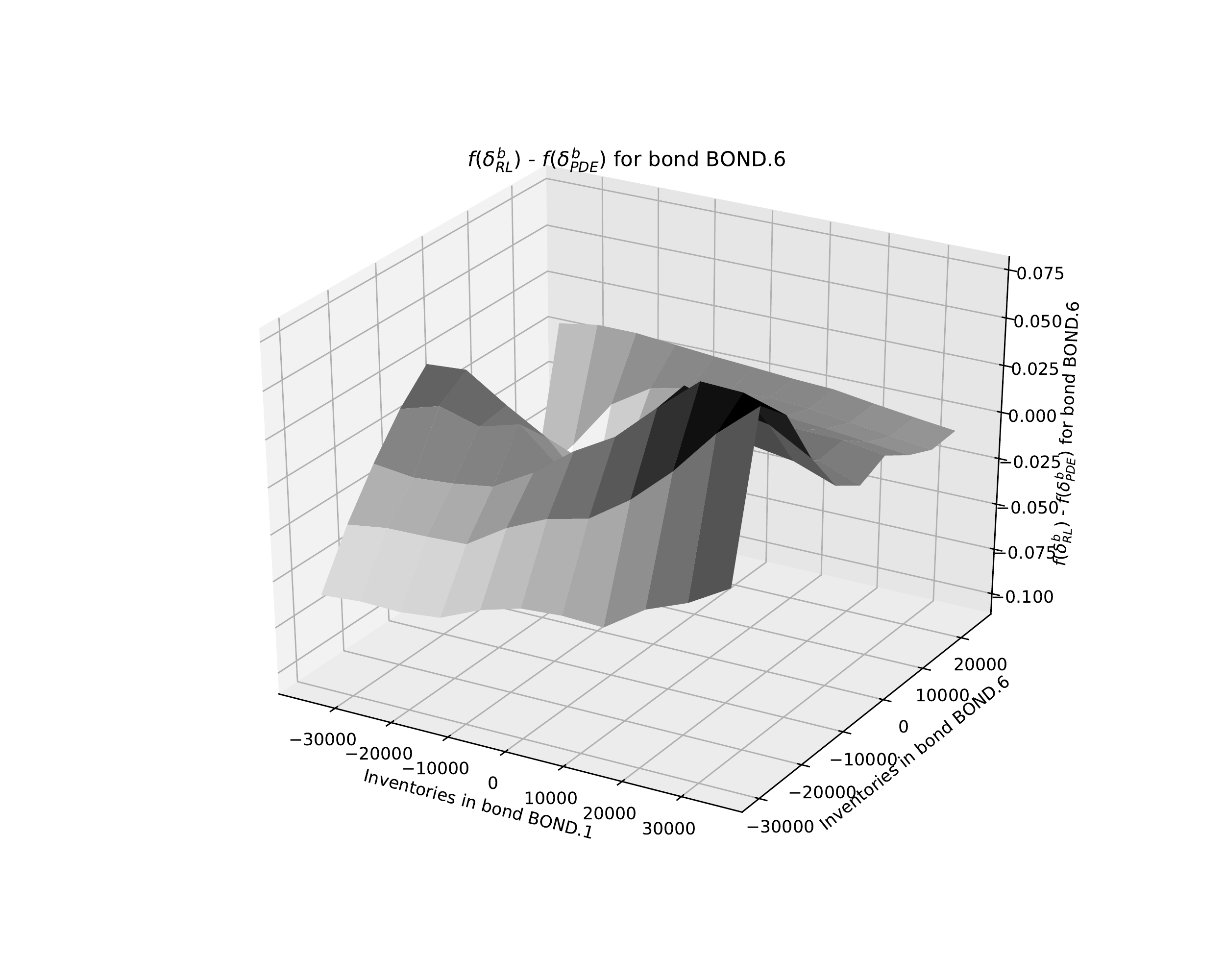}
  \caption{Comparison of the probabilities to trade obtained with the two methods.}
  \label{acth_2d_deltas}
\end{figure}

\begin{figure}[H]
  \centering
  \includegraphics[width=0.48\textwidth]{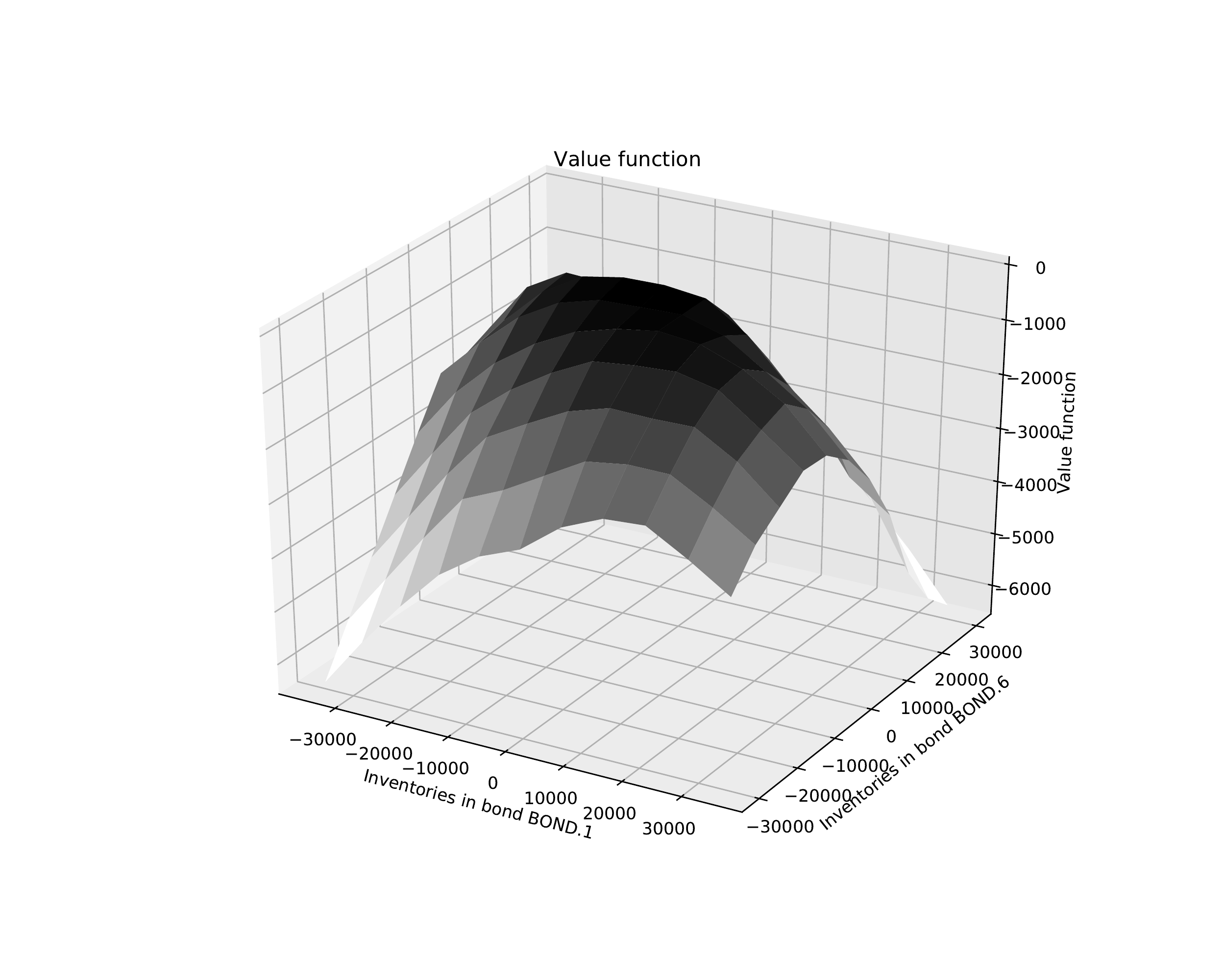}
  \includegraphics[width=0.48\textwidth]{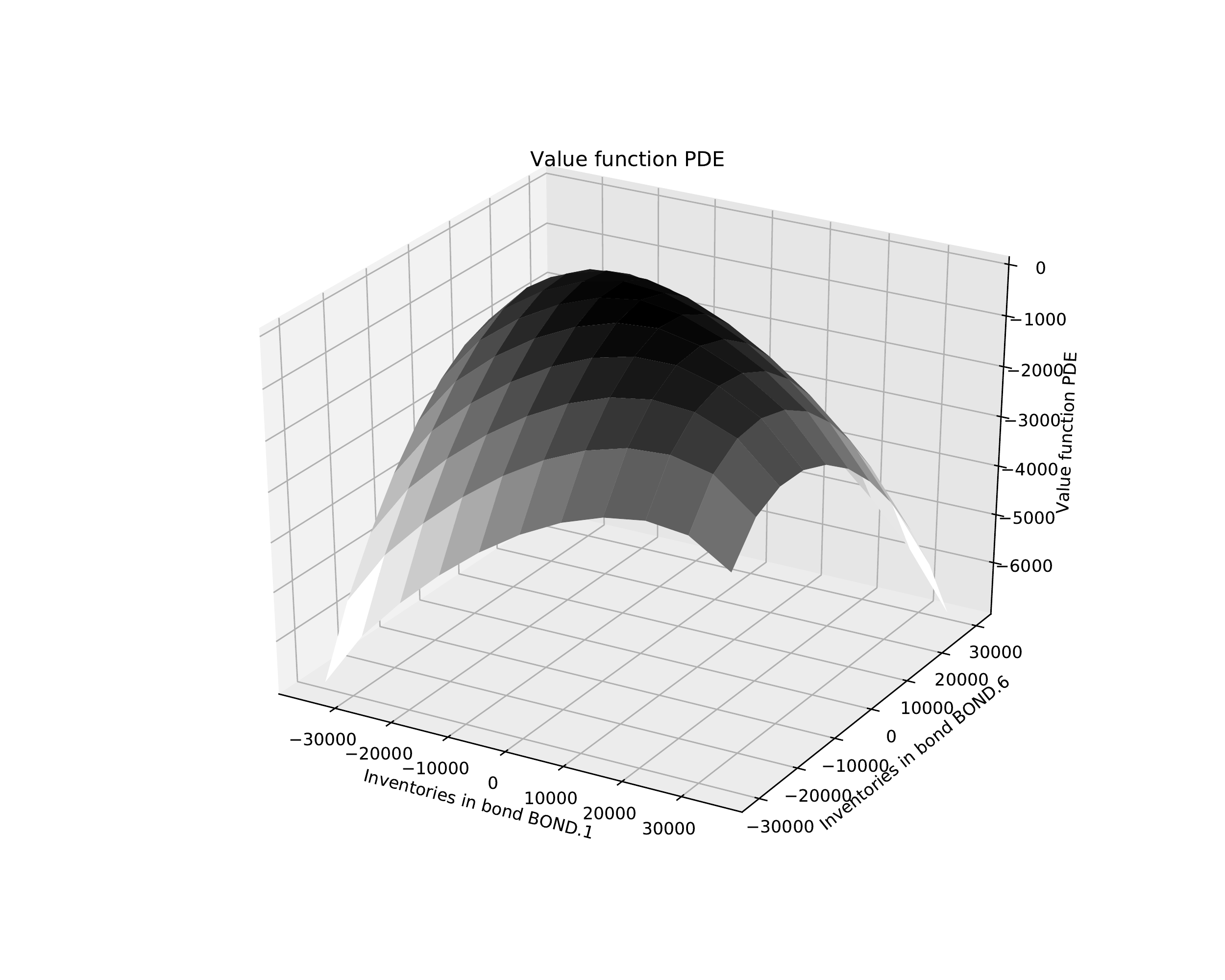}\\
  \vspace{-7mm}
  \includegraphics[width=0.48\textwidth]{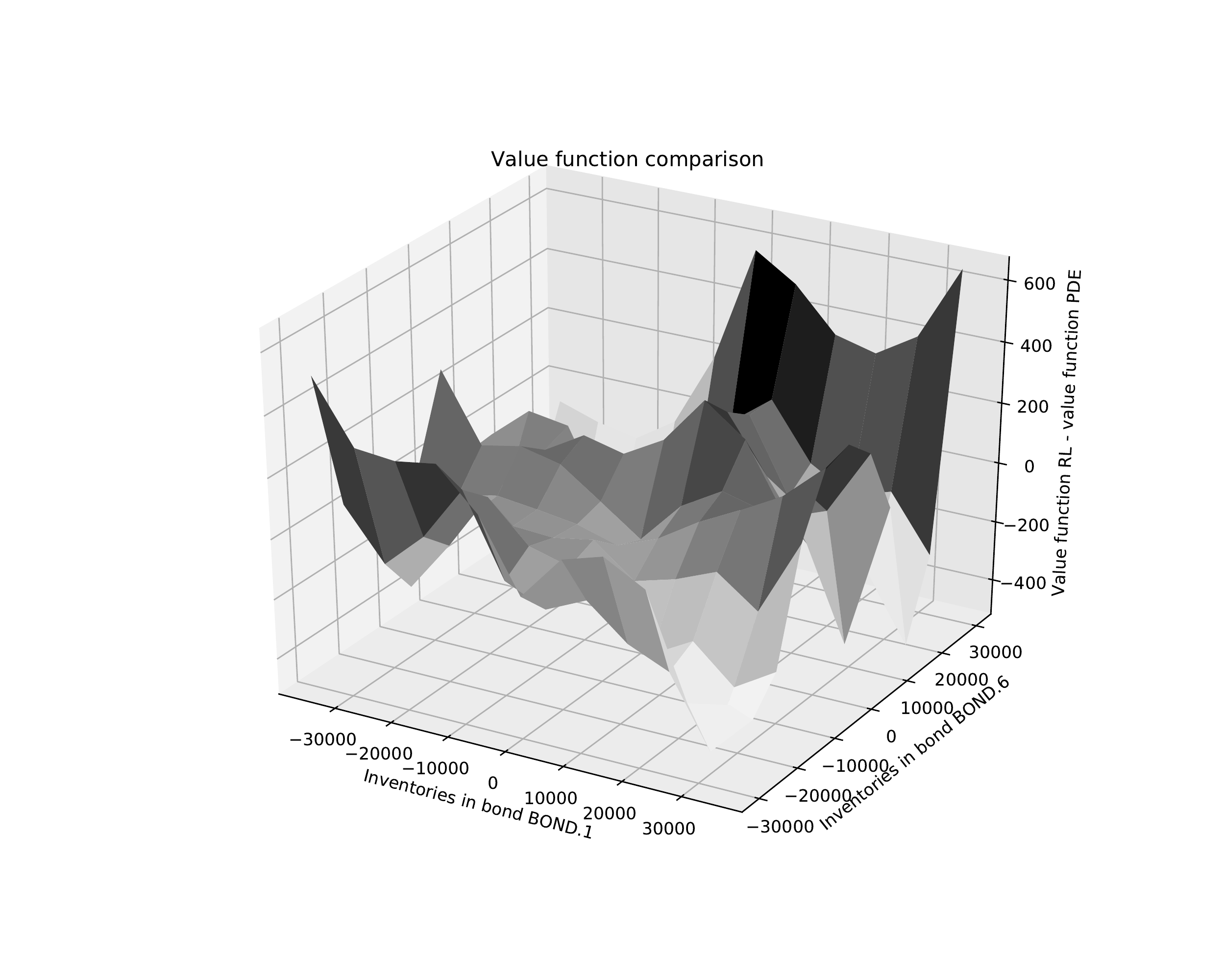}
  \caption{Top left: Optimal value function obtained with our RL algorithm. Top right: Optimal value function obtained with our finite difference method. Bottom: Comparison of the two.}
  \label{value_functions}
\end{figure}

In Figure \ref{rl_2d_deltas} and \ref{th_2d_deltas}, we plotted the optimal (bid) quotes computed with our two methods -- the difference between the two in terms of probability to trade is in Figure \ref{acth_2d_deltas}. The value functions are documented in Figure \ref{value_functions}.

\vspace{-7mm}
\vspace{-7mm}

We see in Figures \ref{rl_2d_deltas} and \ref{th_2d_deltas} that the optimal bid quotes depend on the current inventory in both the bond that is requested and the other bond. Because the correlation between the two bonds is positive, the optimal bid quote functions are increasing with respect to both coordinates. This means that being long one bond reduces the willingness to buy and increases the willingness the sell the other bond.\\

Regarding value functions, we see that correlations are taken into account as the plots of the value functions are not plots of separable functions.\\

It is interesting to notice that the approximations obtained with the two methods are different, even though the average rewards per RFQ are similar. It is noteworthy that the more extreme the inventories, the more different the quotes and the value of the value function. This may be linked to the fact that we learn more slowly at points that are seldom visited in Monte-Carlo simulations.\\

Let us now consider another 2-bond case where bonds have low correlation. We have chosen BOND.18 and BOND.20, as the correlation between their price variations is $7\%$. We used the same algorithm parametrization as in the previous 2-bond case.\\

\begin{figure}[H]
  \centering\includegraphics[width=0.40\textwidth]{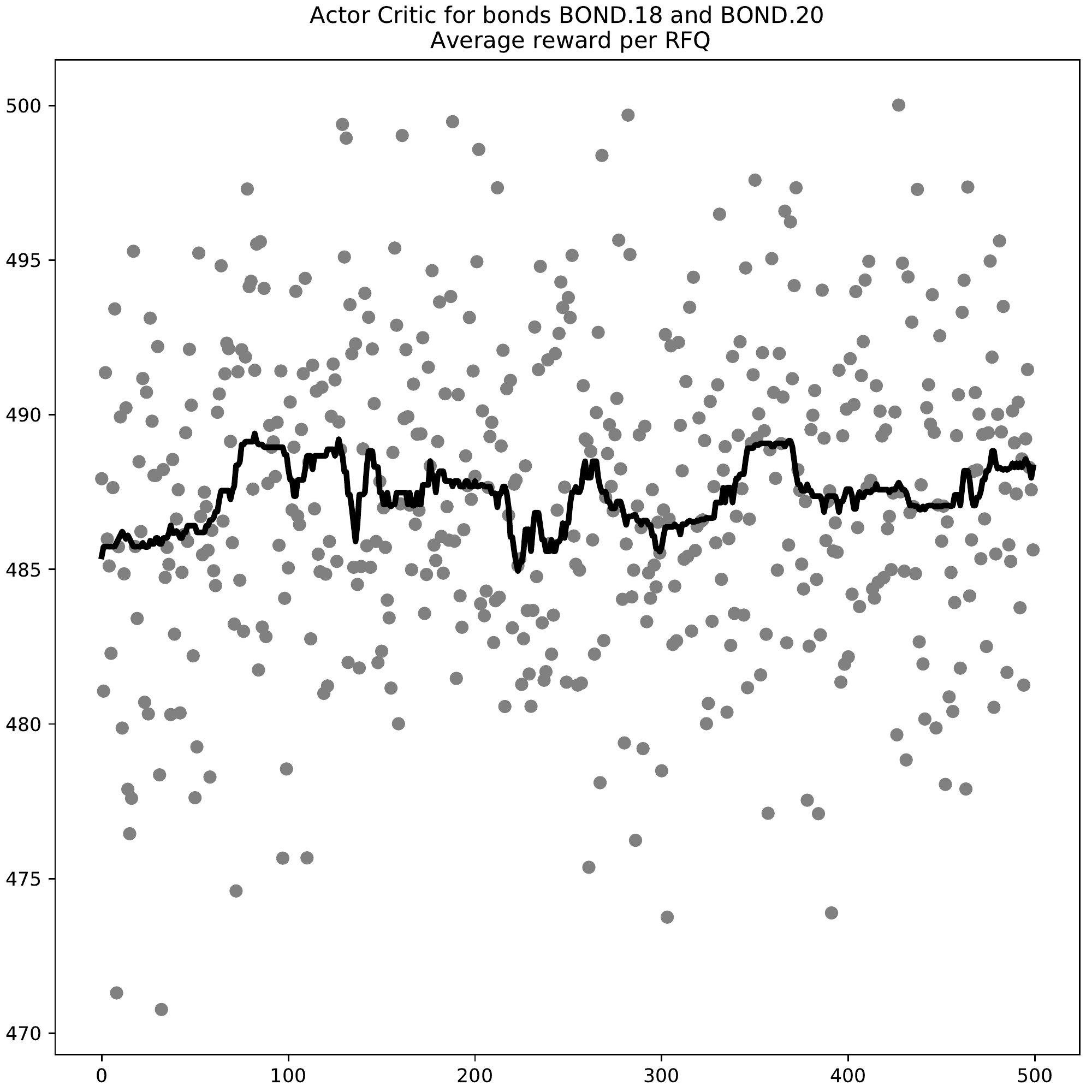}
  \caption{Average reward per RFQ -- Learning process.}
  \label{vf_lc1}
\end{figure}

\begin{figure}[H]
  \centering
\includegraphics[width=0.55\textwidth]{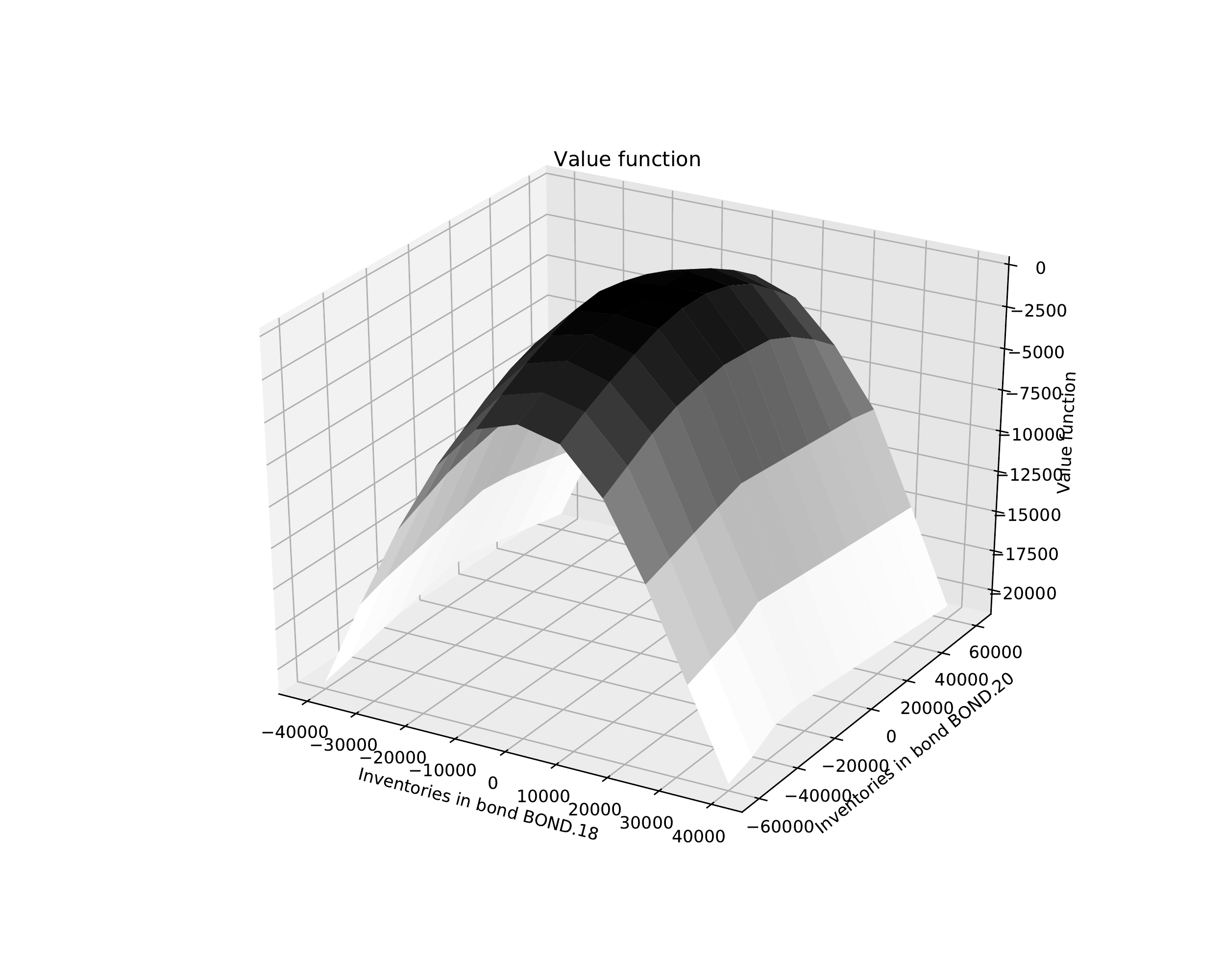}
\caption{Optimal value function obtained with our RL algorithm.}
\label{vf_lc2}
\end{figure}

The learning curve and the value function obtained by our RL algorithm are plotted in Figures \ref{vf_lc1} and \ref{vf_lc2} respectively.\\
\vspace{-3mm}

We clearly see that the algorithm does not learn anything. This is expected because there is almost nothing to learn as the algorithm starts from the quotes obtained by our RL algorithm in the single-bond cases (i.e. ignoring correlation). For the sake of completeness, the average reward per RFQ obtained using the optimal quotes computed with the finite difference method is $490.3$, almost in line with the values of Figure \ref{vf_lc2}. We also see that the value function looks separable.\\

\subsubsection{8-bond case}

In addition to cases with two bonds we consider our target cases where the market maker takes account of his position in a large number of bonds. We consider cases with 8 and 20 bonds. Of course, in these cases, there is no possibility of comparison to a solution based on finite differences. The goal is therefore to verify that the RL algorithm enables to obtain a strategy with a higher average reward per RFQ than that obtained without accounting for the correlation structure between bonds.\\

Let us start with the case of the 8 most volatile bonds in our set of bonds (see Section \ref{dataset}).\\

We considered $\gamma = 5\cdot10^{-2}$ and $r = 10^{-4}$ as above. Risk limits were chosen equal to $5$ times the RFQ size at the beginning of the learning process and were increased every $500$ steps by the RFQ size for each bond until the maximum risk limits equal to $10$ times the RFQ size were reached (this is the reverse Matryoshka dolls principle), except for BOND.5 -- due to its low liquidity and high volatility -- for which we chose a risk limit of $5$ times the associated RFQ size. For the critic and the actor, we considered neural networks with $2$ hidden layers and $18$ nodes in each of these layers with ReLU activation functions. As above, the final layer of each neural network contains one node and the activation function is affine in the case of the critic and sigmoid in the case of the actor. For the pre-training we used the quotes obtained by our RL algorithm in the single-bond case for each bond.  For the learning phase we considered $3000$ steps, i.e. $3000$ steps of TD learning and $3000$ steps of policy improvement for each of the 8 bonds. At each step we carried out $1$ rollout of length $10000$ starting from a zero inventory and $100$ additional rollouts of length $100$ starting from a random inventory. The noise $\epsilon$ in each rollout is distributed uniformly in $[-0.05, 0.05]$ and we chose the probability limit $\nu = 0.005$. The learning rate for the critic is $\eta=5\cdot10^{-8}$ and we used mini-batches of size $50$. The learning rate for the actor is  $\tilde{\eta} = 0.01$ and we used mini-batches of size $50$. \\

\begin{figure}[H]
  \centering
  \includegraphics[width=0.48\textwidth]{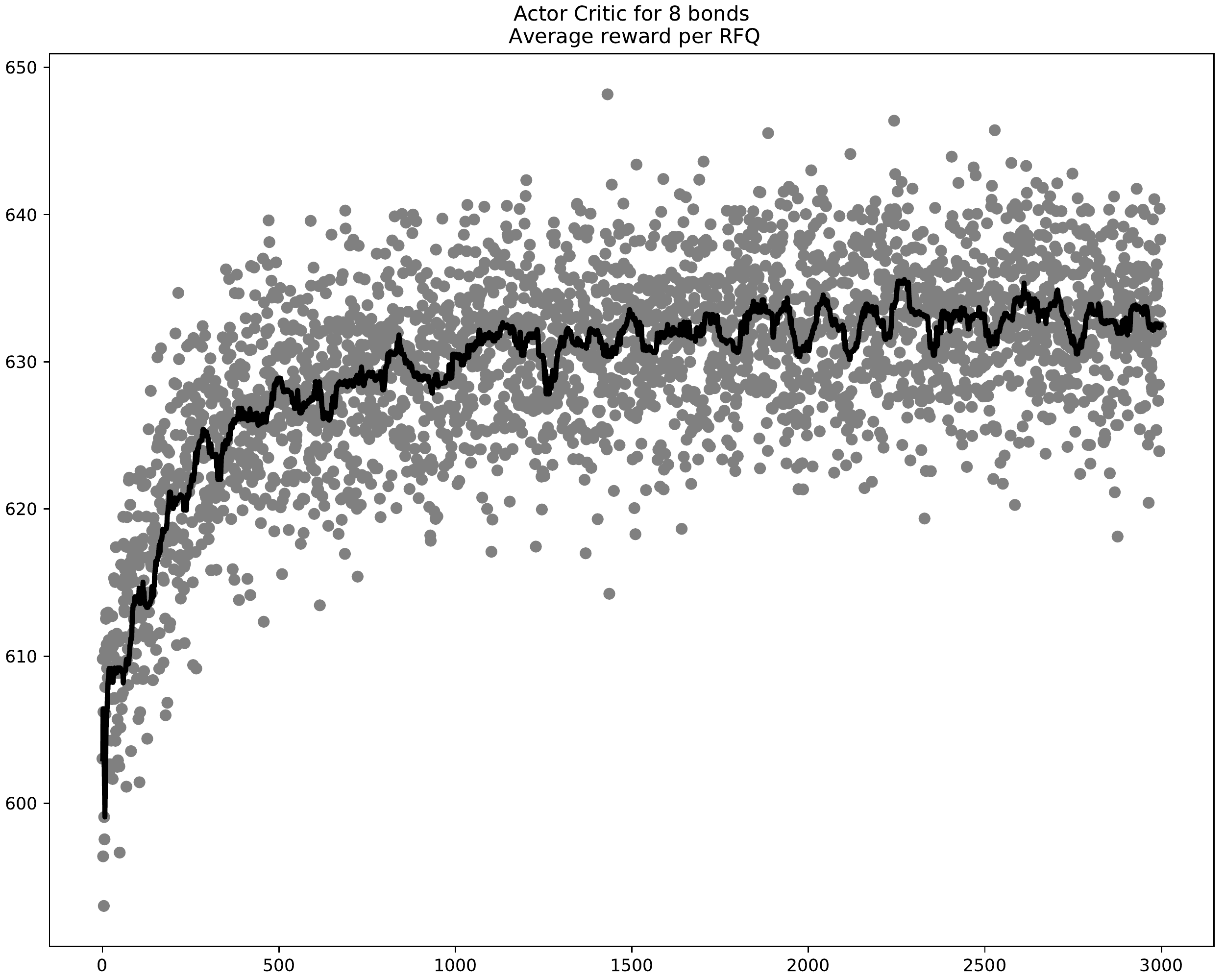}\\
  \caption{Average reward per RFQ -- Learning process for the 8-bond case.}
  \label{rl_8d_sqrt}
\end{figure}

In Figure \ref{rl_8d_sqrt} we see the learning curve in terms of average reward per RFQ. We see that there is a sharp increase in average reward per RFQ over the first 500 steps and that the algorithm then continues to improve progressively the quoting strategy. This important improvement is linked to the correlation structure of the 8 chosen bonds and to the fact that we chose volatile bonds, for which the possibility of hedging provides significant value.

\subsubsection{20-bond case}

Let us now come to the 20-bond case for which our RL algorithm can be applied, but not the finite difference method.\\

We considered $\gamma = 5\cdot10^{-2}$ and $r = 10^{-4}$ as above. Risk limits were chosen equal to $5$ times the RFQ size at the beginning of the learning process and were increased every $500$ steps by the RFQ size for each bond until the maximum risk limits equal to $10$ times the RFQ size were reached, except for BOND.5 for which we chose a risk limit of $5$ times the associated RFQ size, as above. For the critic and the actor, we considered neural networks with $2$ hidden layers and $30$ nodes in each of these layers with ReLU activation functions. As above, the final layer of each neural network contains one node and the activation function is affine in the case of the critic and sigmoid in the case of the actor. For the pre-training we used the quotes obtained by our RL algorithm in the single-bond case for each bond.  For the learning phase we considered $5000$ steps, i.e. $5000$ steps of TD learning and $5000$ steps of policy improvement for each of the 20 bonds. At each step we carried out $1$ rollout of length $10000$ starting from a zero inventory and $100$ additional rollouts of length $100$ starting from a random inventory. The noise $\epsilon$ in each rollout is distributed uniformly in $[-0.05, 0.05]$ and we chose the probability limit $\nu = 0.005$. The learning rate for the critic is $\eta=5\cdot10^{-8}$ and we used mini-batches of size $70$. The learning rate for the actor is  $\tilde{\eta} = 0.01$ and we used mini-batches of size $50$. \\

\begin{figure}[H]
  \centering
  \includegraphics[width=0.4\textwidth]{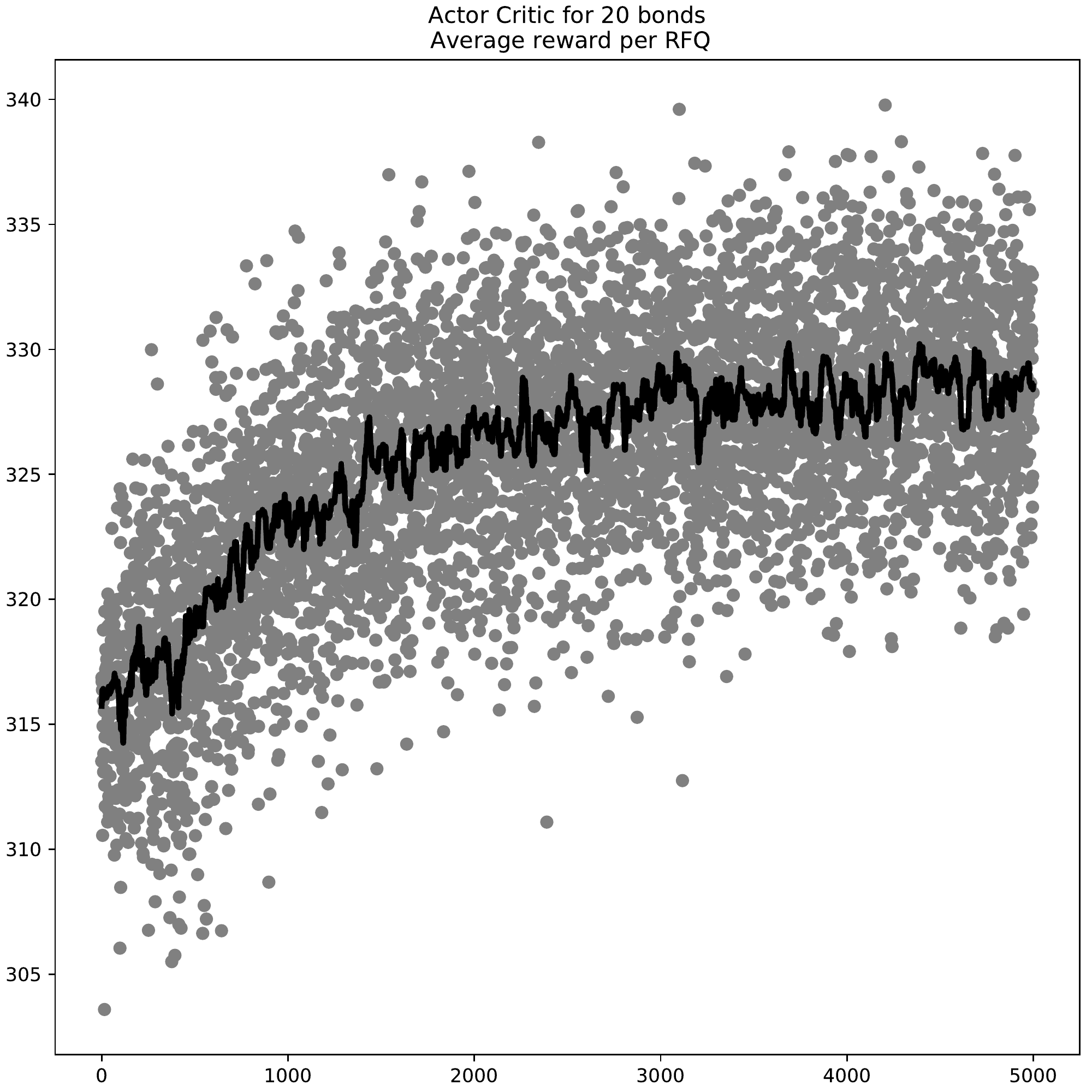}\\
  \caption{Average reward per RFQ -- Learning process for the 20-bond case.}
  \label{rl_20d_sqrt}
\end{figure}

In Figure \ref{rl_20d_sqrt} we see the evolution of the average reward per RFQ during the learning process.

As in the above cases, we clearly see that correlations are taken into account.\\

As a conclusion, we clearly see that our algorithm scales to dimension 20. Furthermore, it is important to mention that (i) the relevant size for the neural networks seems to scale linearly with the number of bonds, and so does the number of steps to reach the maximum average reward per RFQ, and (ii) the same hyperparameters (learning rate, batch size, rollout size, etc.) can be used independently of the dimension.\\

\subsection{Results in the case $\psi(q) = \frac 12 \gamma q'\Sigma q$}
\label{sqr}
We now come to the case of a penalty function of the form $$\psi(q) = \frac 12 \gamma q'\Sigma q.$$ This penalty, proportional to the instantaneous variance of the MtM value of the portfolio, is the penalty function used in many market making models such as those of Cartea \emph{et al.} \cite{carteabook} or in Guéant \cite{gueant}. In many cases, the expected utility framework with CARA utility function used in the original paper of Avellaneda and Stoikov \cite{avellaneda} or in Guéant \emph{et al.} \cite{glft} can also be reduced to the maximization of the expected PnL minus a penalty of the above form up to a change in the intensity functions (see \cite{manziuk}). If quadratic penalties are often considered in academic papers, they are known to be strong (maybe too strong) and to penalize large inventories a lot more than the penalty function of Section~\ref{sqrt}.

\subsubsection{Single-bond cases}

As in the previous case, we start with examples of single-bond market making and compare the results obtained by using the PDE method of the appendix and the results obtained with our RL algorithm. In what follows we consider $\gamma=2\cdot10^{-5}$ and $r=10^{-4}$. The risk limits were set to $5$ times the size of RFQs.\\

\begin{figure}[H]
  \centering
  \includegraphics[width=0.88\textwidth]{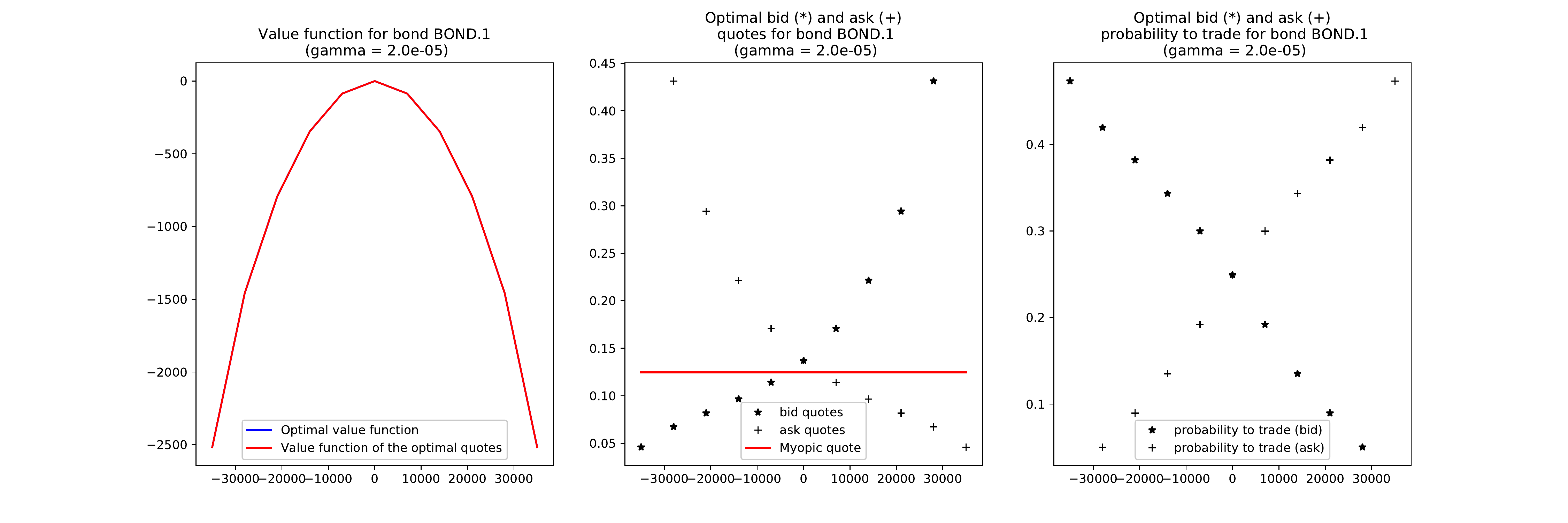}\\
  \includegraphics[width=0.88\textwidth]{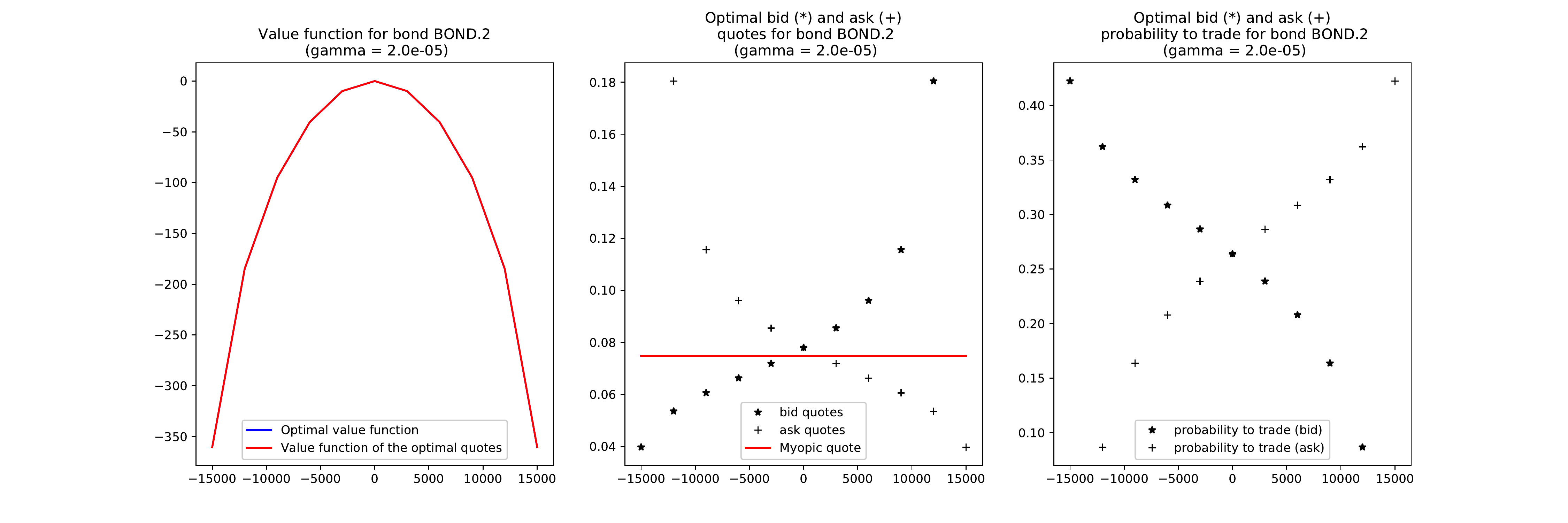}\\
  \includegraphics[width=0.88\textwidth]{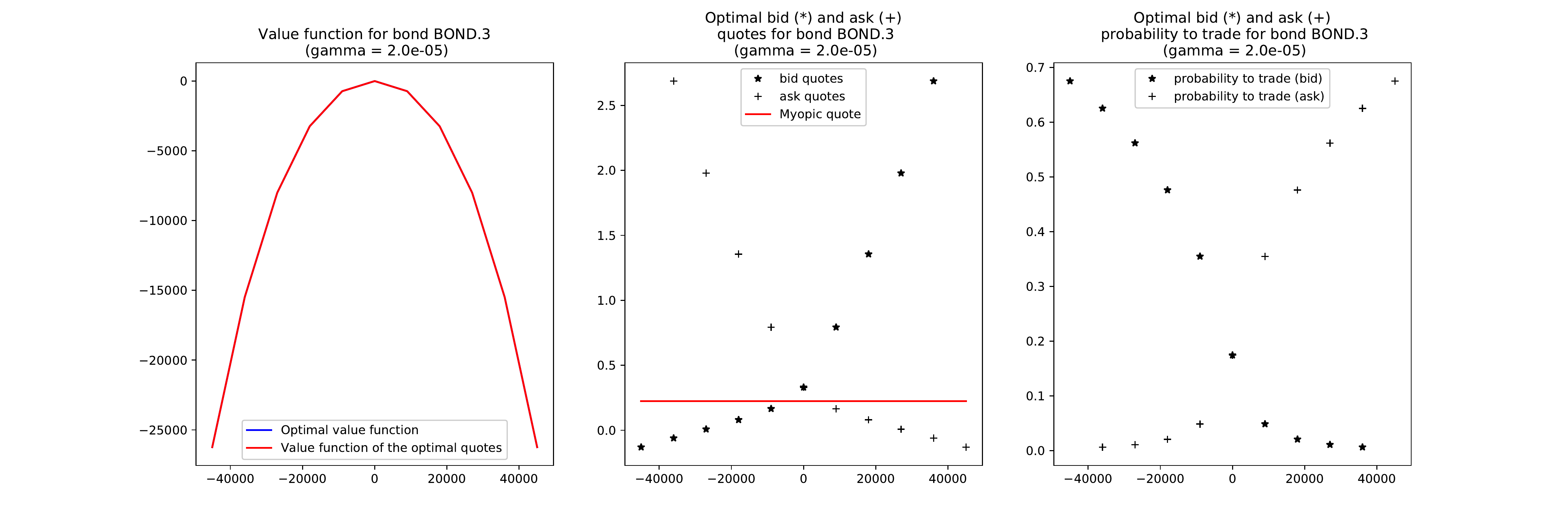}\\
  \includegraphics[width=0.88\textwidth]{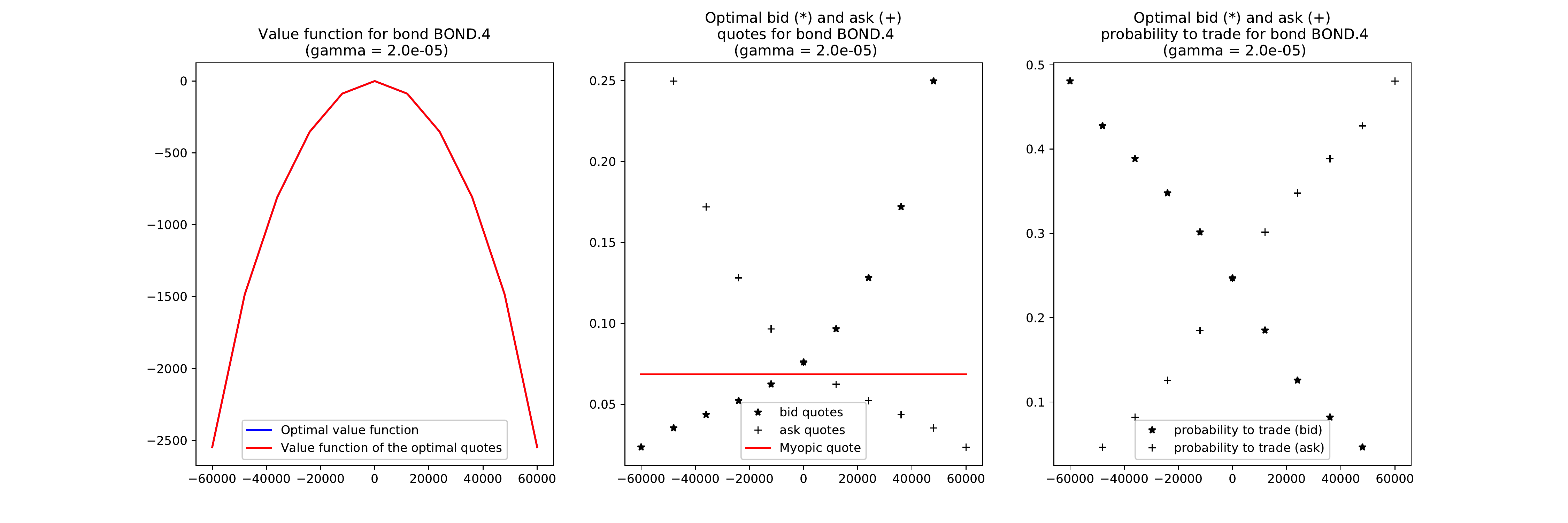}
  \caption{Value functions, optimal quotes, and optimal probabilities to trade with the finite difference approach.}
  \label{pde_square_1}
\end{figure}

\begin{figure}[H]
  \centering
  \includegraphics[width=0.88\textwidth]{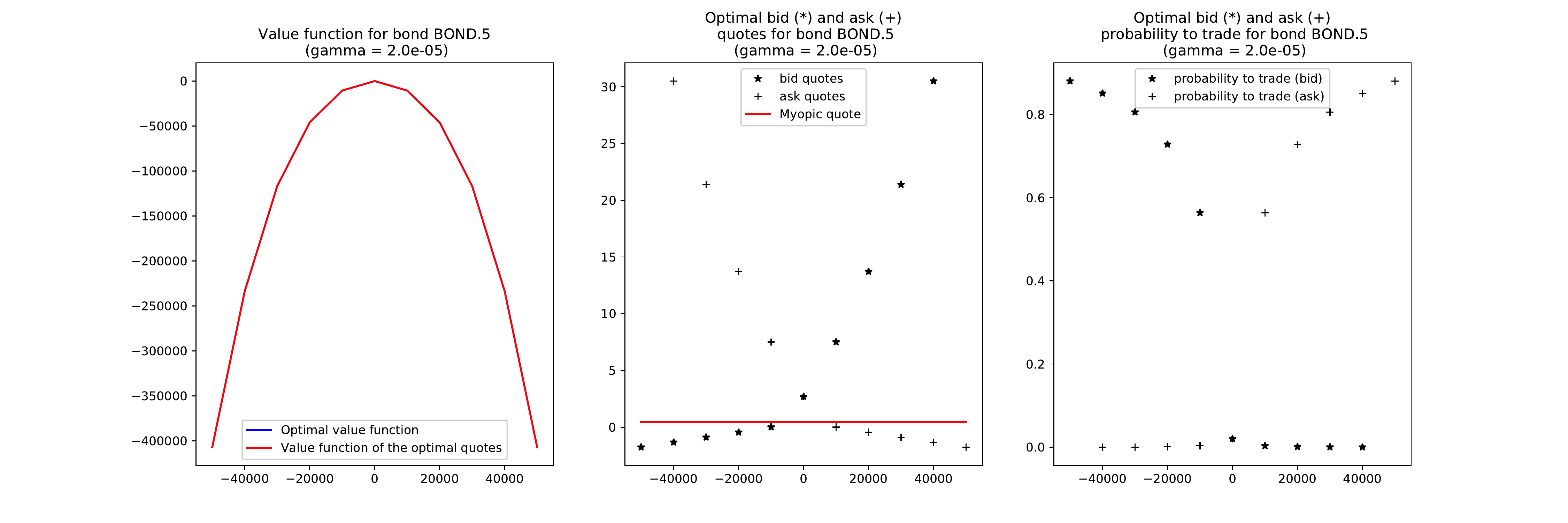}\\
  \includegraphics[width=0.88\textwidth]{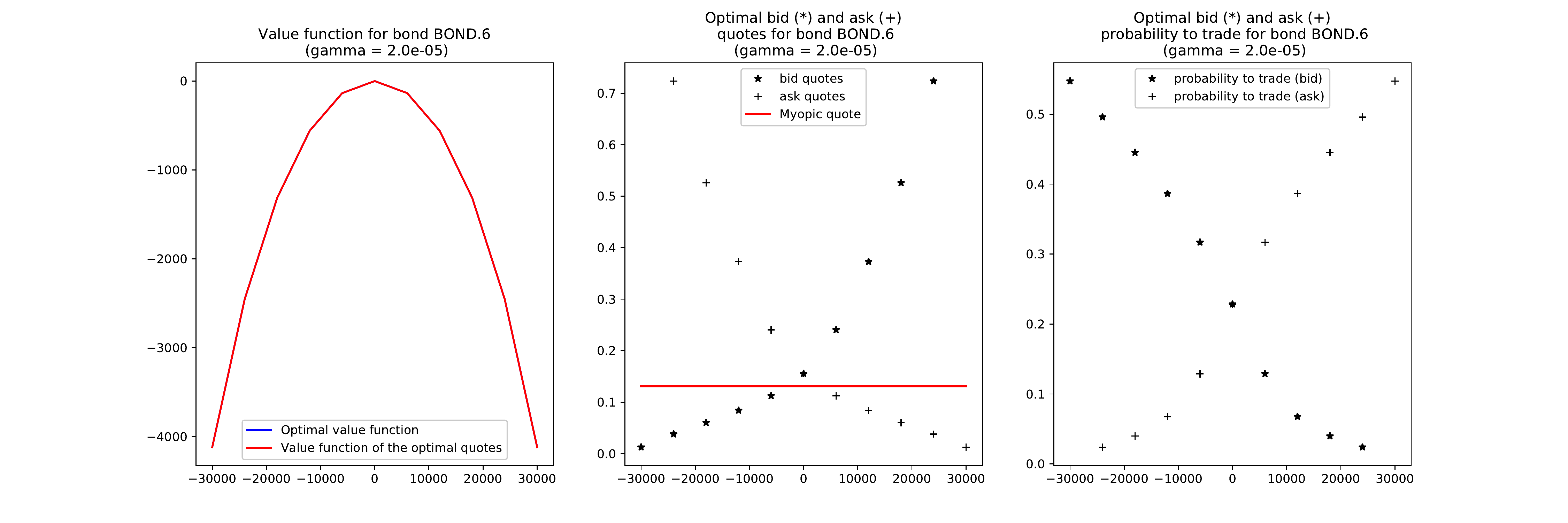}\\
  \includegraphics[width=0.88\textwidth]{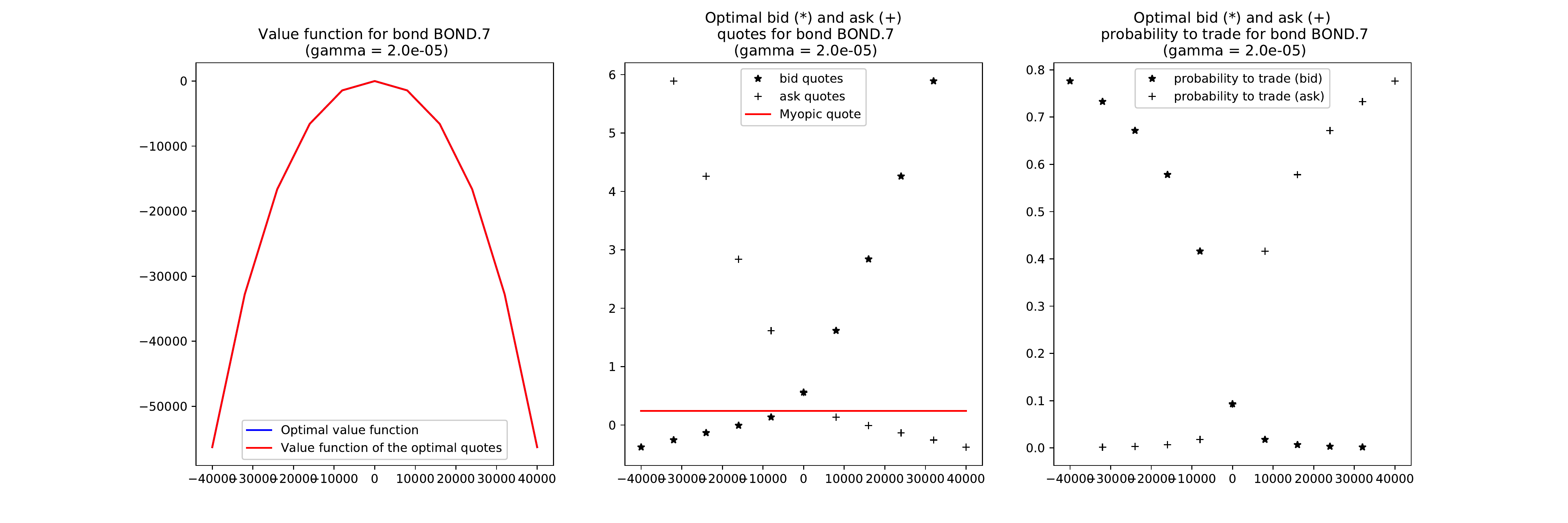}\\
  \includegraphics[width=0.88\textwidth]{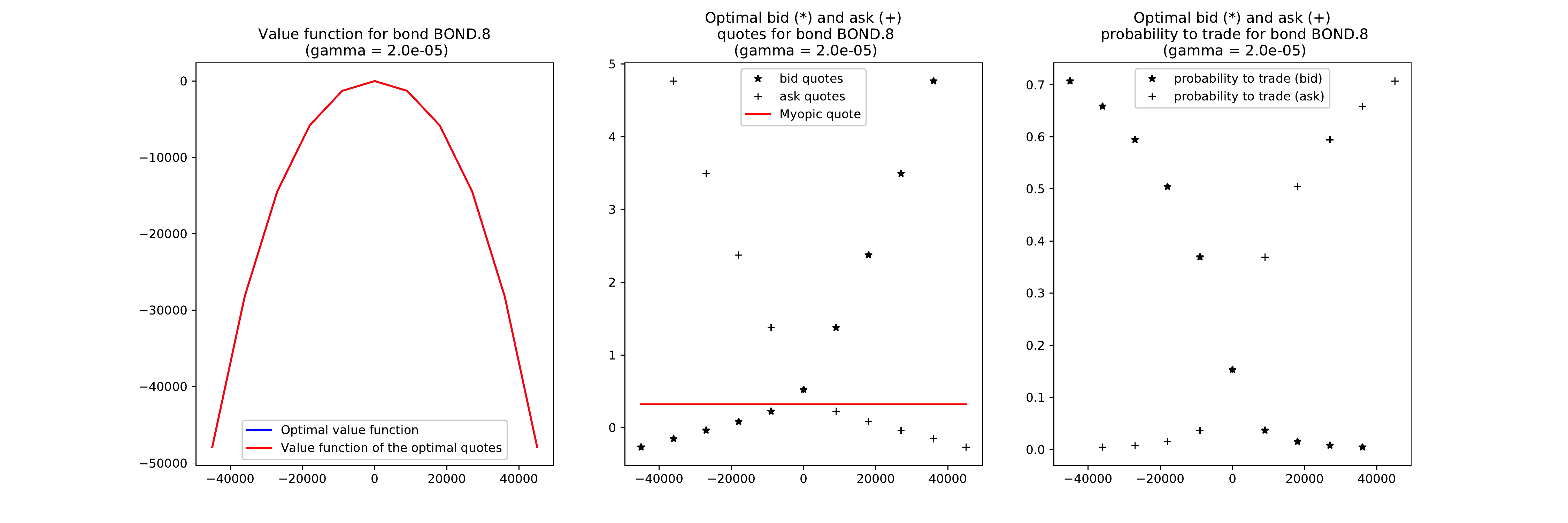}
  \caption{Value functions, optimal quotes, and optimal probabilities to trade with the finite difference approach.}
  \label{pde_square_2}
\end{figure}

\begin{figure}[H]
  \centering
  \includegraphics[width=0.88\textwidth]{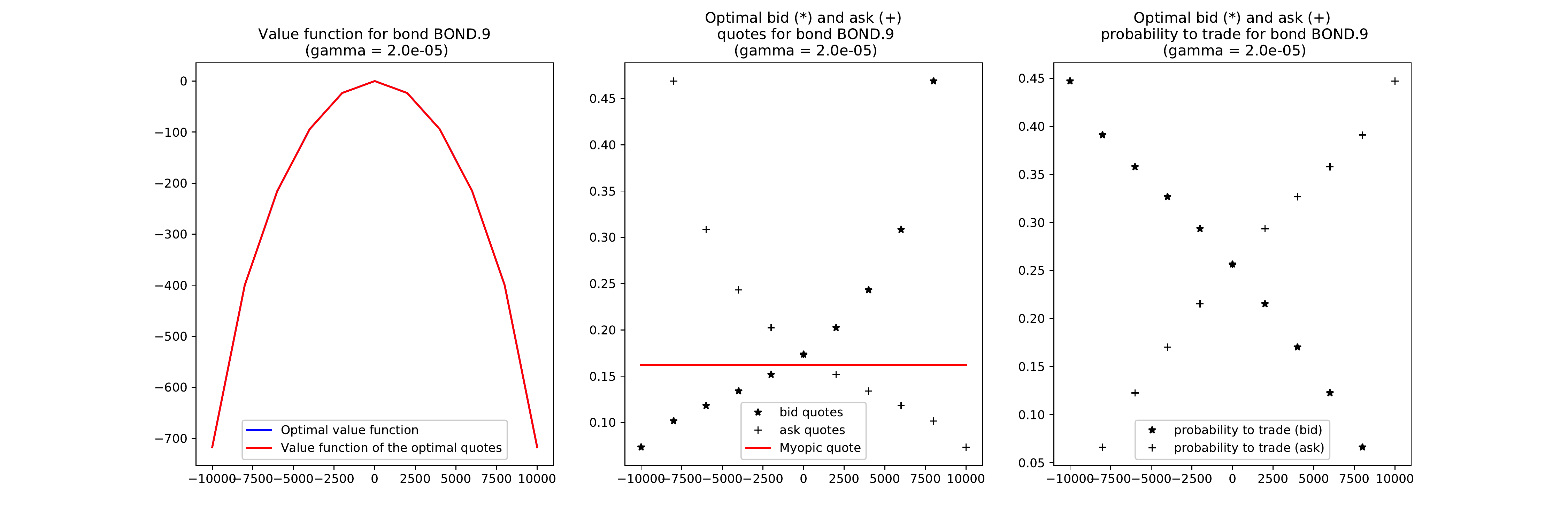}\\
  \includegraphics[width=0.88\textwidth]{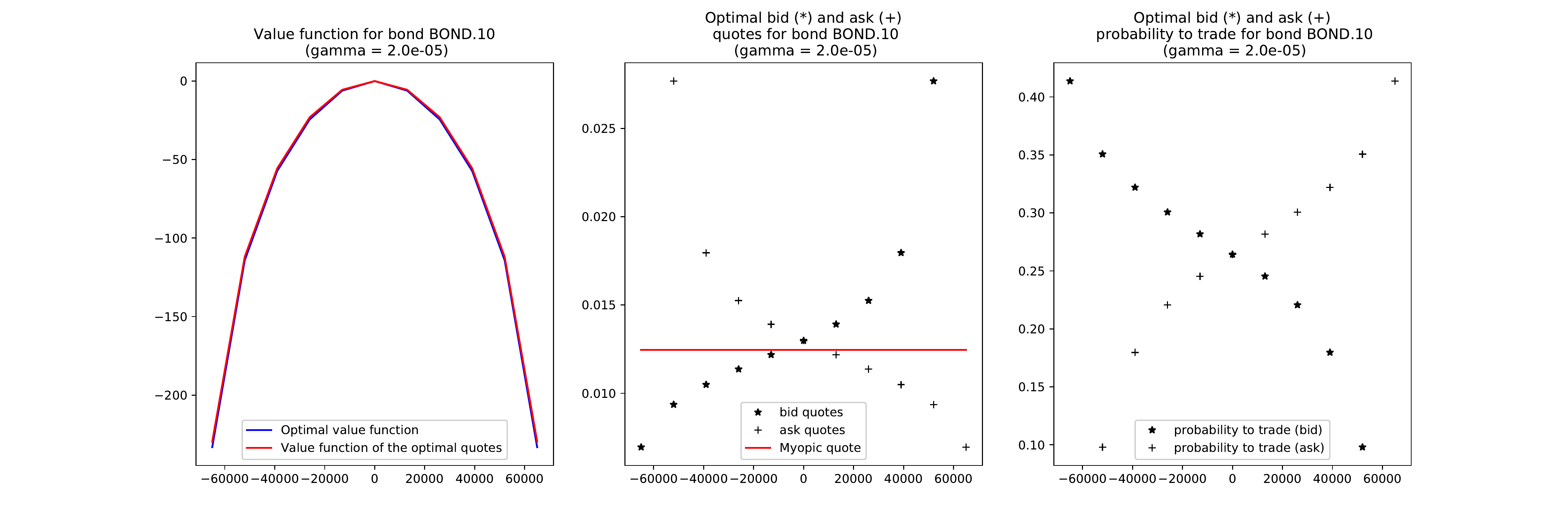}\\
  \includegraphics[width=0.88\textwidth]{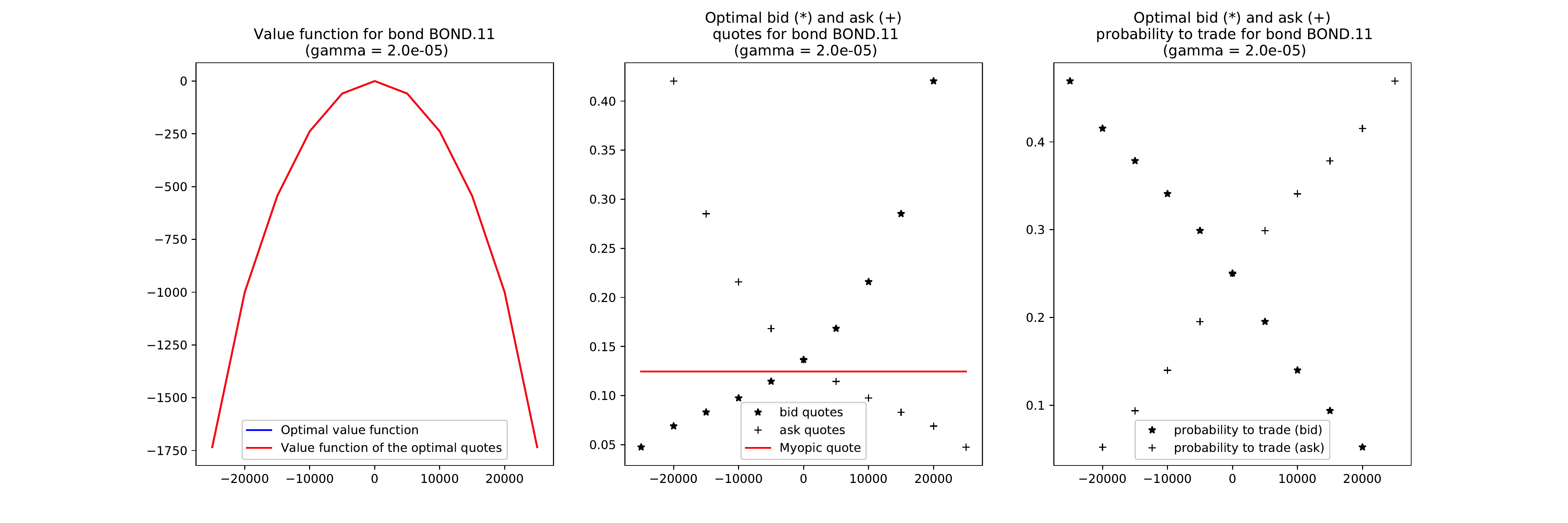}\\
  \includegraphics[width=0.88\textwidth]{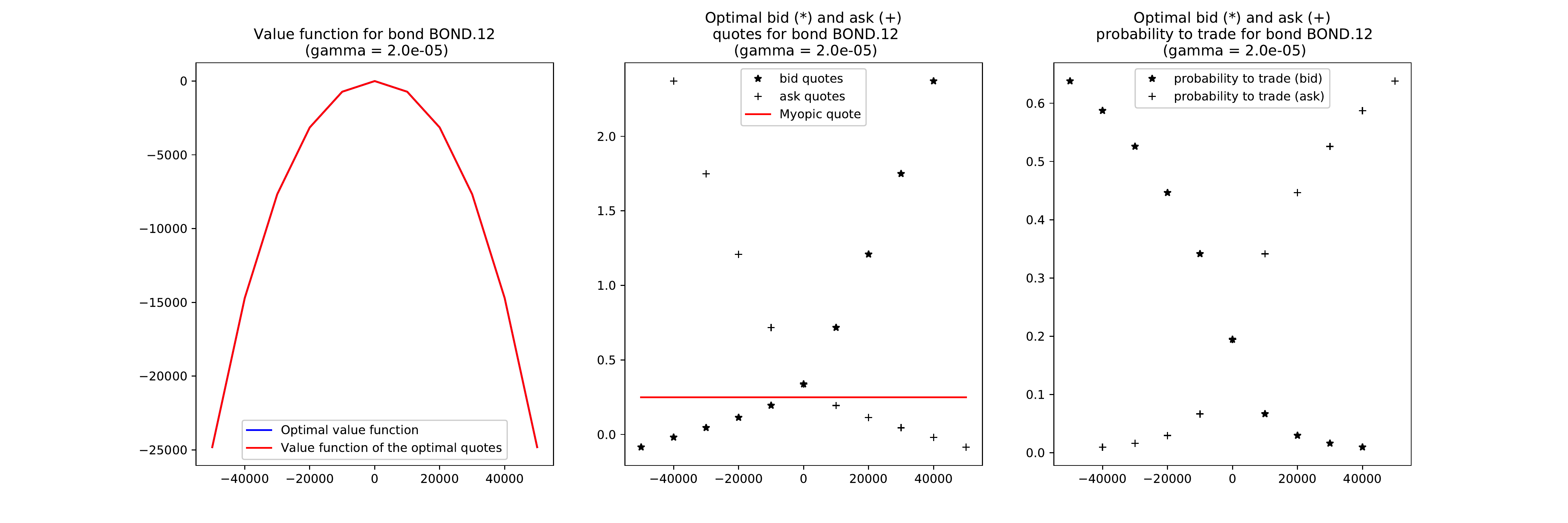}
  \caption{Value functions, optimal quotes, and optimal probabilities to trade with the finite difference approach.}
  \label{pde_square_3}
\end{figure}

\begin{figure}[H]
  \centering
  \includegraphics[width=0.88\textwidth]{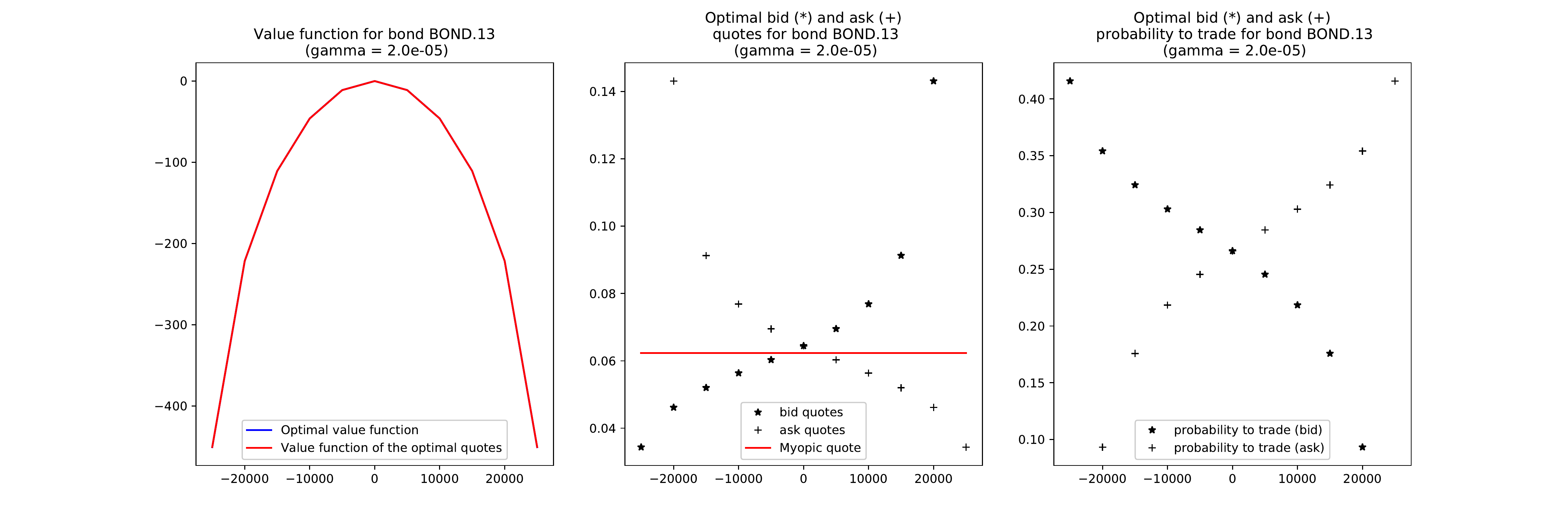}\\
  \includegraphics[width=0.88\textwidth]{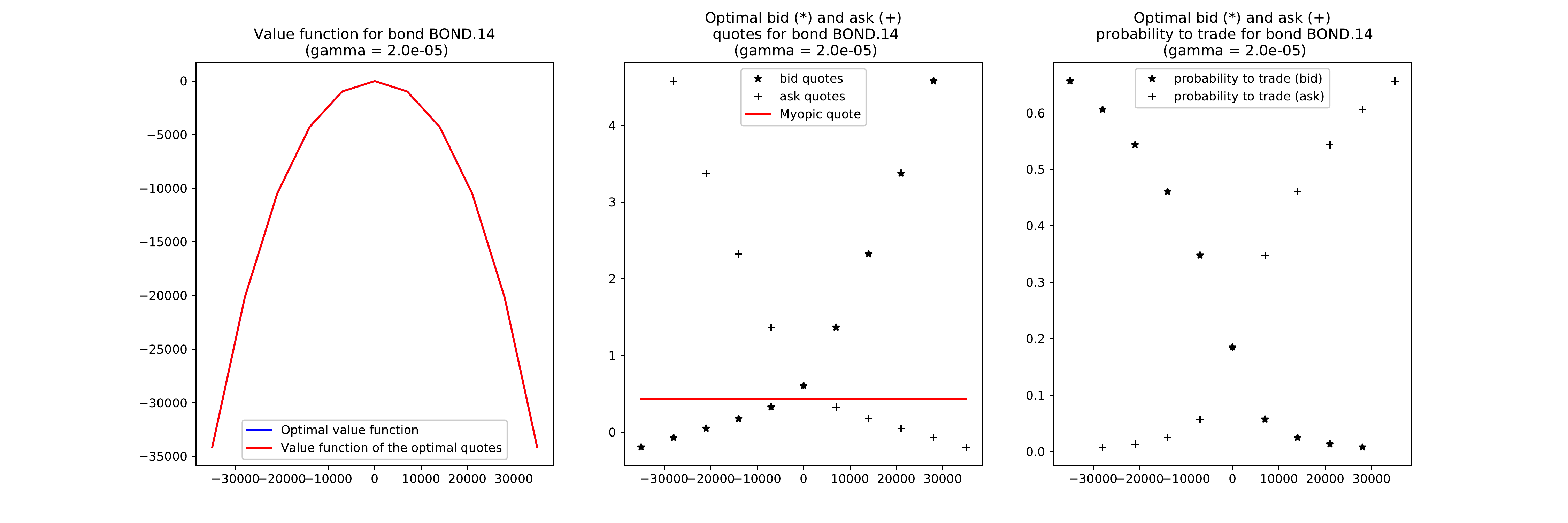}\\
  \includegraphics[width=0.88\textwidth]{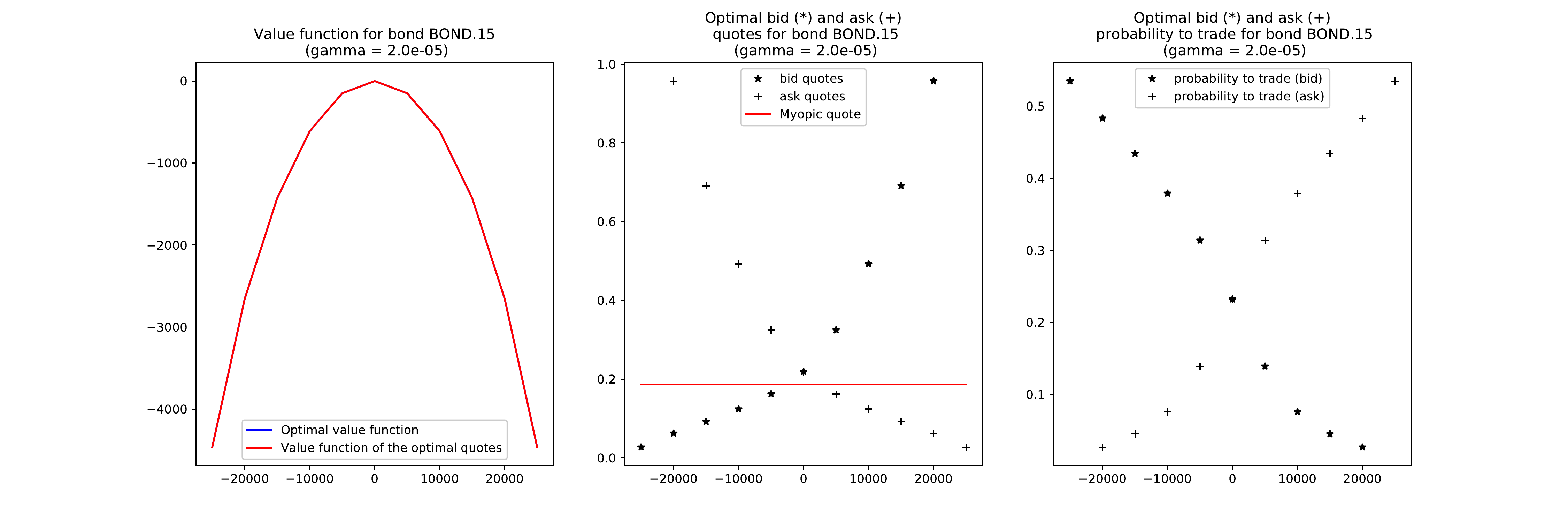}\\
  \includegraphics[width=0.88\textwidth]{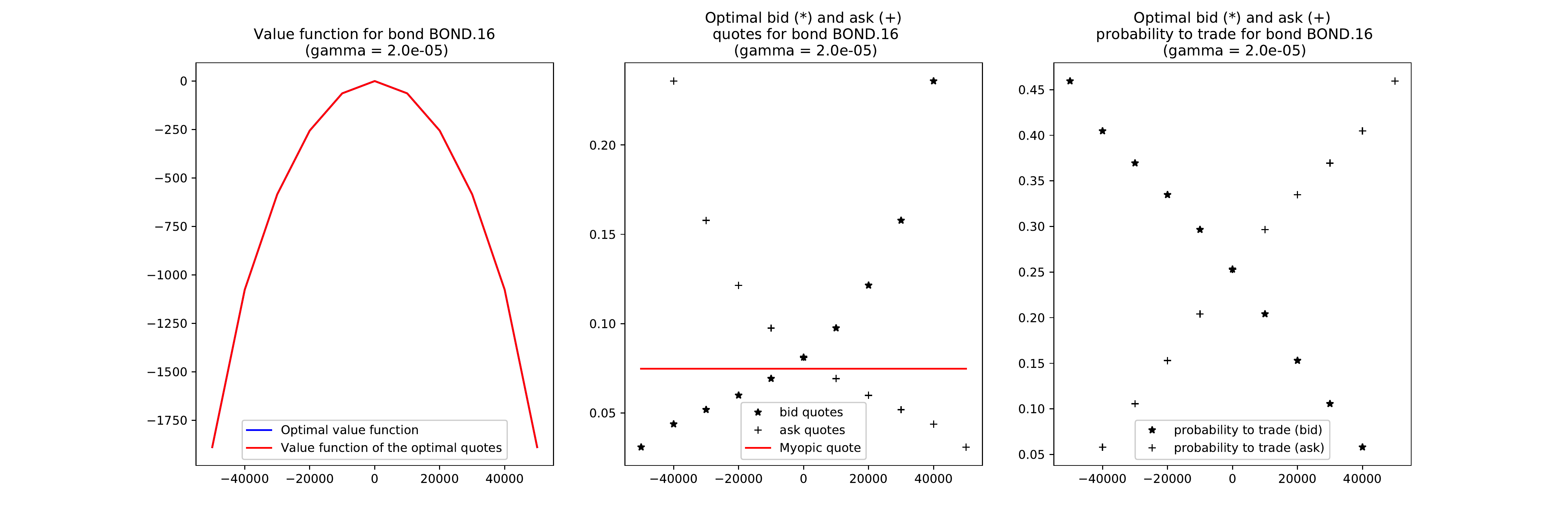}
  \caption{Value functions, optimal quotes, and optimal probabilities to trade with the finite difference approach.}
  \label{pde_square_4}
\end{figure}

\begin{figure}[H]
  \centering
  \includegraphics[width=0.88\textwidth]{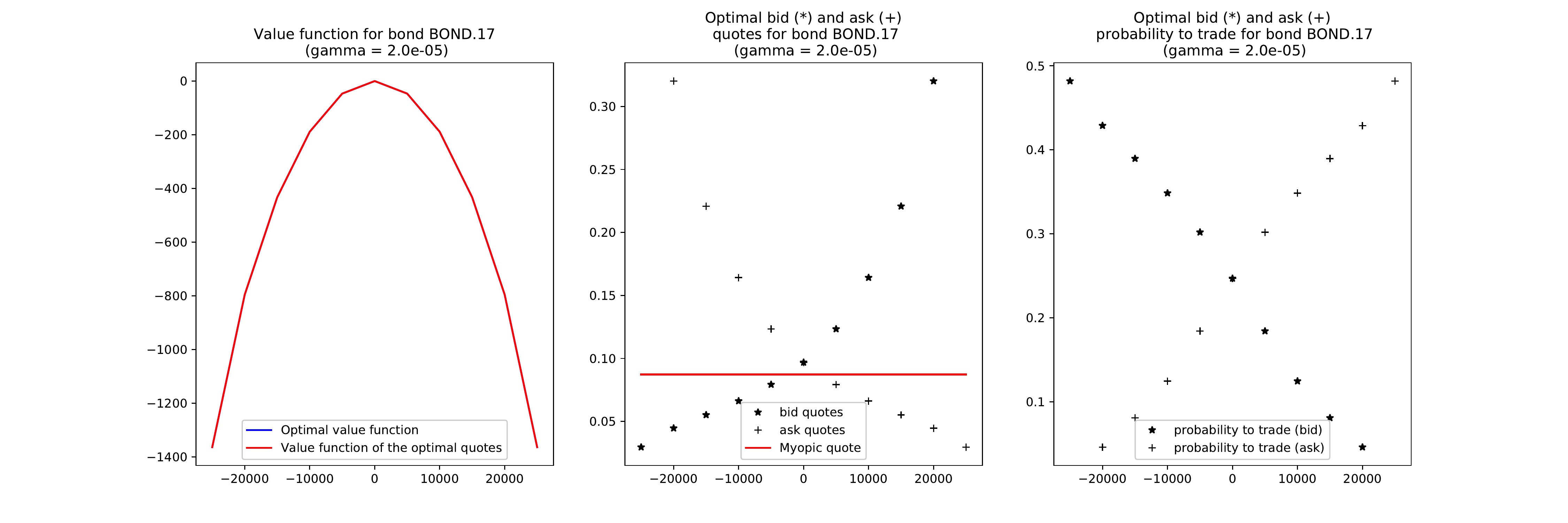}\\
  \includegraphics[width=0.88\textwidth]{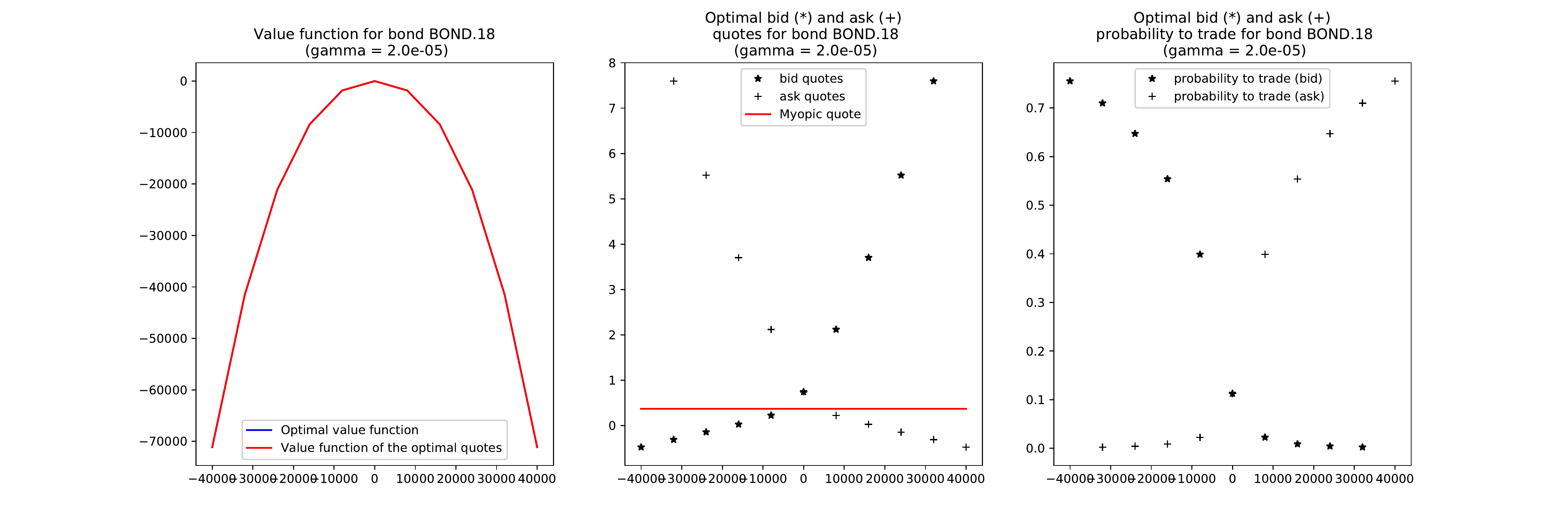}\\
  \includegraphics[width=0.88\textwidth]{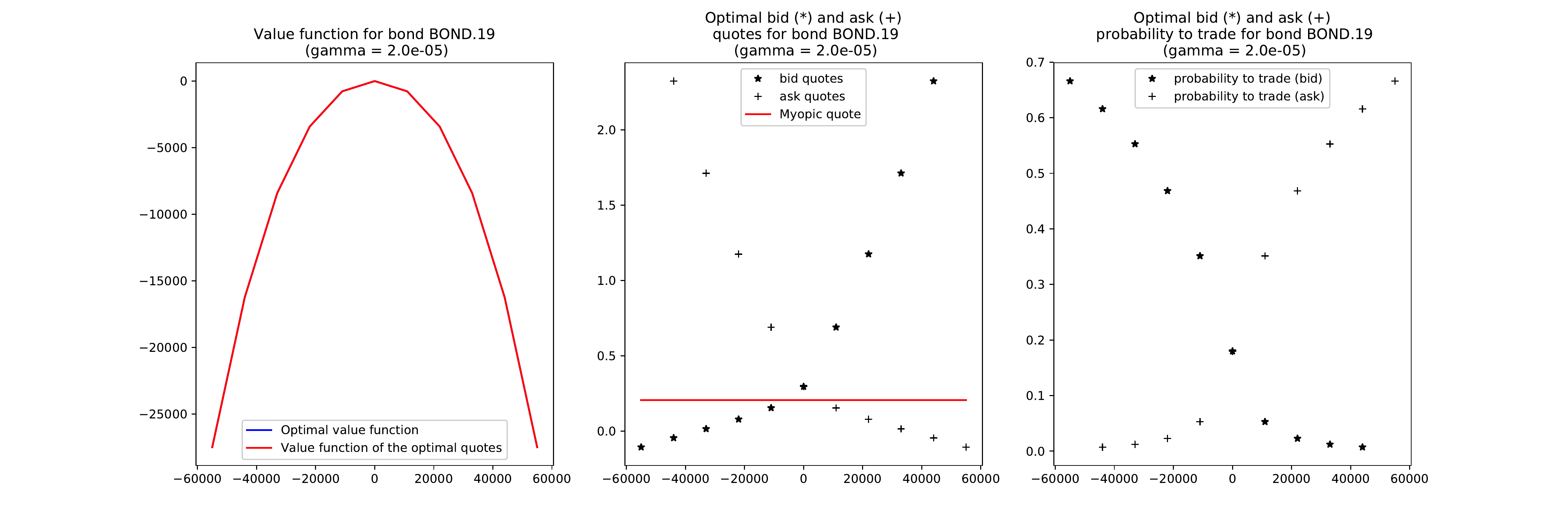}\\
  \includegraphics[width=0.88\textwidth]{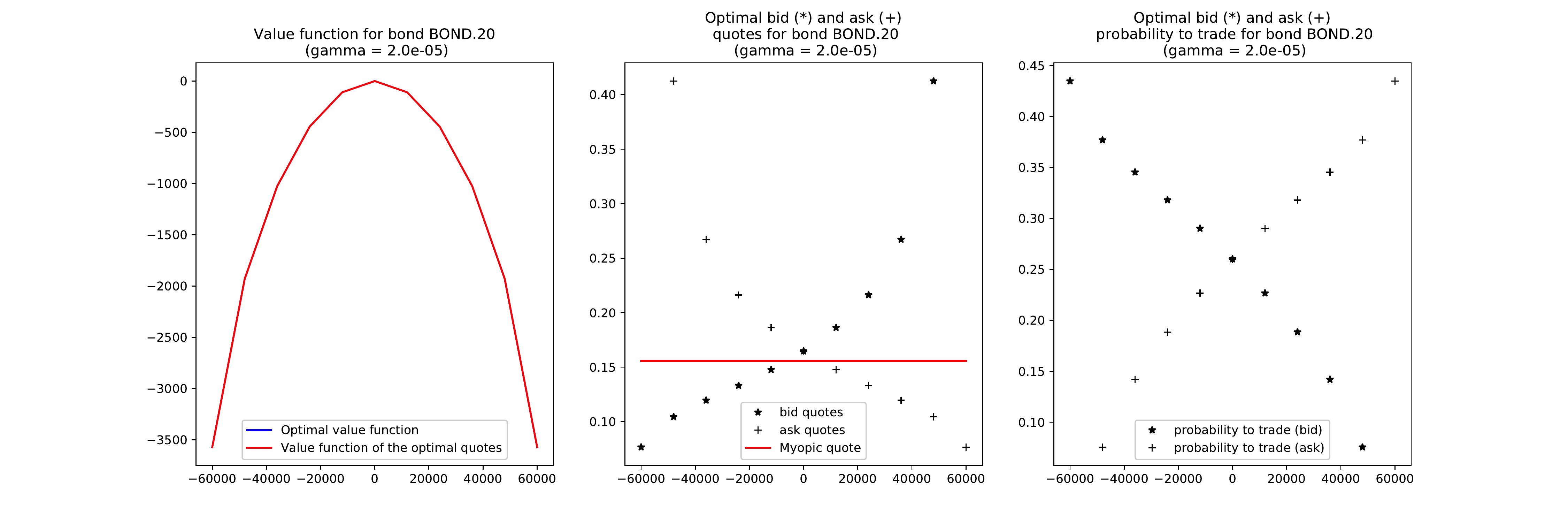}
  \caption{Value functions, optimal quotes, and optimal probabilities to trade with the finite difference approach.}
  \label{pde_square_5}
\end{figure}

As above, for the finite difference approach, we used interpolations of the Hamiltonian functions. The results are documented in Table \ref{R_mean_table_sqr} and Figures \ref{pde_square_1}, \ref{pde_square_2}, \ref{pde_square_3}, \ref{pde_square_4}, and \ref{pde_square_5}.

\begin{table}[H]
\centering
\begin{tabular}{|c|c|}
  \hline
  Bond identifier& Average reward per RFQ \\
  \hline
BOND.1&213.8\\
BOND.2&59\\
BOND.3&404\\
BOND.4&203.1\\
BOND.5&302.2\\
BOND.6&182.6\\
BOND.7&270.2\\
BOND.8&522.7\\
BOND.9&83.2\\
BOND.10&43.2\\
BOND.11&156.1\\
BOND.12&520.8\\
BOND.13&83.1\\
BOND.14&602.2\\
BOND.15&224.3\\
BOND.16&188\\
BOND.17&109.6\\
BOND.18&464.8\\
BOND.19&439.8\\
BOND.20&489\\
  \hline
\end{tabular}
\caption{Average reward per RFQ for the optimal quotes computed with the finite difference method.}
\label{R_mean_table_sqr}
\end{table}

\newpage
For our RL algorithm, we considered again $\gamma = 2\cdot10^{-5}$ and $r = 10^{-4}$. The initial risk limits were set to $3$ times the size of RFQ and gradually augmented with the help of the previously described reverse Matryoshka dolls principle to $5$ times the size of RFQ by the end of the learning process. For both the critic and the actor, we considered neural networks with 2 hidden layers and 10 nodes in each of these layers with ReLU activation functions. The final layer of each neural network contains one node and the activation function is affine in the case of the critic and sigmoid in the case of the actor. As far as pre-training is concerned, we used myopic quotes as described above. Regarding the learning phase, for most bonds we considered 50 steps of the algorithm, i.e. 50 steps of TD learning and 50 steps of policy improvement. For 5 bonds we decided to increase this number of steps to either 100, 150, or 200. At each step, we carried out 1 rollout of length 10000 starting from a zero inventory and 100 additional rollouts of length 100 starting from a random inventory. The noise $\epsilon$ in each rollout is distributed uniformly in $[-0.05, 0.05]$ and we chose the probability limit $\nu = 0.005$. The learning rate for the critic is $\eta = 5\cdot 10^{-8}$ and we used mini-batches of size $70$. The learning rate for the actor is $\tilde{\eta} = 0.01$ and we used mini-batches of size $50$.\\

\begin{figure}[H]
  \centering
  \includegraphics[width=0.8\textwidth]{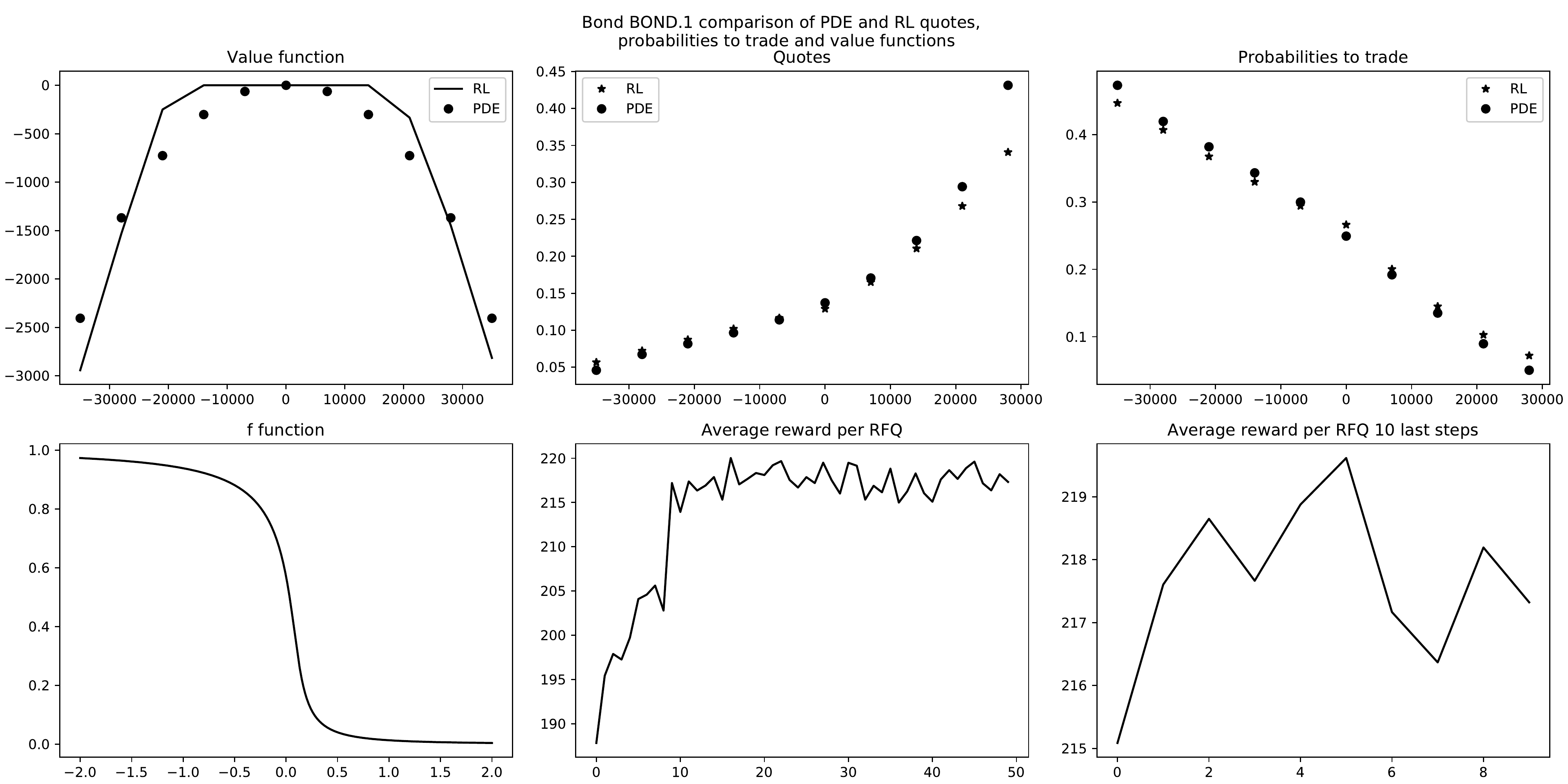}\\
  \caption{Comparison between the two numerical methods.}
  \label{comp_sqr_1}
\end{figure}

\begin{figure}[H]
  \centering
  \includegraphics[width=0.8\textwidth]{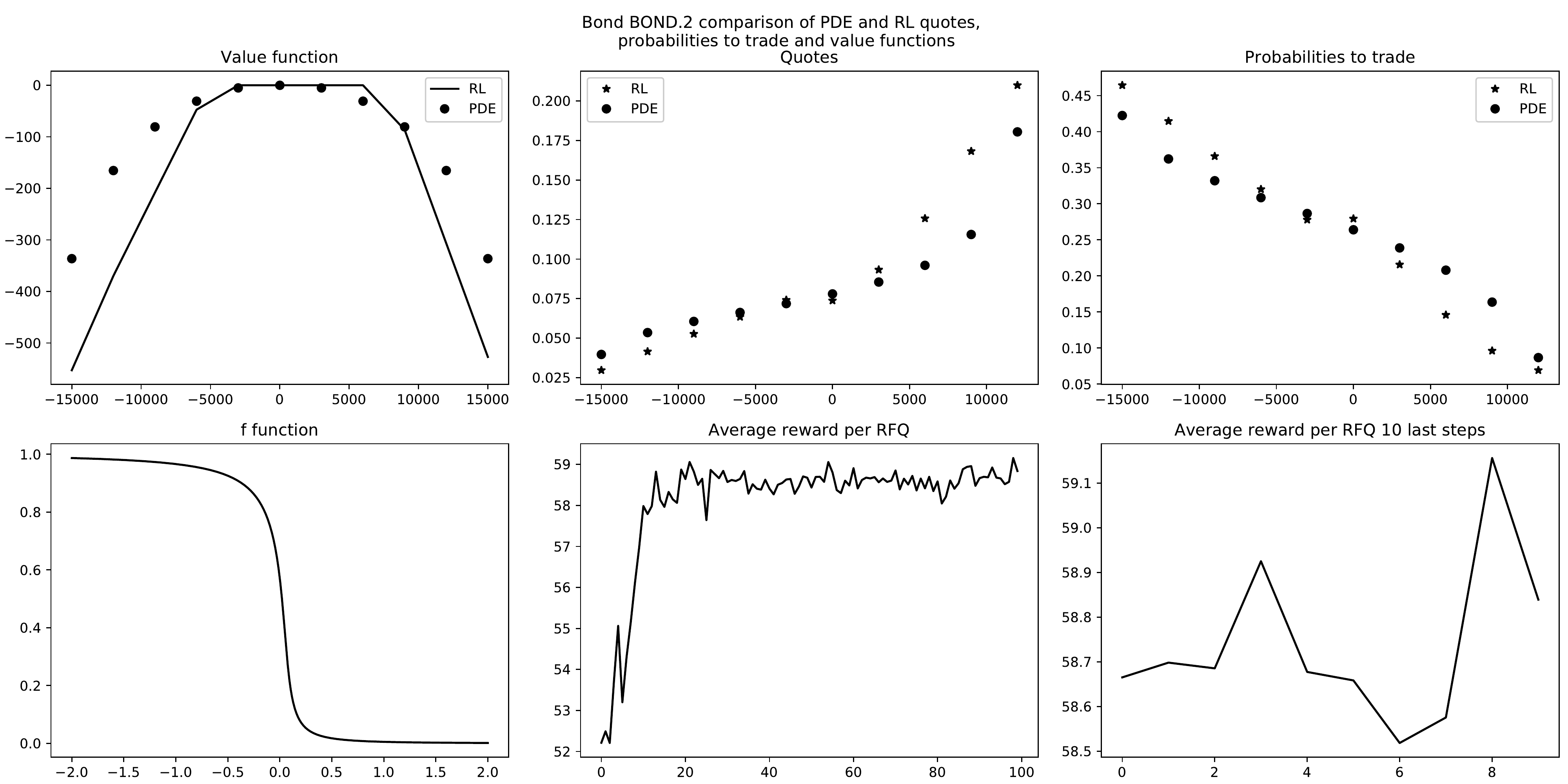}\\
  \includegraphics[width=0.8\textwidth]{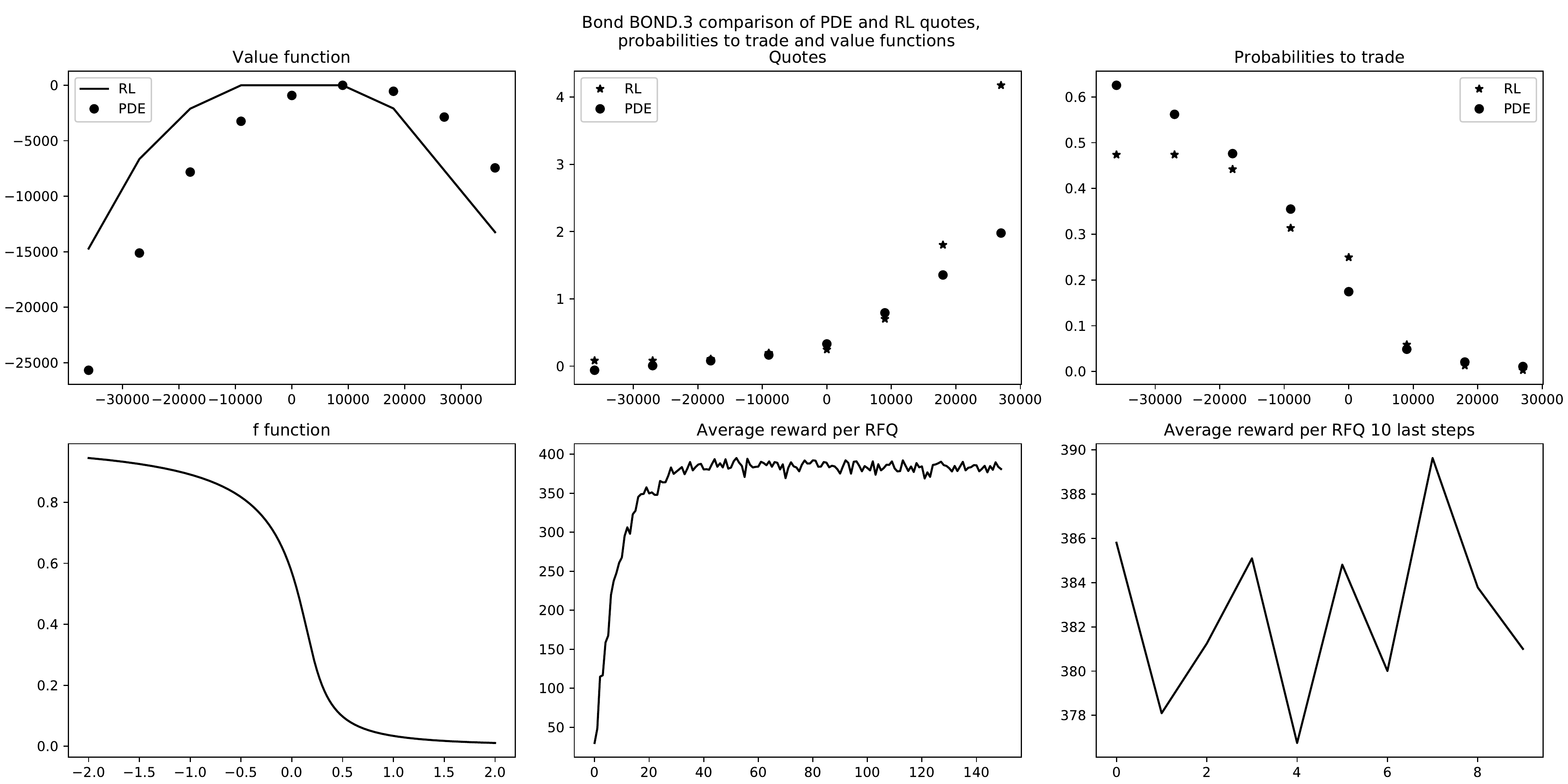}\\
  \includegraphics[width=0.8\textwidth]{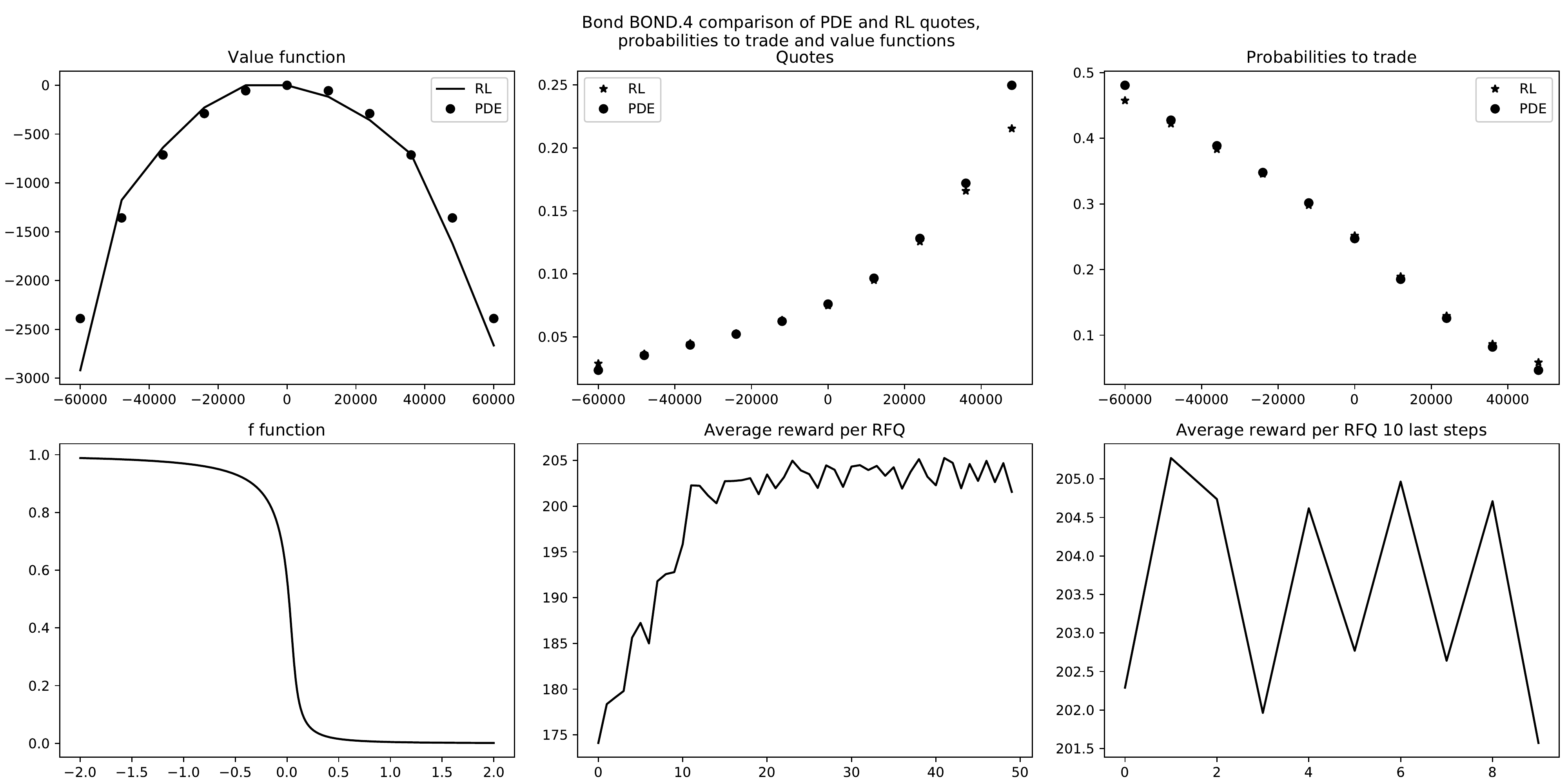}\\
  \caption{Comparison between the two numerical methods.}
  \label{comp_sqr_2}
\end{figure}

\begin{figure}[H]
  \centering
  \includegraphics[width=0.8\textwidth]{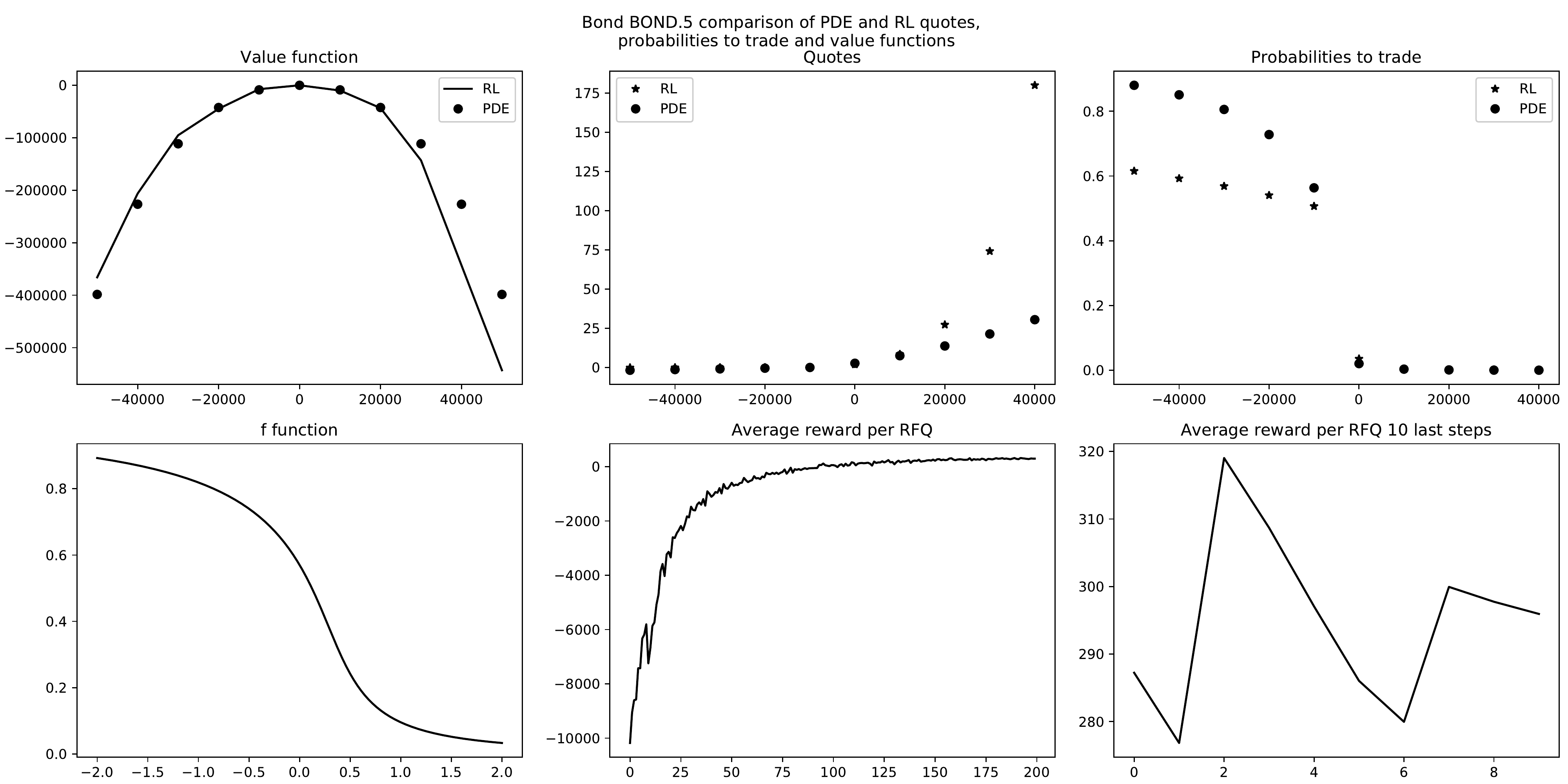}\\
  \includegraphics[width=0.8\textwidth]{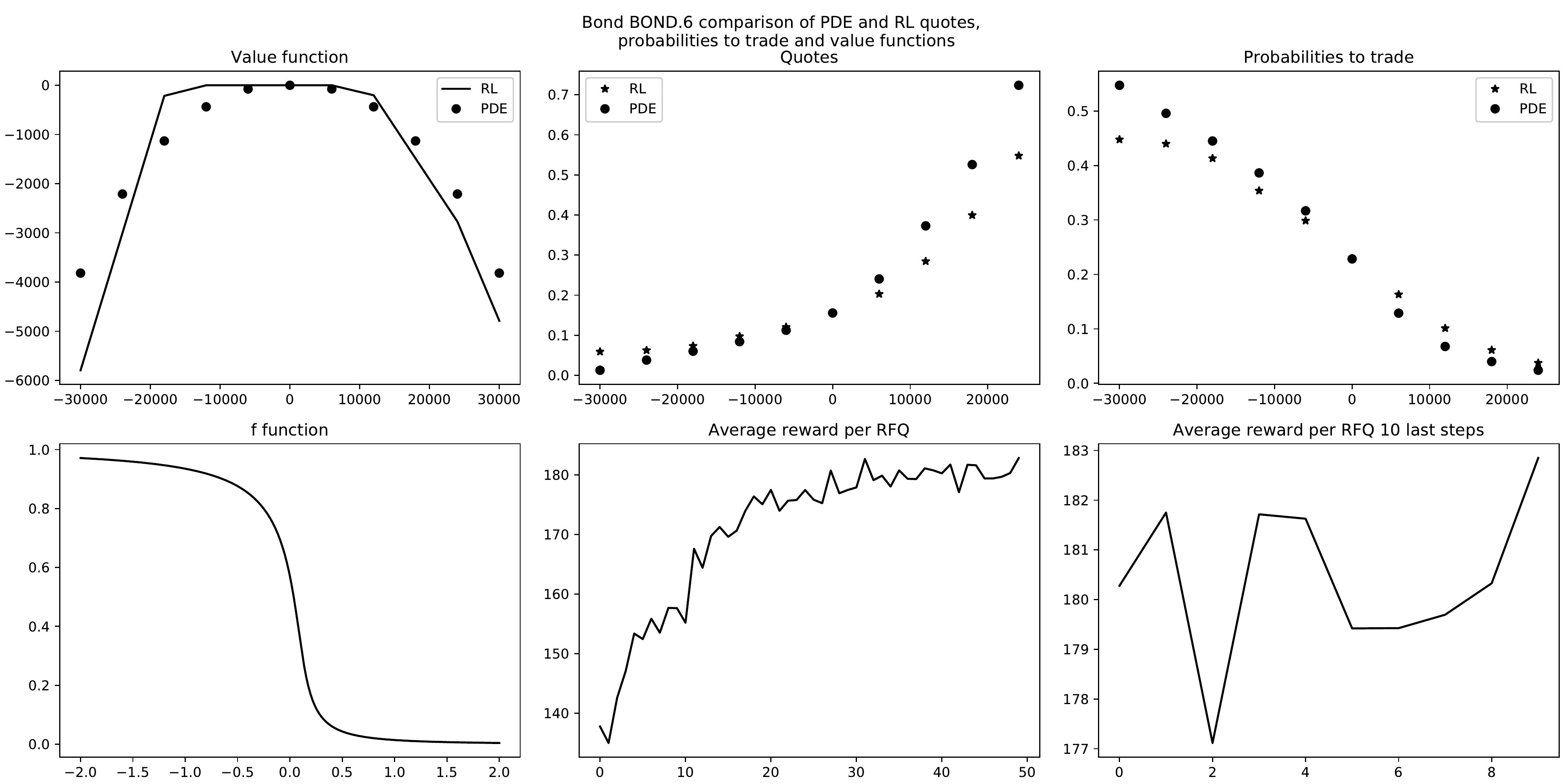}\\
  \includegraphics[width=0.8\textwidth]{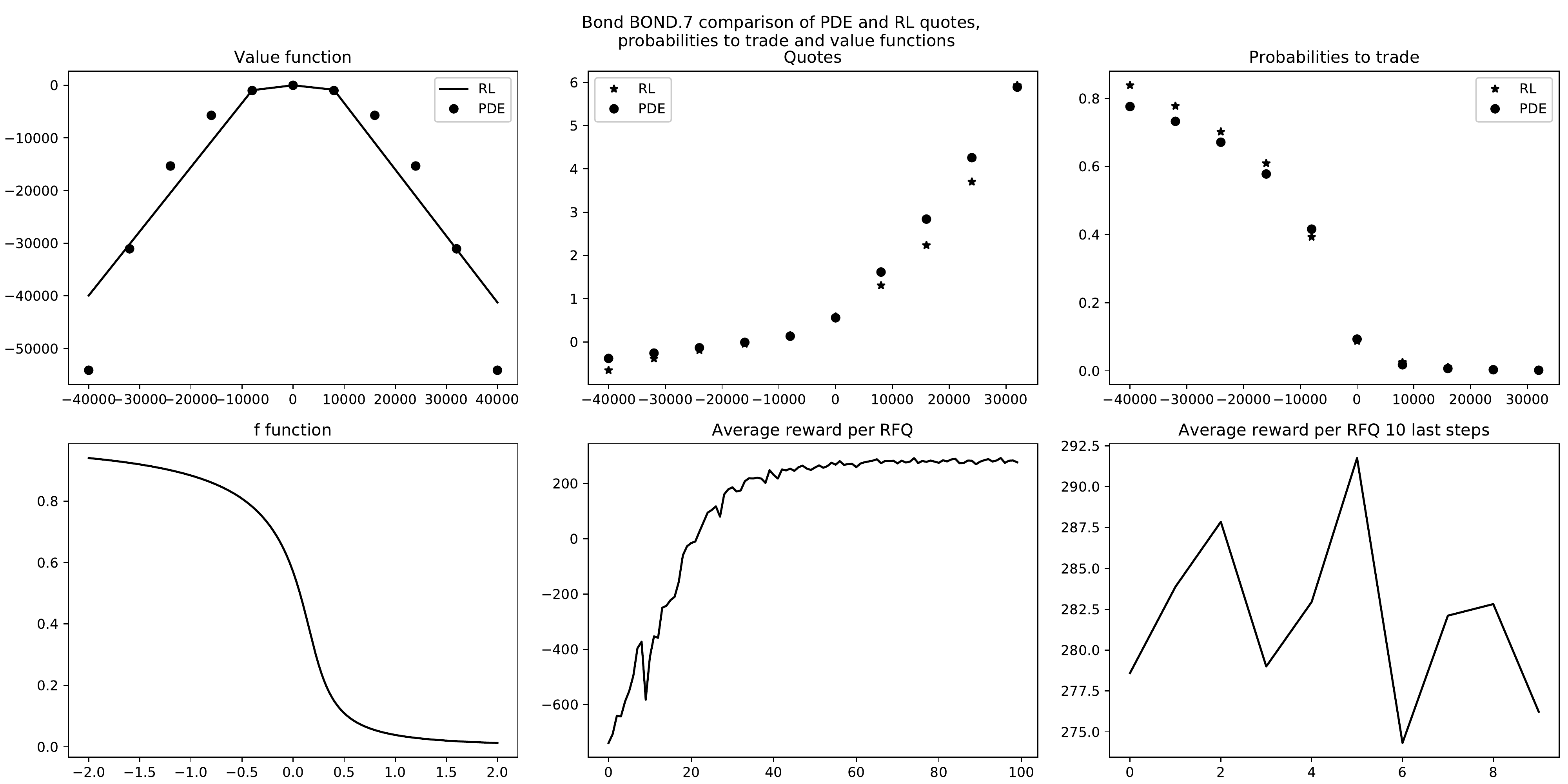}\\
  \caption{Comparison between the two numerical methods.}
  \label{comp_sqr_3}
\end{figure}

\begin{figure}[H]
  \centering
  \includegraphics[width=0.8\textwidth]{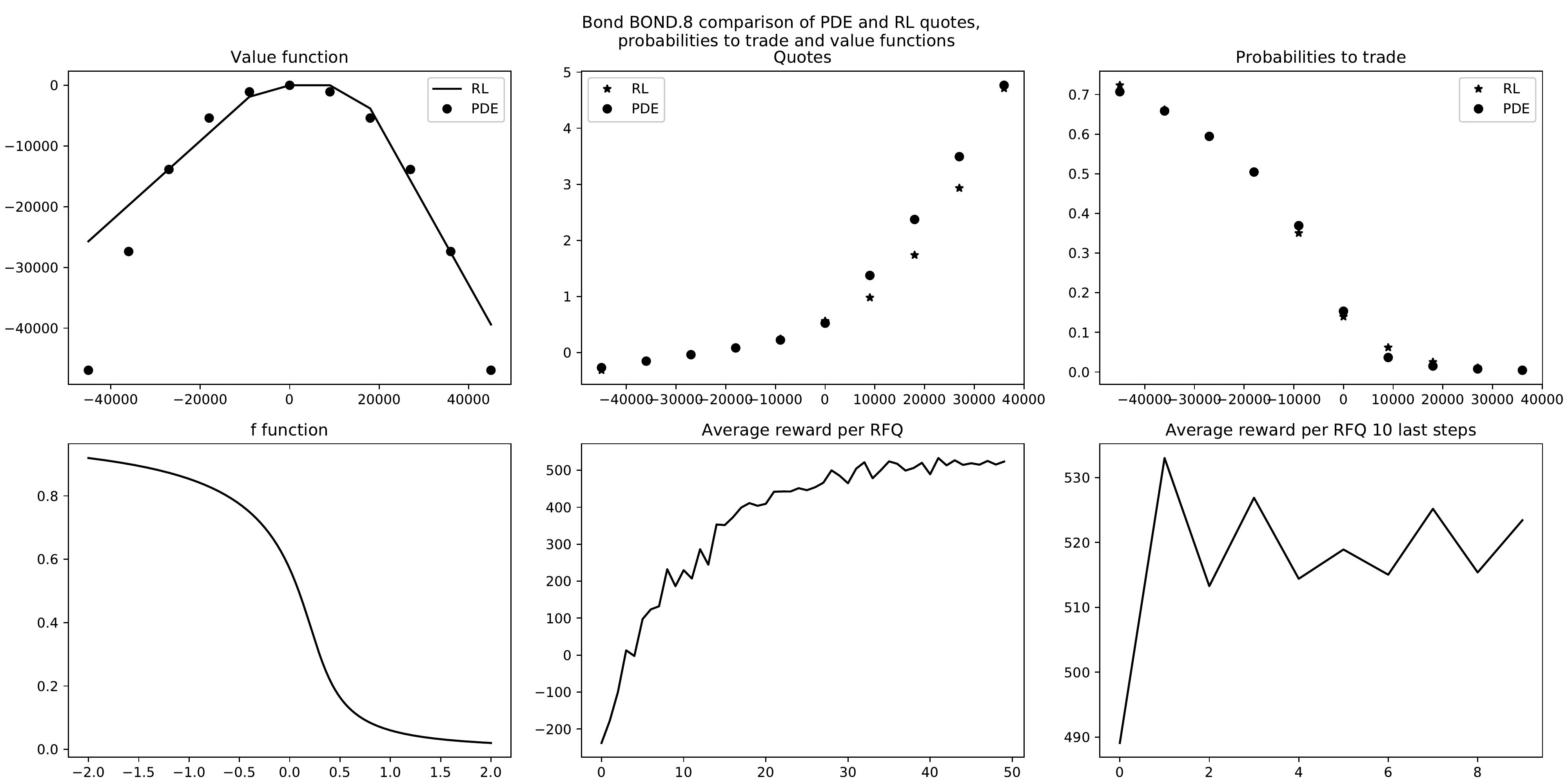}\\
  \includegraphics[width=0.8\textwidth]{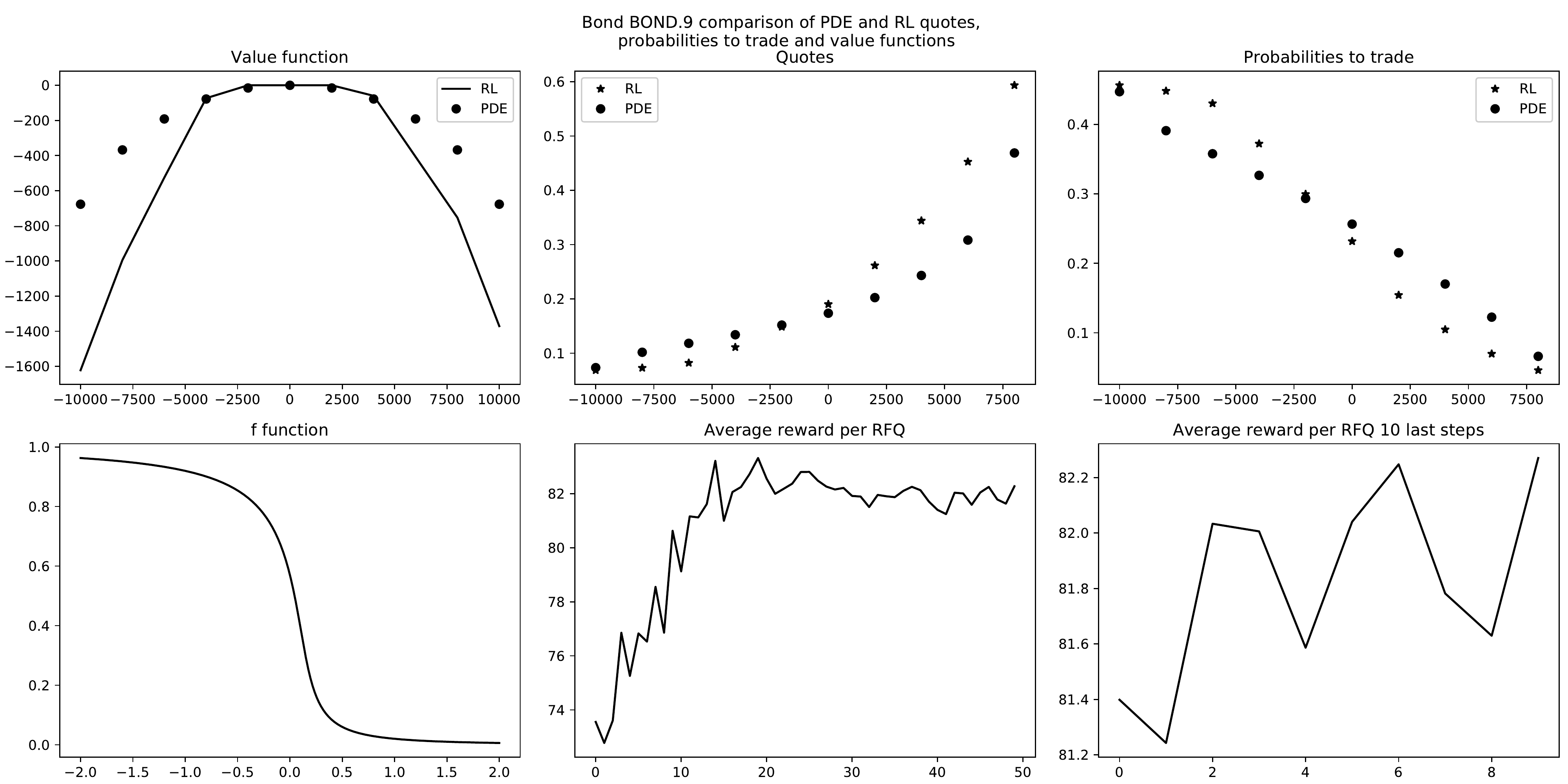}\\
  \includegraphics[width=0.8\textwidth]{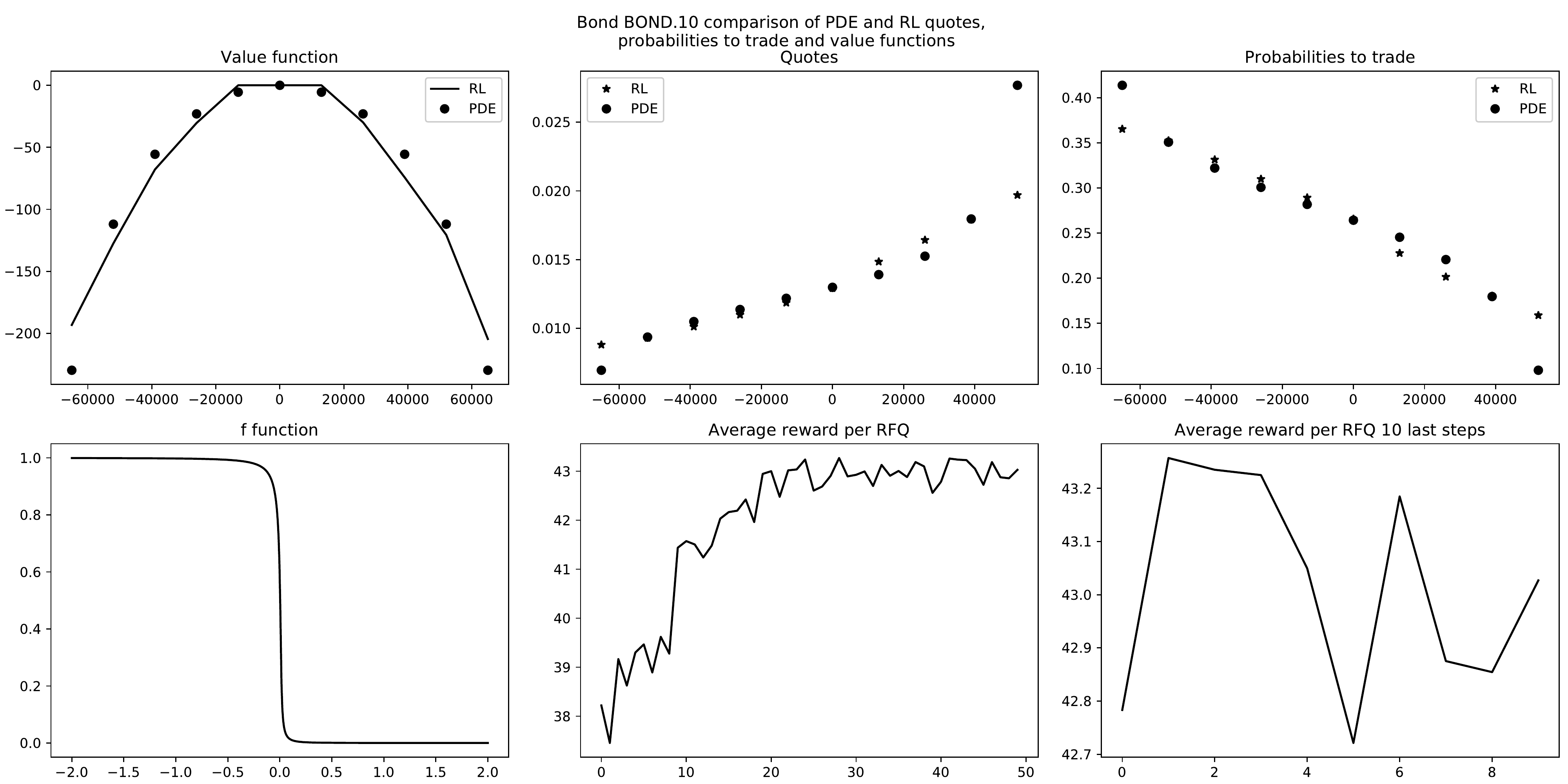}\\
  \caption{Comparison between the two numerical methods.}
  \label{comp_sqr_4}
\end{figure}

\begin{figure}[H]
  \centering
  \includegraphics[width=0.8\textwidth]{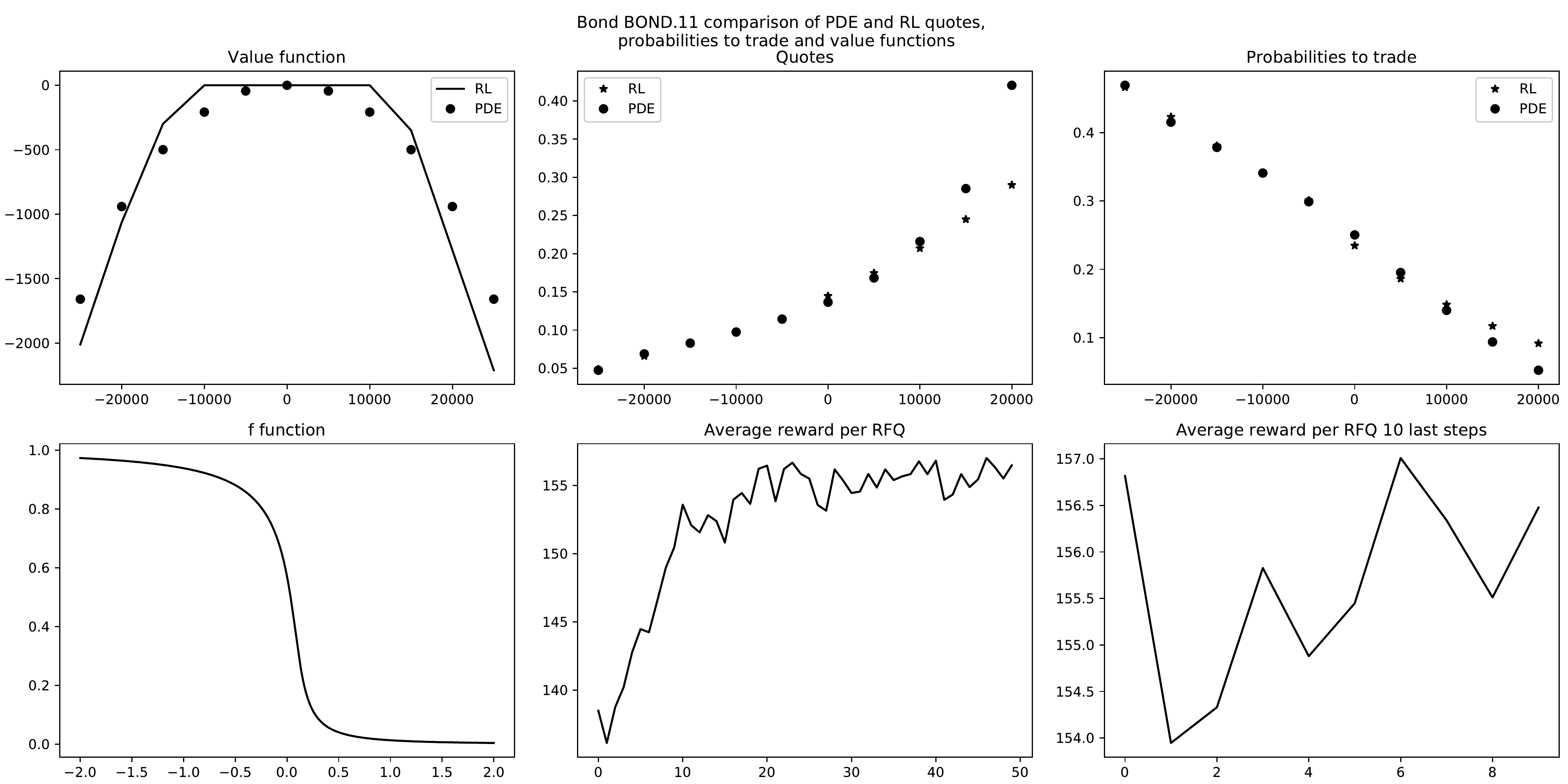}\\
  \includegraphics[width=0.8\textwidth]{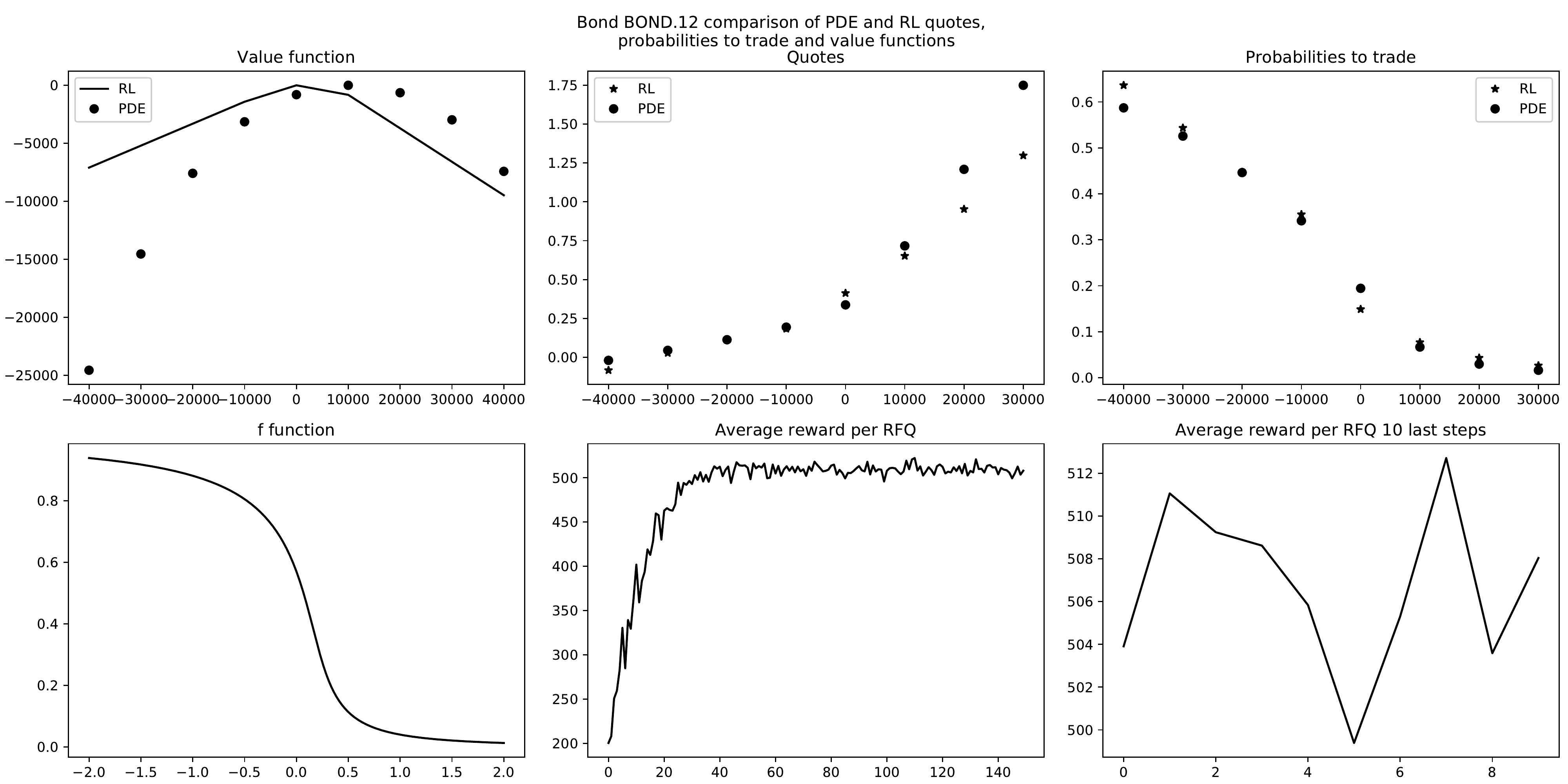}\\
  \includegraphics[width=0.8\textwidth]{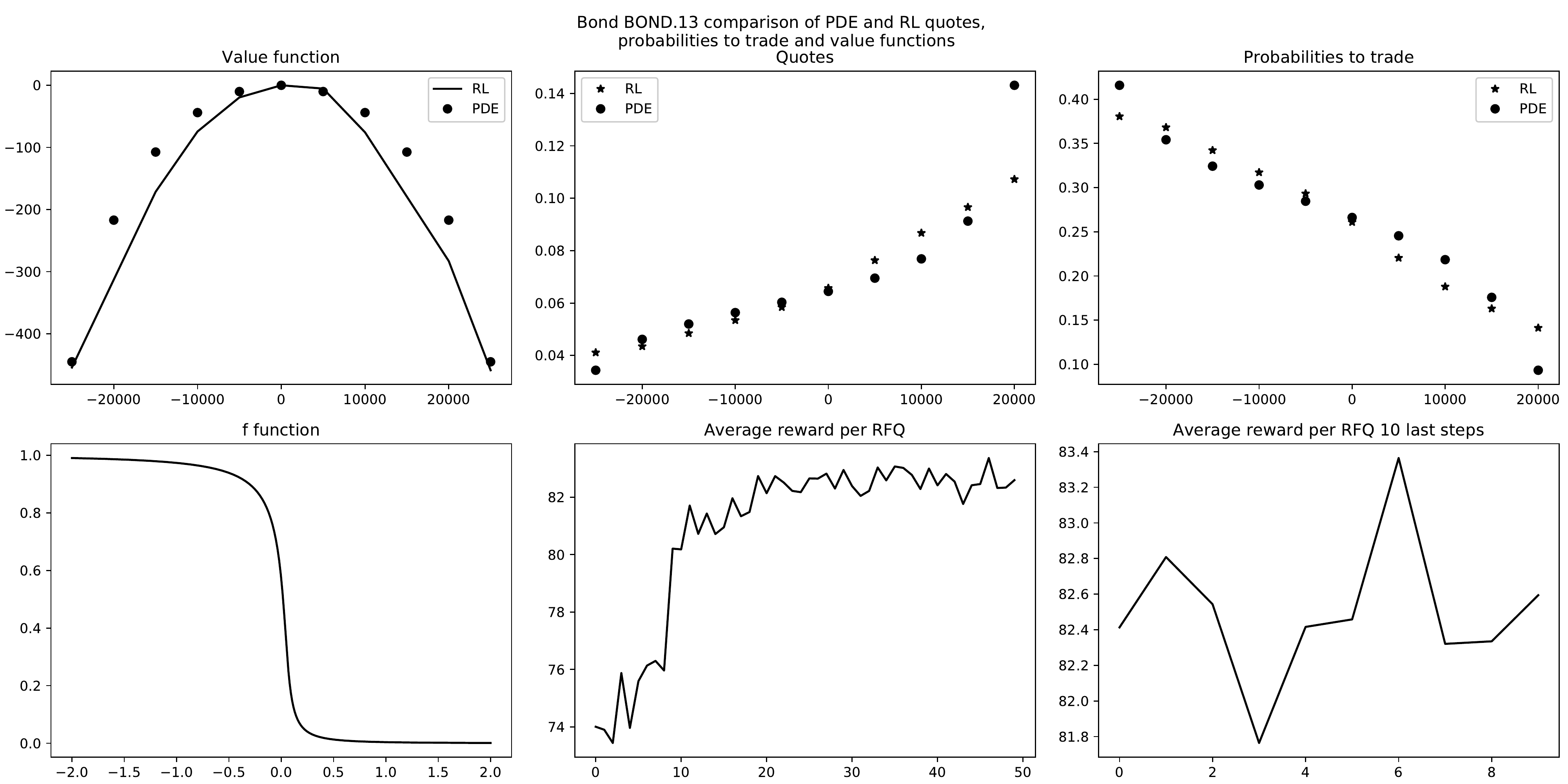}\\
  \caption{Comparison between the two numerical methods.}
  \label{comp_sqr_5}
\end{figure}

\begin{figure}[H]
  \centering
  \includegraphics[width=0.8\textwidth]{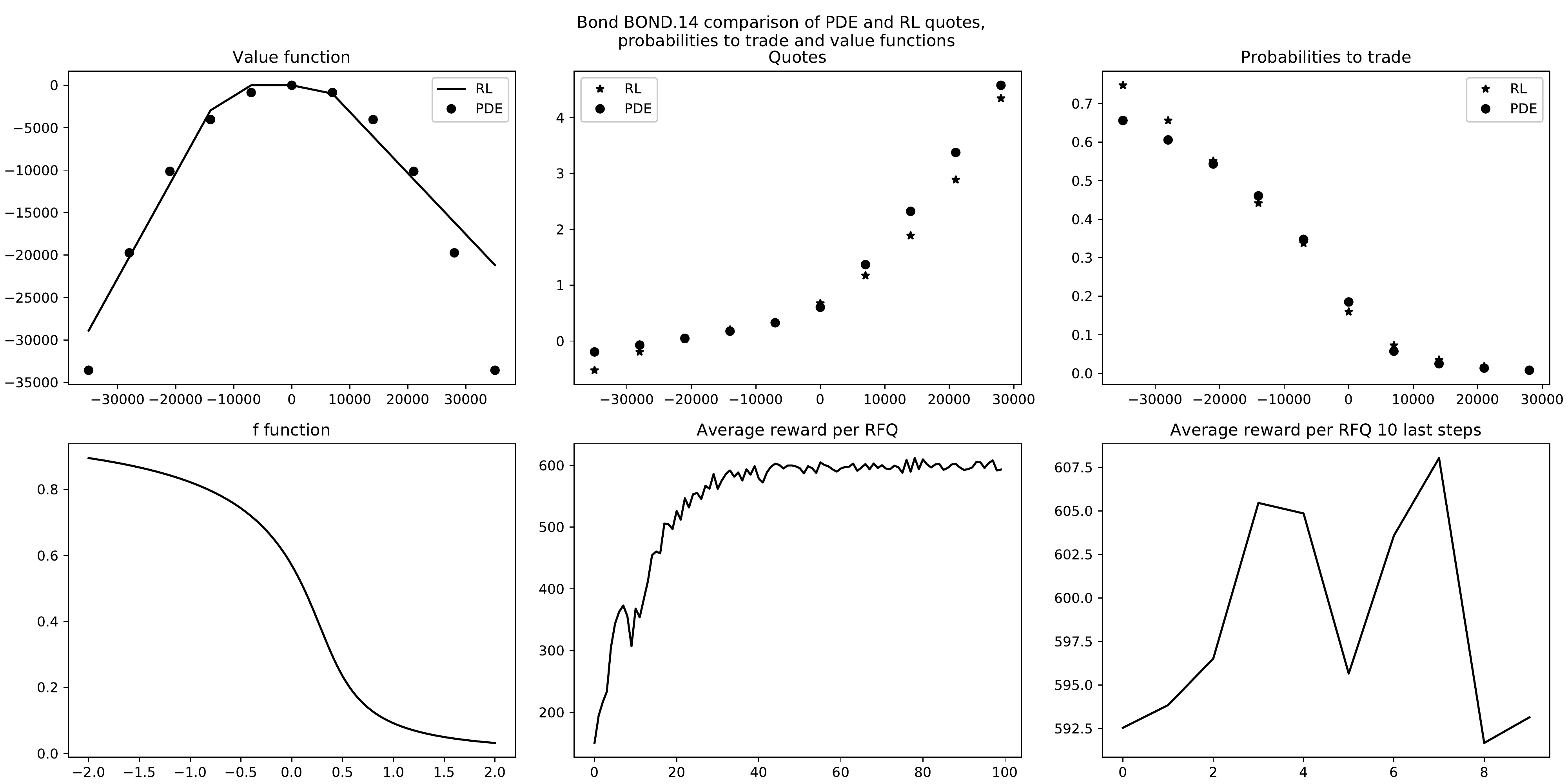}\\
  \includegraphics[width=0.8\textwidth]{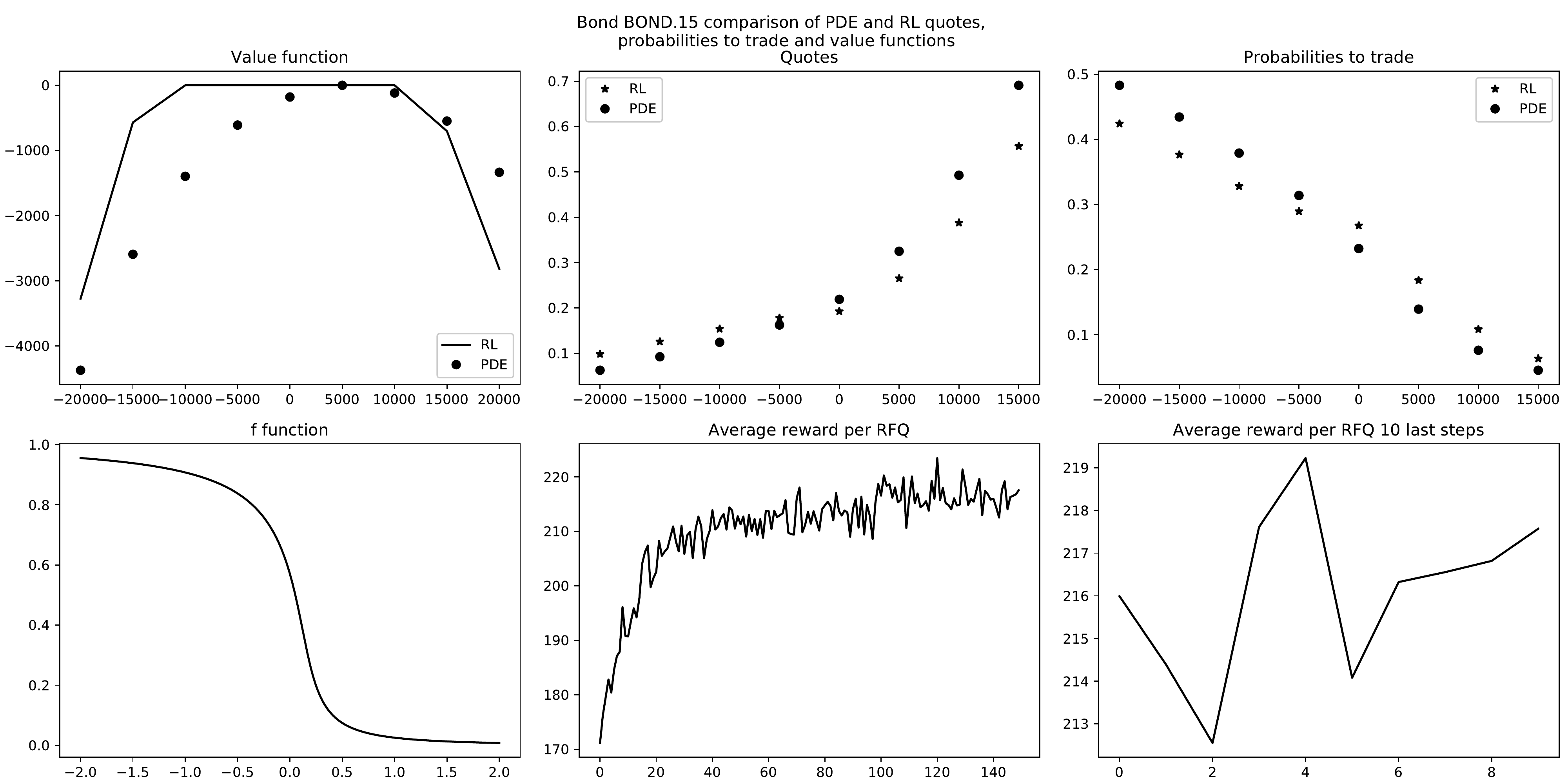}\\
  \includegraphics[width=0.8\textwidth]{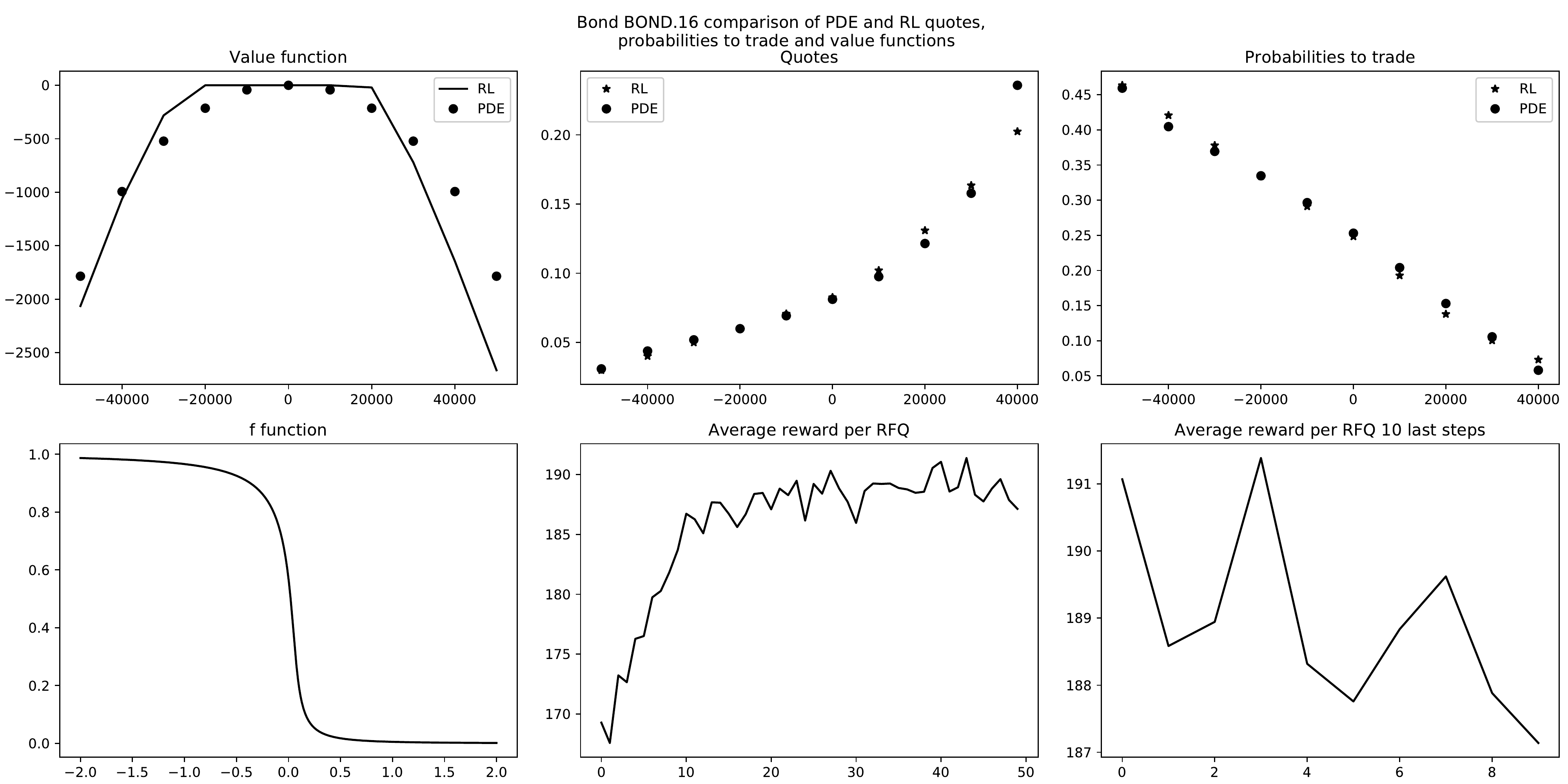}\\
  \caption{Comparison between the two numerical methods.}
  \label{comp_sqr_6}
\end{figure}

\begin{figure}[H]
  \centering
  \includegraphics[width=0.8\textwidth]{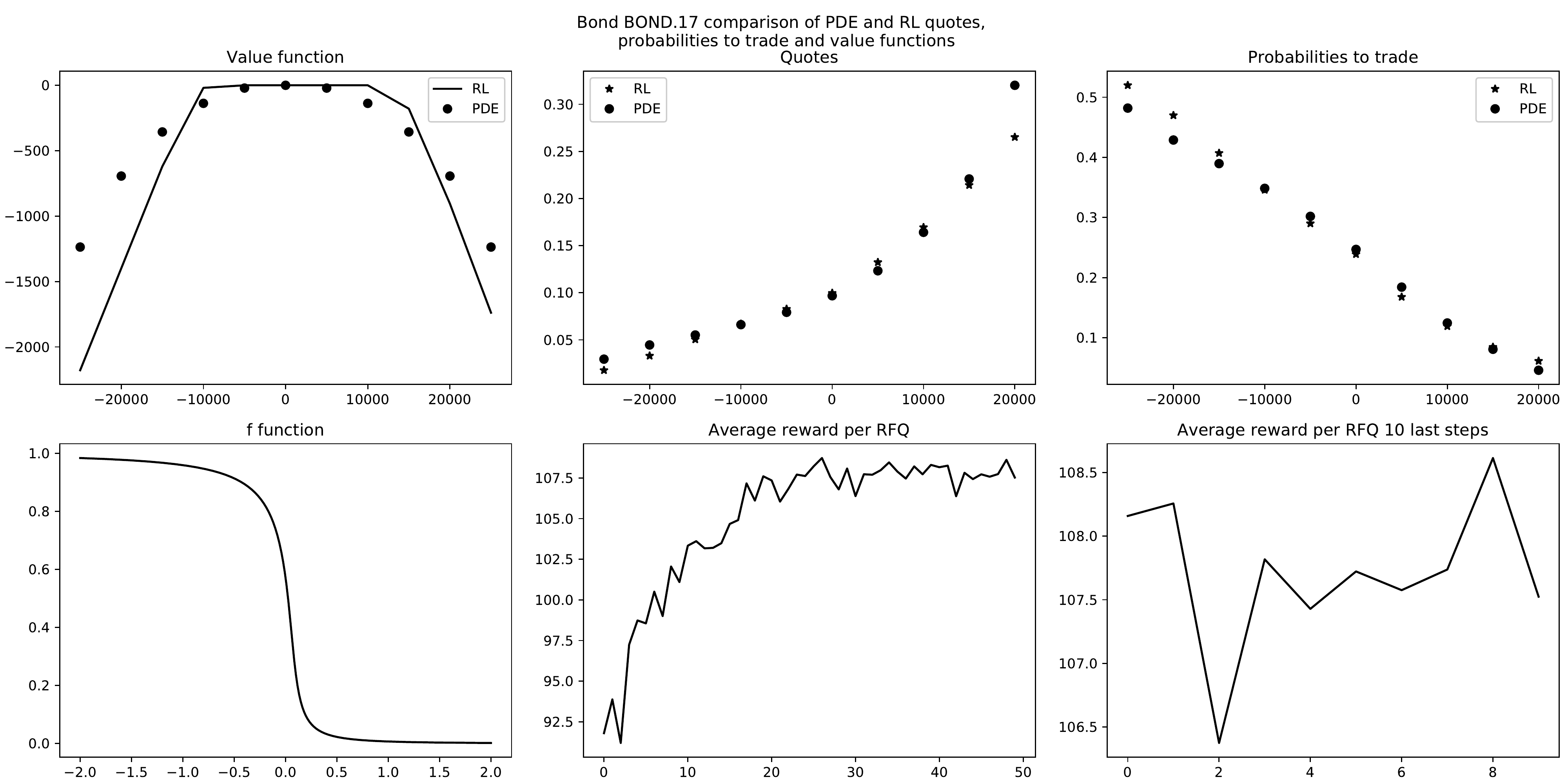}\\
  \includegraphics[width=0.8\textwidth]{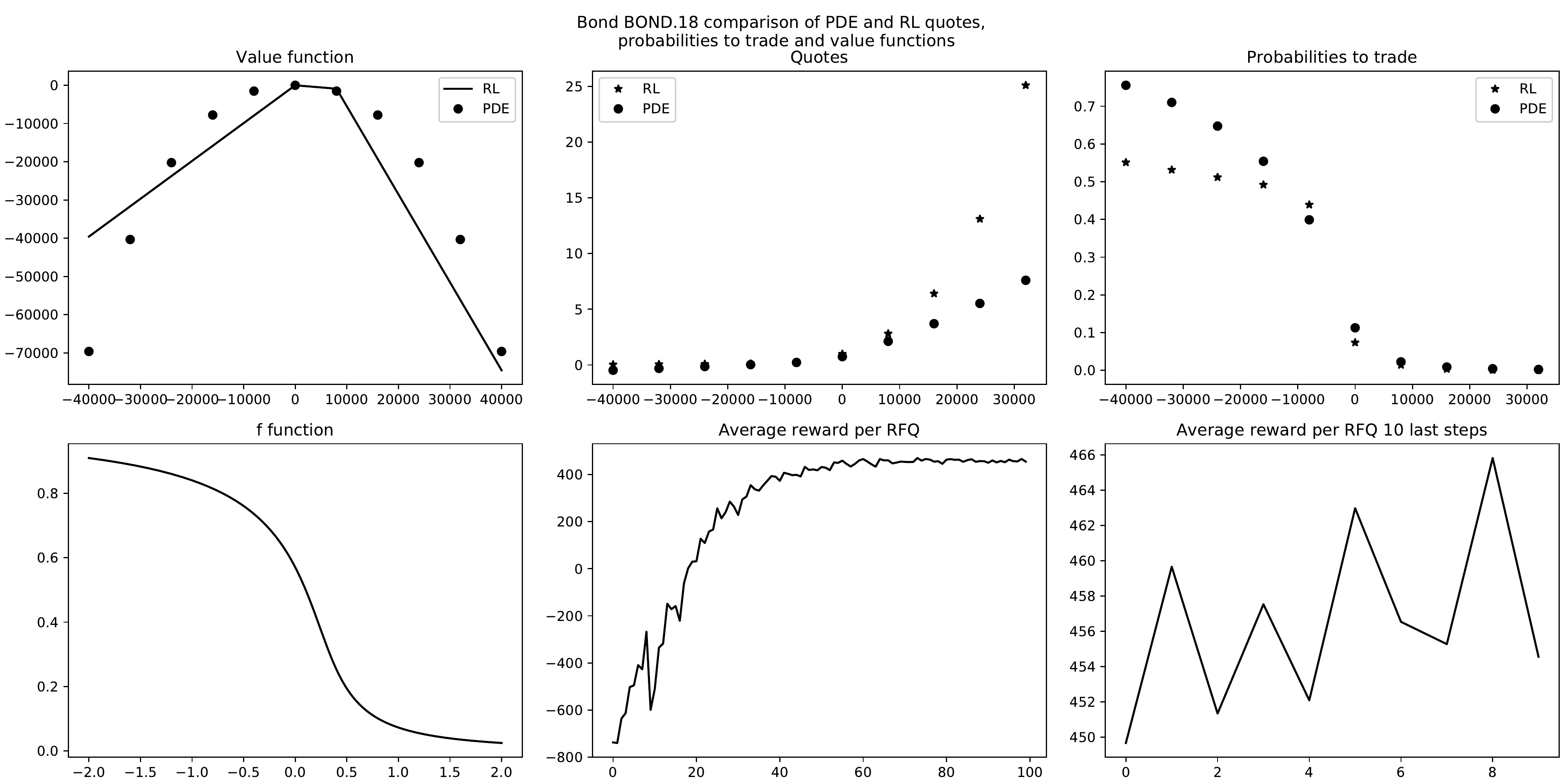}\\
  \includegraphics[width=0.8\textwidth]{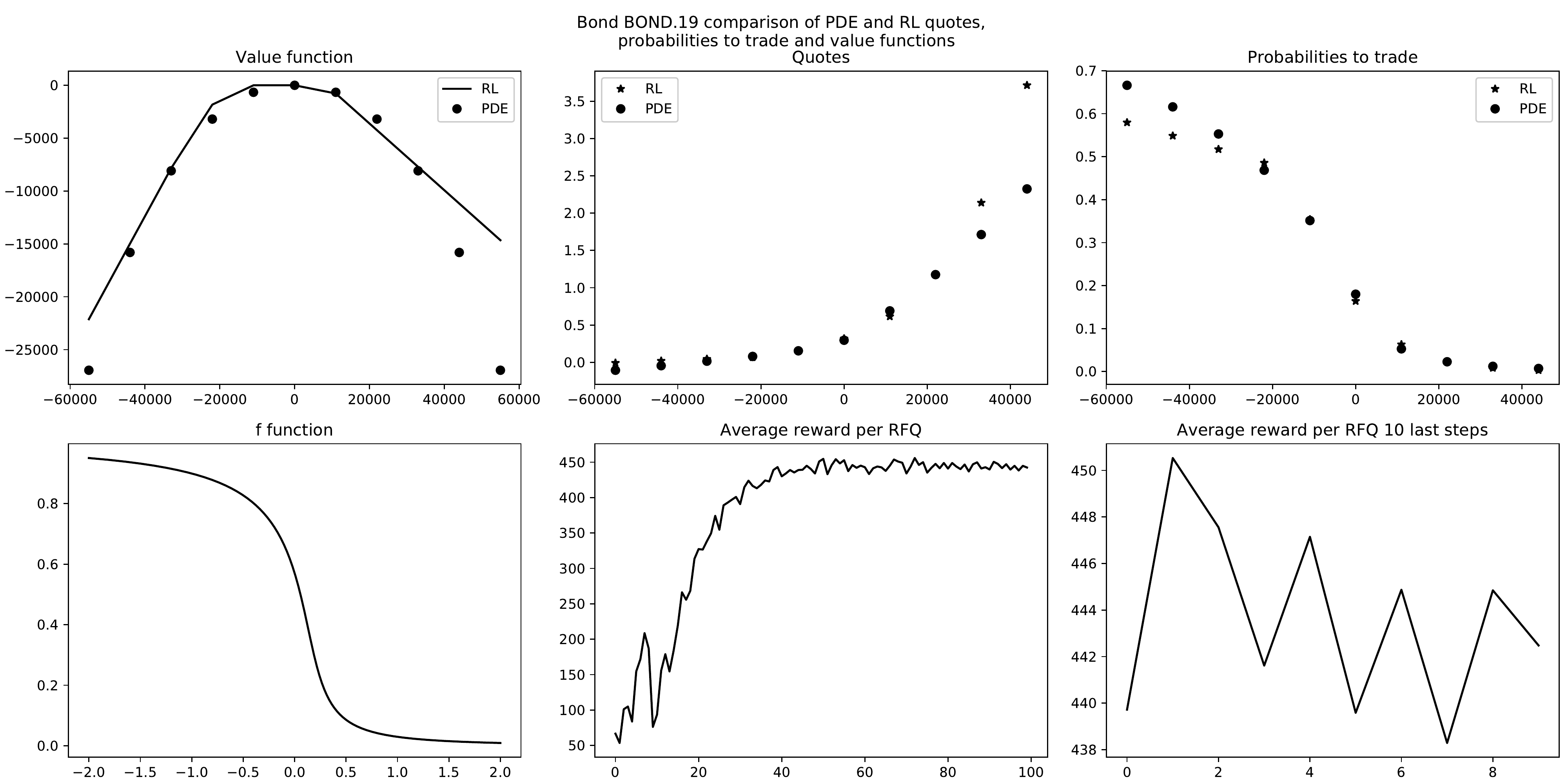}\\
  \caption{Comparison between the two numerical methods.}
  \label{comp_sqr_7}
\end{figure}

\begin{figure}[H]
  \centering
  \includegraphics[width=0.8\textwidth]{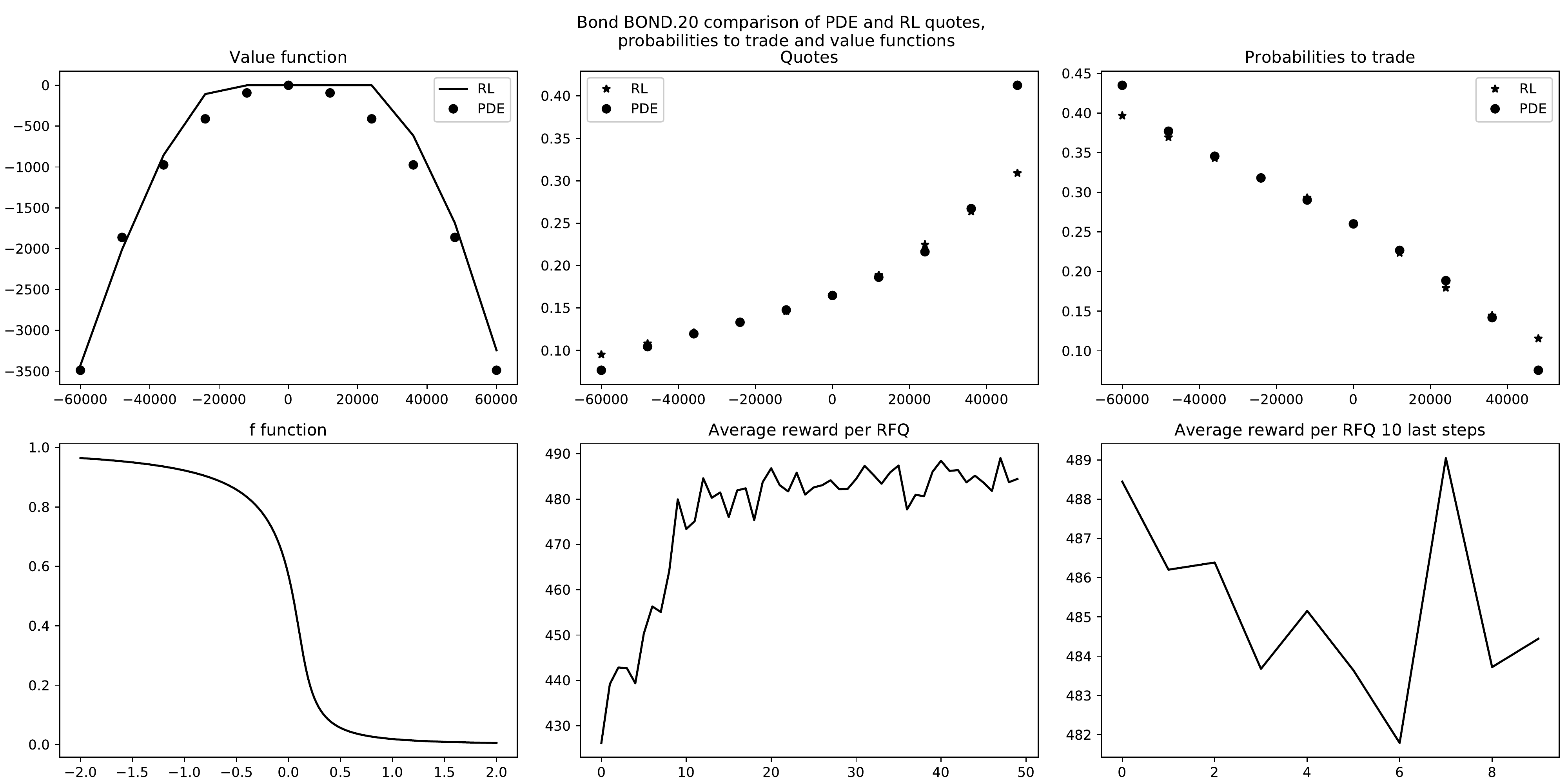}\\
  \caption{Comparison between the two numerical methods.}
  \label{comp_sqr_8}
\end{figure}

The results are shown in Figures \ref{comp_sqr_1}, \ref{comp_sqr_2}, \ref{comp_sqr_3}, \ref{comp_sqr_4}, \ref{comp_sqr_5}, \ref{comp_sqr_6}, \ref{comp_sqr_7}, and \ref{comp_sqr_8} along with the comparison between the finite difference method and our RL algorithm.\\


As it can be easily seen, the results with the two methods are often very close in terms of probability to trade (though sometimes not in terms of quotes and value functions) and they are always comparable when it comes to the average reward per RFQ (see the values in the bottom right panel of the figures and the values in Table 5). It should be noticed that the quality of the approximation in the case of the penalty function considered in this section is not as good as in the case of the penalty function of Section \ref{sqrt}, especially for large values of inventories. This effect is related to the fact that the variance penalty is much stronger than the standard deviation penalty which makes large inventory holdings more unfavorable and therefore more rarely visited over the course of the learning process. Nevertheless, in terms of performance, our algorithm is competitive for the single-bond case. The question is now to see how it scales to the multi-bond case for the variance penalty function.

\subsubsection{2-bond cases}

As above, in order to validate our RL approach in the multi-bond case, we first consider a case with two bonds and compare the approximation obtained with the RL approach to the approximation obtained with the finite difference approach. We want indeed to verify whether our RL approach captures the influence of the correlation between bond prices.\\

We start with the case of a market maker in charge of BOND.1 and BOND.6.\\

For our RL algorithm, we considered $\gamma = 2\cdot10^{-5}$ and $r = 10^{-4}$. The risk limits were set to $3$ times the RFQ size at the beginning of the learning process and were increased every $50$ steps by the RFQ size for each bond according to the previously described reverse Matryoshka dolls principle until the maximum risk limits equal to $5$ times the RFQ size were reached. For the critic and the actor, we considered neural networks with $2$ hidden layers and $12$ nodes in each of these layers with ReLU activation functions. Again, the final layer of each neural network contains one node and the activation function is affine in the case of the critic and sigmoid in the case of the actor. For the pre-training we used the quotes obtained by the RL algorithm in the single-bond case for each bond.  For the learning phase we considered $500$ steps, i.e. $500$ steps of TD learning and $500$ steps of policy improvement for each bond. At each step we carried out $1$ rollout of length $10000$ starting from a zero inventory and $100$ additional rollouts of length $100$ starting from a random inventory. The noise $\epsilon$ in each rollout is distributed uniformly in $[-0.05, 0.05]$ and we chose the probability limit $\nu = 0.005$. The learning rate for the critic is $\eta=1\cdot10^{-8}$ and we used mini-batches of size $70$. The learning rate for the actor is  $\tilde{\eta} = 0.01$ and we used mini-batches of size $50$. \\

\begin{figure}[H]
  \centering
  \includegraphics[width=0.45\textwidth]{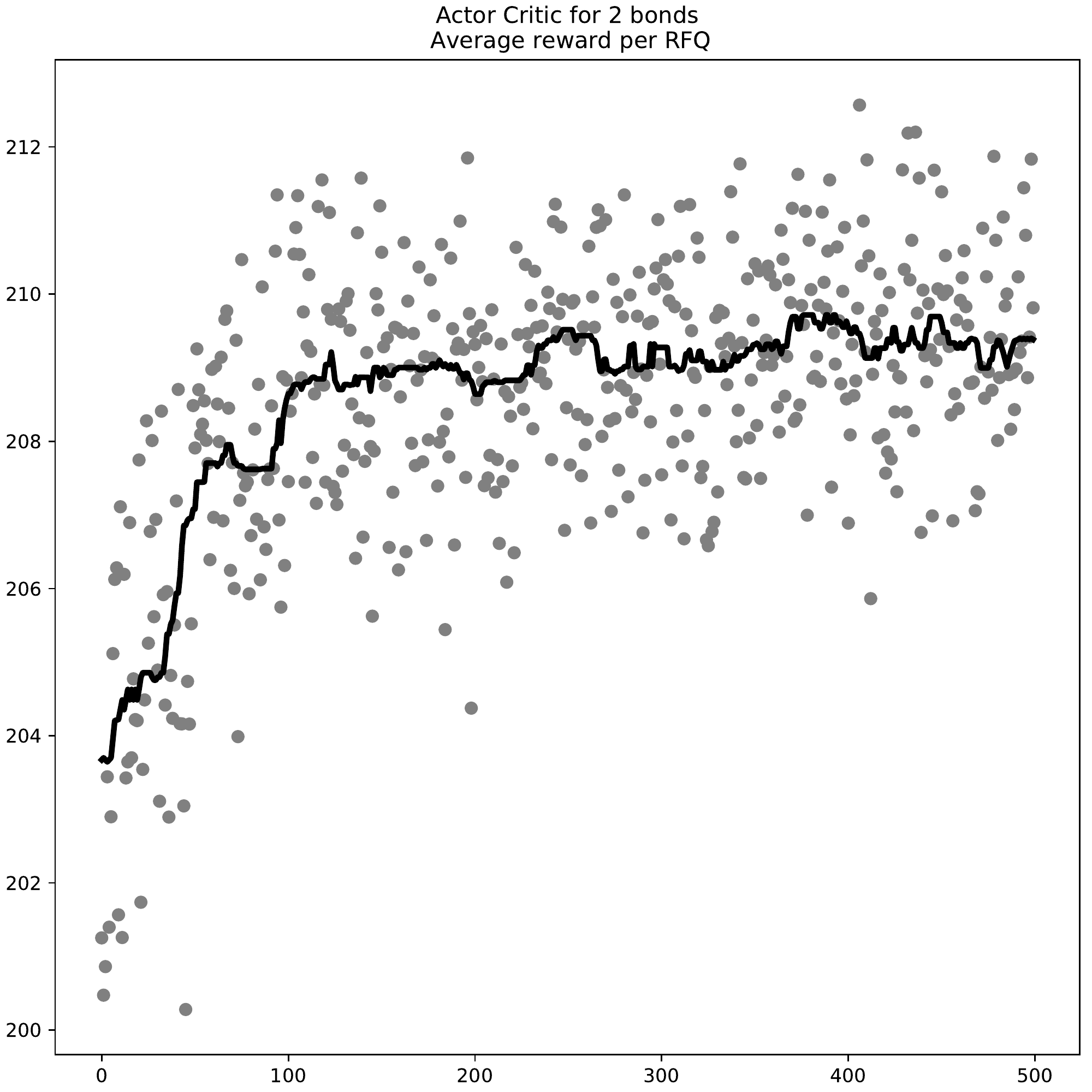}\\
  \caption{Average reward per RFQ -- Learning process for the 2-bond case.}
  \label{rl_2d_sqr}
\end{figure}

The learning curve of the algorithm is plotted in Figure \ref{rl_2d_sqr}.
\vspace{-3mm}

The average reward per RFQ increases during the learning phase. This means that our algorithm takes account of the correlation between the two bonds. Interestingly, the average reward per RFQ obtained when using the optimal quotes computed with the finite difference method of the appendix is~$210.1$. This is almost in line with the average reward per RFQ obtained with our RL algorithm ($\simeq 209$).\\

\newpage

\begin{figure}[H]
  \centering
  \includegraphics[width=0.48\textwidth]{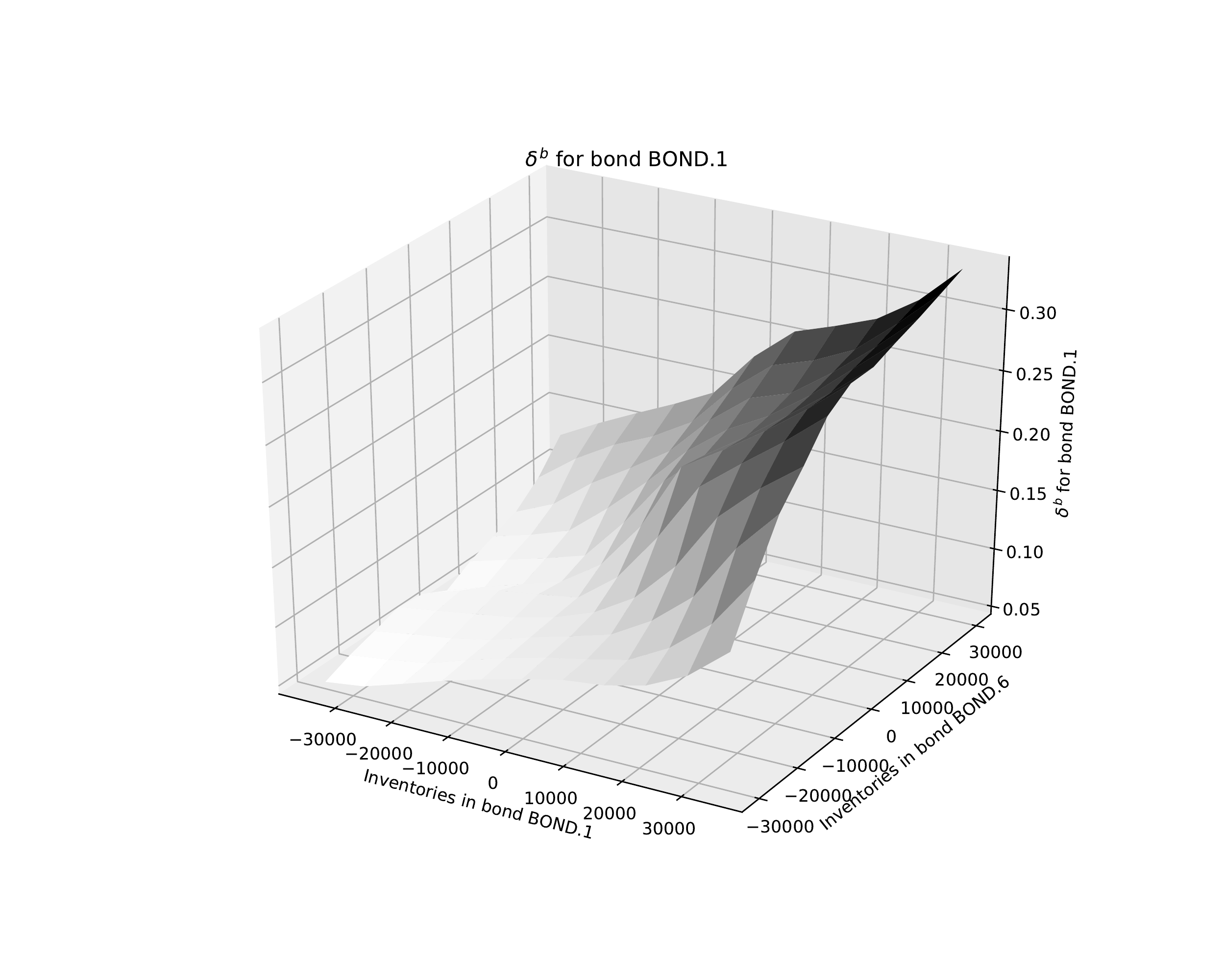}
  \includegraphics[width=0.48\textwidth]{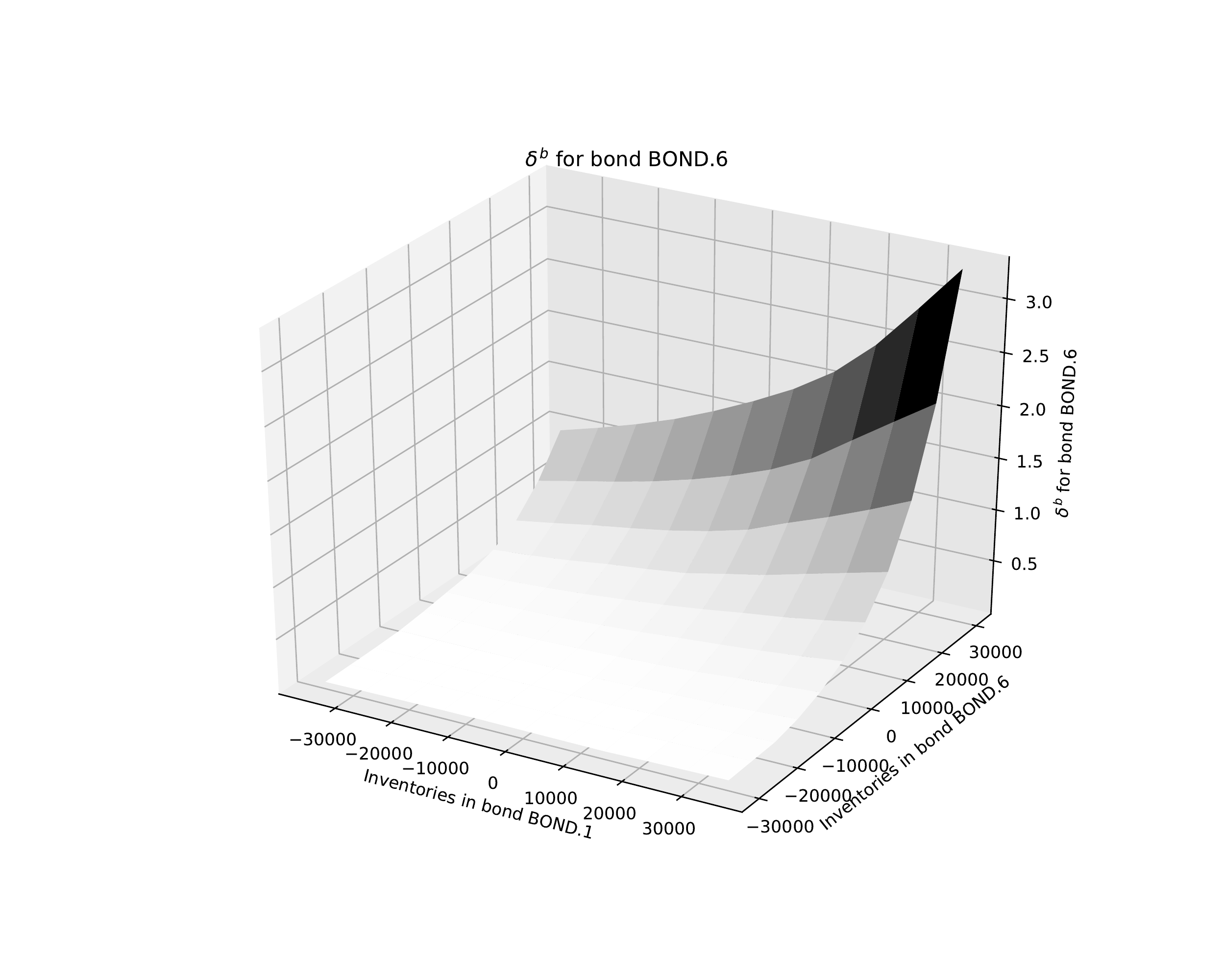}
  \caption{Optimal quotes obtained with our RL algorithm.}
  \label{rl_2d_deltas_sqr}
\end{figure}

\begin{figure}[H]
  \centering
  \includegraphics[width=0.48\textwidth]{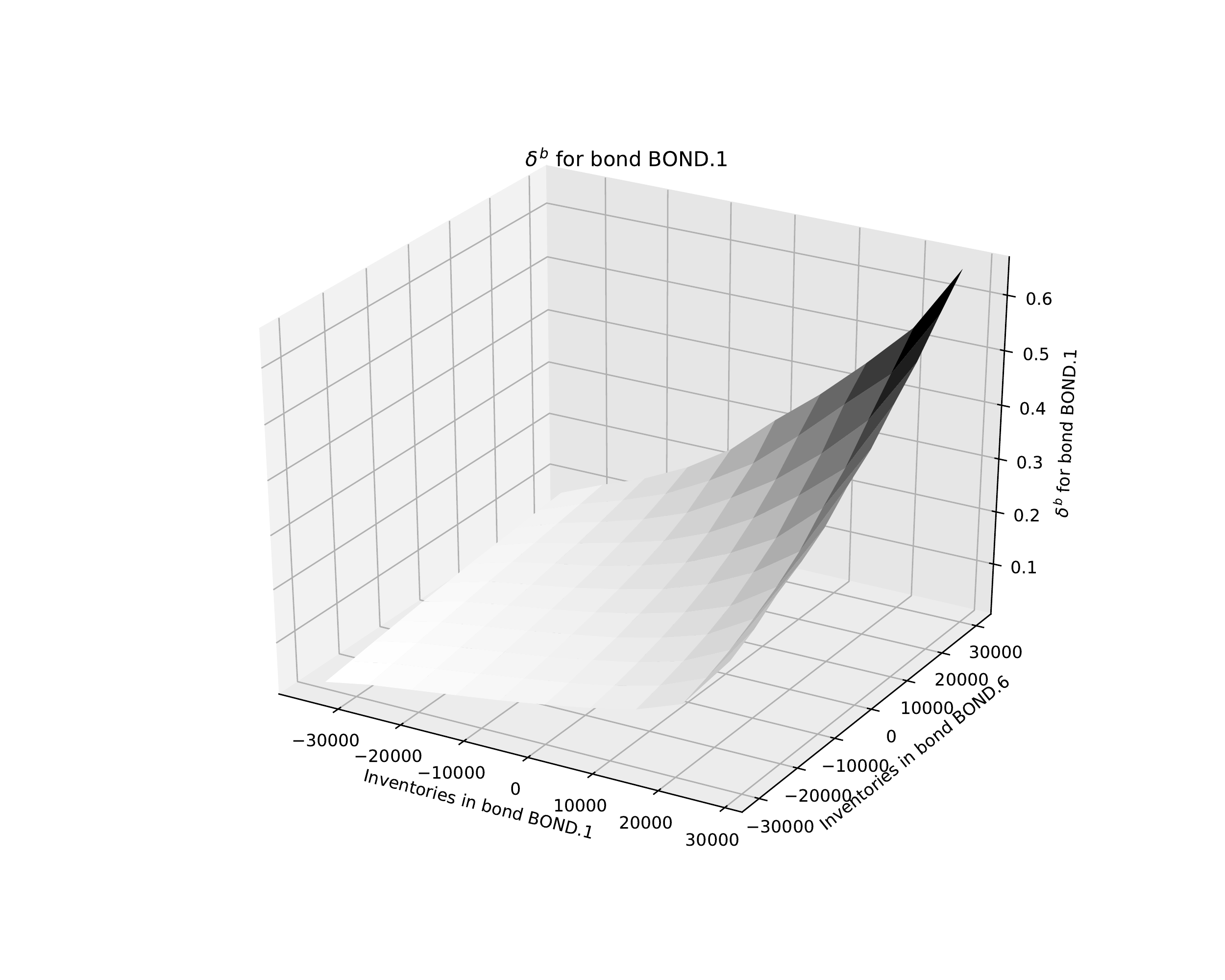}
  \includegraphics[width=0.48\textwidth]{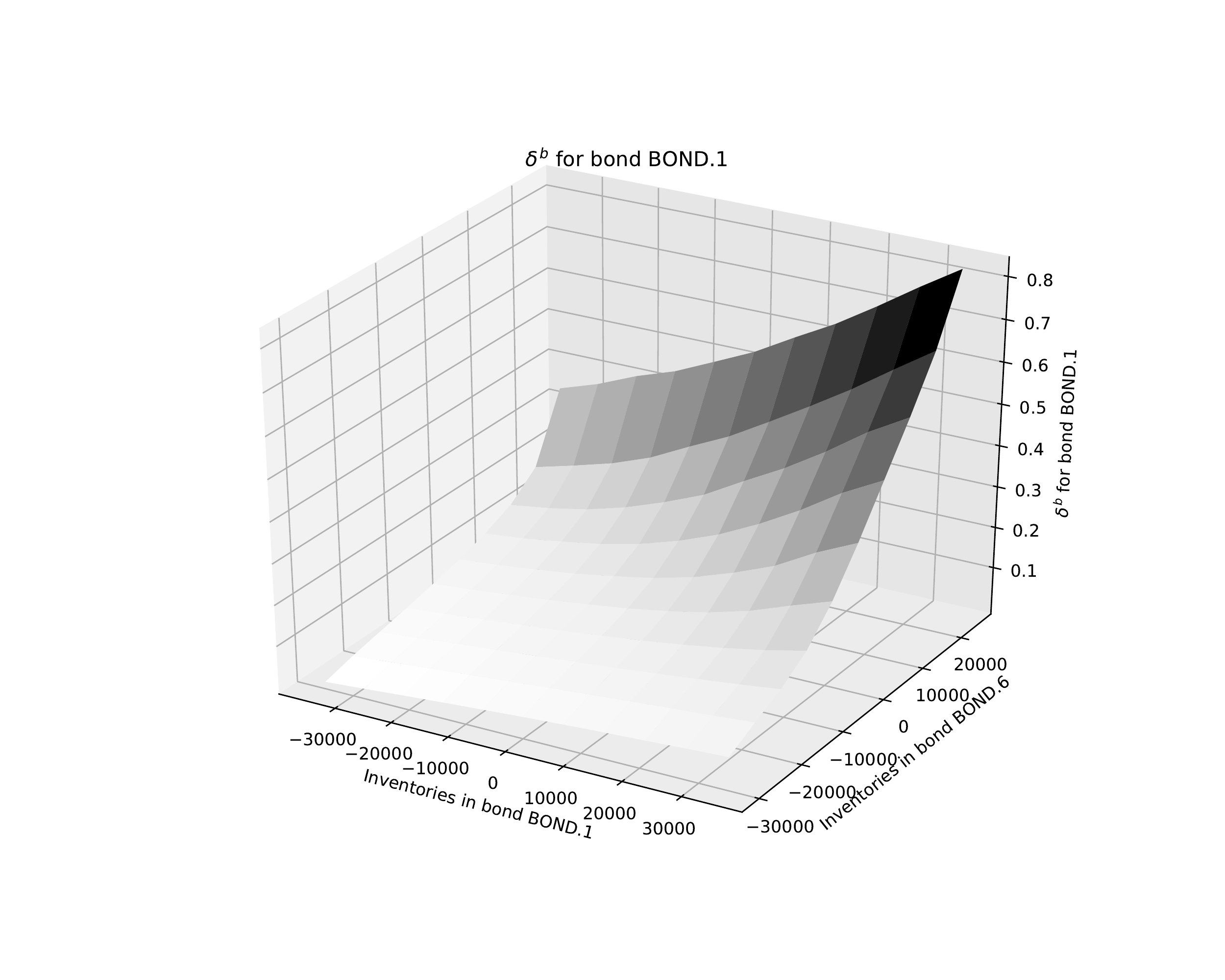}
  \caption{Optimal quotes obtained with our finite difference scheme.}
  \label{th_2d_deltas_sqr}
\end{figure}

We can see in our 2-bond case that the quotes are impacted by the correlation between bonds. In Figure \ref{rl_2d_deltas_sqr}, we plotted the optimal (bid) quotes computed with our RL algorithm. The optimal (bid) quotes computed with our finite difference method are in Figure \ref{th_2d_deltas_sqr}.

We see that the bid quotes are increasing with respect to the inventory in both bonds, as expected because of the positive correlation.\\

\begin{figure}[H]
  \centering
  \includegraphics[width=0.45\textwidth]{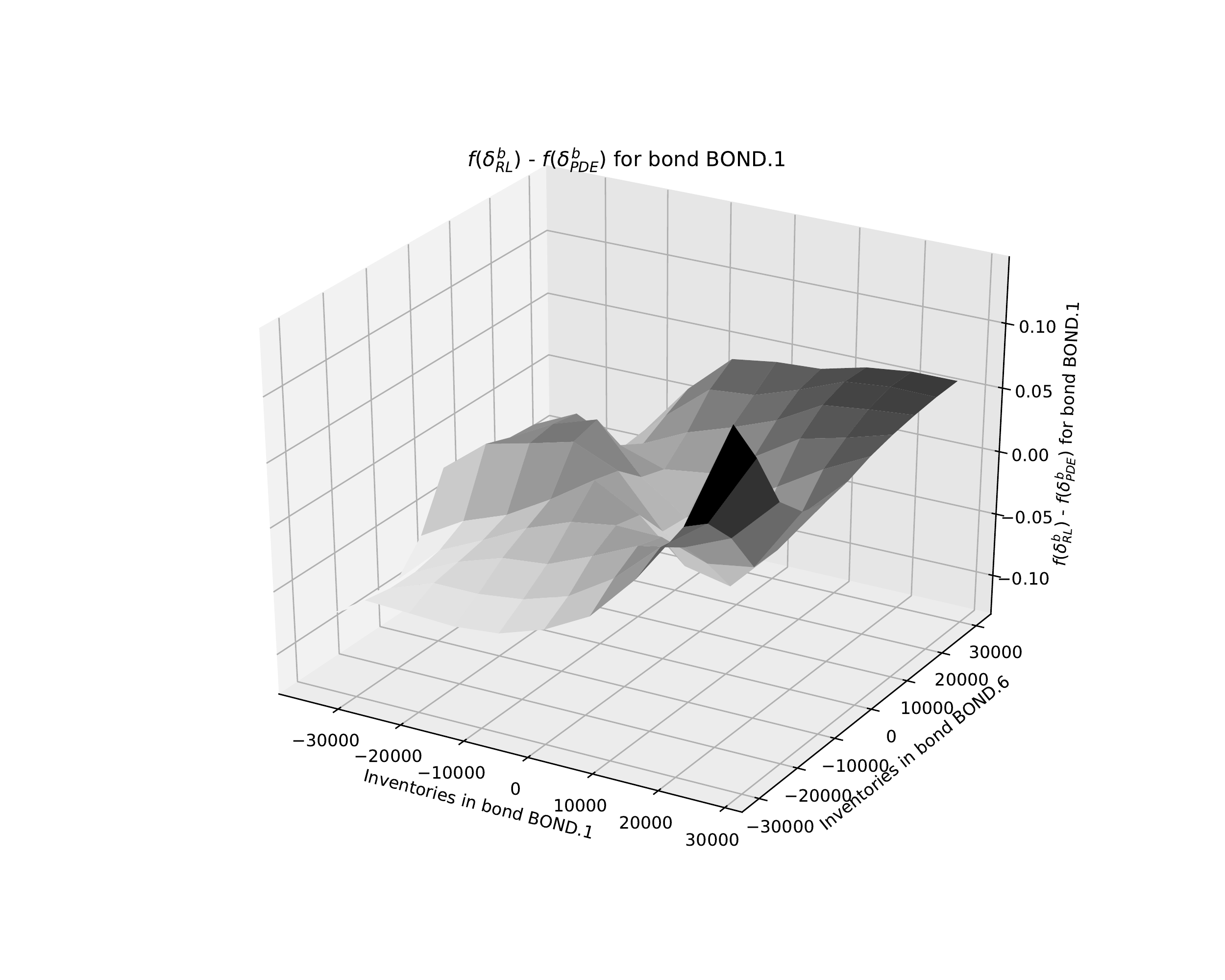}
  \includegraphics[width=0.45\textwidth]{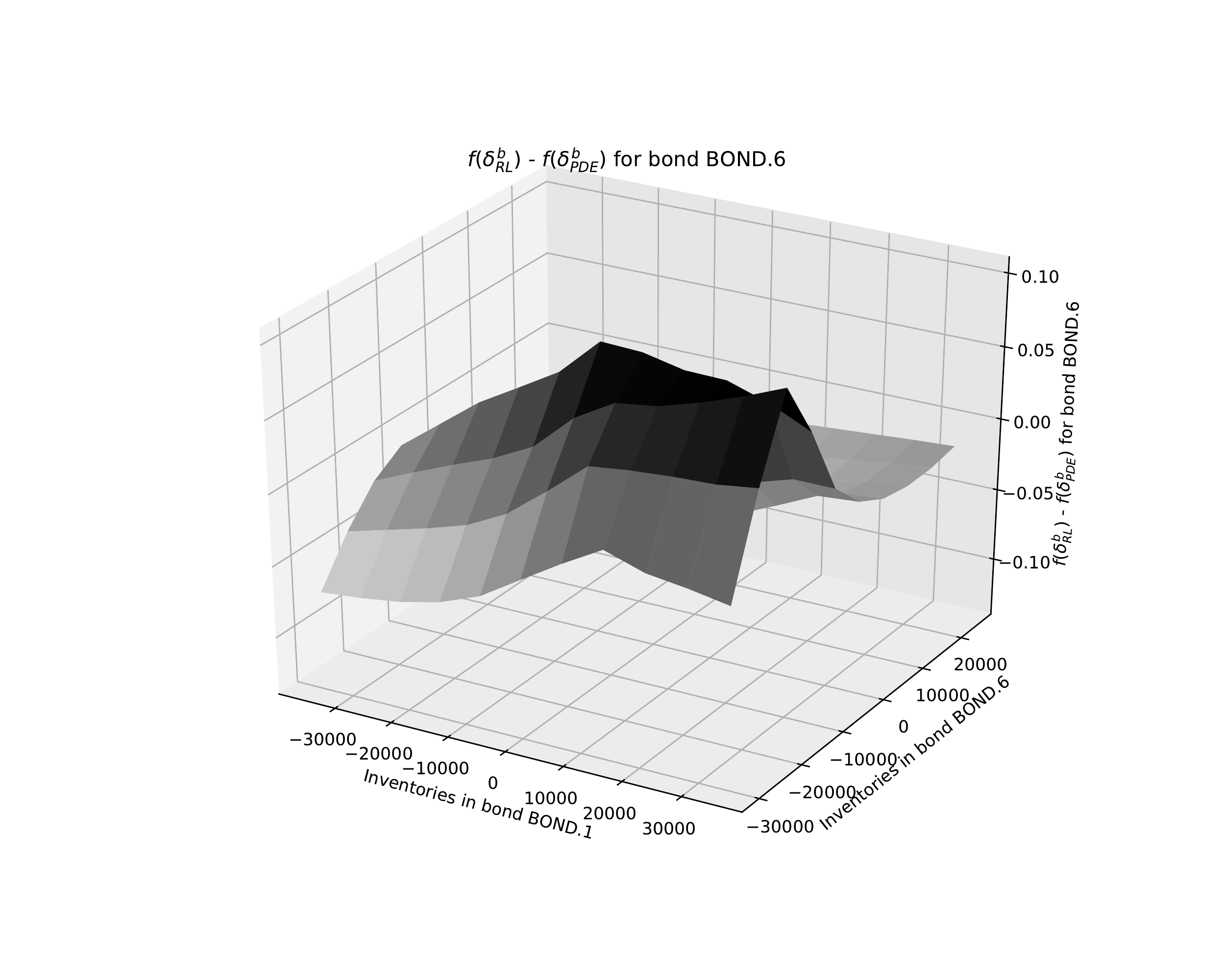}
  \caption{Comparison of the probabilities to trade obtained with the two methods.}
  \label{acth_2d_deltas_sqr}
\end{figure}

As above, we also document the difference between the two approaches in terms of probability to trade in Figure \ref{acth_2d_deltas_sqr}. We see that the difference is rather small, except for inventories that are seldom visited by our RL algorithm.\\

\vspace{-3mm}
\begin{figure}[H]
  \centering
  \includegraphics[width=0.45\textwidth]{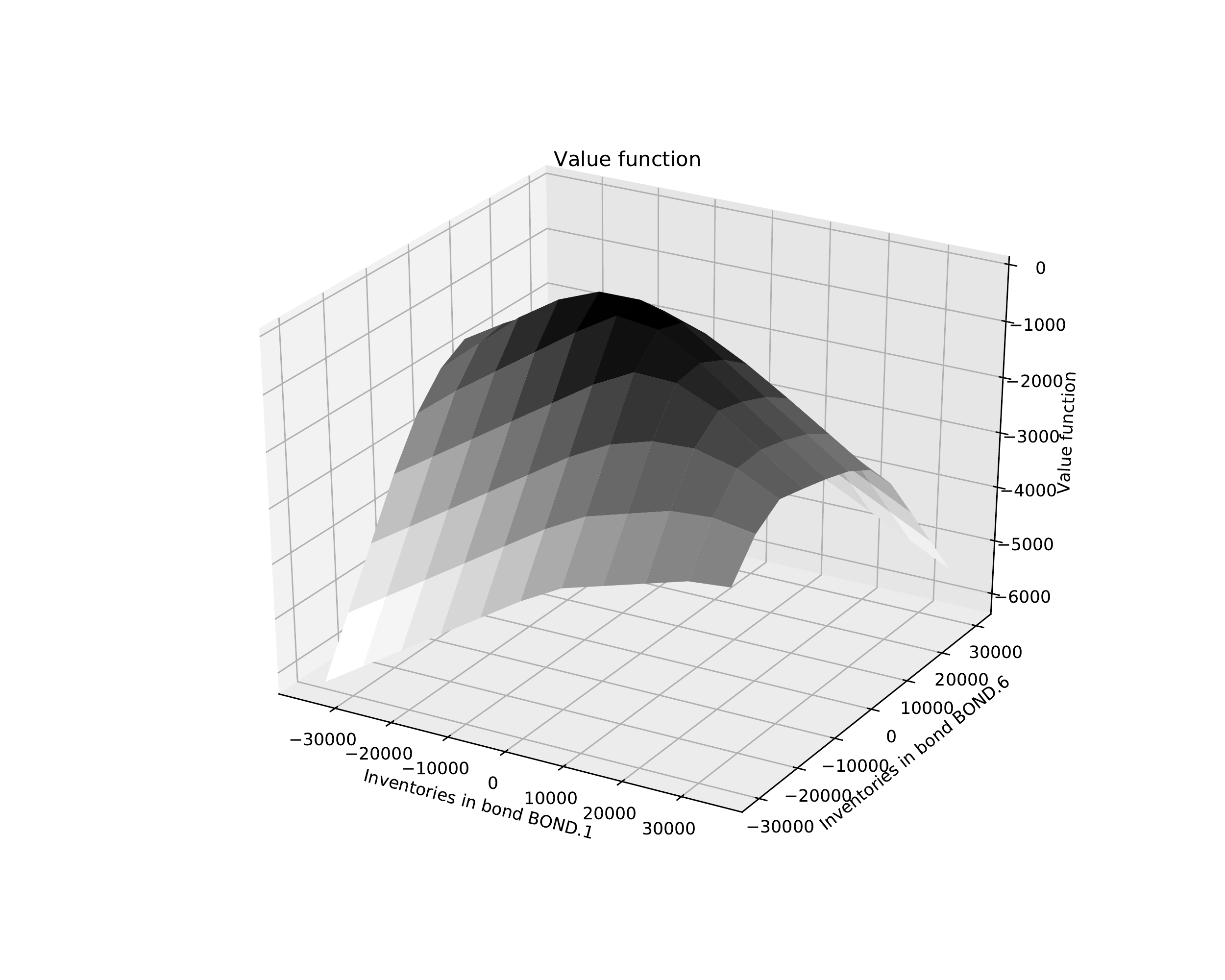}
  \includegraphics[width=0.45\textwidth]{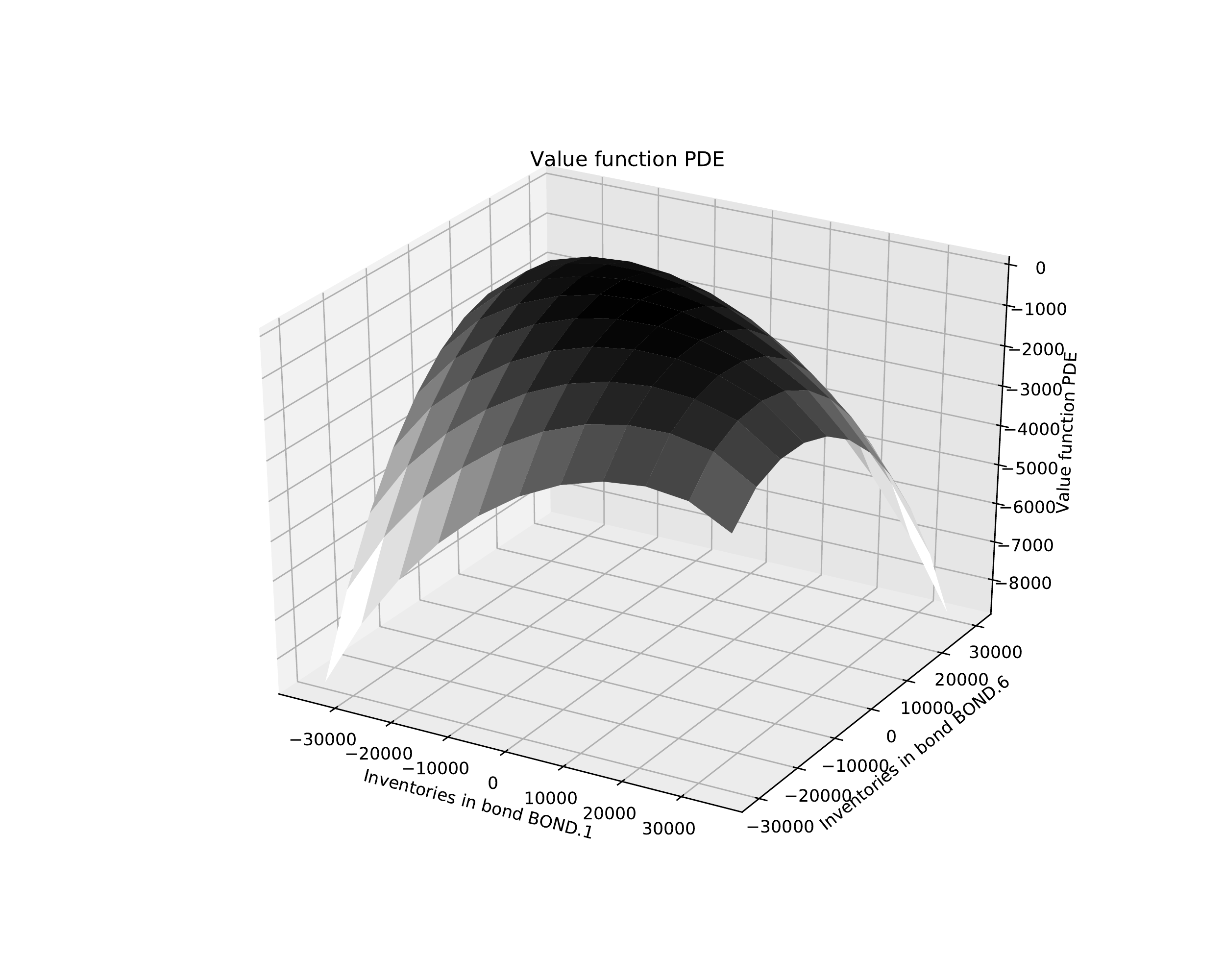}\\
  \includegraphics[width=0.45\textwidth]{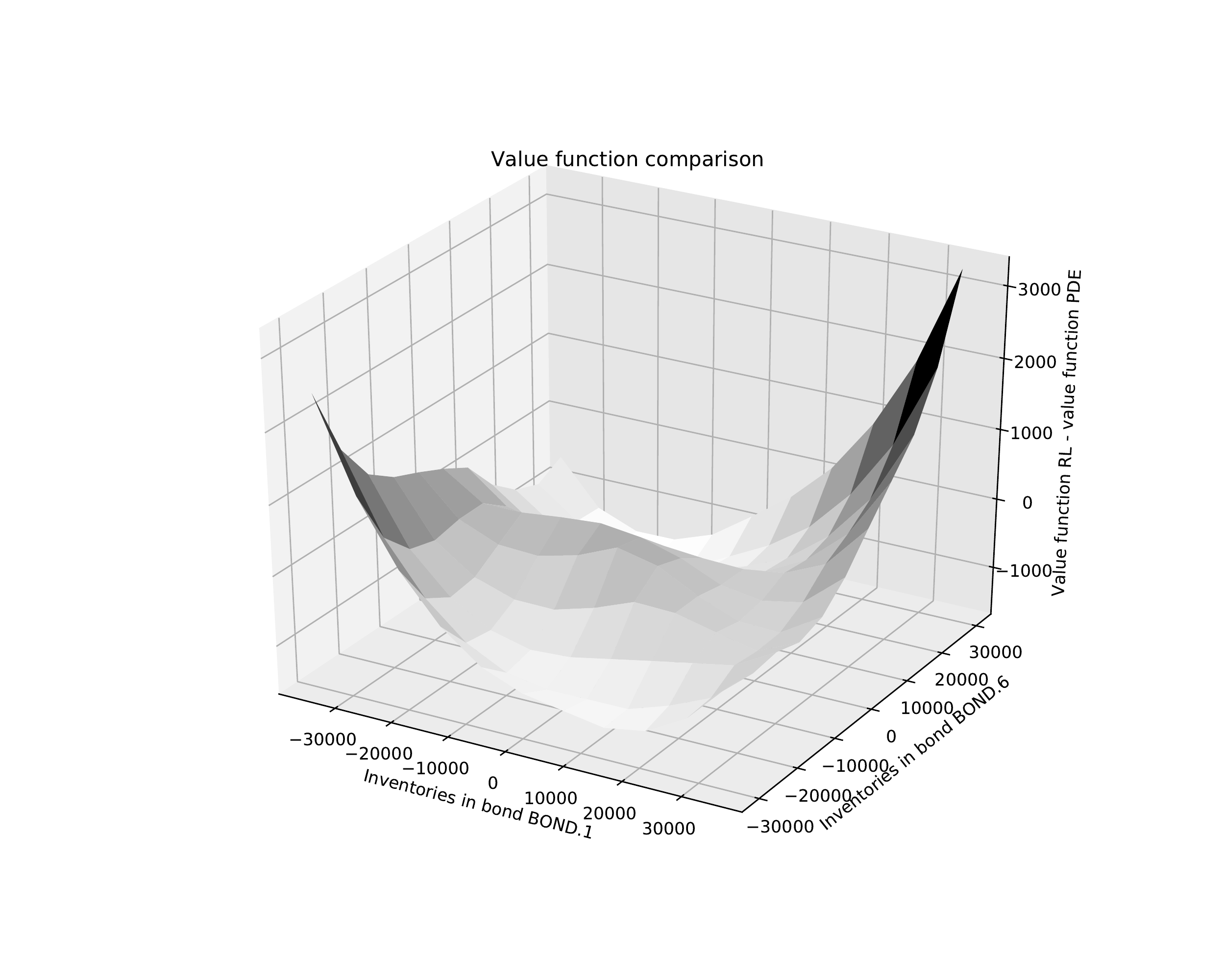}
  \includegraphics[width=0.45\textwidth]{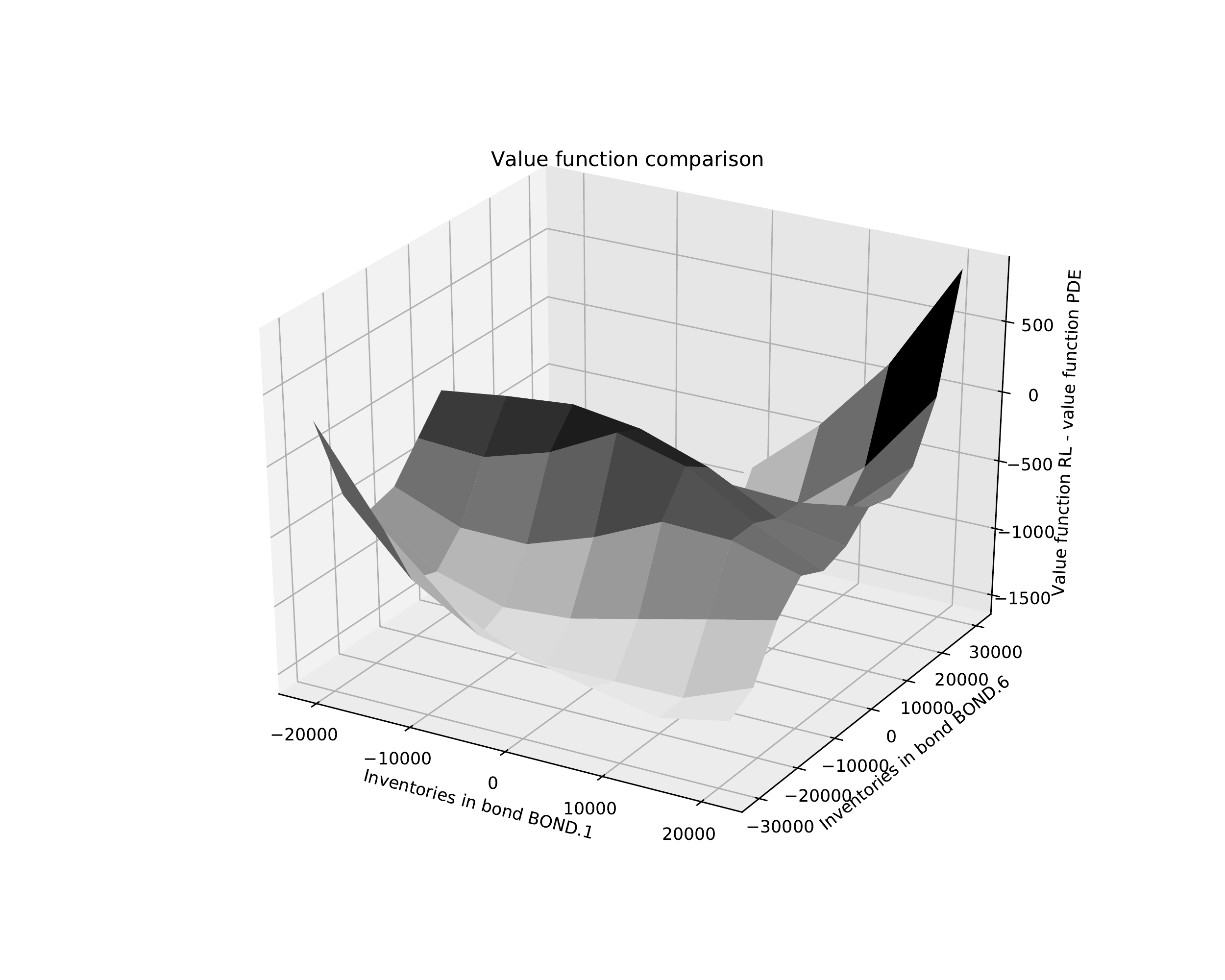}
  \caption{Top: Optimal value function obtained with our RL algorithm and the one obtained with our finite difference method. Bottom: Comparison between the two and zoom on the center of the surface.}
  \label{rl_2d_value_function_sqr}
\end{figure}
\newpage
The optimal value functions associated with our RL algorithm and finite difference method, and the difference between the two approaches are plotted in Figure \ref{rl_2d_value_function_sqr}.\\

As in a single-bond case we can see here that there is a discrepancy for large inventories. This is a consequence of the quadratic penalty function that makes the algorithm visit less often states corresponding to large inventories.\\

What is important to notice is also that the value function is clearly not separable.\\

\subsubsection{8-bond case}

In addition to cases with two bonds we consider our target cases where the market maker takes account of his position in a large number of bonds.\\

Let us start with the same 8-bond case as above.\\

We considered $\gamma = 2\cdot10^{-5}$ and $r = 10^{-4}$ as above. Risk limits were chosen equal to $3$ times the RFQ size at the beginning of the learning process and were increased every $500$ steps by the RFQ size for each bond until the maximum risk limits equal to $5$ times the RFQ size were reached. For the critic and the actor, we considered neural networks with $2$ hidden layers and $18$ nodes in each of these layers with ReLU activation functions. As above, the final layer of each neural network contains one node and the activation function is affine in the case of the critic and sigmoid in the case of the actor. For the pre-training we used the quotes obtained by our RL algorithm in the single-bond case for each bond.  For the learning phase we considered $3000$ steps, i.e. $3000$ steps of TD learning and $3000$ steps of policy improvement for each of the 8 bonds. At each step we carried out $1$ rollout of length $10000$ starting from a zero inventory and $100$ additional rollouts of length $100$ starting from a random inventory. The noise $\epsilon$ in each rollout is distributed uniformly in $[-0.05, 0.05]$ and we chose the probability limit $\nu = 0.005$. The learning rate for the critic is $\eta=1\cdot10^{-8}$ and we used mini-batches of size $70$. The learning rate for the actor is  $\tilde{\eta} = 0.001$ and we used mini-batches of size $50$. \\

\begin{figure}[H]
  \centering
  \includegraphics[width=0.48\textwidth]{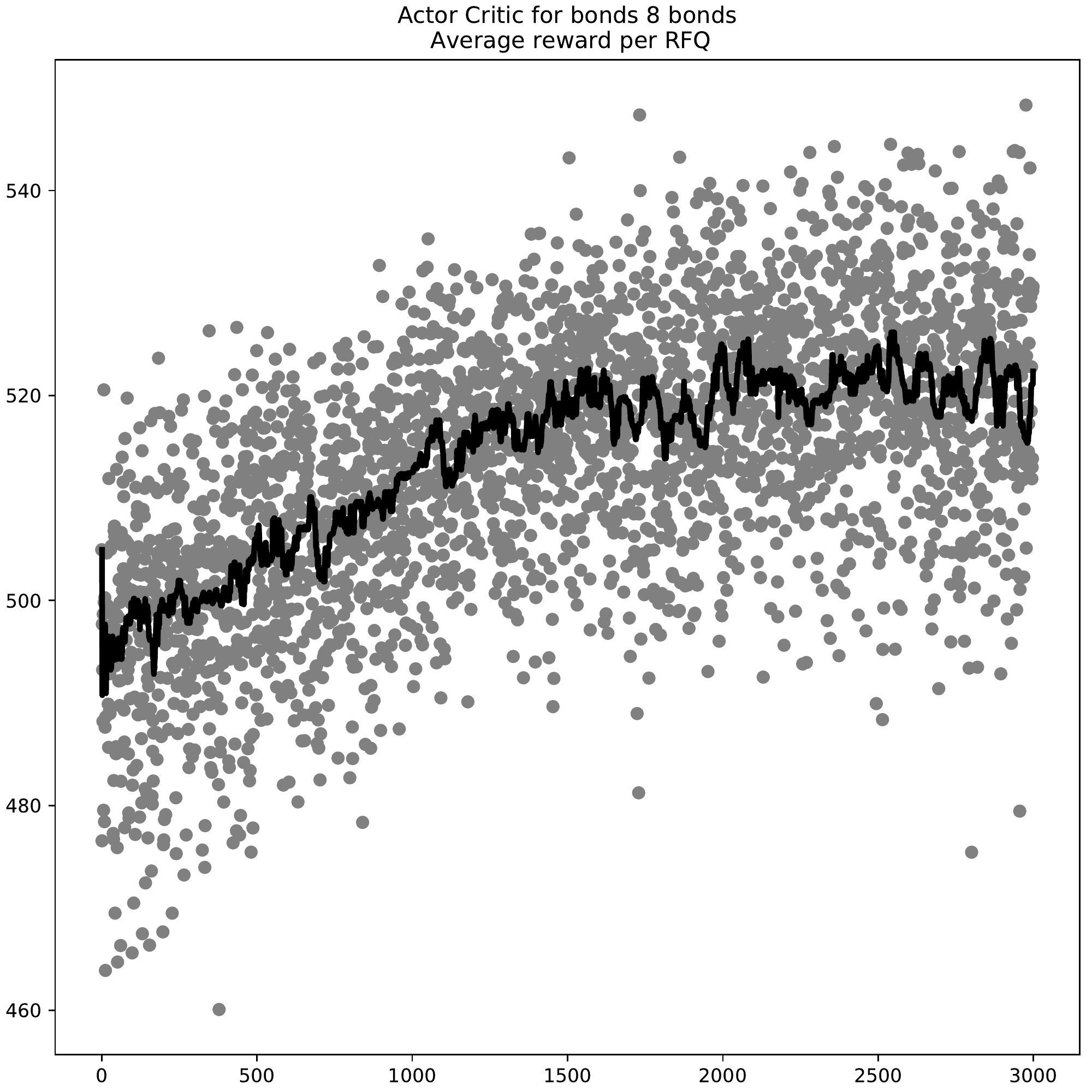}\\
  \caption{Average reward per RFQ -- Learning process for the 8-bond case.}
  \label{rl_8d_sqr}
\end{figure}

In Figure \ref{rl_8d_sqr} we plotted the learning curve in terms of average reward per RFQ. As in the above case, we see that the correlation structure is well taken into account as the average reward per RFQ goes from around 495 to around~520.\\

\subsubsection{20-bond case}

Let us now come to the 20-bond case.\\

We used our RL algorithm with $\gamma = 2\cdot10^{-5}$ and $r = 10^{-4}$. The risk limits were set to $3$ times the RFQ size at the beginning of the learning process and were increased every $500$ steps by the RFQ size for each bond according to the previously described reverse Matryoshka dolls principle until the maximum risk limits were reached ($10$ times the RFQ size for all bonds, except  for BOND.5 -- due to its low liquidity and high volatility -- for which we chose a risk limit of $5$ times the associated RFQ size). For the critic and the actor, we considered neural networks with $2$ hidden layers and $30$ nodes in each of these layers with ReLU activation functions. Again, the final layer of each neural network contains one node and the activation function is affine in the case of the critic and sigmoid in the case of the actor. For the pre-training we used the quotes obtained by our RL algorithm in the single-bond case for each bond. For the learning phase, we considered $4000$ steps, i.e. $4000$ steps of TD learning and $4000$ steps of policy improvement for each of the 20 bonds. At each step we carried out $1$ rollout of length $10000$ starting from a zero inventory and $100$ additional rollouts of length $100$ starting from a random inventory. The noise $\epsilon$ in each rollout is distributed uniformly in $[-0.05, 0.05]$ and we chose the probability limit $\nu = 0.005$. The learning rate for the critic is $\eta=1\cdot10^{-8}$ and we used mini-batches of size $70$. The learning rate for the actor is  $\tilde{\eta} = 0.01$ and we used mini-batches of size $50$. \\

\begin{figure}[H]
  \centering
 \includegraphics[width=0.48\textwidth]{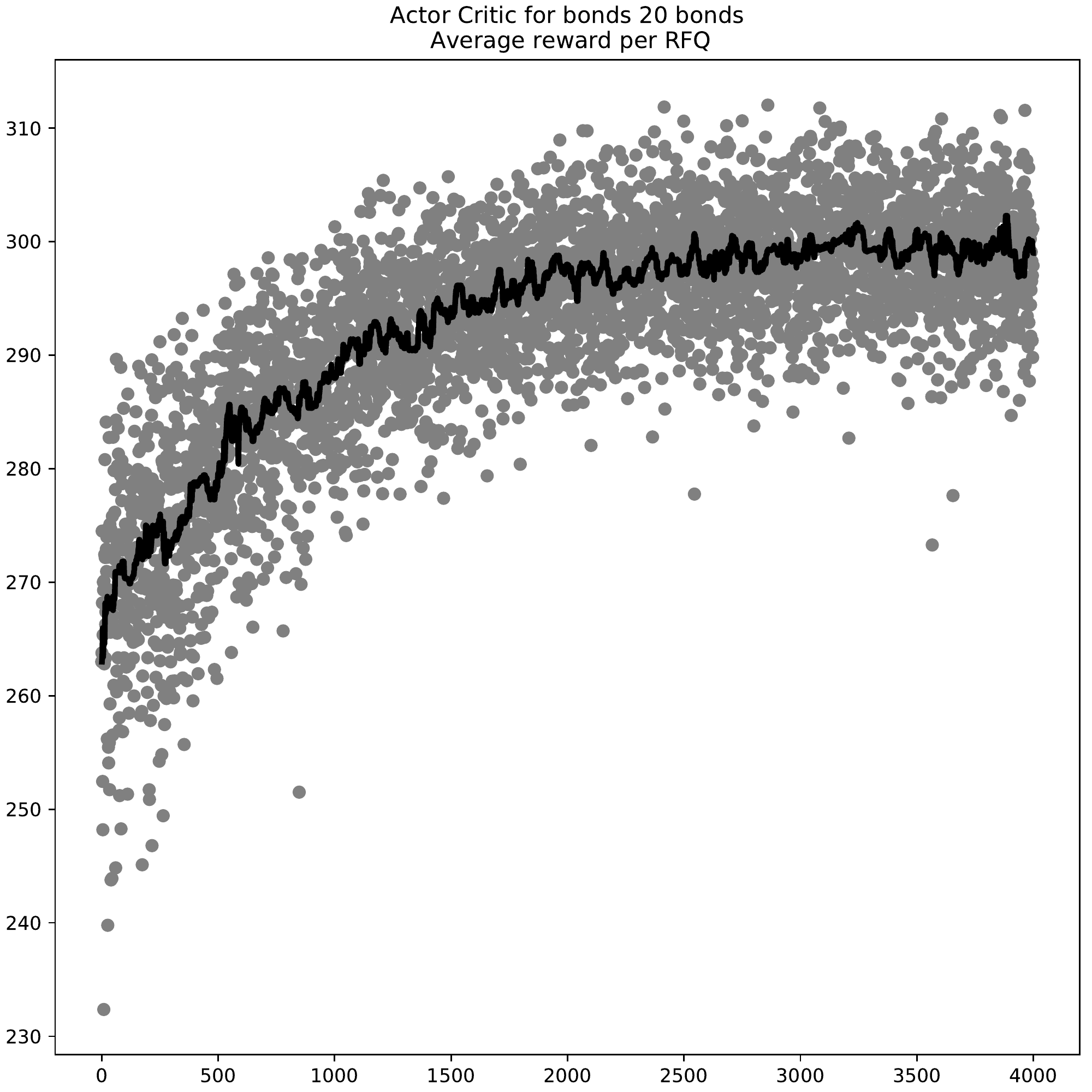}\\
  \caption{Average reward per RFQ -- Learning process for the 20-bond case.}
  \label{rl_20d_sqr}
\end{figure}

In Figure \ref{rl_20d_sqr} we see the evolution of the average reward per RFQ over the course of the learning process.\\

We clearly see that the correlation structure is taken into account.\\

\subsection{The variant with one neural network for all bonds}
\label{one_nn}
So far, we presented examples where the RL algorithm was implemented with one neural network for the value function and $d$ neural networks for the actor, i.e. one for each bond. Although optimization of actor networks can be parallelized in the case of $d$ neural networks for the actor, one can prefer to optimize over a single neural network for the actor to make the method more scalable in terms of memory space. \\

For this reason, we consider now a new implementation of our algorithm with a single-network actor. As discussed in Section \ref{ac_approach}, the inputs of the actor network is now a $2d$-dimensional vector: the first $d$ components are the inventories, the other $d$ is a one-hot vector (of size $d$) to code the bond for which the network has to output a quote.\\

The advantage of this approach is obviously its scalability, but the price to pay for this improvement is that we cannot normalize actor updates as in the case of the multiple-network actor. As a consequence, we need to reduce the learning rate and that affects the time needed for learning.\\

In what follows, we only consider the case of the penalty function used in Section \ref{sqrt}. As above, we consider cases with 2, 8, and 20 bonds.\\

\subsubsection{2-bond case}
As above, we start with a 2-bond case with BOND.1 and BOND.6.\\

For our RL algorithm, we considered $\gamma = 5\cdot10^{-2}$ and $r = 10^{-4}$. We considered risk limits equal to $5$ times the RFQ size and kept them unchanged during learning. For the critic and the actor, we considered neural networks with $2$ hidden layers and $12$ nodes in each of these layers with ReLU activation functions. Again, the final layer of each neural network contains one node and the activation function is affine in the case of the critic and sigmoid in the case of the actor. For the pre-training phase, we used for each bond the quotes obtained by our RL algorithm in the single-bond case.  For the learning phase we considered $500$ steps, i.e. $500$ steps of TD learning and $500$ steps of policy improvement. At each step we carried out $1$ rollout of length $10000$ starting from a zero inventory and $100$ additional rollouts of length $100$ starting from a random inventory. The noise $\epsilon$ in each rollout is distributed uniformly in $[-0.05, 0.05]$ and we chose the probability limit $\nu = 0.005$. The learning rate for the critic is $\eta=5\cdot10^{-8}$ and we used mini-batches of size $70$. The learning rate for the actor is  $\tilde{\eta} = 0.01$ and we used mini-batches of size $50$.

\begin{figure}[H]
  \centering
  \includegraphics[width=0.45\textwidth]{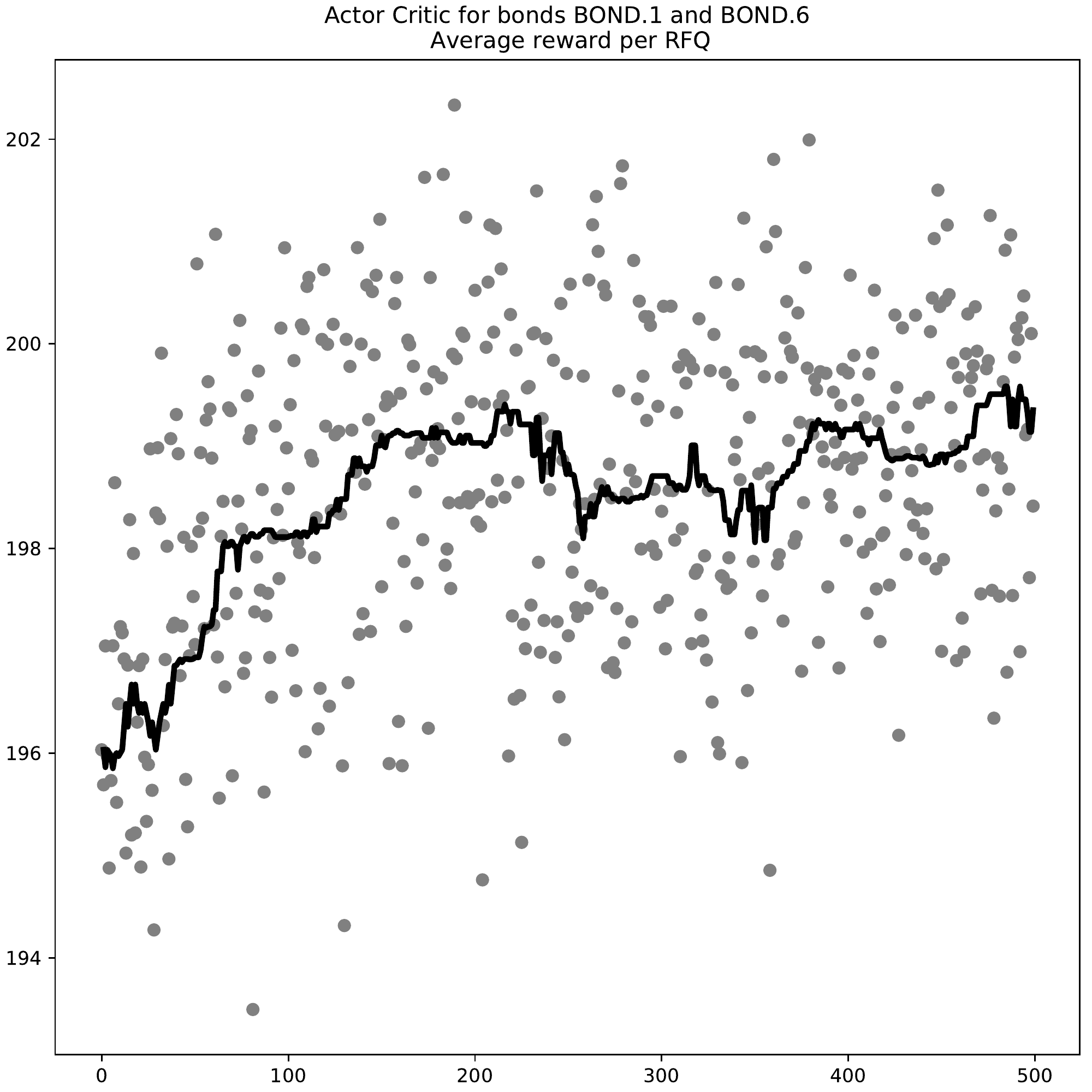}\\
  \caption{Average reward per RFQ -- Learning process for the 2-bond case (single-network actor).}
  \label{rl_2d_single}
\end{figure}

The learning curve of the algorithm is plotted in Figure \ref{rl_2d_single}.\\

The average reward per RFQ obtained with this approach is above $198.0$, in line with the value obtained with the multi-network actor and the finite difference method.\\

\begin{figure}[H]
  \centering
  \includegraphics[width=0.48\textwidth]{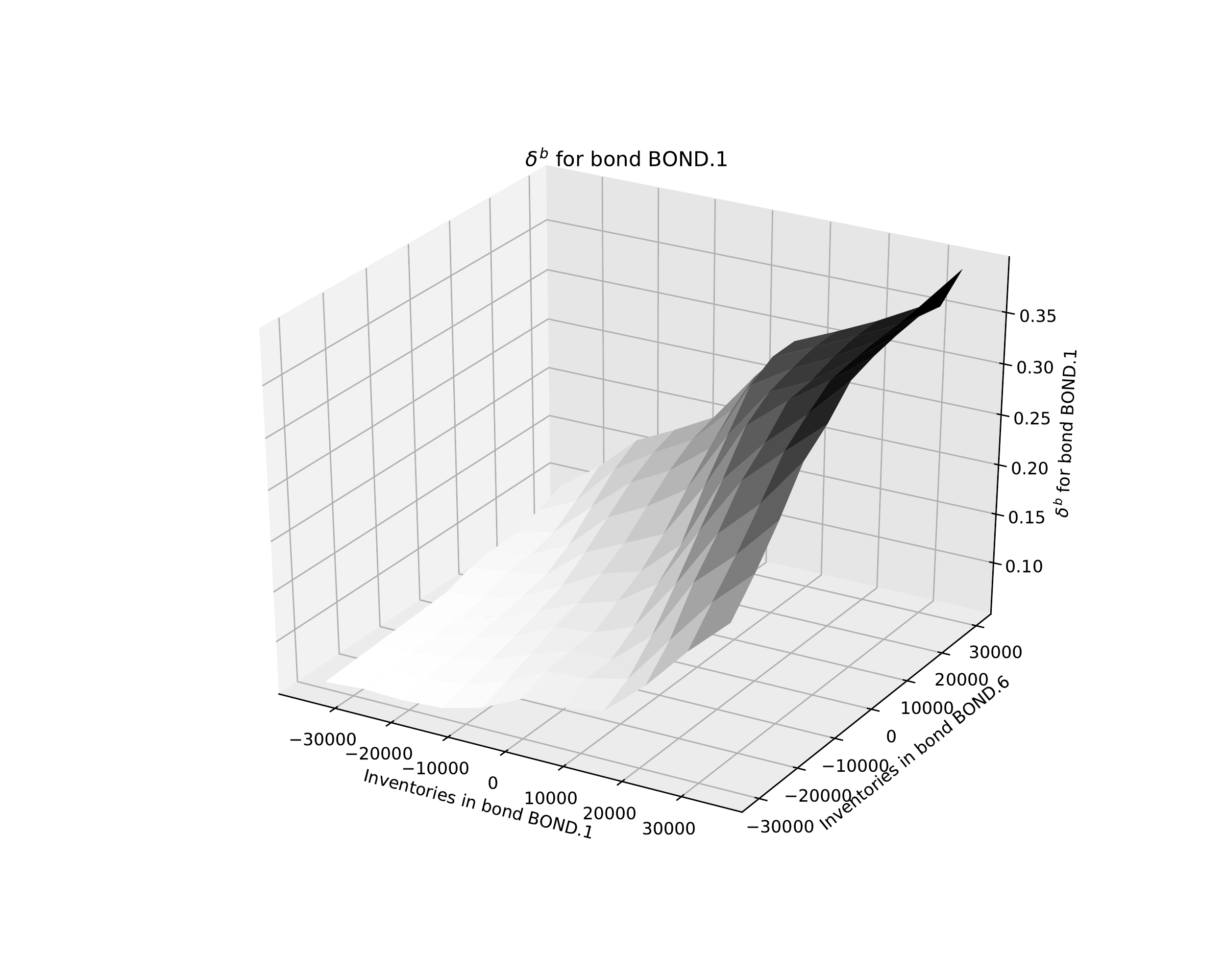}
  \includegraphics[width=0.48\textwidth]{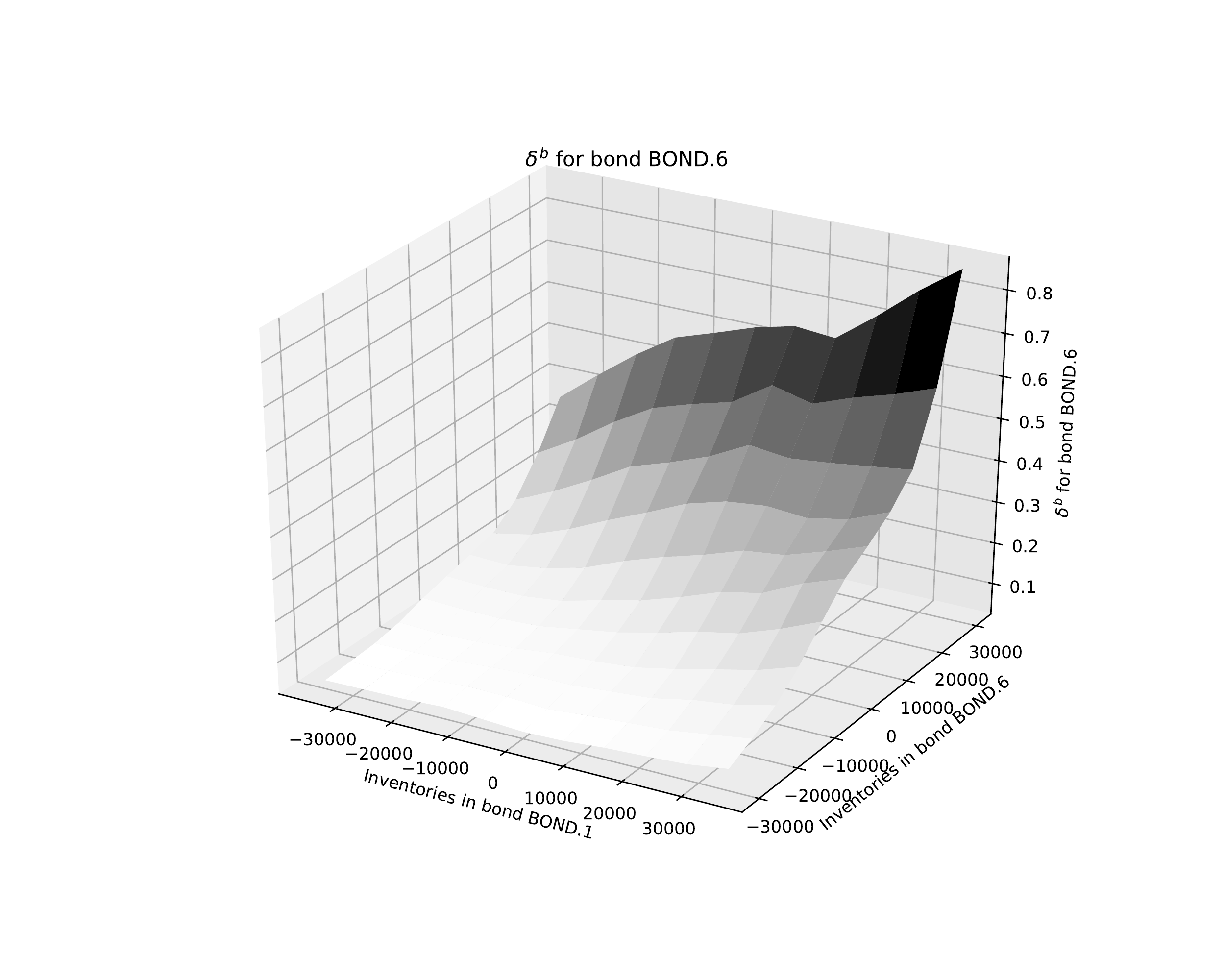}
  \caption{Optimal bid quotes obtained with our RL algorithm (single-network actor).}
  \label{rl_2d_deltas_single}
\end{figure}

\begin{figure}[H]
  \centering
  \includegraphics[width=0.45\textwidth]{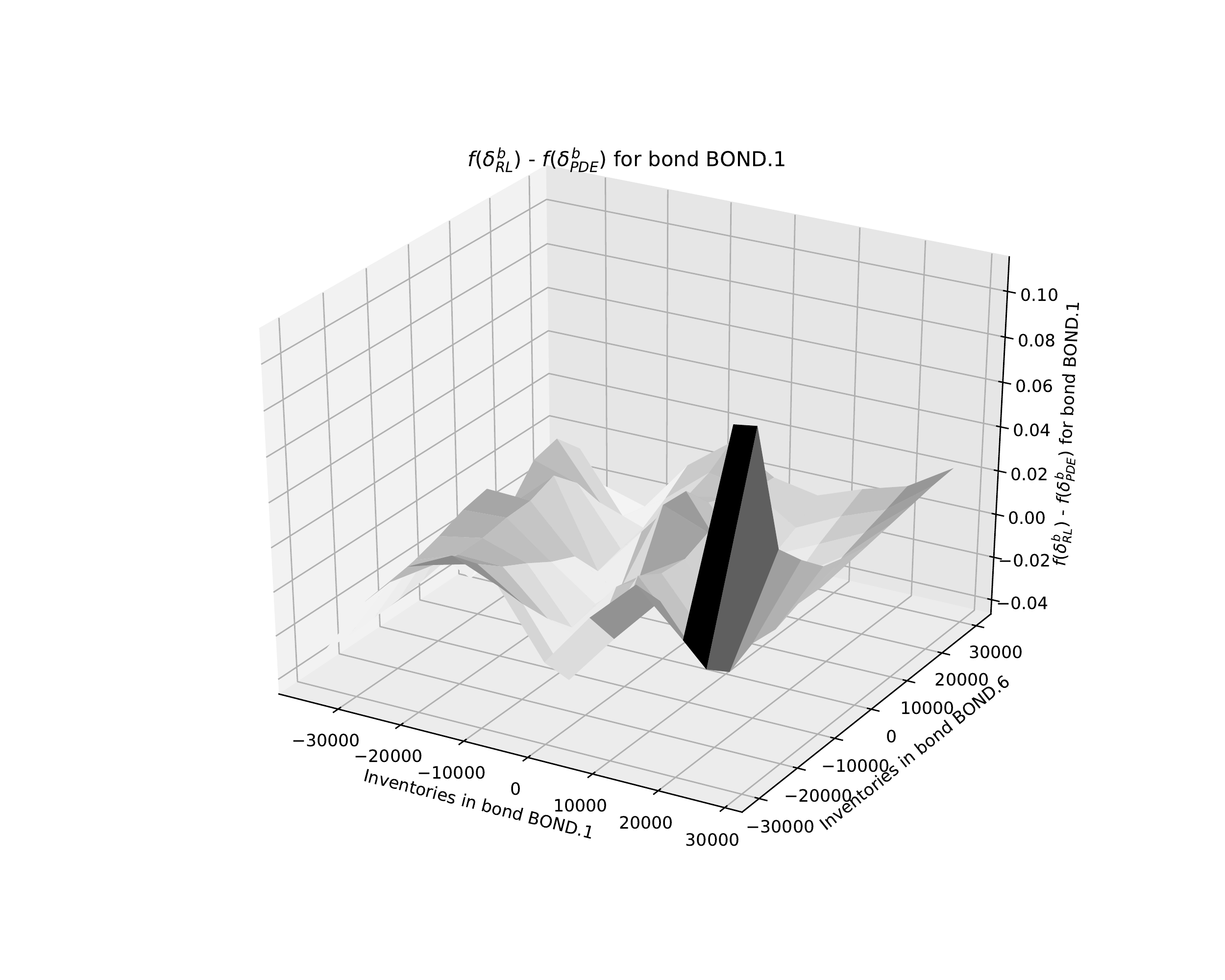}
  \includegraphics[width=0.45\textwidth]{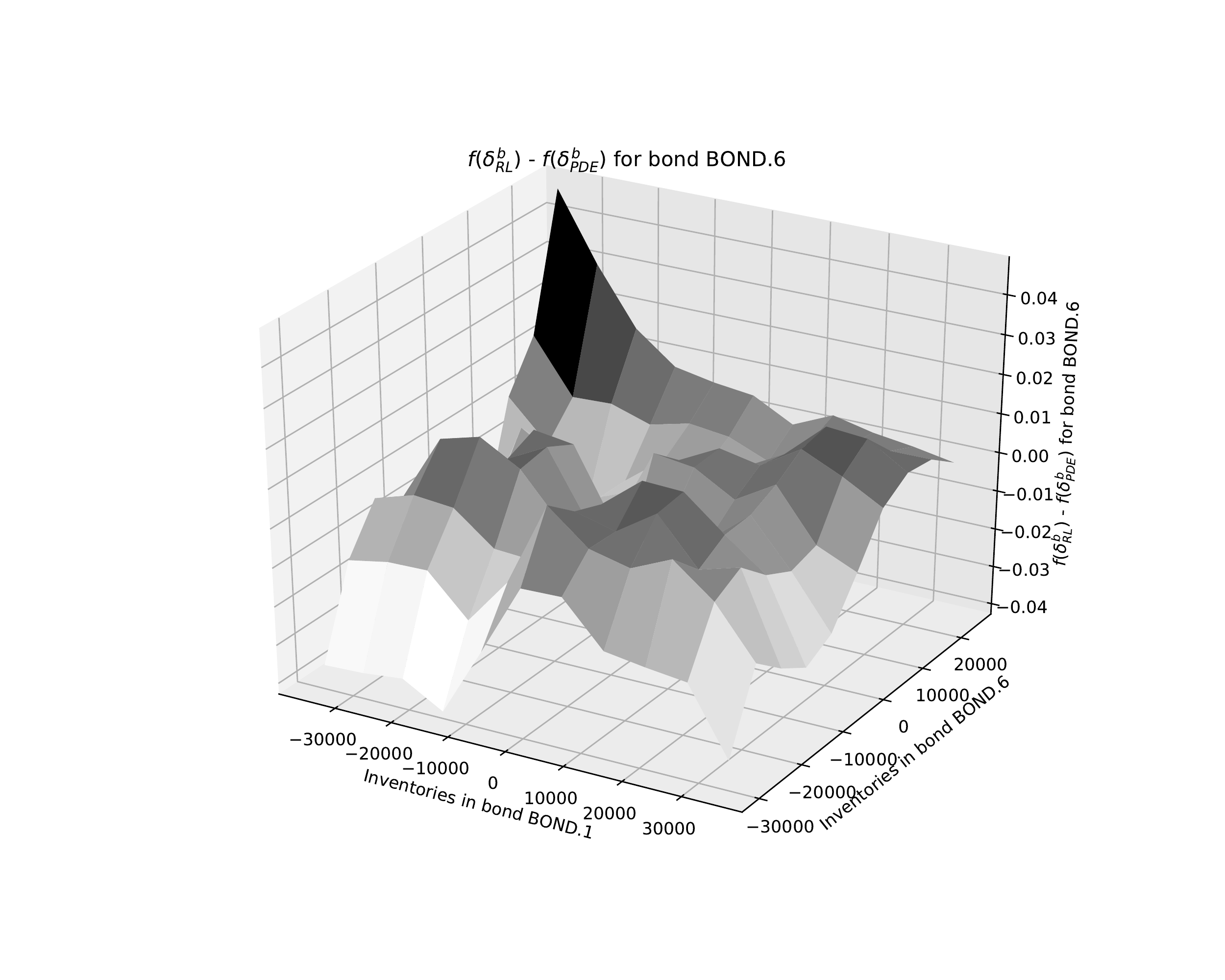}
  \caption{Comparison of the probabilities to trade obtained with the two methods.}
  \label{acth_2d_deltas_sqrt_single}
\end{figure}

In Figure \ref{rl_2d_deltas_single}, we plotted the optimal (bid) quotes computed with our RL single-network actor algorithm -- the difference between these quotes and the those obtained with the finite difference method of the appendix, in terms of probability to trade, is plotted in Figure \ref{acth_2d_deltas_sqrt_single}.

We see that the difference in terms of probability is less than 8\% for BOND.1 and less than 4\% for BOND.6 (and the large differences concern inventories close to risk limits only).\\

\begin{figure}[H]
  \centering
  \includegraphics[width=0.48\textwidth]{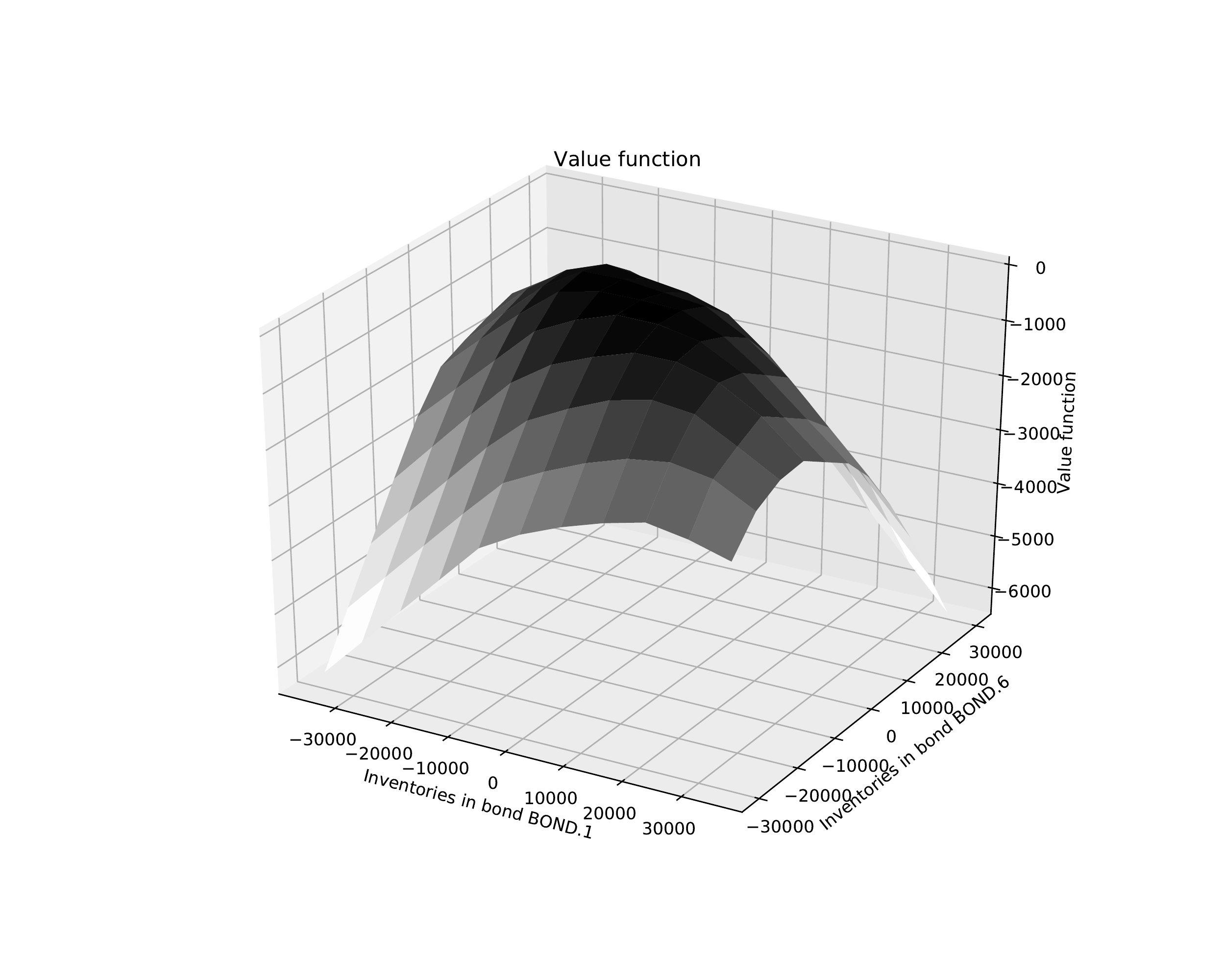}
  \includegraphics[width=0.48\textwidth]{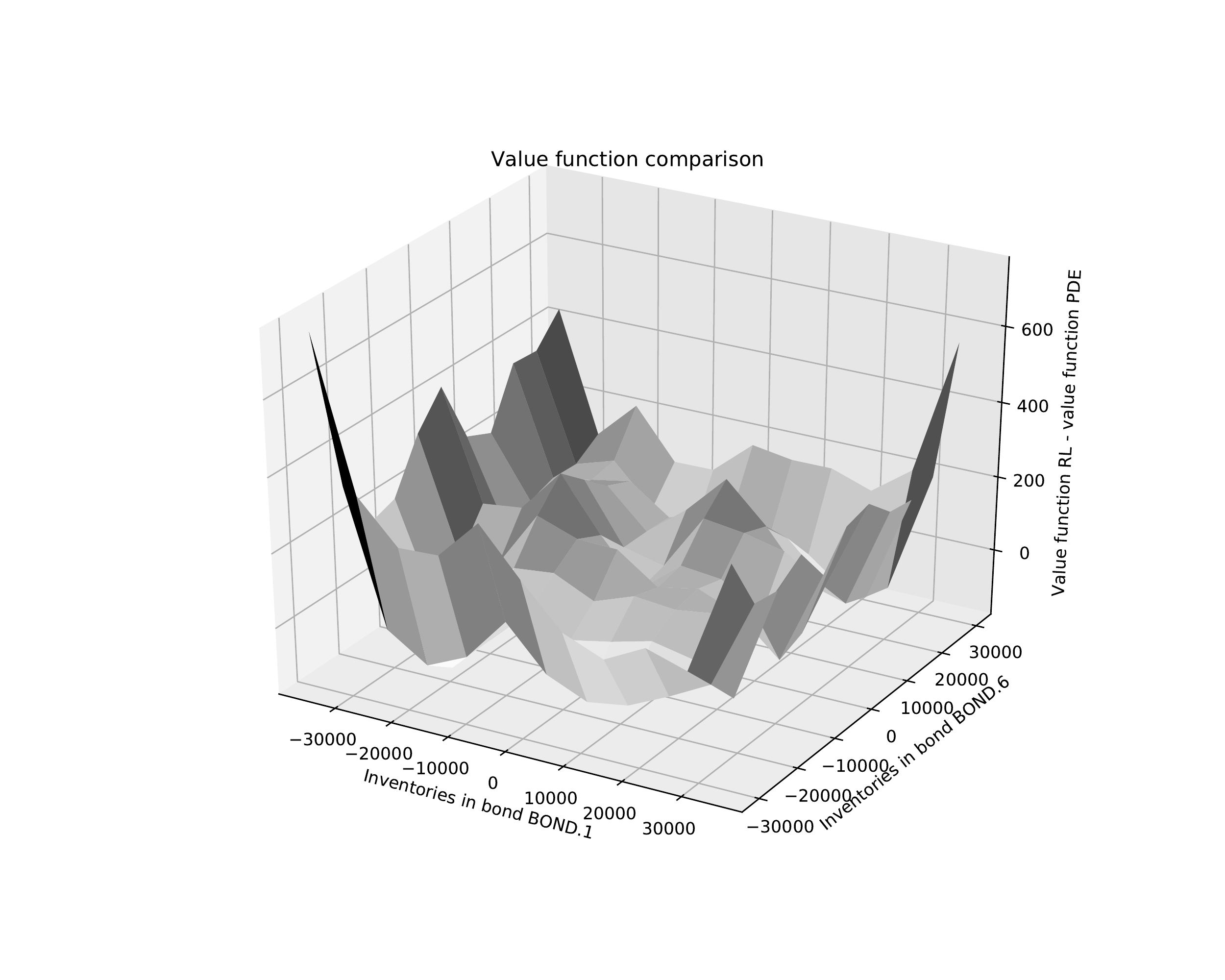}
  \caption{Left: Optimal value function obtained with our RL algorithm. Right: Comparison of the optimal value function obtained with our RL algorithm to the optimal value function obtained with the finite difference method.}
  \label{value_functions_single}
\end{figure}

The value function approximated with our RL method and its comparison with the one obtained with the finite difference method are depicted in Figure~\ref{value_functions_single}.

\subsubsection{8-bond case}

Let us now come to the same 8-bond case as in the previous sections.\\

We considered $\gamma = 5\cdot10^{-2}$ and $r = 10^{-4}$ as above. Risk limits were chosen equal to $5$ times the RFQ size at the beginning of the learning process and were increased every $500$ steps by the RFQ size for each bond until the maximum risk limits equal to $10$ times the RFQ size were reached (this is the reverse Matryoshka dolls principle), except  for BOND.5 -- due to its low liquidity and high volatility -- for which we chose a risk limit of $5$ times the associated RFQ size. For the critic and the actor, we considered neural networks with $2$ hidden layers and $28$ nodes in each of these layers with ReLU activation functions. As above, the final layer of each neural network contains one node and the activation function is affine in the case of the critic and sigmoid in the case of the actor. For the pre-training we used the quotes obtained by our RL algorithm in the single-bond case for each bond.  For the learning phase we considered $3000$ steps, i.e. $3000$ steps of TD learning and $3000$ steps of policy improvement for each of the 8 bonds. At each step we carried out $1$ rollout of length $10000$ starting from a zero inventory and $100$ additional rollouts of length $100$ starting from a random inventory. The noise $\epsilon$ in each rollout is distributed uniformly in $[-0.05, 0.05]$ and we chose the probability limit $\nu = 0.005$. The learning rate for the critic is $\eta=5\cdot10^{-9}$ and we used mini-batches of size $100$. The learning rate for the actor is  $\tilde{\eta} = 0.001$ and we used mini-batches of size $100$. \\

\begin{figure}[H]
  \centering
  \includegraphics[width=0.48\textwidth]{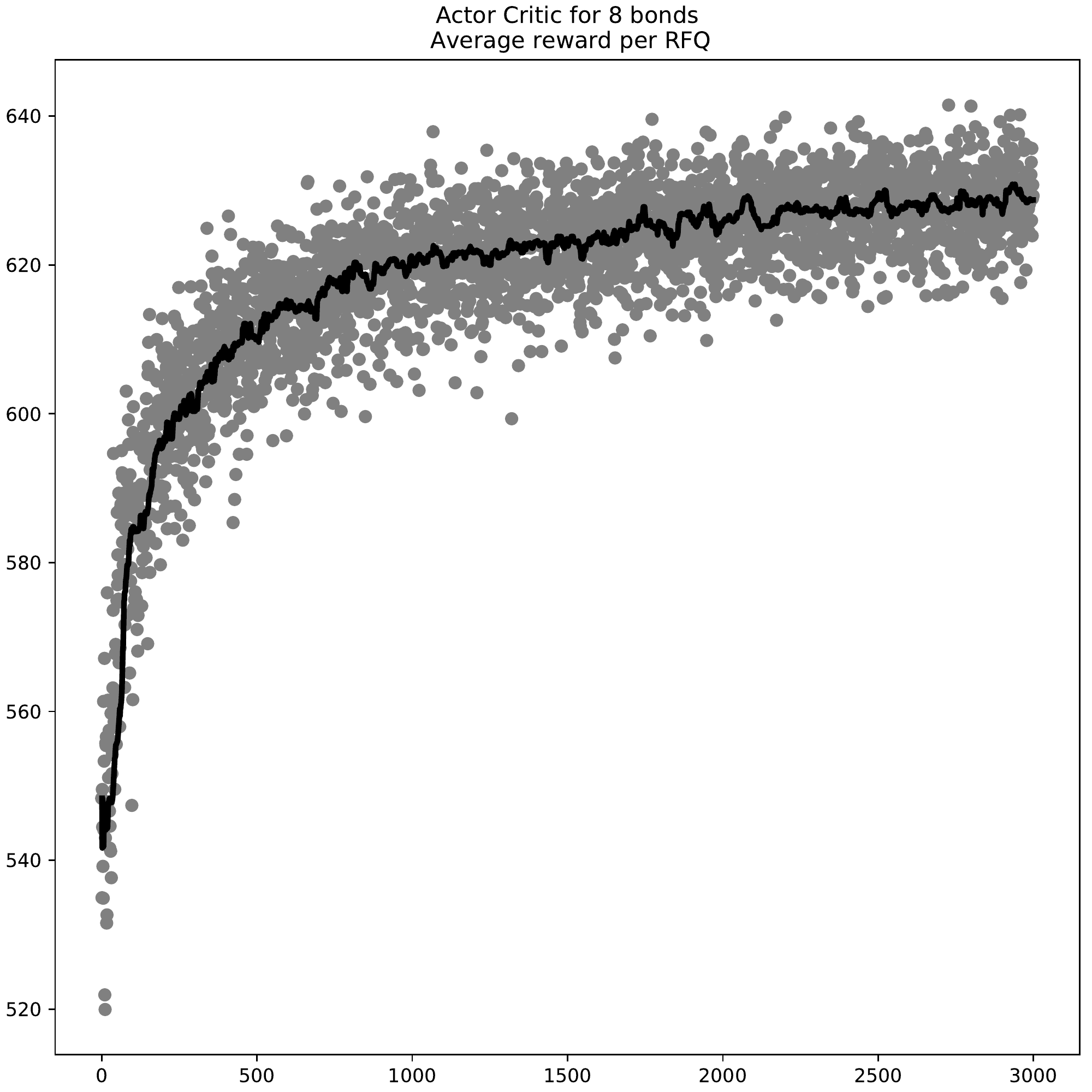}\\
  \caption{Average reward per RFQ -- Learning process for the 8-bond case.}
  \label{rl_8d_sqrt_single}
\end{figure}

The learning curve in terms of average reward per RFQ is plotted in Figure~\ref{rl_8d_sqrt_single}.

We see that the average reward per RFQ reached by the single-network method is around $630$, in line with the results obtained before with the multi-network variant. Interestingly, the initial average reward per RFQ (after pre-training) is lower in the single-network case than in the multi-network case, in spite of an extended pre-training period. This is a sign that training is more complex in the single-network case because of the very structure of the neural network.\\

\subsubsection{20-bond case}

Now, let us move to the 20-bond case.\\

We considered $\gamma = 5\cdot10^{-2}$ and $r = 10^{-4}$ as above. Risk limits were chosen equal to $5$ times the RFQ size at the beginning of the learning process and were increased every $200$ steps by the RFQ size for each bond until the maximum risk limits equal to $10$ times the RFQ size were reached (this is the reverse Matryoshka dolls principle), except  for BOND.5 -- due to its low liquidity and high volatility -- for which we chose a risk limit of $5$ times the associated RFQ size. For the critic and the actor, we considered neural networks with $2$ hidden layers and $30$ nodes in each of these layers for the critic and $300$ for the actor, with ReLU activation functions. As above, the final layer of each neural network contains one node and the activation function is affine in the case of the critic and sigmoid in the case of the actor. For the pre-training we used the quotes obtained by our RL algorithm in the single-bond case for each bond.  For the learning phase we considered $20000$ steps, i.e. $20000$ steps of TD learning and $20000$ steps of policy improvement for each of the 20 bonds. At each step we carried out $1$ rollout of length $5000$ starting from a zero inventory and $100$ additional rollouts of length $50$ starting from a random inventory. The noise $\epsilon$ in each rollout is distributed uniformly in $[-0.05, 0.05]$ and we chose the probability limit $\nu = 0.005$. The learning rate for the critic is $\eta=5\cdot10^{-8}$ and we used mini-batches of size $50$. The learning rate for the actor is  $\tilde{\eta} = 5\cdot10^{-4}$ and we used mini-batches of size $50$. \\

\begin{figure}[H]
  \centering
  \includegraphics[width=0.48\textwidth]{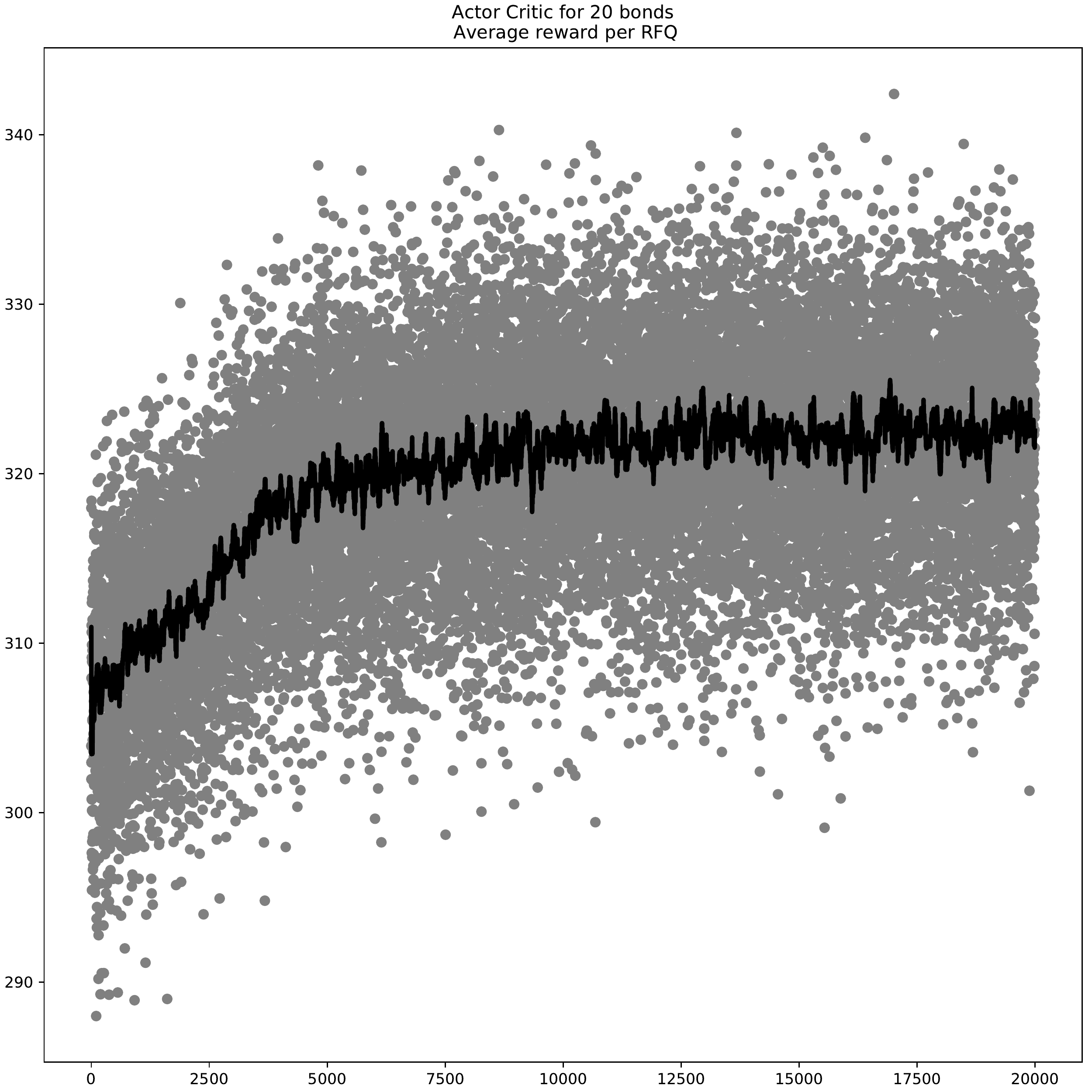}\\
  \caption{Average reward per RFQ -- Learning process for the 20-bond case.}
  \label{rl_20d_sqrt_single}
\end{figure}

The learning curve in terms of average reward per RFQ is plotted in Figure~\ref{rl_20d_sqrt_single}.\\

As in the 8-bond case, we see that the initial point (after pre-training) is lower than in the multi-network case, confirming the difficulty to train the neural network, even for supervised learning. Nevertheless, we see that the correlation structure is taken into account, but we reach a value slightly lower than in the multi-network case.\\

\section*{Conclusion}

The problem of solving numerically the equations characterizing the optimal bid and ask quotes of a market maker in charge of a large number of assets cannot be solved with classical methods based on grids (e.g. finite difference methods). In this paper we proposed a model-based actor-critic-like algorithm involving deep neural networks to address this issue. We showed that our method is consistent with existing ones in the case of low-dimensional problems, and, unlike classical methods, scalable to portfolios of a few dozens of bonds, hence beating the curse of dimensionality. Our method can be adapted to cover multiple extensions, such as the presence of variable RFQ sizes or the choice of a reduced-size state space consisting of risk factors.

\section*{Appendix: Numerical scheme for the Hamilton-Jacobi-Bellman equation}

The numerical approximation of the solution $\tilde{\theta}^*_{r}$ of Eq. \eqref{theta} with finite difference schemes is a classical problem in numerical analysis.\\

A natural idea consists in considering an implicit scheme with operator splitting for the equation
$$0 = \partial_t \tilde{\theta}_{r}(t,q) - r \tilde{\theta}_{r}(t,q) - \psi(q)$$
$$+ \sum_{i=1}^d 1_{q^i<Q^i} H^{i,b}\left(\frac{\tilde{\theta}_{r}(t,q) - \tilde{\theta}_{r}(t,q+\Delta^ie^i)}{\Delta^i}\right)$$
\begin{equation*}
+ \sum_{i=1}^d 1_{q^i>-Q^i} H^{i,a}\left(\frac{\tilde{\theta}_{r}(t,q) - \tilde{\theta}_{r}(t,q-\Delta^ie^i)}{\Delta^i}\right)
\end{equation*}
defined over $(t,q) \in [0,T]\times \mathcal{Q}$ with an arbitrary terminal condition, and then to consider the value of the function $\tilde{\theta}_r$ at time $t=0$ as an approximation $\tilde{\theta}^*_{r}$.\\

More precisely, we consider a time discretization $t_0=0, \ldots, t_K=T$ of $[0,T]$, and start from a space discretization of the terminal condition at time $t_K=T$. Then, for going from an approximation $\hat{\theta}_{k+1}$ of $\tilde{\theta}_{r}(t_{k+1},\cdot)$ to an approximation $\hat{\theta}_{k}$ of $\tilde{\theta}_{r}(t_{k},\cdot)$, where $\tau = t_{k+1} - t_k$, we consider the following scheme (solved iteratively for $y_0, \ldots, y_d$ with Newton's methods when necessary):
$$\frac{\hat{\theta}_{k+1} - y_{0}(q)}{\tau} - r y_0(q) - \psi(q) = 0,$$
and, $\forall i \in \{1, \ldots, d\},$
$$ \frac{y_{i-1}(q) - y_{i}(q)}{\tau} + 1_{q^i<Q^i} H^{i,b}\left(\frac{y_{i}(q) - y_{i}(q+\Delta^ie^i)}{\Delta^i}\right)$$$$ + 1_{q^i>-Q^i} H^{i,a}\left(\frac{y_{i}(q) - y_{i}(q-\Delta^ie^i)}{\Delta^i}\right) = 0,$$
and eventually $\hat{\theta}_{k} = y_d$.\\

This scheme is a convergent monotone scheme but it requires a grid in space and is therefore not scalable to large values $d$ of the number of bonds.

\end{document}